\documentclass{article}
\usepackage{latexsym}
\usepackage{amsmath}
\usepackage{amssymb}
\usepackage{amsfonts}
\usepackage{pdflscape}

\usepackage{setspace}

\usepackage[pdftex]{color}
\definecolor{myred}{rgb}{0.6,0,0}

\newcounter{smallarabics}
\newenvironment{arabicenumerate}
{\begin{list}{{\normalfont\textrm{(\arabic{smallarabics})}}}
  {\usecounter{smallarabics}\setlength{\itemindent}{0cm}
   \setlength{\leftmargin}{5ex}\setlength{\labelwidth}{4ex}
   \setlength{\topsep}{0.75\parsep}\setlength{\partopsep}{0ex}
   \setlength{\itemsep}{0ex}}}
{\end{list}}

\newcounter{smallroman}
\newenvironment{romanenumerate}
{\begin{list}{{\normalfont\textrm{(\roman{smallroman})}}}
  {\usecounter{smallroman}\setlength{\itemindent}{0cm}
   \setlength{\leftmargin}{5ex}\setlength{\labelwidth}{4ex}
   \setlength{\topsep}{0.75\parsep}\setlength{\partopsep}{0ex}
   \setlength{\itemsep}{0ex}}}
{\end{list}}

\newcommand{\ben}{\begin{arabicenumerate}}  
\newcommand{\een}{\end{arabicenumerate}}  

\def\init{\setcounter{equation}{0}}

\newtheorem{theoreme}{Theorem }[section]
\newtheorem{proposition}[theoreme]{Proposition}
\newtheorem{lemma}[theoreme]{Lemma}
\newtheorem{definition}[theoreme]{Definition}

\newtheorem{remark}[theoreme]{Remark}
\newtheorem{example}[theoreme]{Example}
\newcommand{\beq}{\begin{equation}}
\newcommand{\eeq}{\end{equation}}

\newcommand{\bex}{\begin{example}}
\newcommand{\eex}{\end{example}}

\def\bel{\begin{lemma}}
\def\eel{\end{lemma}}
\def\bet{\begin{theoreme}}
\def\eet{\end{theoreme}}
\def\bed{\begin{definition}}
\def\eed{\end{definition}}
\def\ber{\begin{remark}}
\def\eer{\end{remark}}

\def\rr{{\mathbb R}}
\def\zz{{\mathbb Z}}
\def\cc{{\mathbb C}}

\def\app{{\rm app}}
\def\chir{{\rm chir}}
\def\E{{\rm E}}
\def\Born{{\rm Born}}
\def\coupl{\lambda}
\def\HP{{\rm HP}}
\def\amp{{\rm amp}}
\def\gi{{\rm gi}}
\def\rv{{\rm vec}}
\def\+{{(+)}}
\def\-{{(-)}}

\def\lpar {{(\!(}}
\def\rpar {{)\!)}}

\def\Coul{{\rm Coul}}
\def\Max{{\rm Max}}
\def\Eul{{\rm Eul}}
\def\Lor{{\rm Lor}}
\def\KG{{\rm KG}}
\def\GL{{\rm GL}}
\def\Pr{{\rm Pr}}

\def\fB{{\mathfrak B}}
\def\fA{{\mathfrak A}}

\def\fF{{\mathfrak F}}
\def\Texp{{\rm Texp}}

\def\part{{\rm par}} 
\def\Int{{\rm Int}}

\def\Heis{{\rm HP}}
\def\Im{{\rm Im}}
\def\Re{{\rm Re}}
\def\lg{{\rm lg}}
\def\slim{{\rm s-}\lim}
\def\wlim{{\rm w-}\lim}

\def\H{{\rm H}}

\def\cpl{{\rm cpl}}

\def\bar{\overline}

\def\rot{{\rm rot}}

\def\div{{\rm div}}
\def\alg{{\rm alg}}

\def\c0inf{C_0^\infty}

\def\s{{\rm s}}
\def\sa{{\rm s/a}}

\def\ren{{\rm ren}}

\def\ret{{\rm ret}}
\def\adv{{\rm adv}}
\def\sp{{\rm sp}}

\def\proof{{\bf Proof.}\ \ }

\def\sc{{\rm sc}}

\def\cZ{{\mathcal Z}}
\def\cY{{\mathcal Y}}
\def\cV{{\mathcal V}}
\def\cL{{\mathcal L}}
\def\cS{{\mathcal S}}
\def\cQ{{\mathcal Q}}
\def\cE{{\mathcal E}}
\def\cD{{\mathcal D}}
\def\cU{{\mathcal U}}

\def\cO{{\mathcal O}}
\def\cC{{\mathcal C}}
\def\cW{{\mathcal W}}
\def\cR{{\mathcal R}}
\def\fF{{\mathfrak F}}

\def\fin{{\rm fin}}

\def\CCR{{\rm CCR}}

\def\a{{\rm a}}

\def\fr{{\rm fr}}

\def\i{{\rm i}}

\def\Dom{{\rm Dom}}

\def\AUT{{\rm Aut}}

\def\loplus{\mathop{\oplus}\limits}

\def\lpi{\mathop{\Pi}\limits}
\def\sgn{{\rm sgn}}
\def\id{{\rm id}}
\def\Tr{{\rm Tr}}
\def\T{{\rm T}}

\def\Ker{{\rm Ker}}

\def\tr{{\rm tr}}

\def\qed{$\Box$\medskip}

\def\12{\frac{1}{2}}
\def\14{\frac{1}{4}}

\def\supp{{\rm supp}}
\def\dd{{\mathbb D}}

\def\e{{\rm e}}
\def\D{{\rm D}}

\def\d{{\rm d}}

\def\bbbone{{\mathchoice {\rm 1\mskip-4mu l} {\rm 1\mskip-4mu l}
{\rm 1\mskip-4.5mu l} {\rm 1\mskip-5mu l}}}
\def\one{\bbbone}
\def\cH{{\mathcal H}}

\def\can{{\rm can}}

\def\loplus{\mathop{\oplus}\limits}

\def\lpi{\mathop{\Pi}\limits}
\def\sgn{{\rm sgn}}
\def\id{{\rm id}}
\def\Tr{{\rm Tr}}

\def\Ker{{\rm Ker}}

\def\tr{{\rm tr}}

\def\cj{{ j}}
\def\cP{{\mathcal P}}
\def\cJ{{\mathcal J}}
\def\cT{{\mathcal T}}
\def\cF{{\mathcal F}}

\def\cX{{\mathcal X}}
\def\I{{\rm I}}

\def\X{{\rm X}}

\def\R{{\rm R}}

\def\12{\frac{1}{2}}

\def\supp{{\rm supp}}
\def\dd{{\mathbb D}}

\def\e{{\rm e}}

\def\d{{\rm d}}

\def\cH{{\mathcal H}}

\def\P{{\rm P}}

\def\bep{\begin{proposition}}
\def\eep{\end{proposition}}

\def\b{{\rm b}}

\def\s{{\rm s}}

\newcommand{\mat}[4]{\left[\begin{array}{cc}#1 &#2  \\ #3 &#4 \end{array}\right]}
\def\ext{{\rm ext}}
\newcommand{\nowastrona}{}

\def\CARal{{\rm C\hskip 0.25 em \hbox{\raise 1.72 ex 
\hbox{$\scriptscriptstyle\rm al$}\kern -0.57 em A}R}}

\def\t{{\scriptscriptstyle\#}}
\def\otimesal{\mathop{\hbox{\raise 1.5 ex
  \hbox{$\scriptscriptstyle\rm al$}
\kern -0.92 em \hbox{$\otimes$}}}}
\def\oplusal{\mathop{\hbox{\raise 1.5 ex
  \hbox{$\scriptscriptstyle\rm al$}
\kern -0.92 em \hbox{$\oplus$}}}}
\def\Gammal{\hbox{\raise 1.68 ex 
\hbox{$\scriptscriptstyle\rm al$}\kern -0.50 em $\Gamma$}}
\def\Bal{\hbox{\raise 1.68 ex 
\hbox{$\scriptscriptstyle\rm  al$}\kern -0.50 em $B$}}
\def\CARal{{\rm C\hskip 0.25 em \hbox{\raise 1.72 ex 
\hbox{$\scriptscriptstyle\rm al$}\kern -0.57 em A}R}}
\def\t{{\scriptscriptstyle\#}}

\usepackage{rotating}

\begin{document} 

\title{Quantum fields \\
 with classical perturbations}

\author{Jan Derezi\'{n}ski\\
Dept. of Math. Methods in Phys.,\\
 Faculty of Physics, University of Warsaw\\ 
Hoza 74, 00-682 Warszawa, Poland\\ 
email Jan.Derezinski@fuw.edu.pl}
\maketitle

\begin{abstract}
The main purpose of these notes is a review of various models of Quantum Field Theory involving quadratic Lagrangians. We discuss scalar and vector bosons, spin $\frac12$ fermions, both neutral and charged. Beside free theories, we study their interactions with classical perturbations, called, depending on the context, an external
linear source, mass-like term, current  or  electromagnetic
potential. The notes may serve as a first introduction to QFT.
\end{abstract}

\tableofcontents

\setcounter{section}{-1}
\init\section{Introduction}

In these notes we discuss various models of Quantum Field Theory in
1+3 dimensions 
involving quadratic Lagrangians or, equivalently,
quadratic Hamiltonians. 

First of all, we describe basic types of {\em free fields}:
\ben \item {\em neutral scalar bosons},
\item {\em neutral massive  vector bosons} (``{\em massive photons}''),
\item {\em neutral massless  vector bosons} (``{\em massless photons}''),
\item {\em charged scalar bosons},
\item (charged) {\em Dirac fermions},
\item (neutral) {\em Majorana fermions}.
\een

We also consider free fields perturbed by a linear or quadratic
perturbation involving a classical ($c$-number) function.
\begin{arabicenumerate}
\item neutral scalar bosons interacting with  a {\em linear source},
\item neutral scalar bosons  interacting with   a {\em mass-like perturbation},
\item massive
 photons   interacting with   a   {\em classical 4-current},
\item massless
 photons   interacting with   a   {\em classical 4-current},
\item
charged scalar bosons   interacting with   an {\em electromagnetic 4-potential},
\item
Dirac fermions  interacting with   an {\em electromagnetic 4-potential},
\item
Majorana fermions interacting with a  {\em mass-like perturbation}.
\een

All the above models are
(or at least can be) well understood in the non-perturbative sense.
Perturbation theory is not necessary to compute their {\em scattering
 operators} and {\em Green's functions}, which is not the case (at least so far) of truly interacting models.

\nowastrona

\nowastrona

Quantum fields interacting with classical perturbations is a topic with many important applications to realistic physical systems. Therefore, the formalism developed in our text is well motivated physically.

Clearly, many important issues of quantum field theory are outside of the scope of free fields interacting with classical perturbations.
 However, surprisingly many difficult topics can be discussed already on this level.  Therefore, we  believe that our text  has
 pedagogical value, as a kind of an introduction to full quantum field theory.

In our text we stress the deductive character of quantum field
theory. Models that we discuss are quite rigid and built according
to strict principles. Among these principles let us mention the {\em Poincar\'{e} covariance}, the
{\em Einstein causality} and the {\em boundedness of the Hamiltonian from below}. Some of these principles are  encoded
in the {\em Haag-Kastler} and {\em Wightman axioms}. Even if these axioms are often too
restrictive, they
provide useful guidelines.

The only known models for Haag-Kastler or Wightman axioms in 1+3
dimensions are 
  free theories. Their
scattering theory is trivial.
To obtain  interesting physical information one needs {\em interacting
theories}. Unfortunately, interacting theories  are known only perturbatively.

Free theories are the quantizations of covariant 2nd order linear hyperbolic
equations on the Minkowski space. These
 equations can be perturbed by  0th or 1st order terms involving an arbitrary
 space-time functions called, depending on the context, a classical (=external)
linear source, mass-like term, 4-current  or  electromagnetic
4-potential.  We can consider the
quantization of the perturbed equation.
 Such a theory is still essentially  exactly solvable, since the
Hamiltonian is quadratic. It has no Poincar\'{e}{}
covariance. However, it still gives rise to a net of observable algebras
satisfying
 the Einstein causality. 

In our discussion we always start from the study of a {\em classical
theory}, which we discuss  from the
Hamiltonian and Lagrangian point of view. Then we discuss its
quantization. Even though in all the cases we consider the Hamiltonian is
quadratic, its quantization often has various subtle points. In some cases, especially for vector fields, there are several natural approaches to
 quantization, which in the end lead to the same physical
results.  We try to discuss 
 various possible approaches. In our
opinion, the existence of seemingly different formalisms for the same
physical system constitutes one of the most confusing aspects of
quantum field theory.


Classical perturbations that we consider are usually described by smooth space-time functions that decay fast both in space and time.
 In particular, their dynamics is typically described by {\em time-dependent Hamiltonians}. This is a certain minor difficulty, which is often ignored in the literature. We discuss how to modify the usual formalism in order to deal with this problem.

The models that we discuss illustrate many
 problems of interacting theories, such as the {\em ultraviolet problem,}
 the {\em infrared problem} and the {\em gauge invariance.}

The ultraviolet problem means that when we try to define a theory in a naive way some integrals are divergent for large momenta. In the context of our paper this is never due to classical perturbations, which we always assume to be smooth -- the source of ultraviolet divergences is the behavior of propagators.

The ultraviolet problem is already visible when we consider neutral fields with a masslike perturbation or charged fields with
a classical electromagnetic 4-potential. In these systems
classical dynamics exists under rather weak assumptions. However  there are problems with the quantum dynamics.   

In some cases the quantum
dynamics  cannot be implemented on a  Hilbert space. This is 
the case of charged particles (bosons or fermions) in the presence of
 variable spatial components of the  4-potential. On the other hand,  the 
 the scattering
operator exists  under rather weak assumptions
for 4-potential going to zero in the past and future.

Even if we are able to implement
 the classical dynamics or the classical scattering operator, we
 encounter
another
 unpleasant surprise. The
only quantity that is not fixed by the classical considerations is the
phase factor of the scattering operator, written  as $\e^{-\i\cE/\hbar}$, where $\cE$ is usually called the  {\em vacuum energy}.
Computed  naively, it often turns out to be   divergent. In order
to make this phase factor finite it
 is necessary to renormalize the naive expression. This
divergence appears in low order vacuum energy
diagrams. It was first successfully studied by Heisenberg and Euler in the
30's.  A quantity closely related to this phase factor is the {\em 
effective action}, which for a constant field was computed exactly by Schwinger.

The infrared problem means that in the naive theory some integrals are divergent for small momenta.
This problem appears already in non-relativistic quantum mechanics
-- in scattering theory with {\em Coulomb  forces}.
 These forces are {\em long-range}, which
 makes the usual definition of the scattering operator impossible \cite{DG0}. 
Its another manifestation is the appearance of {\em inequivalent
 representations of
canonical commutation relations}, when
 we consider scattering of photons against a
classical 4-current that has a different direction in the past and in the
future \cite{De,DeGe}. Thus,  even in these toy non-relativistic situations 
the usual scattering operator is ill-defined. Therefore, 
it is not surprising that
(much bigger) problems are present eg. in the full QED. One can cope
with the infrared problem by approximating massless photons with 
massive ones 
and restricting computations only to {\em inclusive cross-sections} justified by an
{\em imperfect 
 resolution} of the measuring device \cite{YFS,JR,We1}.

\nowastrona

The expression {\em gauge invariance} has in the context of quantum
field theory several
meanings.

\ben\item The most common meaning, discussed already in the context of classical
electrodynamics, is the fact that if a {\em total derivative} is added to a 4-potential solving
 the Maxwell
equation, then it still solves the Maxwell equations.
Of course, this no longer holds for the {\em Proca equations} -- the massive
generalization of the Maxwell equations. Therefore,
it is often stressed that gauge invariance implies that the photons are
massless. 
\item There exists another meaning of  gauge invariance:
 we can multiply
{\em charged fields} by a {\em space-time dependent phase factor} and compensate it by
changing the external potentials. 

1. and 2.
go together in
the full  QED, which is invariant with respect to these
two 
gauge transformations applied simultaneously.

\nowastrona

\item One often uses the term ``gauge invariance'' in yet
another meaning: To compute the scattering operator we can use various
 {\em (free) photon
propagators}. Equivalently, we have the freedom of choosing a Lagrangian in the path integral formalism. This meaning  applies both to  massive and massless
photons. Some  of these propagators are distinguished, such as the
{\em propagator in the Feynman or the Coulomb
gauge}. (Note, however, that time-ordered $N$-point Green's functions depend on the choice of the propagator).
\een


All these three meanings of  gauge invariance can be illustrated
with models that we consider.

The paper is most of the time rigorous mathematically. In the places where it is not, we believe that many  readers can quite easily  make it rigorous. We try to make the  presentation  of various models parallel by applying, if possible, coherent notation and formalism. This makes our text sometimes repetitious -- we believe that this helps the reader
 to understand  small but often confusing differences between distinct models.

Mathematical language that we use is most of the time elementary. Sometimes we use some  mathematical concepts and facts that are, perhaps,  less commonly known, such as $C^*$-algebras, von Neumann algebras, the Schwartz Kernel Theorem.
The readers unfamiliar with them should not be discouraged -- their role in the article is minor.

Most of the material of this work has been considered in one way or another in the literature. Let us give a brief and incomplete review of references.

On the formal level examples of quantum fields with classical perturbations 
 are discussed in most textbooks
on quantum field theory, see eg. \cite{IZ,JR,Sch,Sr,We1,Wa,BB}.

Linear hyperbolic equations is a well established domain of partial
differential equations, see eg \cite{BGP}.

Axioms of quantum field theory are discussed in \cite{StW,HaaKa,Haa}.

A necessary and sufficient condition for the implementability of
Bogoliubov transformation was given by Shale for bosons  \cite{Sh} and
by Shale and Stinespring for fermions \cite{ShSt}, see also \cite{DeGe}

Problems with implementability of the dynamics of charged particles in
external potentials was apparently first noticed on a heuristic level
in \cite{SSS}. It was  studied rigorously by various authors. In
particular, charged bosons were studied in
\cite{Seiler1972,Lundberg1973,Lundberg1973a,Lundberg1973b,Hochstenbach1976,Asmuss}
and charged  fermions in
\cite{NenciuScharf,KlausScharf,Klaus,Ruijsenaars,DDMS}.
Rigorous discussion of the smeared out local charge for charged fermions is contained in \cite{Lang2}.

The renormalization of the vacuum energy goes back to pioneering
work of \cite{HeEu}. In the mathematically rigorous literature it
leads to the concept of a causal phase
discussed in the fermionic case in \cite{Scharf,G-B}.

The infrared problem goes back to \cite{BN,Ki}, see also \cite{De}.

The Gupta-Bleuler method of quantization of photon fields goes back to
\cite{Gupta,Bleuler}. The $C^*$-algebraic formulation of the
subsidiary condition method is discussed in \cite{TN}.

Rigorous study of vacuum energy for Dirac fermions in a
stationary potential is given in \cite{GHLS}.

A topic that not included
 in these notes are {\em anomalies} in QFT, which to a large extent can be treated in the context of external classical perturbations
\cite{FuSu,Lang1,CDP} 

The notes also treat only dimension 1+3. Note, however, that related problems can be considered in other dimensions. Of particular importance is the case of 1+1 dimension with a large literature, eg. \cite{CR,LangSem}

\bigskip

{\bf Acknowledgements} The origins of these notes go back to a lecture
course on QED at LMU, Munich, which I prepared at the request of
H.Siedentop. The course, with some modifications, was then repated at IHP,
Paris, at the request of M.Lewin, at the  Copenhagen University at the request of
J.P.Solovej, at Warsaw University, and at CRM, Montreal, at the request of V.Jaksic and
R.Seiringer. I am grateful to the colleagues who invited me to give these courses  for their
support and  advice.
Remarks of participants of my courses, among
them of P.Majewski, E.Stockmeier, P.Pickl, D.-A.Deckert, P.Duch and M.Duch are also appreciated. I profited also from the criticism of E.Langmann and anonymous referees.

I am  grateful to the experts in quantum field
theory, K.Pachucki, I.B.Bia{\l}ynicki-Birula, K.Meissner, V.Shabaev and P.Chankowski,  who 
answered my various questions.

Last but not least,
 I profited greatly from my long time collaboration with C.G\'erard.

My research  was supported in part by the National Science Center
(NCN) grant No. 2011/01/B/ST1/04929.  
\nowastrona
\init\section{Basic concepts}
\def\l{{\rm l}}

\subsection{Minkowski space}
\subsubsection{Coordinates in Minkowski space}

The coordinates of the {\em Minkowski space} $\rr^{1,3}$ will be
typically denoted by $x^\mu$,
$\mu=0,1,2,3$.
By definition, the Minkowski space  is  the vector space $\rr^{4}$ equipped with the canonical
pseudo-Euclidean form of signature
$(-+++)$
\[
 g_{\mu\nu}x^\mu x^\nu=-(x^0)^2+\sum_{i=1}^3(x^i)^2.\]
(Throughout these notes the  velocity of light has the value  $1$ and we use the {\em Einstein summation convention}).
We use metric tensor $[g_{\mu\nu}]$ to lower  indices and its inverse $[g^{\mu\nu}]$ to raise indices:
\[x_\mu=g_{\mu\nu}x^\nu,\ \ \ x^\mu=g^{\mu\nu}x_\nu.\]

For a function $\rr^{1,3}\ni x\mapsto f(x)$, 
we will sometimes use various kind of notation for partial derivatives:
\[\frac{\partial f(x)}{\partial x^\mu}=\partial_{x^\mu} f(x)=\partial_\mu f(x)=f_{,\mu}(x)
.\]

\nowastrona

 Writing
$\rr^3$ we will typically denote
 the {\em spatial part} of the Minkowski space obtained by
setting $x^0=0$. If $x\in\rr^{1,3}$, then $\vec x$ will denote the projection
of $x$ onto $\rr^3$. Latin letters $i,j,k$ will sometimes denote the spatial
indices of a vector. Note that $x_i=x^i$.

 $\epsilon^{ijk}$ denotes the 3-dimensional Levi-Civita tensor (the
fully antisymmetric tensor satisfying $\epsilon^{123}=1$). 

For a vector field $\rr^3\ni \vec x\mapsto\vec A(\vec x)$ we define its {\em divergence} and {\em rotation} in the standard way:
\[\div\vec A=\partial_iA^i,\ \ \ (\rot \vec A)^i=\epsilon^{ijk}\partial_j A_k.\] We write $\vec\partial \vec A$ as the shorthand for the tensor $\partial_i A_j$, moreover,
\[\big(\vec\partial \vec A\big)^2:=\sum_{ij}\big(\partial_i A_j\big)^2.\]

On $\rr^{1,3}$ we have the standard Lebesgue measure denoted $\d x$. The
notation $\d\vec{x}$ will be used for the Lebesgue measure on
$\rr^3\subset\rr^{1,3}$.

We will often write $t$ for $x^0=-x_0$. The time derivative will be often denoted by a dot:
\[\dot f(t)=\frac{\partial f(t)}{\partial t}=\partial_tf(t)=\frac{\partial f(
x^0)}{\partial x^0}=\partial_0f(x^0)=f_{,0}(x^0).\]

$\theta(t)$ will denote the Heaviside function. We set
$|t|_+:=\theta(t)|t|$.

\nowastrona
\subsubsection{Causal structure}

 A nonzero vector $x\in \rr^{1,3}$ is called 
\begin{eqnarray*}
\hbox{\em timelike}&\hbox{if} &
x_\mu x^\mu<0,\\
\hbox{\em causal}&\hbox{if} &
x_\mu x^\mu\leq 0,\\
\hbox{\em lightlike}&\hbox{if}& 
x_\mu x^\mu=0,\\
\hbox{\em spacelike}&\hbox{if} &
x_\mu x^\mu>0.\end{eqnarray*}
A causal vector $x$ is called 
\begin{eqnarray*}
\hbox{\em future oriented}&\hbox{if} &
x^0>0,\\
\hbox{\em past oriented}&\hbox{if} &
x^0<0.
\end{eqnarray*}

\nowastrona

The set of future/past oriented causal vectors is called the 
{\em future/past light cone} and denoted $J^\pm$.
 We set $J:=J^+\cup J^-$. 

If $\cO\subset\rr^{1,3}$, its {\em causal shadow} is defined as
$J(\cO):=\cO+J$. We also define its {\em future/past shadow}
$J^\pm(\cO):=\cO+J^\pm.$

Let $\cO_i\subset\rr^{1,3}$, $i=1,2$. We will write $\cO_1\times\cO_2$ iff
$J(\cO_1)\cap\cO_2=\emptyset$, or equivalently, $\cO_1\cap
J(\cO_2)=\emptyset$. We then say that $\cO_1$ and $\cO_2$ are {\em spatially
separated.}  

\nowastrona

A function on $\rr^{1,3}$ is called {\em space-compact} 
iff there exists a compact $K\subset\rr^{1,3}$ such that $\supp
f\subset J(K)$. It is called {\em future/past   space-compact} 
iff there exists a compact $K\subset\rr^{1,3}$ such  that $\supp
f\subset J^\pm(K)$.

  The set of space-compact smooth functions will
be denoted
 $C_\sc^\infty(\rr^{1,3})$.
  The set of future/past space-compact smooth functions will
be denoted
 $C_{\pm\sc}^\infty(\rr^{1,3})$.

\nowastrona

\subsubsection{Fourier transform}

The definition of the Fourier transform of $\rr^3\ni \vec x\mapsto f(\vec x)$
will be standard:
\[\cF f(\vec k):=\int\e^{-\i \vec k\cdot\vec x}f(\vec x)\d\vec x.\]
Often, we will drop $\cF$ -- the name of the variable will indicate whether
we use the position or momentum representation:
\[f(\vec k)=\int\e^{-\i \vec k\cdot\vec x}f(\vec x)\d\vec x,\ \ 
f(\vec x)=\frac{1}{(2\pi)^3}\int\e^{\i \vec k\cdot\vec x}f(\vec k)\d\vec k.\]
For the time variable (typically $t$)
 we reverse the sign in the Fourier transform:
\[f(\varepsilon )=\int\e^{\i \varepsilon t}f(t)\d t,\ \ 
f(t)=\frac{1}{2\pi}\int\e^{-\i \varepsilon t}f(\varepsilon)\d \varepsilon.\]

\subsubsection{Lorentz and Poincar\'e{} groups}

The pseudo-Euclidean group $O(1,3)$
is called the {\em full Lorentz group}.
Its connected component of unity is denoted $SO^\uparrow(1,3)$ and called the {\em connected Lorentz group}.

The full Lorentz group contains special elements: the time reversal $\T$ and the space inversion (the parity) $\P$ and the space-time inversion $\X:=\P\T$:
\[\T(x^0,\vec x)=(-x^0,\vec x),\ \  \P(x^0,\vec x)=(x^0,-\vec x),\ \ \X x=-x.\]
It consists of four connected components
\[SO^\uparrow(1,3),\ \T{\cdot}SO^\uparrow(1,3),\ \P{\cdot} SO^\uparrow(1,3),\ \X{\cdot}SO^\uparrow(1,3).\]
$O(1,3)$ has three subgroups of index two: the {\em special Lorentz group} (preserving the spacetime orientation),  the {\em orthochronous Lorentz group} (preserving the forward light cone) and the {\em chiral  Lorentz group} (preserving the parity):
\begin{eqnarray}
 SO(1,3)&=&SO^\uparrow(1,3)\cup \X{\cdot}SO^\uparrow(1,3),\label{gru1}\\
O^\uparrow(1,3)&=&SO^\uparrow(1,3)\cup \P{\cdot}SO^\uparrow(1,3),\label{gru2}\\
O^\chir(1,3)&=&SO^\uparrow(1,3)\cup \T{\cdot}SO^\uparrow(1,3).\label{gru3}
\end{eqnarray}

The affine extension of the full  Lorentz group
$\rr^{1,3}\rtimes O(1,3)$
 is called the {\em full Poincar\'e
group}. Its elements will be typically written as $(y,\Lambda)$.
 We will often write $y$ instead of $(y,\one)$ and $\Lambda$ instead of $(0,\Lambda)$. It is the full symmetry group of the Minkowski space.

Quantum field theory models are often not invariant wrt the full Poincar\'e{} group but one of its subgroups: the {\em connected}, {\em special}, {\em ortochronous} or {\em chiral Poincar\'e{} group}, which have the obvious definitions.

\subsubsection{Double coverings of  Lorentz and Poincar\'e{} groups}
\label{Double coverings}

The full Poincar\'e{} group or one of its subgroups discussed above is sufficient to describe spacetime symmetries on the level observables.
On the level of the Hilbert space one needs to replace it by one of its double coverings.

There exists a unique, up to an isomorphism, 
connected  group $Spin^\uparrow(1,3)$
such that the following  short exact sequence is true:
\beq \one\to\zz_2\to Spin^\uparrow(1,3)\to SO^\uparrow(1,3)\to\one.\label{wqw}\eeq
We  say that
 $Spin^\uparrow(1,3)$ is a {\em connected double covering of $SO^\uparrow(1,3)$}.
The group $Spin^\uparrow(1,3)$ happens to be isomorphic to $SL(2,\cc)$. 


Unfortunately, the theory of double coverings of the group $O(1,3)$ is relatively complicated.
$O(1,3)$ has 8 non-isomorphic  double coverings  that  extend
(\ref{wqw}),
that is groups $G$ such that the following diagram with exact rows and columns commutes:
\beq \begin{array}{ccccccccc}
&&\one&&\one&&\one&&\\
&&\downarrow&&\downarrow&&\downarrow&&\\
 \one&\to&\zz_2&\to& Spin^\uparrow(1,3)&\to &SO^\uparrow(1,3)&\to&\one\\
&&\downarrow&&\downarrow&&\downarrow&&\\
\one&\to&\zz_2&\to &G&\to& O(1,3)&\to&\one,\\
&&\downarrow&&\downarrow&&\downarrow&&\\
&&\one&\to&\zz_2\times\zz_2&\to&\zz_2\times\zz_2&\to&\one\\
&&&&\downarrow&&\downarrow&&\\
&&&&\one&&\one&&
\end{array}
\label{wqw1}\eeq
 (\cite{Stern}  Sect. 3.10). Indeed, let us fix  elements $\tilde\P$, $\tilde\T$ and $\tilde\X$ that cover  $\P$, $\T$, $\X$. We can independently demand that
\beq\tilde\P^2=\pm\one,\ \ \tilde\T^2=\pm\one,\ \ \tilde\X^2=\pm\one.\label{rela}\eeq
To obtain these double coverings we need to take the group $G$ generated by $Spin^\uparrow(1,3)$ and
$\tilde\P$, $\tilde\T$ and $\tilde\X$ 
with the relations (\ref{rela}) and
$\tilde\X:=\tilde\P\tilde\T$. (This defines (\ref{wqw1}) uniquely).
Clearly, (\ref{rela})
 gives $2\cdot2\cdot2=8$  possibilities.

 Among them one has two  distinguished double coverings
$G= Pin_\pm(1,3)$ with the relations
\[\tilde\T_\pm^2=\pm\one,\ \ \ \tilde\P_\pm^2=\pm\one,\ \ \ \tilde\X_\pm^2=-\one.
\]
The elements  $\tilde\P_\pm$, $\tilde\T_\pm$ and $\tilde\X_\pm$ anticommute among themselves. In what follows we will drop $\pm$ from $\tilde\P_\pm$, $\tilde\T_\pm$ and $\tilde\X_\pm$.

Later on we will need the homomorphism $\theta:Pin_\pm(1,3)\to\zz_2=\{1,-1\}$, which is $\theta(\tilde\Lambda)=1$ for orthochronous $\tilde\Lambda$ and
 $\theta(\tilde\Lambda)=-1$ for non-orthochronous $\tilde\Lambda$. It is called the {\em time orientation}.

Each group (\ref{gru1}), (\ref{gru2}) and (\ref{gru3}) has two non-isomorphic double coverings extending (\ref{wqw}).
In particular, we have 
\beq \begin{array}{ccccccccc}
&&\one&&\one&&\one&&\\
&&\downarrow&&\downarrow&&\downarrow&&\\
 \one&\to&\zz_2&\to& Spin^\uparrow(1,3)&\to &SO^\uparrow(1,3)&\to&\one\\
&&\downarrow&&\downarrow&&\downarrow&&\\
\one&\to&\zz_2&\to &Spin(1,3)&\to& SO(1,3)&\to&\one,\\
&&\downarrow&&\downarrow&&\downarrow&&\\
&&\one&\to&\zz_2&\to&\zz_2&\to&\one\\
&&&&\downarrow&&\downarrow&&\\
&&&&\one&&\one&&
\end{array}
\label{wqw2}\eeq
The group $Spin(1,3)$ is contained in both $Pin_+(1,3)$ and  $Pin_-(1,3)$. It is obtained from  $Spin^\uparrow(1,3)$ by adjoining $\tilde \X$ satisfying $\tilde\X^2=-\one$. (The other double covering, obtained by adjoining $\tilde\X$ satisfying $\tilde\X^2=\one$ will not play a role in our considerations).

We also have two    double coverings of $O^\uparrow(1,3)$ extending (\ref{wqw}), one contained in $Pin_+(1,3)$, the other in $Pin_-(1,3)$:
\beq \begin{array}{ccccccccc}
&&\one&&\one&&\one&&\\
&&\downarrow&&\downarrow&&\downarrow&&\\
 \one&\to&\zz_2&\to& Spin^\uparrow(1,3)&\to &SO^\uparrow(1,3)&\to&\one\\
&&\downarrow&&\downarrow&&\downarrow&&\\
\one&\to&\zz_2&\to &Pin_\pm^\uparrow(1,3)&\to& O^\uparrow(1,3)&\to&\one,\\
&&\downarrow&&\downarrow&&\downarrow&&\\
&&\one&\to&\zz_2&\to&\zz_2&\to&\one\\
&&&&\downarrow&&\downarrow&&\\
&&&&\one&&\one&&
\end{array}
\label{wqw3}\eeq
 $Pin_\pm^\uparrow(1,3)$ is 
obtained by adjoining $\tilde\P$ satisfying $\tilde\P^2=\pm\one$. 

Finally, we have two double coverings of
 $O^\chir(1,3)$ extending (\ref{wqw}), one contained in $Pin_+(1,3)$, the other in $Pin_-(1,3)$:
\beq \begin{array}{ccccccccc}
&&\one&&\one&&\one&&\\
&&\downarrow&&\downarrow&&\downarrow&&\\
 \one&\to&\zz_2&\to& Spin^\uparrow(1,3)&\to &SO^\uparrow(1,3)&\to&\one\\
&&\downarrow&&\downarrow&&\downarrow&&\\
\one&\to&\zz_2&\to &Pin_\pm^\chir(1,3)&\to& O^\chir(1,3)&\to&\one,\\
&&\downarrow&&\downarrow&&\downarrow&&\\
&&\one&\to&\zz_2&\to&\zz_2&\to&\one\\
&&&&\downarrow&&\downarrow&&\\
&&&&\one&&\one&&
\end{array}
\label{wqw4}\eeq
$Pin_\pm^\chir(1,3)$ is
obtained by adjoining $\tilde\T$ satisfying $\tilde\T^2=\pm\one$.

Clearly, $\rr^{1,3}\rtimes Pin_\pm(1,3)$ is a double covering of the full Poincar\'e{} group.
Its elements  will be often written as
$(y,\tilde\Lambda)$ and then the corresponding element of
$\rr^{1,3}\rtimes O(1,3)$ will be denoted by
$(y,\Lambda)$. 

\subsubsection{Complex Lorentz groups}
\label{Complex Lorentz groups}

The complexification of $Spin^\uparrow(1,3)$ is called $Spin(4,\cc)$. It is a connected group  isomorphic to $SL(2,\cc)\times SL(2,\cc)$.  We have the embedding
\[ Spin^\uparrow(1,3)\simeq SL(2,\cc)\ni \tilde\Lambda\mapsto (\tilde\Lambda,\bar{\tilde\Lambda})\in
SL(2,\cc)\times SL(2,\cc)\simeq Spin(4,\cc).\]

The group $Spin(1,3)$ is contained in $Spin(4,\cc)$. In particular, the two elements covering the spacetime inversion are represented as follows:
\begin{eqnarray*}
 Spin(1,3)\ni \tilde\X&\mapsto& (\one,-\one)\in
 Spin(4,\cc),\\
 Spin(1,3)\ni -\tilde\X&\mapsto& (-\one,\one)\in
 Spin(4,\cc).
\end{eqnarray*}

Every finite dimensional representation of
 $Spin^\uparrow(1,3)$ extends uniquely by holomorphic continuation to a  representation of the  connected  complex group
 $Spin(4,\cc)$.
This representation can be restricted to a  representation of $Spin(1,3)$. Thus every finite dimensional representation of
 $Spin^\uparrow(1,3)$ has a natural extension to a  
 representation of $Spin(1,3)$.

The complexification of
 $Pin_-(1,3)$ coincides with the complexification of  $Pin_+(1,3)$. It is denoted $Pin(4,\cc)$.

It will also be  useful to introduce a group that we will call $Pin_\ext(1,3)$. It is the real subgroup of  $Pin(4,\cc)$ that  is generated by  $Pin_-(1,3)$ and $\i\one$, and also by $Pin_+(1,3)$ and $\i\one$. We have
\beq \begin{array}{ccccccccc}
&&\one&&\one&&\one&&\\
&&\downarrow&&\downarrow&&\downarrow&&\\
 \one&\to&\zz_2&\to& Spin^\uparrow(1,3)&\to &SO^\uparrow(1,3)&\to&\one\\
&&\downarrow&&\downarrow&&\downarrow&&\\
\one&\to&\zz_4&\to &Pin_\ext(1,3)&\to& O(1,3)&\to&\one.\\
&&\downarrow&&\downarrow&&\downarrow&&\\
\one&\to&\zz_2&\to&\zz_2\times\zz_2\times\zz_2&\to&\zz_2\times\zz_2&\to&\one\\
&&\downarrow&&\downarrow&&\downarrow&&\\
&&\one&&\one&&\one&&
\end{array}
\label{wqw1-}\eeq

\subsection{General concepts of quantum field theory}
\subsubsection{Quantum mechanics}
\label{Quantum mechanics}

Pure quantum  states are described by normalized vectors in a Hilbert space.
 In typical situations the dynamics is generated by a
 bounded from below self-adjoint operator called the {\em Hamiltonian}. It does not affect any physical
predictions if we subtract from the Hamiltonian the infimum of its spectrum. The Hamiltonian has often a ground state. The ground state is typically nondegenerate.

It will be convenient to  formalize these properties. 
\bed We will say that 
$\cH,H,\Omega$ satisfy   the {\em standard requirements of quantum
  mechanics (QM)}  if
\ben
\item $\cH$ is a Hilbert space;
\item $H$ is a positive self-adjoint operator on $\cH$ (called 
the  Hamiltonian);
\item $\Omega$ is a normalized eigenvector of $H$ with eigenvalue $0$;
\item $\Omega$ is nondegenerate as an eigenvector of $H$.
\een
\label{standard}\eed

\nowastrona
\subsubsection{Time reversal}

If $R$ is a unitary operator $R$ reversing the time, that is, satisfying
\[R\e^{-\i tH}R^{-1}=\e^{\i tH},\]
then $RH R^{-1}=-H$. Therefore, if  $H$ is positive, then $H=0$. Hence unitary operators are not appropriate for the time reversal invariance.

Following Wigner, by a {\em time reversal} operator we will mean an anti-unitary operator $T$ satisfying
\[T\e^{-\i tH}T^{-1}=\e^{\i tH}.\]
We have then $THT^{-1}=H$, which is compatible with the positivity of $H$.

Let us review some concepts and notation related to 
linear and especially anti-linear operators, motivated by their applications  as Wigner's time reversal. Consider the complex vector space  $\cW=\cc^n$. Let $\rho$ be a linear operator on $\cW$.  Then there exists a matrix $[\rho_{ab}]$ such that
\beq (\rho w)_a=\sum_b\rho_{ab}w_b,\label{mat1}\eeq
where $w=[w_a]\in\cW$. We will call $[\rho_{ab}]$ the {\em matrix of $\rho$}.
Note that it is natural to denote the operator and its matrix by the same symbol. In particular, the matrix of the product of linear operators is simply the product of their matrices.

Let $\kappa$ be an  antilinear operator  on $\cW$. Then there exists a matrix $[\kappa_{ab}]$ such that
\beq(\kappa w)_a=\sum_b\kappa_{ab}\bar w_b,\label{mat2}\eeq
where, as usual, the bar denotes the complex conjugation. We will say that $[\kappa_{ab}]$ is the {\em matrix of $\kappa$}.
 Unfortunately, it is dangerous to use the same letter for an antilinear operator and its matrix, even if we will sometimes do so, as in (\ref{mat2}).  The matrix of the  product of two antilinear operators is in general not the product of their matrices -- one needs to put the bar in an appropriate place.

 Sometimes it will be convenient to denote linear transformations on $\cW$  by $L_1(\cW)$ instead of the usual  $L(\cW)$. Then antilinear transformations will be denoted by $L_{-1}(\cW)$. 

Let $G$ be a group equipped with a homomorphism $\theta:G\to\zz_2=\{1,-1\}.$
It yields an obvious partition of $G$: \[G=G_1\cup G_{-1}.\]
 We will say that  $G\ni g\mapsto \pi(g)$ is a {\em $\theta$-linear} representation on $\cW$ if we have a pair of maps
\begin{eqnarray}
G_1\ni g&\mapsto&\pi(g)\in L_1(\cW),\label{pada1}\\
G_{-1}\ni g&\mapsto&\pi(g)\in L_{-1}(\cW),\label{pada2}
\end{eqnarray}
which together form a representation of $G$.
One can write (\ref{pada1}) and (\ref{pada2}) more compactly:
\begin{eqnarray}
G\ni g&\mapsto&\pi(g)\in L_{\theta(g)}(\cW).\label{pada3}
\end{eqnarray}

Suppose that $\cW$ is equipped with a scalar product.
Sometimes it will be convenient to denote unitary transforations on $\cW$ by $U_1(\cW)$ instead of the usual $U(\cW)$. Then anti-unitary transformations will be denoted by $U_{-1}(\cW)$. 
We say that   $G\ni g\mapsto \pi(g)$ is a {\em $\theta$-unitary} representation on $\cW$ if we have a pair of maps
\begin{eqnarray}
G_1\ni g&\mapsto&\pi(g)\in U_1(\cW),\label{pada4}\\
G_{-1}\ni g&\mapsto&\pi(g)\in U_{-1}(\cW),\label{pada5+}
\end{eqnarray}
which together form a representation of $G$.
Again, (\ref{pada4}) and (\ref{pada5+}) can be written more compactly:
\begin{eqnarray}
G\ni g&\mapsto&\pi(g)\in U_{\theta(g)}(\cW).\label{pada6}
\end{eqnarray}

 \subsubsection{Relativistic
quantum mechanics}

Relativistic covariance of a quantum system described by a Hilbert space $\cH$ is expressed by choosing a
strongly continuous unitary
representation of the double cover of the connected Poincar\'{e}{} group
\beq\rr^{1,3}\rtimes Spin^\uparrow(1,3)\ni (y,\tilde\Lambda)\mapsto
U(y,\tilde\Lambda)\in U(\cH).\label{unito}\eeq
We will denote the self-adjoint generator of space-time translations
by
  $P=(P^0,\vec
P)$. $P^0=H$ is  the {\em Hamiltonian}. $\vec P$ is called  the {\em momentum}. Thus
\[U((t,\vec y),\one)=\e^{-\i tH+\i \vec y\vec P}.\]
(We assume that the Planck constant $\hbar$ equals $1$).

Representations of  $Spin^\uparrow(1,3)$ can be divided into two
categories. Integer spin representations induce a representation of
 $SO^\uparrow(1,3)$, and half-integer representations do not.
The projections
\[\frac12\big(\one+U(0,-\one)\big), \ \hbox{resp.}\  
\frac12\big(\one-U(0,-\one)\big)\]
project onto the spaces
of representations of integer, resp.  half-integer spin. We will
write
\[I:=U(0,-\one).\]
Obviously, $PI=IP$. Anticipating the connection of spin and statistics we will call $I$
the {\em fermionic parity}. 
Denote the $*$-automorphism defined by $U(y,\tilde\Lambda)$ by $\cU_{(y,\tilde\Lambda)}$
\[\cU_{(y,\tilde\Lambda)}(A):=U(y,\tilde\Lambda)AU(y,\tilde\Lambda)^*.\]
Restricted to the commutant of $I$
\[\{I\}':=\{A\in B(\cH)\ :\ IA=AI\}\]
$\cU_{(y,\tilde\Lambda)}=\cU_{(y,-\tilde\Lambda)}$, and thus we obtain a  representation of the Poincar\'e{} group:
\[\rr^{1,3}\rtimes SO^\uparrow(1,3)\ni(y,\Lambda)\mapsto\cU_{(y,\Lambda)}\in\AUT\big(\{I\}'\big).\]

\bed The following conditions will be called the {\em basic requirements of
  relativistic quantum mechanics (RQM)}:
\ben
\item {\em Existence of a Poincar\'{e}{} invariant vacuum}:
There exists a (normalized) vector $\Omega$ invariant with respect to 
$\rr^{1,3}\rtimes Spin^\uparrow(1,3)$.
\item {\em Spectral condition}: The joint spectrum of the energy-momentum operator is contained in the forward light cone, that is,
$\sp(P)\subset J^+$.
\item {\em Uniqueness of the vacuum}:
The vector $\Omega$ is unique up to a phase factor.
\item {\em Integer and half-integer spin states live in separate
 superselection sectors}: Observables are contained in $\{I\}'$.
\een
\label{standard1}\eed

Note that  conditions (1)-(3) imply the standard requirements of QM.

More precisely,  (2) implies $H\geq0$. Conversely, the Poincar\'{e}{} invariance
 and the boundedness from below of $H$ implies (2).

(2) implies also that $\Omega$ is the {\em ground state of $H$}. (3) implies that
this ground state is unique. 

Obviously,  $I\Omega=\Omega$.


\nowastrona

\begin{remark} Sometimes the expression {\em relativistic quantum mechanics} is used for the theory of relativistic linear  hyperbolic equations,  such as the Klein-Gordon and Dirac  equation. For the Klein-Gordon equation this is certainly incorrect. This is a classical equation -- in particular, it does not have a natural interpretation in terms of a unitary dynamics on a Hilbert space. In our terminology Dirac equation is also a classical equation -- its unitary dynamics is non-physical because the Hamiltonian is unbounded from below. 
\end{remark}

\nowastrona

\subsubsection{Haag-Kastler axioms for observable algebras}

We still need some postulates that express the idea of causality.
In the mathematical physics 
literature one can find two kinds of axioms that try to formalize this concept: the {\em Haag-Kastler}
and the {\em Wightman axioms}.
 Even though the Wightman axioms were formulated earlier,
it is  more natural to start with the Haag-Kastler axioms. 

\bed We keep the  basic requirements of RQM.

 In addition,
to each open bounded set $\cO\subset\rr^{1,3}$ we associate a 
{\em von Neumann algebra}
 $\fA(\cO)\subset \{I\}'$.
 We will say that the family $\fA(\cO)$,
$\cO$ open in $\rr^{1,3}$, is a {\em net of observable algebras} satisfying the
  {\em Haag-Kastler axioms} if  
the following conditions hold:
\ben\item {\em Isotony}: $\cO_1\subset\cO_2$ implies $\fA(\cO_1)\subset\fA(\cO_2) $.
\item {\em Poincar\'e{} covariance}: for $(y,\tilde\Lambda)\in \rr^{1,3}\rtimes Spin^\uparrow(1,3)$, we have 
\[\cU_{(y,\tilde\Lambda)}\big(\fA(\cO)\big)
=\fA\big((y,\Lambda)\cO\big).\]
\item {\em Einstein causality}:
Let $\cO_1\times\cO_2$. Then  
\[A_i\in\fA(\cO_i),\  i=1,2,\  \hbox{implies}\ 
A_1A_2=A_2A_1.\]
%
\een\eed
\nowastrona

Self-adjoint elements of 
the algebras $\fA(\cO)$ are supposed to describe {\em observables in
  $\cO$}.
 This means
that in principle an observer contained in $\cO$ can perturb the dynamics by
a self-adjoint operator from $\fA(\cO)$, and only from
$\fA(\cO)$.


\ber One can ask why
 von Neumann algebras are used in the Haag-Kastler axioms to describe sets of
observables. We would like to argue that it is a natural choice.

 Suppose we weaken the
Haag-Kaster axioms as follows: We replace
the family of von Neumann 
algebras $\fA(\cO)$ by arbitrary sets $\fB(\cO)$ of self-adjoint
elements of $B(\cH)$, and otherwise we keep the axioms unchanged. Then, if we
set $\fA(\cO):=\fB(\cO)''$ (which obviously contain $\fB(\cO)$),  we obtain a family of von Neumann algebras
satisfying the usual Haag-Kastler axioms. In particular, to see that
the Einstein causality still holds, we use the following easy fact:

Let $\fB_1$, $\fB_2$, be two $*$-invariant subsets of $B(\cH)$
such that \[A_1\in\fB_1,\ A_2\in\fB_2\ \ \hbox{
  implies}\ \ A_1A_2=A_2A_1.\] Set $\fA_1:=\fB_1''$, $\fA_1:=\fB_1''$. Then
\[ A_1\in\fA_1,\ A_2\in\fA_2\ \hbox{ implies }\ \ A_1A_2=A_2A_1.\]
\eer

\nowastrona

\subsubsection{Haag-Kastler axioms for field algebras}

It is often natural to consider nets of algebras containing not only
observables, but also other operators that can be useful to construct
observables. 
 They are called {\em field algebras} and satisfy a slightly modified version of Haag-Kastler axioms.

\bed  We assume the  basic requirements of RQM. 
We say that a family
of von Neumann algebras
 $\fF(\cO)\subset B(\cH)$ associated to
bounded open subsets $\cO$ of $\rr^{1,3}$ is a {\em net of field algebras} in the
sense of {\em Haag-Kastler axioms} if the following conditions hold:
\ben\item[(1)'] {\em Isotony}: $\cO_1\subset\cO_2$ implies $\fF(\cO_1)\subset\fF(\cO_2) $.
\item[(2)'] {\em Poincar\'e{} covariance}: for $(y,\tilde\Lambda)\in \rr^{1,3}\rtimes Spin^\uparrow(1,3)$, we have 
\[\cU_{(y,\tilde\Lambda)}\big(\fF(\cO)\big)
=\fF\big((y,\Lambda)\cO\big).\]
\item[(3)'] {\em Twisted Einstein causality}. Let $\cO_1\times\cO_2$. Then  
\[A_i\in\fF(\cO_i),\  A_i=(-1)^{j_i}IA_iI,\ \ i=1,2,\  \hbox{implies}\ 
A_1A_2=(-1)^{j_1j_2}A_2A_1.\] 
\item[(4)'] {\em Cyclicity}:
$\Big(\bigcup\limits_{\cO}\fF(\cO)\Big)\Omega$ is dense in $\cH$.
\een
\eed

The main reason for introducing the twisted Einstein causality is the
need to accommodate anticommuting fermionic fields. 
Clearly, if the net
$\fF(\cO)$,
$\cO\subset\rr^{1,3}$ satisfies the
Haag-Kastler axioms for 
 field algebras, then the net of their {\em fermionic even subalgebras}
\[\fF_0(\cO):=\{B\in\fF(\cO)\ :\ IBI=B\}, \ \ \cO\subset\rr^{1,3},\]
satisfies the
Haag-Kastler axioms for 
 observable algebras.

Note that in our formulation the decomposition $\cH=\cH_0\oplus\cH_1$
given by the operator $I$ plays a double role. 
\ben \item It
describes the decomposition of the Hilbert space into integer and
half-integer spin representations.
\item In the Einstein causality axiom, block-diagonal  operators have the bosonic character and block-off-diagonal
  operators have the fermionic character. 
\een
A priori it is not obvious that these two properties should give the
same decomposition. However, one can show that it is natural to assume
from the beginning that this is the case. This is the content the
theorem about the {\em connection of the spin and statistics}, described
eg. in \cite{StW}.

Setting $\tilde\Lambda=-\one$ in Axiom (2)' shows that the
  bosonic/fermionic superselection rule is local, ie., 
  $I\fF(\cO)I=\fF(\cO)$  for all $\cO$.
\subsubsection{Global symmetries}
\label{Global symmetries}

Field algebras can be used to describe {\em global
symmetries}.

 Suppose that a group $G$ has a unitary representation on the Hilbert $\cH$:
\[G\ni g\mapsto R(g)\in U(\cH)\]
We assume that $R(g)$, $g\in G$, commute with $U(y,\tilde\Lambda)$ and leave invariant $\Omega$.
This implies that $I$ commutes with $R(g)$.
Let $\cR_g$ denote the automorphism defined by $R(g)$:
\[\cR_g(A):=R(g)AR(g)^{-1},\ \ A\in B(\cH).\]
We define the {\em gauge invariant subalgebras}
\[\fF_\gi(\cO)=\{B\in\fF_0(\cO)\ :\ \cR_g(B)=B,\ g\in G\}\]
or, equivalently,
\[\fF_\gi(\cO)=\fF_0(\cO)\cap\{R(g)\ :\ g\in G\}'.\]
Then the net  $\cO\mapsto\fF_\gi(\cO)$ satisfies then the Haag-Kastler
axioms for observable algebras.

{\subsubsection{Neutral quantum fields}}

In practical computations of quatum field theory the information is encoded in
{\em quantum fields}.
Some of these fields are (formally) Hermitian, and then they are called {\em neutral fields}. Some of them are not -- they are usually called {\em charged fields}. We will first consider only neutral fields. Charged fields will be discussed later.

Neutral fields are typically denoted by
 $\rr^{1,3}\ni x\mapsto \hat\phi_a(x)$, where $a=1,\dots,n$
 enumerates the ``internal degrees of freedom'', eg. the 
species of particles and
 the value of their spin projected
on a distinguished axis. 
Some of the fields are bosonic, some are fermionic. They commute or
anticommute for spatially separated points, 
  which
is expressed by the commutation/anticommutation relations
\[[\hat\phi_a(x),\hat\phi_b(y)]_\pm=0,\ \ (x-y)^2>0.\]

\nowastrona

One can try to interpret neutral quantum fields as
``operator valued tempered distributions'', which become (possibly unbounded)
self-adjoint operators
when smeared out with real Schwartz test functions.
  We can
organize the internal degrees of freedom 
of neutral fields
into a finite dimensional vector
space $\cV=\rr^n$. 
Thus for any $f=(f_a)\in \cS(\rr^{1,3},\rr^n)$ we obtain a
{\em smeared out quantum field}, which is the operator
\beq \hat\phi[f]:=\sum_a\int f_a(x)\hat\phi_a(x)\d x.\label{lne3}\eeq

\nowastrona

\subsubsection{Wightman axioms for neutral fields}
\label{Wightman axioms for neutral fields}

Let us now formulate the Wightman axioms for neutral fields.

\bed We assume that  the basic requirements of RQM are satisfied.

 $\cV$ is a finite dimensional real  vector
space
 equipped with a representation
\beq 
Spin^{\uparrow}(1,3)\ni\tilde\Lambda\mapsto \sigma (\tilde\Lambda)\in L(\cV).\label{REPI}\eeq
We have a unique decomposition $\cV=\cV_0\oplus\cV_1$.
where $\cV_0$, resp. $\cV_1$ is  the space of integer spin,
resp. half-integer spin.

\nowastrona

We suppose that $\cD$ is a dense subspace of $\cH$ containing $\Omega$
and we have a map 
\beq \cS(\rr^{1,3},\cV)\ni f\mapsto \hat\phi[f]\in L(\cD)\label{line}\eeq
 satisfying the following conditions:

\ben
\item {\em Continuity}: For any $\Phi,\Psi\in\cD$,
\beq \cS(\rr^{1,3},\cV)
\ni f\mapsto (\Phi|\hat\phi[f]\Psi)\label{line1}\eeq
is continuous.
\item
{\em Poincar\'e{} covariance}: 
 for $(y,\tilde\Lambda)\in \rr^{1,3}\rtimes Spin^\uparrow(1,3)$ we have 
\[\cU_{(y,\tilde\Lambda)}\big(\hat\phi[f]\big)
=\hat\phi\left[\sigma (\tilde\Lambda)f\circ(y,\Lambda)^{-1}\right].\]
\nowastrona
\item {\em Einstein causality}: Let  ${\nolinebreak \supp f_1\times\supp f_2=\emptyset}$, where $f_i$ have values in $\cV_{j_i}$,
$i=1,2$. Then \[\hat\phi[f_1]\hat\phi[f_2]=(-1)^{j_1j_2}\hat\phi[f_2]\hat\phi[f_1].\]
\item 
{\em Cyclicity of the vacuum}: Let $\fF^\alg$ denote the algebra of polynomials generated by $\hat \phi[f]$. Then $\fF^\alg\Omega$
 is dense in $\cH$.
\item {\em Hermiticity}: For any $\Phi,\Psi\in\cD$,
\[(\Phi|\hat\phi[f]\Psi)=(\hat\phi[ f]\Phi|\Psi).\] 
\een
\eed




In what follows a map (\ref{line}) satisfying Axiom (1) will be called an
{\em operator valued distribution}.
 By saying that it is cyclic we will
mean that it satisfies Axiom (4).

Setting $\tilde\Lambda=-\one$ in Axiom (2),  we see that
 $f\in\cS(\rr^{1,3},\cV_j)$ implies
\[\hat\phi[ f]=(-1)^{j}I\hat\phi[f] I.\]

\subsubsection{Relationship between Haag-Kastler and Wightman axioms}

``Morally", Wightman axioms are stronger than the Haag-Kastler axioms. In
fact, let $\fF^\alg(\cO)$ be the algebra of polynomials in $\hat\phi[f]$ with $\supp f\subset \cO$, which can be treated as a $*$-subalgebra of $L(\cD)$.
Then the family 
  $\cO\mapsto \fF^\alg(\cO)$ is almost a net of field algebras
and  $\cO\mapsto \fF_0^\alg(\cO)$ is almost a net of observable algebras
in the sense of the
Haag-Kastler axioms. Unfortunately, elements of $\fF^\alg(\cO)$ are defined
only on 
$\cD$ and not on the whole $\cH$, and often do not extend to bounded operators
on $\cH$.

\nowastrona

We know that the fields  $\hat\phi[f]$  are {\em Hermitian}
  on $\cD$. Suppose they
are {\em essentially self-adjoint}. Then their closures are self-adjoint
operators on $\cH$. We  could consider the von Neumann algebra $\fF(\cO)$
generated by bounded functions of $\hat\phi[f]$,  $\supp f\subset
\cO$. Let $\fF_0(\cO)$ be its fermionic even part.  Then  there is still no guarantee that
 the net $\cO\mapsto \fF_0(\cO)$ satisfies the Haag-Kastler axioms:
we are not sure whether the Einstein causality holds. 

To see this we recall  that there are serious problems with commutation of
unbounded operators \cite{RS1}. One says that two self-adjoint operators commute (or
strongly commute) if all their spectral projections commute.
There exist however
examples of pairs of two self-adjoint operators $A$, $B$ and a
subspace $\cD\subset \Dom A\cap\Dom B$ with 
the following property:
\ben \item $A$ and $B$ preserve $\cD$ and are essentially self-adjoint on $\cD$.
\item $A$ and $B$ commute on $\cD$.
\item  $A$ and $B$ do not commute strongly.
\item $\cD$ is dense.
\een

More about what is known about the relationship between the  Haag-Kastler and Wightman axioms the reader can find in \cite{Araki}, Sect. 4.9.

\nowastrona
\subsubsection{Global symmetries in the Wightman formalism}
\label{Global-Wight}

In the  Wightman formalism we can encode global symmetries.
Let $G$ be a group acting with the unitary representation $U(g)$ and let $\cU_g$  the corresponding $*$-automorphism, as described 
in Subsect. \ref{Global symmetries}. Suppose in addition that $g$ acts on $\cV$
such that $\cR_g(\hat\phi[f]):=\hat\phi[g f]$, or in the unsmeared notation
\[\cR_g\big(\hat\phi_a(x)\big)=\sum_b g_{ab}\hat\phi_b(x),\]
where $g$ commutes with
 $\sigma(\tilde\Lambda)$.

$\cR_g$ can be interpreted as a $*$-automorphism of the polynomial algebra $\fF^\alg$.
We set $\fF_\gi^\alg(\cO)$ to be the  subalgebra of fixed points of the
 action of $G$ on $\fF_0^\alg(\cO)$. 
One could argue that this $*$-algebra should describe observables in $\cO$.

Note that what we described is  a  {\em global symmetry}
and not  a {\em local gauge invariance}. (In the older literature
sometimes the former is called the {\em gauge invariance of the first kind} and the latter the
 {\em gauge invariance of the second kind}). 
Satisfactory 
treatment of local gauge invariance, even Abelian,  in the framework of Wightman axioms seems to be problematic. In fact, a convenient description of gauge fields seems to require a space with an indefinite scalar product. This goes beyond the usual Wightman axioms and poses serious technical problems \cite{Wigh}.

\nowastrona

Haag-Kastler axioms seem to provide a satisfactory general framework for  quantum
field theory on a flat spacetime, also for theories with local gauge invariance. Their weakness is the abstractness and great
generality. For instance, we do not see how to recognize that a given family of
algebras satisfying Haag-Kastler axioms corresponds to a theory with   local
gauge invariance. (There exists, however, a beautiful theory developed by
Doplicher-Haag-Roberts that allows us to recognize   global
symmetries.)

Wightman axioms 
 seem more concrete.
However, they have flaws. As we mentioned earlier,
 they seem to be incompatible with the
local gauge invariance.

In any case, both Haag-Kastler and Wightman 
 axioms are useful as guiding principles for quantum
field theory.

\nowastrona

{\subsubsection{Charged fields}}


Sometimes, instead of Hermitian fields one uses
a pair of fields
 $\rr^{1,3}\ni
x\mapsto\hat\psi_a(x),\hat\psi_a^*(x)$, $a=1,\dots,m$. We will call them {\em charged fields}. 
One assumes that after smearing with complex test functions
 \begin{eqnarray*}
\hat\psi[h]&:=&\sum_a\int\bar{ h_a(x)}\hat\psi_a(x)\d x,\\
\hat\psi^*[h]&:=&\sum_a\int h_a(x)\hat\psi_a^*(x)\d x,\end{eqnarray*}
 one obtains  linear operators on $\cD$ Hermitian cojugate to one another.
\nowastrona

One can  organize species
 of charged fields into  a
 {\em complex}
space  $\cW=\cc^m$.

Clearly, for any charged field $\psi_a$, by setting
\begin{eqnarray*}
\hat\phi_{a,\R}(x):=\frac{1}{\sqrt2}(\hat\psi_a(x)+\hat\psi_a^*(x)),\\
\hat\phi_{a,\I}(x):=\frac{1}{\i\sqrt2}(\hat\psi_a(x)-\hat\psi_a^*(x))\end{eqnarray*}
 we obtain a pair of neutral fields.
Thus introducing charged fields to the Wightman axioms is essentially only a notational change, which, as we will see, is convenient for describing  $U(1)$ symmetries.

\subsubsection{Wightman axioms for neutral and charged fields}

The modified Wightman axioms that admits  both neutral and charged fields are very similar to the Wightman axioms for neutral fields described
in Subsubsect.
\ref{Wightman axioms for neutral fields}.
It would be boring to state them in full detail. In fact, almost all statements 
from the Wightman axioms for neutral fields remain a part of the new axioms.
 The only exception is Axiom (4) about the cyclicity of the vacuum, which needs to be replaced by a new one.
Below we will  list the additional elements that need to be added. We indicate by (...)  the places where  appropriate statements from Subsubsect.
\ref{Wightman axioms for neutral fields}
should be inserted. 

\bed
(...) We assume that $\cW$ is a finite dimensional complex  vector
space
 equipped with a representation
\beq 
Spin^{\uparrow}(1,3)\ni\tilde\Lambda\mapsto \tau (\tilde\Lambda)\in L(\cW).\label{REPI+}\eeq
We have a unique decomposition $\cW=\cW_0\oplus\cW_1$.
where $\cW_0$, resp. $\cW_1$ is  the space of integer spin,
resp. half-integer spin.

\nowastrona

(...) We have  maps 
\beq \cS(\rr^{1,3},\cW)\ni h\mapsto \hat\psi[h],\hat\psi^*[h]\in L(\cD).\label{line+}\eeq

\ben
\item {\em Continuity}: (...)
\beq \cS(\rr^{1,3},\cW)
\ni h\mapsto (\Phi|\hat\psi[h]\Psi)
\label{line1+}\eeq
is continuous.
\item
{\em Poincar\'e{} covariance}: (...)
\[\cU_{(y,\tilde\Lambda)}\big(\hat\psi[h]\big)
=\hat\psi\left[\tau (\tilde\Lambda)h\circ(y,\Lambda)^{-1}\right].\]
\nowastrona
\item {\em Einstein causality}: (...) 
 Let  $\supp h_1\times\supp h_2=\emptyset$, where $h_i$ have values in $\cW_{j_i}$,
$i=1,2$. Then (...) \begin{eqnarray*}\hat\phi[f_1]\hat\psi[h_2]&=&(-1)^{j_1j_2}\hat\psi[h_2]\hat\phi[f_1],\\
\hat\psi[h_1]\hat\psi[h_2]&=&(-1)^{j_1j_2}\hat\psi[h_2]\hat\psi[h_1],\\
\hat\psi[h_1]\hat\psi^*[h_2]&=&(-1)^{j_1j_2}\hat\psi^*[h_2]\hat\psi[h_1].
\end{eqnarray*}
\item 
{\em Cyclicity of the vacuum}: Let $\fF^\alg$ denote the algebra of polynomials in $\hat \phi[f]$,  $\hat \psi[h]$ and  $\hat \psi^*[h]$. 
Then $\fF^\alg\Omega$
 is dense in $\cH$.
\item {\em Hermiticity}: (...)
\[(\Phi|\hat\psi[h]\Psi)=(\hat\psi^*[ h]\Phi|\Psi).\]  
\een
\eed

It will be convenient to reformulate the  axiom about the Poincar\'e{} invariance in terms of the unsmeared fields:
\begin{eqnarray}
\cU_{(y,\tilde\Lambda)}\big(\hat\phi_a(x)\big)&=&
\sum_b\sigma _{ab}^{{-1}}(\tilde\Lambda)\hat\phi_b(\Lambda x+y),\label{sym1a}
\\\cU_{(y,\tilde\Lambda)}\big(\hat\psi_a(x)\big)&=&
\sum_b\tau_{ab}^{{-1}}(\tilde\Lambda)\hat\psi_b(\Lambda x+y).\label{sym2a}
\end{eqnarray}

\nowastrona

\subsubsection{$U(1)$ symmetry}
\label{$U(1)$ symmetry and charge conjugation}

Consider the group $U(1)=\rr/2\pi\zz$. A global $U(1)$ symmetry is usually encoded by dividing fields into neutral $\hat\phi$ and complex $\hat\psi$. Let $\cH_n$ be the closed span of vectors of the form
\[\hat\phi[f_1]\cdots\hat\phi[f_k]\hat\psi^*[h_1]\cdots\hat\phi^*[h_p]\hat\psi[h_1']\cdots\hat\phi[h_q']\Omega,\ \   \ \ \ n=p-q.\]
Note that the cyclicity of vacuum implies that the sum of $\cH_n$ is dense in $\cH$. Assume that $\cH_n$ are mutally orthogonal, so that we have the decomposition $\cH=\loplus_{n\in\zz}\cH_n$. For $\theta\in U(1)$ we define $R(\theta):=\loplus_{n\in\zz}\e^{\i n\theta}$. Clerly, $R(\theta)\Omega=\Omega$ and $U(1)\ni\theta\mapsto R(\theta)$ is a unitary representation commuting with $U(y,\tilde\Lambda)$. Let $\cR_\theta$ be the corresponding $*$-automorphism:
\[\cR_\theta(A)=R(\theta)AR(-\theta).\]
 We then have
\begin{eqnarray*}
\cR_\theta\big(\hat\phi_a(x)\big)&=&\hat\phi_a(x),\\
\cR_\theta\big(\hat\psi_a(x)\big)&=&\e^{-\i\theta}\hat\psi_a(x),\\
\cR_\theta\big(\hat\psi_a^*(x)\big)&=&\e^{\i\theta}\hat\psi_a^*(x).
\end{eqnarray*}
Thus we have an example of a global symmetry, as in Subsect. \ref{Global-Wight}.

\subsubsection{Charge conjugation}

Let $C$ be a unitary operator such that $C\Omega=\Omega$. Let  $\cC$ be the corresponding $*$-automorphism:
\[\cC(A):=C AC^{-1}.\]
We say that it  is a {\em charge conjugation} if it satisfies
\begin{eqnarray}
\cC\big(\hat\phi_a(x)\big)&=&\sum_b\alpha_{ab}^{{-1}}\hat\phi_b(x),\label{ssym2}\\
\cC\big(\hat\psi_a(x)\big)&=&\sum_b\kappa_{ab}^{{-1}}\hat\psi_b^*( x)
,\label{ssym3}\end{eqnarray}
and hence
\begin{eqnarray}\cC\big(\hat\psi_a^*(x)\big)&=&\sum_b\bar\kappa_{ab}^{{-1}}\hat\psi_b( x)
,\label{ssym3a}\end{eqnarray}
where $\alpha$ and $\kappa$ are some matrices on $\cV$ and $\cW$. We will  also assume that
\beq\alpha^4=\one,\ \ 
(\kappa\bar\kappa)^2=\one,\label{ssym4}\eeq
so that $C^4=\one$.

We have
\[\cC\cR_\theta=\cR_{-\theta}\cC,
\]
which is the reason for  the name charge conjugation.

Note that $C$ is linear, even though $\cC$  acts on fields antilinearly.

\subsubsection{Parity invariance}

 Recall that  the Wightman  axioms involve the connected Lorentz group 
$Spin^\uparrow(1,3)$. In particular, we have  representations 
\begin{eqnarray}Spin^\uparrow(1,3)\ni\tilde\Lambda&\mapsto&\sigma(\tilde\Lambda)\in L(\cV),\label{uniti1}\\
Spin^\uparrow(1,3)\ni\tilde\Lambda&\mapsto&\tau(\tilde\Lambda)\in L(\cW),\\
\label{uniti2}
\rr^{1,3}\rtimes Spin^\uparrow(1,3)\ni (y,\tilde\Lambda)&\mapsto&
U(y,\tilde\Lambda)\in U(\cH).\label{uniti3}\end{eqnarray}

Chose $+$ or $-$. Replace the group $Spin^\uparrow(1,3)$ in the Wightman axioms  by
 $Pin_\pm^\uparrow(1,3)$, so that we have the representations
\begin{eqnarray}Pin_\pm^\uparrow(1,3)\ni\tilde\Lambda&\mapsto&\sigma(\tilde\Lambda)\in L(\cV),\label{uniti1p}\\
Pin_\pm^\uparrow(1,3)\ni\tilde\Lambda&\mapsto&\tau(\tilde\Lambda)\in L(\cW),\\
\label{uniti2p}
\rr^{1,3}\rtimes Pin_\pm^\uparrow(1,3)\ni (y,\tilde\Lambda)&\mapsto&
U(y,\tilde\Lambda)\in U(\cH).\label{uniti3p}\end{eqnarray}
 The resulting set of axioms will be called the {\em Wightman axioms of a $P$-invariant theory}.

In particular,  the space inversion (parity) $\tilde\P\in Pin_\pm^\uparrow(1,3)$ is represented in the Hilbert space by the unitary operator $P:= U(\tilde\P)$. 
$\cP:=\cU_{\tilde\P}$
denotes the corresponding automorphism. It acts on the fields as follows:
\begin{eqnarray*}
\cP\big(\hat\phi_a(x^0,\vec x)\big)&=&\sum_b\sigma _{ab}^{{-1}}(\tilde\P)\hat\phi_b(x^0,-\vec x),\\
\cP\big(\hat\psi_a(x^0,\vec x)\big)&=&\sum_b\kappa_{ab}^{{-1}} (\tilde\P)\hat\psi_b(x^0,-\vec x)
.\end{eqnarray*}
We have $P^2=\one$ in the case $+$ and $P^4=\one$ in the case $-$.

Obviously, $P$ is linear and $\cP$
acts on fields linearly.

\subsubsection{Time reversal invariance}

Chose again $+$ or $-$. Let us replace the group $Spin^\uparrow(1,3)$ in the Wightman axioms by
 $Pin_\pm^\chir(1,3)$. We have now representations
\begin{eqnarray}Pin_\pm^\chir(1,3)\ni\tilde\Lambda&\mapsto&\sigma(\tilde\Lambda)\in L(\cV),\label{uniti1t}\\
Pin^\chir(1,3)\ni\tilde\Lambda&\mapsto&\tau(\tilde\Lambda)\in L_{\theta(\tilde\Lambda)}(\cW),
\label{uniti2t}\\
\rr^{1,3}\rtimes Spin^\uparrow(1,3)\ni (y,\tilde\Lambda)&\mapsto&
U(y,\tilde\Lambda)\in U_{\theta(\tilde\Lambda)}(\cH).\label{uniti3t}\end{eqnarray}
Note that we demand that (\ref{uniti2t}) is $\theta$-linear and (\ref{uniti3t}) is $\theta$-unitary. We denote by 
$[\tau_{ab}(\tilde\Lambda)]$ the matrix of $\tau$. For nonorthochronous $\tilde\Lambda$ we  demand that   (\ref{sym2a}) is replaced by
\begin{eqnarray*}
U(y,\tilde\Lambda)\hat\psi_a(x)U(y,\tilde\Lambda)^{-1}&=&
\sum_b\tau_{ab}^{{-1}}(\tilde\Lambda)\hat\psi_b^*(\Lambda x+y).
\end{eqnarray*}
The resulting set of axioms will be called the {\em Wightman axioms of a $T$-invariant theory}.

In particular, the time reversal is implemented by the anti-unitary operator $T:=U(\tilde\T)$. $\cT:=\cU_{\tilde\T}$
denotes the corresponding automorphism. The time reversal acts on the fields as follows
\begin{eqnarray*}
\cT\big(\hat\phi_a(x^0,\vec x)\big)&=&\sum_b\sigma _{ab}^{{-1}}(\tilde\T)\hat\phi_b(-x^0,\vec x),\\
\cT\big(\hat\psi_a(x^0,\vec x)\big)&=&\sum_b\tau_{ab}^{{-1}}(\tilde\T)\hat\psi_b^*(-x^0,\vec x)
.\end{eqnarray*}
We have $T^2=\one$ in the case $+$ and $T^4=\one$ in the case $-$.

Note that $T$ is antilinear and $\cT$ acts on fields antilinearly.

\nowastrona

\subsubsection{The CPT Theorem}

Suppose that we have a theory satisfying  the Wightman axioms (without the $P$ and $T$ invariance). As described in
Subsubsection 
\ref{Complex Lorentz groups}
the representations (\ref{uniti1}) and (\ref{uniti2}) possess  natural extensions
\begin{eqnarray}Spin(1,3)\ni\tilde\Lambda&\mapsto&\sigma(\tilde\Lambda)\in L(\cV),\label{uniti1x}\\
Spin(1,3)\ni\tilde\Lambda&\mapsto&\tau(\tilde\Lambda)\in L(\cW).\label{uniti2x}
\end{eqnarray}
Let us stress that (\ref{uniti2}) is   linear  and not $\theta$-linear!
A deep theorem, called the {\em $CPT$ Theorem}, says that we can extend the representation (\ref{uniti3}) to a $\theta$-unitary representation 
\begin{eqnarray}\rr^{1,3}\rtimes Spin(1,3)\ni (y,\tilde\Lambda)\mapsto
U(y,\tilde\Lambda)\in U_{\theta(\tilde\Lambda)}(\cH),\label{uniti3x}\end{eqnarray}
such that
 (\ref{sym1a}) and (\ref{sym2a}) hold on the whole $Spin(1,3)$. 

In particular, the spacetime inversion is implemented by the anti-unitary operator $X:=U(\tilde \X)$. $\cX:=\cU_{\tilde\X}$
 denotes the corresponding automorphism. Then
\begin{eqnarray*}
\cX\big(\hat\phi_a( x)\big)&=&\sum_b\sigma _{ab}^{{-1}}(\tilde \X)\hat\phi_b(- x),\\
\cX\big(\hat\psi_a( x)\big)&=&\sum_b\tau_{ab}^{{-1}}(\tilde \X)\hat\psi_b(- x)
.\end{eqnarray*}

Note that $X$ is antilinear but $\cX$ acts on the fields linearly.

\subsubsection{The CPT Theorem in a $P$ and $T$-invariant theory}

Let $\epsilon,\delta\in\{+,-\}$. Suppose we have a theory that satisfies the Wigthman axioms with the $P$ invariance described by the group
 $Pin_\epsilon^\uparrow(1,3)$ and the $T$ invariance described by the group  $Pin_\delta^\chir(1,3)$.
By the CPT Theorem the theory is also $Spin(1,3)$ invariant. 
In particular, we have the matrices
$\sigma (\tilde\X)$, $\sigma (\tilde\T)$, $\sigma (\tilde\P)$,
$\tau (\tilde\X)$, $\tau (\tilde\T)$ and $\tau (\tilde\P)$,
We also have the operators $X$, $T$ and $P$. Define 
\begin{eqnarray*}
\alpha&:=&\sigma(\tilde\X)\sigma(\tilde\T^{-1})\sigma (\tilde\P^{-1}),\\
\kappa&:=&\tau(\tilde\X)\bar{\tau(\tilde\T^{-1})}\tau(\tilde\P^{-1}),\\
C&:=&X T^{-1}P^{-1}.
\end{eqnarray*}
Then  $C$ is unitary and the corresponding automorphism
$\cC(A)=CAC^{-1}$ satisfies (\ref{ssym2}) and (\ref{ssym3}).

The groups  $Pin_\epsilon^\uparrow(1,3)$,  $Pin_\delta^\chir(1,3)$ and
 $Spin(1,3)$ can be treated as subgroups of 
$Pin_\ext(1,3)$.  Let $G$ denote the group generated by these three groups together. (Clearly, there are three possibilities: $G$ is $Pin_\ext(1,3)$,
 $Pin_-(1,3)$ or  $Pin_+(1,3)$).
Assume that there exists representations
\begin{eqnarray*}G\ni\tilde\Lambda&\mapsto&\sigma(\tilde\Lambda)\in L(\cV),\\
G\ni\tilde\Lambda&\mapsto&\tau(\tilde\Lambda)\in L_1(\cW)\cup L_{-1}(\cW),
\end{eqnarray*} 
that extends  the representations (\ref{uniti1p}) and  (\ref{uniti2p}) of
 $Pin_\epsilon^\uparrow(1,3)$,  (\ref{uniti1t}) and  (\ref{uniti2t}) of $Pin_\delta^\chir(1,3)$ and (\ref{uniti1x}) and  (\ref{uniti2x}) of
 $Spin(1,3)$. (If such a representation exists, it is clearly unique).

Now $\X\T^{-1}\P^{-1}\in O(1,3)$ is the identity. Therefore, 
$\tilde\X\tilde\T^{-1}\tilde\P^{-1}\in G$  is one of the four elements of $Pin_\ext(1,3)$ covering the identity. All of them raised to the fourth power are the identity. Therefore, the matrices $\alpha$, $\kappa$ satisfy
\[\alpha^4=\one,\ \ 
(\kappa\bar\kappa)^2=\one,\]
 which is the condition (\ref{ssym4}).
 Thus $C$ is an example of a charge conjugation according to the definition of Subsubsect. \ref{$U(1)$ symmetry and charge conjugation}.

 Obviously,  $X=CPT$. This is explains the name of the CPT Theorem. (Let us stress, however,  that the theorem holds also if the theory is not $P$ and $T$ invariant, so that we cannot write $X=CPT$).

\subsubsection{$N$-point Wightman and Green's functions}

For simplicity, in this subsubsection we use  Wightman axioms for neutral fields. They  allow us to define a multilinear map
\begin{eqnarray}\nonumber
 \cS(\rr^{1,3},\cV)\times\cdots\times \cS(\rr^{1,3},\cV)&&\\
\ni (f_N,\dots,f_1)&\mapsto&
(\Omega|\hat\phi[f_N]\cdots\hat\phi[f_1]\Omega)\in\cc,\label{poap}
\end{eqnarray} which is separately
continuous in its arguments.
 By the {\em Schwartz Kernel Theorem} \cite{Gel,RS1},
 (\ref{poap}) can be extended to a linear map
\[\cS\left((\rr^{1,3})^N,\cV^{\otimes N}\right) \ni F\mapsto
\int W(x_N,\dots,x_1)F(x_N,\dots,x_1)\d x_N\cdots\d x_1,
\] where $\rr^{(1,3)N}\ni(x_N,\dots,x_1)\mapsto
W(x_N,\dots,x_1)$ is a tempered distribution on 
$\rr^{(1,3)N}$ with values in the space dual to $\cV^{\otimes N}$, called the
{\em $N$-point Wightman function}, so that (\ref{poap}) equals
\[\int W(x_N,\dots,x_1)f_N(x_1)\cdots f_1(x_1)\d x_N\dots\d x_1.\]

\nowastrona

 From the point of view of the Wightman axioms,
the collection of Wightman functions $W_N$, $N=0,1,\dots$, 
contains all the information about a given
quantum field theory.
In particular,
\begin{eqnarray*}&&\left(\hat\phi[f_N]\cdots \hat\phi[f_1]\Omega|
\hat\phi[g_M]\cdots\hat\phi[g_1]\Omega\right)\\
&=&\int W(y_1,\dots,y_N,x_M,\dots,x_1)\\&&\times f_1(x_1)\cdots f_N(x_N)
 g_M(y_M)\cdots g_1(y_1)\d x_1\cdots \d x_N\d y_M\cdots \d y_1.
\end{eqnarray*}
This is expressed in the so called {\em Wightman Reconstruction
  Theorem} \cite{StW}.


\nowastrona

In practical computations Wightman functions are not often used. Much more frequent are  the so-called  {\em (time-ordered) 
Green's functions}. Their
formal definition is as follows:
\begin{eqnarray}\label{gree} &&\langle\hat\phi(x_N)\cdots\hat\phi(x_1)\rangle
\\&:=&\sum_{\sigma\in S_N}\sgn_\epsilon(\sigma)
\theta\left(x_{\sigma(N)}^0-x_{\sigma(N-1)}^0\right)
\cdots \theta\left(x_{\sigma(2)}^0-x_{\sigma(1)}^0\right)W(x_{\sigma(N)},\dots,x_{\sigma(1)})
,\nonumber\end{eqnarray}
where $\sgn_\a(\sigma)$ is the sign of the permutation
of the fermionic elements among $N,\dots,1$.

Note that
 we multiply a distribution with a discontinuous function  in (\ref{gree}), which strictly
 speaking is illegal. Disregarding this problem,  Green's functions are
 covariant 
 due to the
commutativity/anticommutativity of fields at spacelike separations.

\subsection{General scattering theory}

\subsubsection{Time ordered exponential}
\label{Time ordered exponential}
We will often use the formalism of time-dependent Hamiltonians. In
this subsection we describe the main concepts of this formalism.

Assume that $I$ is a unitary involution.
 (In applications, $I$ will be the
fermionic parity operator).
 We call an operator $B$
{\em even}, resp. {\em odd}, if $B=\pm IBI$. Such operators will be called {\em of pure parity}.

Let $t\mapsto B_k(t),\dots,B_1(t)$ be time dependent
 operators of pure parity. 
 Let $t_n,\dots, t_1$ be pairwise distinct. We define the {\em
time-ordered product  of $B_n(t_n),$..., $B_i(t_1)$} by 
\[\T\left(B_n(t_n)\cdots B_1(t_1)\right):=
\sgn_\a(\sigma) B_{\sigma_n}(t_{\sigma_n})\cdots B_{\sigma_1}(t_{\sigma_1}),\]
where $(\sigma_1,\dots,\sigma_n)$ is the permutation such that 
 $t_{\sigma_n}\geq\cdots \geq t_{\sigma_1}$ and $\sgn_\a(\sigma)$ is the sign of
this permutation restricted to the odd elements among $B_n,\dots,B_1$.

Consider a  family of self-adjoint operators
\beq t \mapsto H(t).\label{schr}\eeq We will assume that $H(t)$ are even.
(\ref{schr}) will be called the  {\em  the Schr\"odinger picture  Hamiltonian}. 
For $t_+>t_-$, we define the {\em  time-ordered exponential}
\begin{eqnarray}\label{dynam}
&&\Texp\left(-\i\int_{t_-}^{t_+} H(t)\d t\right)\\&:=&
\sum_{n=0}^\infty(-\i)^n\mathop{\int\cdots\int}\limits_{t_+\geq t_n\geq\cdots\geq
  t_1\geq t_-}H(t_n)\cdots H(t_1)\d
t_n\cdots\d t_1\notag\\
&\!\!\!=&
\!\!\!\!\!\sum_{n=0}^\infty(-\i)^n\int_{t_-}^{t_+}\cdots\int_{t_-}^{t_+}\frac{1}{n!}\T\left(H(t_n)\cdots H(t_1)\right)\d
t_n\cdots\d t_1.\notag\end{eqnarray}
For brevity, we will write 
$ U(t_+,t_-)$ for (\ref{dynam}) and call it the {\em dynamics generated by $t\mapsto H(t)$.}
(Of course, if $H(t)$ are unbounded, the above definition should be
viewed only as a heuristic indication how to define the family of  unitary
operators $U(t_+,t_-)$. In general, in  most of this subsection we are not very precise about the boundedness of operators, limits, etc.)

 We also set $U(t_-,t_+):=U(t_+,t_-)^{-1}$.

Clearly, if $H(t)=H$, then $U(t_+,t_-)=\e^{-\i(t_+-t_-)H}$.
\nowastrona

\subsubsection{Heisenberg picture}
\label{Heisenberg picture}

Let $A$ be an operator.
Its {\em evolution in the  Heisenberg picture} is
\beq A(t):=U(0,t)A U(t,0).\label{heisen}\eeq
Equivalently,  $A(t)$ is the solution of
\begin{eqnarray}
\frac{\d}{\d t}A(t)&=&\i\left[H_{\HP}(t),A(t)\right],\\
A(0)&=&A,\label{heisen6}
\end{eqnarray}
where the {\em  Hamiltonian in the Heisenberg picture}
 is defined as
\beq t\mapsto H_{\HP}(t):=U(0,t)H(t)U(t,0).\label{heisen1}\eeq
Thus a quantum dynamics can be described by two time-dependent Hamiltonians:
(\ref{schr}) and (\ref{heisen1}). If they do not depend on time, they coincide.

Let us note that a similar distinction exists in classical dynamical
systems.
Consider a flow on $\rr^d$ given by the equation
\[\frac{\d}{\d t}x(t)=v\big(t,x(t)\big).\]
For any initial condition $x(0)=x_0\in\rr^d$, we obtain a solution $\rr\ni
t\mapsto x(t,x_0)$. Thus any time dependent observable $F$ has two descriptions:
\begin{eqnarray}
\rr\times\rr^d\ni(t,x)&\mapsto &F(t,x),\label{euler}\\
\rr\times\rr^d\ni(t,x_0)&\mapsto &F\big(t,x(t,x_0)\big).\label{lagrange}
\end{eqnarray}
In fluid dynamics,
 (\ref{euler}) is sometimes called the {\em Eulerian}  description,
and (\ref{lagrange})  the {\em Lagrangian } description.

In classical mechanics the phase space  is described by coordinates $(\phi,\pi)
\in\rr^m\times\rr^m$. 
The time evolution  is described by
the Hamilton equations
\begin{eqnarray*}
\dot\phi(t)&=&
\partial_\pi H\big(t,\phi(t),\pi(t)\big),
\\\dot\pi(t)
&=&-\partial_\phi H\big(t,\phi(t),\pi(t)\big).
\end{eqnarray*}
For any initial condition $\big(\phi_0,\pi_0\big)\in\rr^m\times\rr^m$, we obtain a solution of the Hamilton equations. Similarly as in the quantum case,  we have two kinds of the classical Hamiltonian:
\begin{eqnarray}
\rr\times\rr^m\times\rr^m\ni(t,\phi,\pi)&\mapsto &H(t,\phi,\pi),\label{euler1}\\
\rr\times\rr^m\times\rr^m\ni(t,\phi_0,\pi_0)&\mapsto &H\big(t,\phi(t,\phi_0,\pi_0),\pi(t,\phi_0,\pi_0)\big).\label{lagrange1}
\end{eqnarray}
(\ref{euler1}) is the Hamiltonian in the Eulerian description and  (\ref{lagrange1})  is the Hamiltonian in the Lagrangian description. 
The former is the analog of the Schr\"odinger picture and the latter of the Heisenberg picture. If they do not depend on time, they coincide.

We will use the classical  Hamiltonian in the Lagrangian description and the quantum Hamiltonian in the Schr\"odinger picture as the standard ones.

\subsubsection{Time-dependent perturbations}
\label{Time-dependent perturbations}

Our time-dependent Hamiltonians will usually have the form
\[H(t):=H_\fr+ \lambda V(t),\] where
 $H_\fr$ is a self-adjoint operator and
$\rr\ni t\mapsto V(t)$
is a family of self-adjoint operators. 
We define the {\em evolution in the interaction picture} and
the {\em interaction picture Hamiltonian}:
\begin{eqnarray*}U_\Int(t_+,t_-)&:=&\e^{\i t_+H_\fr}U(t_+,t_-)\e^{-\i
    t_-H_\fr},\\H_\Int(t)&:=&\lambda\e^{\i tH_\fr}V(t)\e^{-\i tH_\fr}.\end{eqnarray*}
Note that
\begin{eqnarray*}U_\Int(t_+,t_-)
&=&\Texp\left(-\i\coupl\int_{t_-}^{t_+}H_\Int(t)\d t\right).\end{eqnarray*}
We define the {\em scattering operator} by
\begin{eqnarray}
S&:=&\lim_{t_+,-t_-\to\infty} U_\Int(t_+,t_-)\notag\\
&=&\Texp\left(-\i\int_{-\infty}^{\infty}H_\Int(t)\d t\right).\label{scatte1}\end{eqnarray}
We also introduce the {\em M{\o}ller operators}
\begin{eqnarray}
S^-&:=&\lim_{t\to\infty} U(0,-t)\e^{\i tH_\fr }=\lim_{t\to\infty}U_\Int(0,-t)\notag\\
&=&\Texp\left(-\i\int_{-\infty}^{0}H_\Int(t)\d t\right),\label{scatte2}\\
S^+&:=&\lim_{t\to\infty} U(0,t)\e^{-\i tH_\fr }=\lim_{t\to\infty} U_\Int(0,t)
\notag\\
&=&\Texp\left(-\i\int_{0}^{\infty}H_\Int(t)\d t\right)^*.\label{scatte3}
\end{eqnarray}
Clearly, $S=S^{+*}S^-$.

\nowastrona

For any operator $A$ we distinguish now two Heisenberg pictures -- wrt. the {\em full dynamics} (\ref{heisen}) and wrt. the {\em free dynamics}:
\beq A_\fr (t):=\e^{\i tH_\fr }A\e^{-\i tH_\fr }.\label{hei1}\eeq
Equivalently,  $A_\fr(t)$ is the solution of
\begin{eqnarray*}
\frac{\d}{\d t}A_\fr(t)&=&\i\left[H_\fr,A_\fr(t)\right],\\
A_\fr(0)&=&A,
\end{eqnarray*}

\subsubsection{Time ordered Green's functions}
\label{Time ordered Green's functions}

Assume that $H_\fr $ and $V(t)$
are even.
Let $\Phi_\fr $ be a fixed even vector with $H_\fr \Phi_\fr =0$, which we will call the {\em vacuum}.
(In our applications, $\Phi_\fr$ will be always the ground state of $H_\fr$.)
Let $A_k,\dots, A_1 $ be operators of fixed parity.
The {\em free time-ordered Green's functions} are defined as
\begin{eqnarray}\nonumber
&& \langle A_{k,\fr}(t_k)\cdots A_{1,\fr}(t_1)\rangle_\fr\\
&:=&\left(\Phi_\fr |\T\bigl(A_{k,\fr}(t_k)\cdots
  A_{1,\fr}(t_1)\bigr)\Phi_\fr \right),\label{las-0} 
\end{eqnarray}
where $A_{i,\fr}(t)$ are $A_i$ in the Heisenberg picture for the free dynamics (\ref{hei1}).

Suppose that there exist
\beq \Phi^\pm:
=\lim_{t\to\pm\infty}U(0,t)\Phi_\fr .\label{pembroke}\eeq
 The {\em interacting  time-ordered Green's functions} are defined as
\begin{eqnarray}\nonumber
&&\langle  A_k(t_k)\cdots A_1(t_1)\rangle\\
&:=&\left(\Phi^+|\T\bigl(A_{k}(t_k)\cdots
  A_{1}(t_1)\bigr)\Phi^-\right),\label{las} 
\end{eqnarray}
where $A_i(t)$ are  $A_i$ in the Heisenberg picture for the full dynamics (\ref{heisen}).

Let $t\mapsto f(t)=(f_1(t),\dots,f_k(t))$ be an $n$tuple of  functions. If $A_i$ is even, we simply assume that $f_i(t)$ has real values, if $A_i$ is odd, the values of $f_i(t)$ are (anticommuting) Grassmann numbers.
The {\em generating function} is defined as
\begin{eqnarray*}\label{gener}
Z(f)&=&\sum_{N=0}^\infty\frac{(-\i)^N}{N!}\sum_{i_1}\int f(t_1)\d t_1
\cdots\sum_{i_N}\int f(t_N)\d t_N\langle A_{i_N}(t_N)\cdots A_{i_1}(t_1)\rangle.
\notag\end{eqnarray*}
Note that the generating function  is  the vacuum expectation value of a certain scattering operator:
\beq Z(f)=(\Phi_\fr |S(f)\Phi_\fr),\eeq
where $S(f)$ is the scattering operator (\ref{scatte1}) with $\lambda V(t)$ replaced by \[\lambda V(t)+\sum\limits_{i}f_i(t)A_i.\]

\nowastrona
 We can
express  interacting Green's functions by the free ones:
\begin{eqnarray}\label{gell-}
\langle A_k(t_k)\cdots A_1(t_1)\rangle
&=&\sum_{n=0}^\infty\frac{(-\i\lambda)^n}{n!}\int_{-\infty}^\infty\d
  s_n\cdots\int_{-\infty}^\infty\d s_1 \\
&&\times \langle
V_\fr(s_n,s_n)\cdots V_\fr(s_1,s_1)A_{k,\fr}(t_k)\cdots A_{1,\fr}(t_1)\rangle_\fr,\nonumber
\end{eqnarray}
where $V_\fr(s,t):=\e^{\i tH_\fr}V(s)\e^{-\i tH_\fr}.$

 \subsubsection{Adiabatic switching and the energy shift}

In most of our notes we concentrate on time dependent perturbations that decay sufficiently fast in the past and future. Such perturbations lead to a relatively simple scattering theory, described in Subsubsection 
\ref{Time-dependent perturbations} and \ref{Time ordered Green's functions}.

One would like also to consider the case of time-independent perturbations. In fact, let $V$ be a (time-independent) self-adjoint perturbation.
In this and the following two subsubsections we are interested in the time-independent Hamiltonian $H:=H_\fr +\lambda V$. 

 It is often convenient to extract information about $H$ from the time-dependent formalism.
This can be done by introducing the so called {\em adiabatical switching} invented by Gell-Mann and Low.

In this subsubsection we describe how to compute the energy shift using adiabatic switching. We  follow quite closely  \cite{Mo}.

Let $\epsilon>0$. We define
$V_{\epsilon}(t):=\e^{-\epsilon |t|}V$. We will write
\[H_\epsilon(t):=H_\fr +\lambda V_\epsilon\]
 for the corresponding time-dependent
Hamiltonian. We also introduce the corresponding $H_{\epsilon\Int}$, 
 $U_\epsilon(t_+,t_-)$, $U_{\epsilon\Int}(t_+,t_-)$,
$S_\epsilon^\pm$, 
$S_\epsilon$.

\bep We have
\begin{eqnarray}
&&\i\epsilon\lambda\partial_\lambda U_\epsilon(t_+,t_-)\label{toz1}\\&=&
\begin{cases}
H_\epsilon(t_+)U_\epsilon(t_+,t_-)-U_\epsilon(t_+,t_-)H_\epsilon(t_-),&
0\geq t_+\geq t_-;\\
-H_\epsilon(t_+)U_\epsilon(t_+,t_-)+U_\epsilon(t_+,t_-)H_\epsilon(t_-),&
t_+\geq t_-\geq0;\end{cases}\notag\\[3ex]&&
\i\epsilon\lambda\partial_\lambda U_{\epsilon\Int}(t_+,t_-)\label{toz2}\\
&=&
\begin{cases}
H_{\epsilon\Int}(t_+)U_{\epsilon\Int}(t_+,t_-)-U_{\epsilon\Int}(t_+,t_-)H_{\epsilon\Int}
(t_-),&
0\geq t_+\geq t_-;\notag\\
-H_{\epsilon\Int}(t_+)U_{\epsilon\Int}(t_+,t_-)+U_{\epsilon\Int}(t_+,t_-)
H_{\epsilon\Int}(t_-),&
t_+\geq t_-\geq0;\end{cases}\end{eqnarray}
\begin{eqnarray}
\pm\i\epsilon\lambda\partial_\lambda S_\epsilon^\pm&=& H S_\epsilon^\pm-
 S_\epsilon^\pm H_\fr;\label{toz3}\\
\i\epsilon\lambda\partial_\lambda S_\epsilon&=&H_\fr S_\epsilon+
 S_\epsilon H_\fr-2S_\epsilon^{+*}HS_\epsilon^-
.\label{toz4}
\end{eqnarray}
\eep
\nowastrona

\proof Display the dependence on $\lambda$ by writing
$U_{\epsilon,\lambda}(t_+,t_-)$. For $t_+\geq t_-\geq0$ we have
\[U_{\epsilon,\lambda}(t_+,t_-)=U_{\epsilon,\lambda\e^{\theta\epsilon}}
(t_++\theta,t_-+\theta).\]
Hence,
\begin{eqnarray*}
0&=&
\frac{\d}{\d\theta}U_{\epsilon,\lambda\e^{\theta\epsilon}}
(t_++\theta,t_-+\theta)\Big|_{\theta=0}\\
&=&\epsilon\lambda\partial_\lambda U_{\epsilon,\lambda}(t_+,t_-)
+\frac{\d}{\d t_+}U_{\epsilon,\lambda}(t_+,t_-)+
\frac{\d}{\d t_-}U_{\epsilon,\lambda}(t_+,t_-)\\
&=&\epsilon\lambda\partial_\lambda U_{\epsilon,\lambda}(t_+,t_-)
-\i H_{\epsilon,\lambda}(t_+)U_{\epsilon,\lambda}(t_+,t_-)+
\i U_{\epsilon,\lambda}(t_+,t_-)H_{\epsilon,\lambda}(t_-).
\end{eqnarray*} This proves (\ref{toz1}), from which the remaining identities
follow. 
\qed

\nowastrona

 Assume that $\Phi_\fr$ is an eigenvector of $H_\fr$ with
$H_\fr\Phi_\fr=E_\fr\Phi_\fr$. 
Set
\begin{eqnarray*}
\Phi_\epsilon^\pm&:=&\frac{S_\epsilon^\pm\Phi_\fr}{(\Phi_\fr|
S_\epsilon^\pm\Phi_\fr)},\\
E_\epsilon^\pm&:=&\frac{(\Phi_\fr|HS_\epsilon^\pm\Phi_\fr)}
{(\Phi_\fr|S_\epsilon^\pm\Phi_\fr)}
=(\Phi_\fr|H\Phi_\epsilon^\pm)
.\end{eqnarray*}

\bep
\begin{eqnarray}\label{pq1}
(H-E_\fr\pm\i\epsilon\lambda\partial_\lambda)S_\epsilon^\pm\Phi_\fr&=&0,\\
\label{pq2}
\mp\i\epsilon\lambda\partial_\lambda\log(\Phi_\fr|S_\epsilon^\pm\Phi_\fr)
&=&E_\epsilon^\pm
-E_\fr,\\
\left(H-E_\epsilon^\pm\pm\i\epsilon\lambda\partial_\lambda\right)
\Phi_\epsilon^\pm&=&0.\label{pq3}
\end{eqnarray}
\eep
\nowastrona

\proof Applying (\ref{toz3}) to $\Phi_\fr$ yields (\ref{pq1}). We scalar multiply it  with
$\Phi_\fr$ obtaining (\ref{pq2}). Combining 
 (\ref{pq1}) with (\ref{pq2}) gives  (\ref{pq3}). \qed


In the following theorem we argue that the 
adiabatic switching often  allows us to compute an eigenvector of $H$ and its
eigenvalue.

\bet
\ben\item Assume that\beq
\hbox{ there exist a nonzero }\ \  
\lim\limits_{\epsilon\searrow0}\Phi_\epsilon^\pm.\label{gell-1}\eeq Then there exist
\begin{eqnarray*}
E_\GL^\pm&:=&\lim\limits_{\epsilon\searrow0}
E_\epsilon^\pm,\\
\Phi_\GL^\pm&:=&
\lim\limits_{\epsilon\searrow0}|(\Phi_\fr|S_\epsilon^\pm\Phi_\fr)|
\Phi_\epsilon^\pm.\end{eqnarray*}
\item Supose in addition that
\beq \lim\limits_{\epsilon\searrow0}\epsilon
\lambda\partial_\lambda \Phi_\epsilon^\pm=0.\label{gell-2}\eeq
Then \[H\Phi_\GL^\pm=E_\GL^\pm\Phi_\GL^\pm.\]
\een\label{gell-mann}\eet

\nowastrona
\proof 
(1) The existence of $E_\GL^\pm$ is immediate. Next we note that 
$(\Phi_\epsilon^\pm|\Phi_\epsilon^\pm)=|(\Phi_\fr|S_\epsilon^\pm\Phi_\fr)|^{-2}$. Hence
$\lim\limits_{\epsilon\searrow0}|(\Phi_\fr|S_\epsilon^\pm\Phi_\fr)|$ exists. 
 This implies
the existence  of $\Phi_\GL^\pm$. 

By (\ref{pq3}), we have
\[(H-E_\GL^\pm)\lim\limits_{\epsilon\searrow0}
\Phi_\epsilon^\pm=0.\] 
This implies (2). \qed

In the remaining part of this subsubsection we assume that (\ref{gell-1}) and (\ref{gell-2}) are true. 

Suppose now that
 for
small enough $\lambda_0>0$ and $|\lambda|<\lambda_0$, $H_\lambda$ has a unique
nondegenerate eigenvalue close to $E_\fr$, which we denote $E_\lambda$, depending continuously on $\lambda$, with $E_0=E_\fr$.
 If $E_{\GL,\lambda}^\pm$ also depends continuously on $\lambda$, then we
see that $E_{\GL,\lambda}^+=E_{\GL,\lambda}^-=E_\lambda$
 and $\Phi_{\GL,\lambda}^+$ is proportional to $\Phi_{\GL,\lambda}^-$. Note that both
 $\Phi_{\GL,\lambda}^+$ and $\Phi_{\GL,\lambda}^-$ are normalized, hence they may differ only
 by a phase
 factor.



\nowastrona

In what follows we simply assume that
\beq  E:=E_\GL^+=E_\GL^-,\ \ 
\Phi:=\Phi_\GL^+=\e^{\i\alpha}\Phi_\GL^-.\label{gell}\eeq
 is true (even if the argument given above does not apply).

\bel \label{bele}Let $B$ be an operator. Then
\beq (\Phi | B\Phi)=
\lim_{\epsilon\searrow0}
\frac{(S_\epsilon^+\Phi_\fr|BS_\epsilon^-\Phi_\fr)}{(\Phi_\fr|S_\epsilon\Phi_\fr)}.
\label{ist}\eeq\eel
\proof
The right hand side of (\ref{ist}) equals
\begin{eqnarray*}
\lim_{\epsilon\searrow0}
\frac{(S_{\epsilon}^+\Phi_\fr|BS_{\epsilon}^-\Phi_\fr)}
{(S_{\epsilon}^+\Phi_\fr|S_{\epsilon}^-\Phi_\fr)}
&=&
\lim_{\epsilon\searrow0}
\frac{(\Phi_\epsilon^+|B\Phi_\epsilon^-)}
{(\Phi_\epsilon^+|\Phi_\epsilon^-)}
=
\frac{(\Phi_\GL^+|B\Phi_\GL^-)}{(\Phi_\GL^+|\Phi_\GL^-)}.
\end{eqnarray*}
 \qed

\nowastrona

The following theorem describes the {\em Sucher formula} often used in practical computations of the energy shift.

\bet
\begin{eqnarray}
E-E_\fr&=&\lim_{\epsilon\searrow0}\frac{\i\epsilon\lambda}{2}\partial_\lambda
\log(\Phi_\fr|S_\epsilon\Phi_\fr).
\end{eqnarray}
\eet

\proof 
We sandwich (\ref{toz4}) with $\Phi_\fr$ and divide with
$(\Phi_\fr|S_\epsilon\Phi_\fr)$ obtaining
\[-\i\epsilon\lambda\partial_\lambda\log(\Phi_\fr|S_\epsilon\Phi_\fr)=2E_\fr
-2\frac{(S_{\epsilon}^+\Phi_\fr|H S_{\epsilon}^-\Phi_\fr)}
{(S_{\epsilon}^+\Phi_\fr|S_{\epsilon}^-\Phi_\fr)}.\]
The last term, by Lemma
\ref{bele}, converges to $2E$.
\qed

Note that the right hand side of the Sucher formula may have a nonzero
imaginary part. In this case we expect that it describes a resonance close to $E_\fr$.

\subsubsection{Adiabatic switching and Green's functions}

Recall that if $A$ is an operator then $A(t)=\e^{\i tH}A\e^{-\i tH}$.
To define the interacting Green's functions we fix a vector $\Phi$, which is a bound state of $H$, and we set
\begin{eqnarray*}
\langle A_k(t_k)\cdots A_1(t_1)\rangle
&:=&\left(\Phi|\T\bigl(A_{k}(t_k)\cdots
  A_{1}(t_1)\bigr)\Phi\right).
\end{eqnarray*}

\nowastrona

Suppose  that $H=H_\fr+\lambda V$. The {\em  Gell-Mann and Low Theorem about Green's functions} allows us to express interacting  Green's functions by the free ones:

\bet Suppose that (\ref{gell-1}), (\ref{gell-2}) and (\ref{gell}) are true, so that we can apply the results of the previous subsubsection. Then
\begin{eqnarray}
\langle A_{k}(t_k)\cdots
  A_{1}(t_1)\rangle\label{las0} 
&=&\lim_{\epsilon\searrow0}
\frac{1}{(\Phi_\fr|S_\epsilon \Phi_\fr)}\sum_{n=0}^\infty\frac{(-\i\lambda)^n}{n!}\\
&& \times 
\int_{-\infty}^\infty\d
  s_n\cdots\int_{-\infty}^\infty\d s_1
\langle
V_{\epsilon}(s_n)\cdots V_{\epsilon}(s_1)A_{k}(t_k)\cdots
A_{1}(t_1)\rangle_\fr 
,\nonumber\\(\Phi_\fr|S_\epsilon \Phi_\fr)
&=&
\lim_{\epsilon\searrow0}
\sum_{n=0}^\infty\frac{(-\i\lambda)^n}{n!}\int_{-\infty}^\infty\d
  s_n\cdots\int_{-\infty}^\infty\d s_1\label{las2}
\\
&&\ \ \ \ \ \ \times
\langle
V_{\epsilon}(s_n)\cdots V_{\epsilon}(s_1)\rangle_\fr.\nonumber
\end{eqnarray}
\eet

\nowastrona

\proof (\ref{las2}) follows from (\ref{scatte1}) applied to $U_\epsilon$.

Let us prove (\ref{las0}). Let $t_k\geq\cdots\geq t_1$. Let $A_{i,\epsilon}(t)$ denote the operator $A_i$ in the Heisenberg picture for the evolution $U_\epsilon$.
The  left-hand side of (\ref{las0}) is
\begin{eqnarray}
&&
\left(\Phi|A_{k}(t_k)\cdots
  A_{1}(t_1)\Phi\right)\notag \\
&=&\lim_{\epsilon\searrow0}
\left(\Phi|A_{k,\epsilon}(t_k)\cdots
  A_{1,\epsilon}(t_1)\Phi\right)\notag \\
&=&
\lim_{\epsilon\searrow0}
\frac{\left(S_\epsilon^+\Phi_\fr|A_{k,\epsilon}(t_k)\cdots
  A_{1,\epsilon}(t_1)S_\epsilon^-\Phi_\fr\right)}
{(\Phi_\fr|S_\epsilon \Phi_\fr)},\label{las00}
\end{eqnarray}
where at the last step we used  Lemma \ref{bele}.

 Let $\langle\cdots\rangle_\epsilon$ denote Green's functions for the dynamics $U_\epsilon$. Then
the numerator of (\ref{las00}) can be written as
\beq
\langle
A_{k,\epsilon}(t_k)\cdots
  A_{1,\epsilon}(t_1)\rangle_\epsilon.\label{las4}\eeq
Applying (\ref{gell-}) to (\ref{las4}) we arrive at (\ref{las0}). \qed

Let us note that on the right of (\ref{las0}) only the free dynamics appears. One can forget about the dynamics $U_\epsilon$, whose use can be treated as a trick. Some authors consider (\ref{las0})  as a (perturbative) definition of Green's functions, forgetting about the auxiliary nonphysical dynamics $U_\epsilon$.
\nowastrona

\subsubsection{Adiabatic scatttering theory}

In our notes we concentrate on  time-dependent Hamiltonians, with perturbations decaying fast as $|t|\to\infty$. For such Hamiltonians the definitions of M{\o}ller and scattering operators given in (\ref{scatte1}), (\ref{scatte2}) and (\ref{scatte3}) work well.

One would also like to consider scattering theory of time-independent Hamiltonians. Unfortunately, in QFT   these definitions typically need to be modified.

In Quantum Mechanics the situation is much better. For  (time-independent) Schr\"odinger Hamiltonians $H:=H_\fr+V(x)$, $H_\fr:=-\Delta$,  (at least with short-range potentials) there exists a very satisfactory scattering theory based on the M{\o}ller operators 
\beq S^\pm:=\slim_{t\to\pm\infty}\e^{\i tH}\e^{-\i tH_\fr}\label{gell1}\eeq
and the scattering operator by $S:=S^{+*}S^-$. 

This approach  almost never works in QFT, even for simple-minded models with classical (time-independent) perturbations considered in these notes. In particular, the limit (\ref{gell1}) almost never exists. 
 One of the reasons is the existence of the ground state for $H_\fr$ in quantum field theory. $H$ has essentially always   a different ground state or no ground state at all. It usually has a different spectrum.

One can try to remedy this problem by introducing adiabatic switching together with a renormalization of the phase,  which is divergent as $\epsilon\searrow0$, as in the following teorem:

\bet   Suppose that (\ref{gell-1}), (\ref{gell-2}) and (\ref{gell}) are true.
\ben\item
 Assume also that there exist
 the {\em adiabatic} or {\em Gell-Mann--Low M{\o}ller operators}
\begin{eqnarray}S_\GL^\pm&:=&\wlim_{\epsilon\searrow0}\frac{|(\Phi_\fr|
S_\epsilon^\pm\Phi_\fr)|}
{(\Phi_\fr|S_\epsilon^\pm\Phi_\fr) }S_\epsilon^\pm,
\label{qrowa}\end{eqnarray}
and
\beq
\wlim_{\epsilon\searrow0}\epsilon\lambda\partial_\lambda
\frac{|(\Phi_\fr|
S_\epsilon^\pm\Phi_\fr)|}
{(\Phi_\fr|S_\epsilon^\pm\Phi_\fr) }S_\epsilon^\pm=0.\label{spla3}\eeq
Then
\begin{eqnarray} S_\GL^\pm (H_\fr-E_\fr)&=&(H-\Re E)S_\GL^\pm,
\label{spla}\\
 S_\GL^\pm\Phi_\fr&=&
\Phi_\GL^\pm.\label{spla1}\end{eqnarray}
\item Define the {\em adiabatic} or  {\em Gell-Mann--Low scattering operator}
\beq
S_\GL=S_\GL^{+*}S_\GL^-.\label{stron}\eeq
Then
\begin{eqnarray*}S_\GL H_\fr&=&H_\fr S_\GL,\\
S_\GL\Phi_\fr&=&\e^{\i\alpha}\Phi_\fr.\end{eqnarray*}
\een
\eet
\proof
Using $\frac{|f|}{f}=\sqrt{\frac{\bar f}{f}}$, we obtain
$\i\partial_\lambda\frac{|f|}{f}=-\frac{|f|}{f}\Re\frac{\i\partial_\lambda
  f}{f}.$
Therefore, setting $f:=(\Phi_\fr|S_\epsilon^\pm\Phi_\fr)$ we compute
\begin{eqnarray*}
\mp\i\epsilon\lambda\partial_\lambda
\frac{|(\Phi_\fr|
S_\epsilon^\pm\Phi_\fr)|}
{(\Phi_\fr|S_\epsilon^\pm\Phi_\fr) }&=&
-\frac{|(\Phi_\fr|
S_\epsilon^\pm\Phi_\fr)|}
{(\Phi_\fr|S_\epsilon^\pm\Phi_\fr) }\Re
\frac{(\Phi_\fr|
(HS_\epsilon^\pm-S_\epsilon^\pm H_\fr)\Phi_\fr)}
{(\Phi_\fr|S_\epsilon^\pm\Phi_\fr) }\\
&=&-
\frac{|(\Phi_\fr|
S_\epsilon^\pm\Phi_\fr)|}
{(\Phi_\fr|S_\epsilon^\pm\Phi_\fr)
}\Re\left(E_\epsilon^\pm-E_\fr\right).\end{eqnarray*}
Therefore,
\begin{eqnarray*}
&&\mp\i\epsilon\lambda\partial_\lambda
\frac{|(\Phi_\fr|
S_\epsilon^\pm\Phi_\fr)|}
{(\Phi_\fr|S_\epsilon^\pm\Phi_\fr) }S_\epsilon^\pm\\&=&
-\frac{|(\Phi_\fr|
S_\epsilon^\pm\Phi_\fr)|}
{(\Phi_\fr|S_\epsilon^\pm\Phi_\fr) }
\left(-HS_\epsilon^\pm+S_\epsilon^\pm H_\fr+\Re(E_\epsilon^\pm-E_\fr)S_\epsilon^\pm\right).
\end{eqnarray*}
Using $\Re E_\fr=E_\fr$, we obtain (\ref{spla}). 
 (\ref{spla1}) follows by definition.
\qed

If the limit in (\ref{spla3}) is strong and not only weak, $S_\GL^\pm$ are unitary. This will be the case  in the examples we consider in our text.
In general, they do not have to be unitary, and one needs to perform an additional {\em wave function renormalization}, which we will not discuss.

\ber Many textbooks use (\ref{gell1}) as the starting point for a derivation of the rules of QFT, eg. \cite{We1}. As  indicated above, this is quite far from being correct. Gell-Mann and Low invented the adiabatic switching as an attempt to make
 this  derivation somewhat more satisfactory. This approach often  works in  QFT models with classical perturbations.\eer

\init\section{Neutral scalar bosons}

In this section we consider the {\em Klein-Gordon equation}
\beq(-\Box+m^2)\phi(x)=0\label{kgo}\eeq
and we quantize the space of its {\em real solutions}.
We study two kinds of interactions: an {\em external linear  source}
\beq(-\Box+m^2)\phi(x)=-j(x),\label{kgo.1}\eeq
and a {\em mass-like perturbation}
\beq(-\Box+m^2)\phi(x)=-\kappa(x)\phi(x).\label{kgo.2}\eeq
\nowastrona

\subsection{Free neutral scalar bosons}

\subsubsection{Special
solutions and Green's functions}
\label{special}

Every function $\zeta$ that solves the (homogeneous) Klein-Gordon equation
\beq(-\Box+m^2)\zeta(x)=0\label{kgo1}\eeq
 can be written as
\begin{eqnarray*}
\zeta(x)&=&
\int\e^{\i kx}g(k)\delta(k^2+m^2)\frac{\d k}{(2\pi)^3}\\
&=&
\sum_{\pm}
\int\frac{\d\vec{k}}{(2\pi)^32\sqrt{\vec k^2+m^2}}g\Big(\pm\sqrt{\vec k^2+m^2},\vec
k\Big) 
\e^{\mp\i x^0\sqrt{\vec
    k^2+m^2} +\i\vec x\vec k},\end{eqnarray*}
where $g$ is a function on the two-sheeted hyperboloid  $k^2+m^2=0$.
\nowastrona
A special role is played by the following  {\em 3 special solutions} of the
  homogeneous 
Klein-Gordon equation.
\ben\item
The {\em positive frequency} or {\em Wightman}, resp. {\em negative frequency} or {\em anti-Wightman solution}:
\begin{eqnarray*}
D^{(\pm)}(x)&=&\pm\i\int\e^{\i kx}\theta(\pm k^0)\delta(k^2+m^2)\frac{\d
    k}{(2\pi)^3}\\
&=&
\pm\i\int\frac{\d\vec{k}}{(2\pi)^32\sqrt{\vec k^2+m^2}}\e^{\mp\i x^0\sqrt{\vec
    k^2+m^2} +\i\vec x\vec k}\\
&=&\frac{1}{4\pi}\sgn x^0\delta(x^2)\\&& \pm
\frac{m\theta(-x^2)}{8\pi\sqrt{-x^2}}H_1^{\mp\sgn x^0}(m\sqrt{-x^2})
\pm\frac{m\i\theta(x^2)}{4\pi^2\sqrt{x^2}}K_1(m\sqrt{x^2}).
\end{eqnarray*}
where $H_1^\pm$ are the Hankel functions and $K_1$ is the MacDonald function
of the 1st order.
\nowastrona
\item
The {\em Pauli-Jordan}  or the {\em commutator function}:
\begin{eqnarray*}
D(x)&=&\i\int\e^{\i kx}\sgn(k^0)\delta(k^2+m^2)\frac{\d
    k}{(2\pi)^3}\\
&=&
\int\frac{\d\vec{k}}{(2\pi)^3\sqrt{\vec k^2+m^2}}\e^{\i \vec x\vec k}
\sin\left( x^0\sqrt{\vec
    k^2+m^2}\right)\\
&=&\frac{1}{2\pi}\sgn x^0\delta(x^2)-
\frac{m\sgn x^0\theta(-x^2)}{4\pi\sqrt{-x^2}}J_1(m\sqrt{-x^2})
,
\end{eqnarray*}
where $J_1$ is the Bessel function
of the 1st order.
$D(x)$ is the unique solution of the 
Klein-Gordon equation satisfying
\[D(0,\vec x)=0,\ \dot{D}(0,\vec x)=\delta(\vec x).\]
We have, $\supp D\subset J$.
\een

\nowastrona

Solutions of
\beq(-\Box+m^2)\zeta(x)=\delta(x),\label{kgo2}\eeq
are called {\em Green's functions} or {\em fundamental solutions} of the Klein-Gordon
equation. 
In particular, let us introduce  the following  3 Green's
functions. \ben\item
The {\em retarded}, resp. {\em advanced Green's function}:
\begin{eqnarray*}
D^{\pm}(x)&=&\int\frac{\e^{\i kx}}{(k^2+m^2\mp\i 0\sgn k^0)}\frac{\d k}{(2\pi)^4}\\
&=&\frac{1}{2\pi}\theta(\pm x^0)\delta(x^2)-
\frac{m\theta(-x^2)\theta(\pm x^0)}{4\pi\sqrt{-x^2}}J_1(m\sqrt{-x^2})
.
\end{eqnarray*}
We have $\supp D^\pm\subset J^\pm$.
In the literature, $D^+(x)$ is usually denoted $D^\ret(x)$ and
 $D^-(x)$ is usually denoted $D^\adv(x)$.

\nowastrona

\item
The {\em causal} or {\em Feynman(-Stueckelberg) Green's function}:
\begin{eqnarray*}
D^{\rm c}(x)&=&\int\frac{\e^{\i kx}}{(k^2+m^2-\i 0)}\frac{\d k}{(2\pi)^4}\\
&=&\frac{1}{4\pi}\delta(x^2)-
\frac{m\theta(-x^2)}{8\pi\sqrt{-x^2}}H_1^-(m\sqrt{-x^2})
+\frac{m\i\theta(x^2)}{4\pi^2\sqrt{x^2}}K_1(m\sqrt{x^2}).
\end{eqnarray*}
\een

\nowastrona

The 
special solutions and Green's functions introduced above are often called {\em propagators}. They satisfy the following relations
\begin{eqnarray*}
\bar{D(x)}=D(x)=-D(-x)&=&D^{(+)}(x)+D^{(-)}(x)\\
&=&D^+(x)-D^-(x),\\
\bar{D^{(-)}(x)}=D^{(+)}(x)&=&-D^{(-)}(-x),\\
\bar{D^+(x)}=D^+(x)=D^-(-x)&=&\theta(x^0)D(x),\\
\bar{D^-(x)}=D^-(x)=D^+(-x)&=&\theta(-x^0)D(x),\\
D^{\rm c}(x)=D^{\rm c}(-x)&=&\theta(x^0)D^{(+)}(x)-\theta(-x^0)D^{(-)}(x). 
\end{eqnarray*}

\nowastrona

Let us prove the last identity.
\begin{eqnarray}
D^{\rm c}(x)&=&\int\frac{\e^{\i kx}}{\big(\vec k^2+m^2-(|k^0|+\i 0)^2\big)}
\frac{\d k}{(2\pi)^4}\notag \\
&=&\frac{1}{(2\pi)^4}\int\frac{\e^{-\i k^0x^0+\i\vec k\vec x}}{2\sqrt{\vec k^2+m^2}\left(\sqrt{\vec
        k^2+m^2}-|k^0|-\i 0\right)}\d k\label{wqe1}\\
&&+\int\frac{\e^{-\i k^0x^0+\i\vec k\vec x}}{2\sqrt{\vec k^2+m^2}\left(\sqrt{\vec
        k^2+m^2}+|k^0|+\i0\right)}\frac{\d k}{(2\pi)^4}.\label{wqe}\end{eqnarray}
In (\ref{wqe}) we can replace $\i0$ with $-\i0$. Then the parts of (\ref{wqe1}) and (\ref{wqe}) with $k^0<0$ are swapped:
\begin{eqnarray}
&=&\int\frac{\e^{-\i k^0x^0+\i \vec k\vec x}}
{2\sqrt{\vec k^2+m^2}\left(\sqrt{\vec
        k^2+m^2}-k^0-\i 0\right)}\frac{\d k}{(2\pi)^4}\notag\\
&&+\int\frac{\e^{-\i k^0x^0+\i \vec k\vec x}}{2\sqrt{\vec
    k^2+m^2}\left(\sqrt{\vec 
        k^2+m^2}+k^0-\i 0\right)}\frac{\d k}{(2\pi)^4},\notag\\
&=&\i\theta(x^0)\int\frac{\e^{-\i \sqrt{\vec k^2+m^2}x^0+\i \vec k\vec
    x}}{2\sqrt{\vec k^2+m^2}}\frac{\d\vec k}{(2\pi)^3}\notag\\
&&+\i\theta(-x^0)
\int\frac{\e^{\i \sqrt{\vec k^2+m^2}x^0+\i \vec k\vec x}}
{2\sqrt{\vec k^2+m^2}}\frac{\d \vec k}{(2\pi)^3}\notag
\\
&=&\theta(x^0)D^{(+)}(x)-\theta(-x^0)D^{(-)}(x).\notag
\end{eqnarray}
where in the last step we  integrate wrt $k^0$ using
\[\int\frac{\e^{-\i x^0 k^0}}{\varepsilon\mp k^0-\i 0}\d k^0=2\pi\i\e^{\mp\i x^0\varepsilon}\theta(\pm x^0).\]

\nowastrona
Let us now prove that $\supp D^+\subset J^+$. By the Lorentz invariance it
suffices to prove that $ D^+$ is zero on the lower half-plane.
We write
\begin{eqnarray*}
D^{+}(x)&=&\int\frac{\e^{\i kx}}{(k^2+m^2-\i 0\sgn k^0)}\frac{\d k}{(2\pi)^4}\\
&=&\int\frac{\e^{-\i k^0x^0+\i \vec k\vec x}}{\big(\vec k^2+m^2-(k^0+\i0)^2\big)}\frac{\d
k^0\d\vec k}{(2\pi)^4}.
\end{eqnarray*}
Next we continuously deform the contour of integration, replacing $ k^0$  by $k^0+\i R$, where $R\in[0,\infty[$. We do not cross any singularities of the integrand
and note that $\e^{-\i x^0(k^0+\i
  R)}$ goes to zero (remember that   $x^0<0$).
\nowastrona

\subsubsection{Space of solutions}

\label{c15.6.1}




\nowastrona

A space-like
 subspace of codimension $1$ will be called a {\em Cauchy subspace}.

Solutions of the {\em Cauchy problem}
 are uniquely parametrized by their {\em Cauchy
data} (the value and the normal derivative on a Cauchy surface).
 They can be expressed by the Cauchy data with help of the Pauli-Jordan
function. 

\vskip -3ex 

\bet \label{curved.01z} Let
$\varsigma,\vartheta \in C_{\rm c}^\infty(\rr^3)$. Then there exists a
unique $\zeta\in C_\sc^\infty(\rr^{1,3})$ that solves
\beq(- \Box+m^2) \zeta=0\label{KG1z}\eeq
with initial conditions $\zeta(0,\vec x)=\varsigma(\vec x),\ \ 
\dot \zeta(0,\vec x)=\vartheta(\vec x).$
It satisfies $\supp \zeta\subset J(\supp \varsigma\cup\supp \vartheta)$
and 
  is given by
\begin{eqnarray}
\zeta(t,\vec x)
&=&\int_{\rr^3}\dot D(t,\vec x-\vec y)\varsigma(\vec y)
\d\vec{y}+\int_{\rr^3}
D(t,\vec x-\vec y)\vartheta(\vec y)\d\vec{y}.\label{qrt}
\end{eqnarray}\eet

\nowastrona

Let $\cY_{\KG}$, resp. $\cc\cY_\KG$  
denote the {\em space of real}, resp. {\em complex, space-compact
  solutions of the Klein-Gordon
equation}.

For $\zeta_1,\zeta_2\in C^\infty(\rr^{1,3})$ we define
\begin{eqnarray}
\cj^\mu(x)=\cj^\mu(\zeta_1,\zeta_2,x)&:=&\partial^\mu \zeta_1(x)\zeta_2(x)
-\zeta_1(x)\partial^\mu \zeta_2(x).\label{curr}\end{eqnarray}
We easily check that 
\[\partial_\mu
\cj^\mu(x)=(\Box -m^2)\zeta_1(x) \zeta_2(x)
-\zeta_1(x)(\Box-m^2) \zeta_2(x), \]
Therefore, if $\zeta_1,\zeta_2\in\cc\cY_\KG$,  then 
\[\partial_\mu \cj^\mu(x)=0.\]One says that $\cj^\mu(x)$ is a {\em
  conserved 4-current}.

\nowastrona

The flux of $\cj^\mu$ across any Cauchy subspace $\cS$
does not depend on its choice. It defines
a {\em symplectic form} on $\cY_{\KG}$
\begin{eqnarray}\nonumber
\zeta_1\omega \zeta_2&=&
\int_\cS \cj^\mu(\zeta_1,\zeta_2,x)\d s_\mu(x)\label{qrt1}\\
\label{symplo}
&=&\int\left(-\dot \zeta_1(t,\vec x)\zeta_2(t,\vec x)+
\zeta_1(t,\vec x)\dot\zeta_2(t,\vec x)\right)\d\vec{x}.
\end{eqnarray}

Clearly, the form (\ref{symplo}) is well defined also if only
$\zeta_2\in\cY_{\KG}$, and $\zeta_1$ is a distributional solution of the
Klein-Gordon equation.

\nowastrona

The Poincar\'{e}{} group $\rr^{1,3}\rtimes O(1,3)$ 
acts on $\cY_{\KG}$ and $\cc\cY_\KG$ by
\[r_{(y,\Lambda)}\zeta(x):=\zeta\left((y,\Lambda)^{-1}x\right).\]
$r_{(y,\Lambda)}$ are symplectic (preserve the symplectic form) for $\Lambda\in  O^\uparrow(1,3)$, otherwise they are antisymplectic (change the sign in front of the symplectic form). 

\nowastrona

The Pauli-Jordan function  $D$ can be used to construct solutions of the
Klein-Gordon equation
parametrized by space-time functions,
 which are especially useful in the axiomatic
formulation of QFT.

\bet \label{curved.2z}\ben \item
For any $f\in C_{\rm c}^\infty(\rr^{1,3},\rr)$, $D*f\in\cY_{\KG}$,
where
 \[D*f(x):=\int D(x-y)f(y)\d y.\] 
\item Every element of
$\cY_{\KG}$ is of this form.
\item
\beq- (D*f_1)\omega (D*f_2)=\int f_1(x)D(x-y)f_2(y)\d x\d y.\label{peierls}\eeq
\item If $\supp f_1\times\supp f_2$, then 
\[(D*f_1)\omega( D*f_2)=0.\]
\een
\eet
The right hand side of (\ref{peierls}) is sometimes called the {\em Peierls
  bracket} of $f_1$ and $f_2$.


\nowastrona
Let us prove (\ref{peierls}). Choose time $t$ later than $\supp f_i$,
$i=1,2$. Then we have $D*f_i=D^+*f_i$. Now
\begin{eqnarray*}
&&-(D*f_1)\omega( D*f_2)\\& =& \int\Big((\dot D^+* f_1)(t,\vec x)(D^+ *f_2)(t,\vec x)
- (D^+* f_1)(t,\vec x)(\dot D^+ *f_2)(t,\vec x)\Big)\d\vec x\\
&=&
\int_{x^0<t}
\Big(-(\Box-m^2)( D^+ *f_1)( x)(D^+ *f_2)( x)\\&&\hskip 4ex+
(D^+* f_1)( x)(\Box-m^2) (D^+* f_2)(x)\Big)\d x\\
&=&
\int
\left( f_1( x)(D^+* f_2)( x)
- (D^+* f_1)( x) f_2(x)\right)\d x\\
&=&
\int
\left( f_1( x)(D^+* f_2)( x)
-  f_1( x) (D^-*f_2)( x)\right)\d x
\ =\  \int
 f_1( x)( D*f_2)( x)\d x.
\end{eqnarray*}

\nowastrona

\subsubsection{Classical fields}

We will also consider the {\em space dual to $\cY_{\KG}$}.
More precisely, we can endow the space  $\cY_{\KG}$ with the standard topology
of $C_{\rm c}^\infty(\rr^3)\oplus C_{\rm c}^\infty(\rr^3)$ 
given by the initial
conditions. The space of real, resp. complex
 continuous functionals on $\cY_{\KG}$
will be denoted by
$\cY_{\KG}^\t$, resp. by $\cc
\cY_{\KG}^\t$. The action of $T\in \cc\cY_{\KG}^\t$ on $\zeta\in\cY_{\KG}$ will be
denoted by
$\langle T|\zeta\rangle,$ and sometimes simply by $T\zeta$.

If $T\in\cc\cY_\KG^\t$, we define $T^*\in\cc\cY_\KG^\t$ by
\[\langle T^*|\zeta\rangle:=\bar{\langle T|\zeta\rangle},\ \ \ \ \zeta\in\cY_\KG.\]
Note that in this context the star does not denote the Hermitian
conjugation (which in our text is the standard meaning of the star).

Let us stress that  the space $\cY_\KG$ is real, which reflects the fact that in this section we consider neutral fields. In the section devoted to charged fields the main role will be played by the complexification of $\cY_\KG$, that is $\cW_\KG:=\cc\cY_\KG$.

\nowastrona

For $x\in\rr^{1,3}$,
$\phi(x)$, $\pi(x)$
 will denote the functionals on $\cY_{\KG}$ given by
\[\langle \phi(x)|\zeta\rangle:=\zeta(x),\ \ \ \ 
\langle \pi(x)|\zeta\rangle:=\dot\zeta(x).\]
They are called {\em classical fields}. Clearly, for any 
$\zeta\in\cY_{\KG}$ we have 
\[(-\Box+m^2)\langle\phi(x)|\zeta\rangle=0.\]
Thus the equation
\beq(-\Box+m^2)\phi(x)=0\label{tauto}\eeq
is a tautology.

On $\cY_{\KG}^\t$ we have the action of the Poincar\'{e}{} group
$(y,\Lambda)\mapsto r_{(y,\Lambda)}^{\t-1}$. Note
that
\[r_{(y,\Lambda)}^{\t-1}\phi(x)=\phi( \Lambda x+y).\]

\nowastrona

Clearly, $\dot\phi(x)=\pi(x)$ and, by (\ref{qrt}),
\beq
\phi(t,\vec x)=\int \dot D(t,\vec x-\vec y)\phi(0,\vec y)\d\vec{y}+
\int  D(t,\vec x-\vec y)\pi(0,\vec y)\d\vec{y}.\label{pzk8}\eeq

By (\ref{qrt1}), the symplectic form  can be written as
\begin{eqnarray*}
\zeta_1\omega \zeta_2
&=&\int\big(-\langle\pi(t ,\vec x)|\zeta_1\rangle
\langle\phi(t ,\vec
x)|\zeta_2\rangle 
+\langle\phi(t ,\vec x)|\zeta_1\rangle
\langle\pi(t ,\vec
x)|\zeta_2\rangle\big)\d\vec{x},
\end{eqnarray*}
or more simply,
\[\omega=\int\phi(t,\vec x)\wedge \pi(t,\vec x)\d\vec{x}.\]
The conserved 4-current can be written as
\[j_\mu(x)= \phi(x)\wedge \partial_\mu\phi(x).\]
\nowastrona

\subsubsection{Poisson brackets}

The symplectic structure on the space $\cY_{\KG}$ leads to 
a {\em Poisson bracket} on
 functions on $\cY_{\KG}$:
\begin{eqnarray}
\{\phi(t,\vec x),\phi(t,\vec y)\}=\{\pi(t,\vec x),\pi(t,\vec y)\}&=&0,\nonumber
\\
\{\phi(t,\vec x),\pi(t,\vec y)\}&=&\delta(\vec x-\vec y).
\label{poisson}\end{eqnarray}
The relations (\ref{poisson}) 
 can be viewed as mnemotechnic identities that yield the correct
Poisson bracket for more regular functions, eg. the smeared out fields in
 (\ref{poi1}) or (\ref{pzk1}) described below.
  Note that formally $\phi(t,\vec x)$ and
 $\pi(t,\vec x)$
generate the algebra of all functions on $\cY_{\KG}$.

Using (\ref{pzk8}) we obtain
\[\{\phi(x),\phi(y)\}=D(x-y).\]
Therefore, the Pauli-Jordan solution is often called the {\em commutator function}.

\subsubsection{Smeared fields}

There are two basic methods to introduce {\em smeared  fields}.

One way  to smear them out is to
 use the {\em pairing given by the symplectic form}. It is convenient to
 allow complex smearing functions paired  antilinearly.
More precisely, for $\zeta\in\cc\cY_{\KG}$ we introduce the functional on
$\cY_{\KG}$ given by 
\[\langle  \phi\lpar \zeta\rpar|\rho\rangle:=\bar\zeta\omega \rho,\ \ \ \rho\in\cY_{\KG}.\]

Note in passing that $\omega$ can be treated as a linear map from $\cY_\KG$ to
$\cY_\KG^\t$, which satisfies
\[-(\omega\bar\zeta)\rho=\bar\zeta\omega\rho.\]
Therefore, a possible alternative notation for 
$ \phi\lpar \zeta\rpar$ is $-\omega\bar\zeta$.

Clearly,
\beq
\phi\lpar\zeta\rpar =\int\left(-\bar{\dot\zeta(t,\vec x)}\phi(t,\vec x)+
\bar{\zeta(t,\vec x)}\pi(t,\vec x)\right)\d\vec{x}.
\label{pzk}\eeq
Note that
\beq
\{\phi\lpar\zeta_1\rpar,\phi\lpar\zeta_2\rpar\}
=\bar\zeta_1\omega \bar\zeta_2.\label{poi1}\eeq

\nowastrona

We can also  smear fields {\em with space-time functions}. For
 $f\in C_{\rm c}^\infty(\rr^{1,3},\rr)$, we set
\[\phi[f]:=\int f(x)\phi(x)\d x.\]
We have
\begin{eqnarray}
\phi[f]&=&\phi\lpar- D*f\rpar,\label{pzk0}\\
\{\phi[f_1],\phi[f_2]\}&=&\int\int f_1(x)D(x-y)f_2(y)\d x\d
y.\label{pzk1}\end{eqnarray} 
To see (\ref{pzk0}), write  an element of $\cY_\KG$ as
 $D*g$ for some $g\in C_{\rm c}^\infty(\rr^{1,3},\rr)$:
\begin{eqnarray*}
\langle\phi\lpar- D*f\rpar| D*g\rangle&=&(D*f)\omega(D*g)=\int f(x)D(x-y)g(y)\d x\d y\\=\ \int
f(x)\langle\phi(x)|D*g\rangle\d x&=&
\langle\phi[f]|D*g\rangle.\end{eqnarray*}

\nowastrona

\nowastrona
\subsubsection{Lagrangian formalism}

In classical mechanics we have  the {\em Hamiltonian formalism}, where the basic object is the phase space equipped with a symplectic form, and the {\em Lagrangian formalism}, where we start from  the configuration space. In classical field theory we  can also use both formalisms.

 In this context, the Hamiltonian approach is often called
 the {\em on-shell formalism}. This means that  the field $\phi(x)$ acts on the space
of solutions of the equations of motion.
In other words, the field $\phi(x)$ that we use in the
Hamiltonian formalism satisfies the equation  (\ref{tauto}) -- one says that it
is {\em on-shell}.

In the Lagrangian formalism one also uses a  classical 
field, which we will denote by
$\phi(x)$, as before. But now, this field     is {\em off-shell}. This
means, we do not enforce any equation on $\phi(x)$. 
One can interpret $\phi(x)$
as the functional on, say, $C^\infty(\rr^{1,3})$ or $C_\sc^\infty(\rr^{1,3})$ such that
$\langle \phi(x)|f\rangle:=f(x)$.

\nowastrona

Using $\phi(x)$ in the off-shell formalism, introduce the {\em Lagrangian density} 
\begin{eqnarray}
\cL(x)=&
-\frac12\partial_\mu\phi(x)\partial^\mu\phi(x)-\frac12 m^2\phi(x)^2.
\label{lagra1}\end{eqnarray}
The {\em Euler-Lagrange equation }
\beq
\partial_{\phi}\cL-
\partial_\mu\frac{\partial \cL}{\partial\phi_{,\mu}}=0
\label{euler-lagrange1}\eeq
yields  the Klein-Gordon equation (\ref{kgo}).

When we go from the Lagrangian to Hamiltonian formalism, we enforce the
{\em on-shell} condition, that is, 
the Euler-Lagrange equation, and we introduce the variable  conjugate to $\phi(x)$:
\begin{eqnarray*}
\pi(x)&:=&\frac{\partial
  \cL}{\partial\phi_{,0}( x)}=\phi_{,0}( x).\end{eqnarray*} 
Then we express everything in terms of $\phi(x)$  and $\pi(x)$.

\subsubsection{Stress-energy tensor}
We can also introduce the {\em stress-energy tensor}
\begin{eqnarray} \label{sfa8}
\cT^{\mu\nu}(x)&:=&-\frac{\partial \cL(x)}{\partial\phi_{,\mu}(x)}\phi^{,\nu}(x)+g^{\mu\nu}\cL(x)\\
&=&
\partial^\mu\phi(x)\partial^\nu\phi(x)
-
g^{\mu\nu}\frac12\left(\partial_\alpha\phi(x)\partial^\alpha\phi(x)+
m^2\phi(x)^2\right). \notag
\end{eqnarray}

It is easy to check that the stress-energy tensor is conserved for a solution of the Klein-Gordon equation (on shell)
\[\partial_\mu \cT^{\mu\nu}(x)=0.\]
We express the stress-energy tensor 
in terms of $\phi(x)$ and $\pi(x)$.
Its components  with the first temporal coordinate
are called the {\em Hamiltonian density} and {\em momentum
  density}:
\begin{eqnarray*}
\cH(x)\ :=\ \cT^{00}(x)&=&\frac12
\left(\pi( x)^2+\big(\vec\partial\phi( x)\big)^2 +m^2\phi(x)^2\right),\\
\cP^i(x)\ :=\ \cT^{0i}(x)&=&-\pi(x)\partial^i
 \phi(x).\end{eqnarray*}
They are examples of quadratic functionals on $\cY_\KG$:
\begin{eqnarray*}
\langle \cH(x)|\zeta\rangle&=&\frac12
\left(\dot\zeta( x)^2+\big(\vec\partial\zeta( x)\big)^2 +m^2\zeta(x)^2\right),\\
\langle \cP^i(x)|\zeta\rangle&=&-\dot\zeta(x)\partial^i
 \zeta(x).\end{eqnarray*}

We introduce  the {\em (total) Hamiltonian} and {\em momentum}:
\begin{eqnarray}\nonumber
H&:=&\int_\cS \cT^{\mu0}(x)\d s_\mu(x)=\int \cH(t,\vec x)\d\vec x,\\
P^i&:=&\int_\cS \cT^{\mu i}(x)\d s_\mu(x)=\int \cP^i(t,\vec x)\d\vec x.\label{sfa9}
\end{eqnarray}
where $\cS$ is any Cauchy subspace.


$H$ and $\vec P$ are
the generators of the time and space translations:
\begin{eqnarray*}
\dot\phi(x)=\{\phi(x),H\},\ \ \ \dot\pi(x)=\{\pi(x),H\},\\
\vec\partial\phi(x)=-\{\phi(x),\vec P\},\ \ \ \vec\partial\pi(x)=-\{ \pi(x),\vec P\}.
\end{eqnarray*}
The observables $H$,
$P^1$, $P^2$ and $P^3$ are in involution. (This means that the Poisson bracket of every pair among these observables  vanishes).
\nowastrona

\nowastrona

\subsubsection{Diagonalization of the equations of motion}



For $\vec k\in\rr^3$, set $\varepsilon=\varepsilon (\vec  k):=\sqrt{\vec k^2+m^2}$ and
 $k:=(\varepsilon(\vec k)
 ,\vec k)$. $k\in\rr^{1,3}$ of this form will be called {\em on shell}.
Define
\begin{eqnarray*}
\phi_t(\vec k)&:=&\int\phi(t,\vec x)\e^{-\i \vec k \vec x}\d\vec x,\\
\pi_t(\vec k)&:=&\int\pi(t,\vec x)\e^{-\i \vec k \vec x}\d\vec x.\end{eqnarray*}
Clearly,
\begin{eqnarray*}\phi_t^*(\vec k)&=&\phi_t(-\vec k),\\
\pi_t^*(\vec k)&=&\pi_t(-\vec k),\\{}
\{\phi_t^*(\vec k),\phi_t(\vec k')\}=\{\pi_t^*(\vec k),\pi_t(\vec k')\}&=&0,\\{}
\{\phi_t^*(\vec k),\pi_t(\vec k')\}&=&
(2\pi)^3\delta(\vec k-\vec k').
\end{eqnarray*}
The equations of motion are
\begin{eqnarray*}
\dot\phi_t(\vec k)&=&\pi_t(\vec k),\\
\dot\pi_t(\vec k)&=&-\varepsilon^2(\vec k)\phi_t(\vec k).
\end{eqnarray*}
For $k$ on shell we set
\begin{eqnarray*}
a_t(k)&=&\frac{1}{\sqrt{(2\pi)^{3}}}\Bigg(
\sqrt{\frac{\varepsilon (\vec k)}{2}}\phi_t(\vec k)
+\frac{\i}{\sqrt{2\varepsilon (\vec k)}}\pi_t(\vec k)\Bigg),\\
a_t^*(k)&=&\frac{1}{\sqrt{(2\pi)^{3}}}\Bigg(
\sqrt{\frac{\varepsilon (\vec k)}{2}}\phi_t^*(\vec k)
-\frac{\i}{\sqrt{2\varepsilon (\vec k)}}\pi_t^*(\vec k)\Bigg).
\end{eqnarray*}

We have the equations of motion
\begin{eqnarray*}
\dot a_t(k)&=&-\i \varepsilon(\vec k)a_t( k),\\
\dot{{ a}}^*_t( k)&=&\i \varepsilon(\vec k)a_t^*( k).
\end{eqnarray*}
We will usually write $a(k)$, $ a^*(k)$ instead of $a_0(k)$, $ a_0^*(k)$, so that
\begin{eqnarray}
a(k):&=&
\int\frac{\d\vec x}{\sqrt{(2\pi)^{3}}}\e^{-\i\vec k\vec x}\Bigg(
\sqrt{\frac{\varepsilon (\vec k)}{2}}\phi(0,\vec x)
+\frac{\i}{\sqrt{2\varepsilon (\vec k)}}\pi(0,\vec x)\Bigg),\label{dou1}
\\
a^*(k)&=&
\int\frac{\d\vec x}{\sqrt{(2\pi)^{3}}}\e^{\i\vec k\vec x}\Bigg(
\sqrt{\frac{\varepsilon (\vec k)}{2}}\phi(0,\vec x)
-\frac{\i}{\sqrt{2\varepsilon (\vec k)}}\pi(0,\vec x)\Bigg)\label{dou2}
. \end{eqnarray}
Thus
\begin{eqnarray}
 a_t(k)&=&\e^{-\i t \varepsilon(\vec k)}a( k),\notag\\
{ a}^*_t( k)&=&\e^{\i t\varepsilon(\vec k)}a^*( k);\notag\\
\{a(k),H\}&=&-\i\varepsilon(\vec k)a( k),\label{qas1}\\
\{a^*(k),H\}&=&\i\varepsilon(\vec k)a^*( k);\label{qas2}\\
\{a(k), a( k')\}=\{ a^*(k), a^*( k')\}&=&0,\label{qas3}\\
\{a( k),a^*( k')\}&=&-\i\delta(\vec k-\vec
k').\label{qas4} \end{eqnarray}

The fields can be written as
%
\begin{eqnarray*}
\phi(x)&=&
\int\frac{\d\vec{k}}{\sqrt{(2\pi)^3}\sqrt{2\varepsilon (\vec k)}}
 \left(\e^{\i kx}a( k)+
\e^{-\i kx}a^*( k)\right),\\
\pi(x)&=&\int\frac{\d\vec{k}\sqrt{\varepsilon(\vec k)}}{\i\sqrt{(2\pi)^3}\sqrt2} \left(\e^{\i
  kx} a(k)-
\e^{-\i kx}a^*(k)\right).
\end{eqnarray*}

\nowastrona

$a(k)$, $ a^*(k)$ 
diagonalize simultaneously the Hamiltonian, momentum and
symplectic form:
\begin{eqnarray*}
H&=&\int\d\vec k \varepsilon (\vec k) a^*(k) a( k),\\
\vec P&=&\int\d\vec k \vec k a^*(k) a( k),\\
\i\omega&=&\int\d\vec k  a^*(k)\wedge a(k).\end{eqnarray*}

With
 $\zeta_1,\zeta_2\in\cY_{\KG}$, the last identity is the shorthand for
\[\i\zeta_1\omega\zeta_2=
\int\left(\bar{\langle a(k)|\zeta_1\rangle}\langle
  a(k)|\zeta_2\rangle
-\langle a(k)|\zeta_1\rangle\bar{\langle a(k)|\zeta_2\rangle}\right)\d\vec{k}.
\]


\nowastrona

\subsubsection{Plane waves}
Let $k\in\rr^{1,3}$ satisfy  $k^0=\varepsilon(\vec k)$.
A {\em plane wave} $|k)$ is defined as
\beq (x|k)=\frac{1}{\sqrt{(2\pi)^3}\sqrt{2\varepsilon(\vec k)}}
\e^{\i kx}.\label{planewave1aa}\eeq
Note that, following Dirac, we denote plane waves using the ``ket notation'' 
$|k)$ when they appear on the right of a bilinear
form.
We also write $(x|k)$ for the evaluation of $|k)$ at the point $x\in\rr^{1,3}$.
Plane waves are solutions of the Klein-Gordon equation which are not space compact.

If a plane wave appears on the left, we employ the ``bra notation'', which implies
 an
additional complex conjugation:
\begin{eqnarray*}
(k|x)=\bar{(x|k)} &=&\frac{1}{\sqrt{(2\pi)^3}\sqrt{2\varepsilon(\vec k)}}
\e^{-\i kx}.\label{planewave1aa-}\\
\end{eqnarray*}
Note that in the neutral case we use only {\em positive frequency} plane waves, corresponding to $k^0>0$.


Let $k=(\varepsilon(\vec k),\vec k)$, $k'=(\varepsilon(\vec k'),\vec k')$.
\begin{eqnarray*}
\i(\bar{k}|\omega|k')=\i(k|\omega|\bar{k'})&=&0,\\
-\i(\bar{k}|\omega|\bar{k'})=\i(k|\omega|k')&=&\delta(\vec k-\vec k').\end{eqnarray*}
$a(k)$ and $a^*(k)$, defined in (\ref{dou1}) and (\ref{dou2}) will be
 called {\em plane wave functionals}. They can be expressed as
\begin{eqnarray*}
a(k)&=&
\i \phi\lpar |k)\rpar\\
&=&
\i\int
\Bigl(\partial_t{(k|}0,\vec x)\phi(0,\vec x)-
(k|0,\vec x) \pi(0,\vec
x)\Bigr)\d \vec x,
\\
a^*(k)
&=&-\i \phi\lpar \bar{|k)}\rpar\\
&=&
-\i\int
\Bigl(\partial_t\bar{(k|0,\vec x)}\phi(0,\vec x)-
\bar{(k|0,\vec x)} \pi(0,\vec
x)\Bigr)\d \vec x
. \end{eqnarray*}



\nowastrona

The fields can be written in terms of plane waves functionals as
%
\begin{eqnarray*}
\phi(x)
&=&
\int\left((x|k)a(k)+\bar{
(x|k)}a^*(k)\right)\d\vec k.
\end{eqnarray*}

So far we used only the real space $\cY_\KG$. We can complexify it and extend $a(k)$ to $\cc\cY_\KG$ by complex linearity.
 Every $\zeta\in\cc\cY_{\KG}$  satisfies
\begin{eqnarray}
\langle a(k)|\zeta\rangle&=&\i(k|\omega\zeta,\label{dfd1}\\
\bar{\langle a(k)|\bar\zeta\rangle}&=&-\i\bar{(k|}\omega\zeta,\label{dfd2}\\
\zeta(x)&=&\int\frac{\d\vec{k}}{\sqrt{(2\pi)^3}\sqrt{2\varepsilon (\vec k)}}
 \left(\e^{\i kx}\langle a( k)|\zeta\rangle +
\e^{-\i kx}\bar{\langle a( k)|\bar\zeta\rangle}\right).\label{pafa}\end{eqnarray}


\nowastrona

\nowastrona

\nowastrona
\subsubsection{Positive frequency space}

 $\cW_\KG^{(\pm)}$ will denote the subspace of $\cc\cY_\KG$
 consisting of {\em positive},
resp. {\em negative frequency
 solutions}, that is,
\begin{eqnarray*}
\cW_\KG^{(+)}&:=&\{g\in\cc\cY_\KG\ :\ 
 \bar{(k|}\omega g=0\},\\
\cW_\KG^{(-)}:=\bar{\cW_\KG^{(+)}}&=&\{g\in\cc\cY_\KG\ :\ 
 (k|\omega g=0\}.
\end{eqnarray*}

\nowastrona

In other words,  $\cW_\KG^{(+)}$ is the subspace of $\cc\cY_\KG$ that consists of functions of the form
\[g(x)=\int\frac{\d\vec{k}}{\sqrt{(2\pi)^3}\sqrt{2\varepsilon (\vec k)}}
\e^{\i kx}\langle a( k)|g\rangle.\]
For $g_1,g_2\in\cW_\KG^{(+)}$ we define the scalar product
\beq(g_1|g_2):=\i\bar g_1\omega g_2=
\int\bar{\langle a(k)|g_1\rangle}\langle a(k)|g_2\rangle\d\vec{k}.\label{wewe}\eeq
The {\em Hilbert space of positive energy solutions}
is denoted
 $\cZ_\KG$, and is the completion of $\cW_\KG^{(+)}$ in this scalar
product.

Note that
\[\langle a(k)|g\rangle =(k|g),\ \ g\in\cZ_\KG.\]
We can identify $\cZ_\KG\simeq L^2(\rr^3)$ and   rewrite (\ref{wewe}) as
\beq(g_1|g_2)=
\int\bar{(k|g_1)}(k|g_2)\d\vec{k}.\label{wewe8}\eeq

$\rr^{1,3}\rtimes  O^\uparrow(1,3)$ leaves $\cZ_\KG$ invariant.

We have a natural identification of $\cY_\KG$ with
$\cW_\KG^{(+)}$. Indeed, if $\zeta\in\cY_\KG$ is given by
(\ref{pafa}), then we can project it onto $\cW_\KG^{(+)}$ obtaining
\beq
\zeta^{(+)}(x)=\int\frac{\d\vec{k}}{\sqrt{(2\pi)^3}\sqrt{2\varepsilon (\vec k)}}
 \e^{\i kx}\langle a( k)|\zeta\rangle .\label{pafa1}\eeq
This identification allows us to define a real scalar product on $\cY_\KG$:
\begin{eqnarray}
\nonumber\langle\zeta_1|\zeta_2\rangle_\cY&:=&\Re(\zeta_1^{(+)}|\zeta_2^{(+)}).
\end{eqnarray}
We can compute explicitly this scalar product:
\begin{eqnarray}\langle\zeta_1|\zeta_2\rangle_\cY
&=&\int\int\dot\zeta_1(0,\vec x)(-\i)D^{(+)}(0,\vec x-\vec y)
\dot\zeta_2(0,\vec y)\d\vec x\d\vec y\label{derive9}\\
&&+
\int\int\zeta_1(0,\vec x)(-\Delta_{\vec x}+m^2)(-\i)D^{(+)}(0,\vec x-\vec y)
\zeta_2(0,\vec y)\d\vec x\d\vec y\nonumber.\end{eqnarray}

\nowastrona

 \subsubsection{Quantization}

Let us describe the quantization of the Klein-Gordon
equation, following the formalism
 of quantization of neutral bosonic systems \cite{DeGe}. 
 We will use the ``hat'' to denote the quantized objects.



\nowastrona

We want to construct $\cH,\hat H,\Omega$ satisfying the standard
requirements of QM  (1)-(3) and a self-adjoint operator valued
distribution \beq
\rr^{1,3}\ni x\mapsto \hat\phi(x),\label{distro}\eeq
such that, with $\hat\pi(x):=\dot{\hat\phi}(x)$,
 \ben
\item 
$(-\Box+m^2)\hat\phi(x)=0,$
\item
$
[\hat\phi(0,\vec x),\hat\phi(0,\vec y)]=[\hat\pi(0,\vec
  x),\hat\pi(0,\vec y)]=0$,\\
$
[\hat\phi(0,\vec x),\hat\pi(0,\vec y)]=\i\delta(\vec x-\vec y).$
\item 
$\e^{\i t\hat H}\hat\phi(x^0,\vec x)\e^{-\i t\hat H}=\hat\phi(x^0+t,\vec x).$
\item 
$\Omega$ is cyclic for $\hat\phi(x)$.\een

\nowastrona

The above problem has a  solution, which is
essentially unique.
Indeed, let $\cH,\hat H, \Omega$, $\rr^{1,3}\ni x\mapsto\hat\phi(x),\hat\pi(x)$ solve the above problem. Decorating (\ref{dou1}) and (\ref{dou2}) with hats leads to the definitions of two operator valued distributions Hermitian conjugate to one another:
\begin{eqnarray}
\hat a(k):&=&
\int\frac{\d\vec x}{\sqrt{(2\pi)^3}}\e^{-\i\vec k\vec x}\Bigg(
\sqrt{\frac{\varepsilon (\vec k)}{2}}\hat\phi(0,\vec x)
+\frac{\i}{\sqrt{2\varepsilon (\vec k)}}\hat\pi(0,\vec x)\Bigg),\label{dou1+}
\\
\hat a^*(k)&=&
\int\frac{\d\vec x}{\sqrt{(2\pi)^3}}\e^{\i\vec k\vec x}\Bigg(
\sqrt{\frac{\varepsilon (\vec k)}{2}}\hat\phi(0,\vec x)
-\frac{\i}{\sqrt{2\varepsilon (\vec k)}}\hat\pi(0,\vec x)\Bigg)\label{dou2+}
. \end{eqnarray} 
Using (2) and (3) we obtain the quantized versions of (\ref{qas1})-(\ref{qas4}): 
\begin{eqnarray}
[\hat a(k),\hat H]&=&\varepsilon(\vec k)\hat a( k),\label{qas1+}\\{}
[\hat a^*(k),\hat H]&=&-\varepsilon(\vec k)\hat a^*( k),\label{qas2+}\\{}
[\hat a(k), \hat a( k')]=[\hat a^*(k), \hat a^*( k')]&=&0,\label{qas3+}\\{}
[\hat a( k),\hat a^*( k')]&=&\delta(\vec k-\vec
k').\label{qas4+} \end{eqnarray}

 $\hat H\Omega=0$  implies $\hat H \hat a(k)\Omega=-\varepsilon(\vec k)\hat a(k)\Omega$. But $\hat H\geq0$. Thus we should assume
 \beq \hat a(k)\Omega=0.\label{anni}\eeq

By (4), $\Omega$ is cyclic for $\hat a(k)$ and $\hat a^*(k)$. Using (\ref{qas1+}), (\ref{qas4+}) and (\ref{anni}) we see that  $\Omega$ is cyclic just for  $\hat a^*(k)$. In other words, 
$\cH$ is spanned by vectors of the form
\[\int\Psi(\vec k_1,\dots,\vec k_n)\hat a^*(k_1)\cdots \hat a^*(k_n)\Omega
\d \vec k_1\cdots\d \vec k_n.\]
Using again (\ref{qas1+}), (\ref{qas4+}) and (\ref{anni}) we see that the scalar product of two vectors $\Psi$, $\Psi'$ is zero if $n\neq n'$, and otherwise it is
\[(\Psi|\Psi')
=\int \bar{\Psi(\vec k_1,\dots,\vec k_n)}
\Psi'(\vec k_1,\dots,\vec k_n)
\d \vec k_1\cdots\d \vec k_n.\]
Therefore, $\cH$ can be identified with 
 $\Gamma_\s\big(L^2(\rr^3)\big)$, $\Omega$ with the Fock
vacuum, $\hat a^*(k)$ with the creation operators in the ``physicist's notation'', the quantum field is
\[\hat\phi(x):=
\int\frac{\d\vec{k}}{\sqrt{(2\pi)^3}\sqrt{2\varepsilon (\vec k)}}
\left(\e^{\i  kx}\hat a( k)+\e^{-\i kx}\hat a^*( k)\right),\]
finally, the quantum Hamiltonian and the momentum are
\begin{eqnarray*}
\hat H&:=&\int \hat a^*( k)\hat a( k)\varepsilon (\vec k)\d\vec{k},\\
\vec {\hat P}&:=&\int \hat a^*( k)\hat a( k)\vec k\d\vec{k}.
\end{eqnarray*}

By  (\ref{wewe}) we can identify
 $L^2(\rr^3)$  with the positive frequency Hilbert space $\cZ_\KG$. Using the   ``mathematician's notation'' on the right we can write
\begin{eqnarray} \hat a^*(k)&=&\hat a^*\big(|k)\big).
\end{eqnarray}


\nowastrona

Note that the whole
$\rr^{1,3}\rtimes  O^\uparrow(1,3)$ is  unitarily implemented on $\cH$
by
 $U(y,\Lambda):=\Gamma\Bigl(r_{(y,\Lambda)}\Big|_{\cZ_\KG}\Bigr)$:
\[U(y,\Lambda)\hat\phi(x)U(y,\Lambda)^*=\hat\phi\bigl((y,\Lambda)x\bigr).\]
This is true even though 
 we only required that  time translations are implemented.

We have
\[ [\hat\phi(x),\hat\phi(y)]=-\i D(x-y).\]

For  $f\in C_{\rm c}^\infty(\rr^{1,3},\rr)$ set
\beq \hat \phi[f]:=\int f(x)\hat\phi(x)\d x.\label{qrt3}\eeq
(\ref{qrt3}) satisfy the 
  Wightman axioms  with
$\cD:=\Gamma_\s^\fin(\cZ_\KG)$. 

For an open  set $\cO\subset \rr^d$ we set
\[\fA(\cO):=\{\exp(\i \hat \phi[f])\ :\ f\in C_{\rm c}^\infty(\cO,\rr)\}''.\]
The algebras $\fA(\cO)$ satisfy the Haag-Kastler axioms.

\nowastrona
\nowastrona
\subsubsection{Quantization in terms of smeared fields}

There exists an alternative equivalent formulation of the quantization
program, which uses  smeared fields instead of point fields, which
may better appeal to some people.

Again, we want to construct $\cH,\hat H,\Omega$ satisfying the standard requirements of QM (1)-(3).
Instead of
(\ref{distro}) we look for a linear function
\[\cY_{\KG}\ni\zeta\mapsto\hat\phi\lpar\zeta\rpar\]
with values in self-adjoint operators such that
\ben\item
\beq
[\hat\phi\lpar\zeta_1\rpar,\hat\phi\lpar\zeta_2\rpar]
=\i\zeta_1\omega \zeta_2.\label{poi1a}\eeq
\item
\[\hat\phi\lpar r_{(t,\vec0)}\zeta\rpar
=\e^{\i t\hat H}\hat\phi\lpar\zeta\rpar\e^{-\i t\hat H}.\]
\item 
$\Omega$ is cyclic for the algebra generated by
$ \hat\phi\lpar\zeta\rpar$.\een

\nowastrona

One can pass between these two versions of the quantization  by
\beq
\hat\phi\lpar\zeta\rpar =\int\left(-\dot\zeta(t,\vec x)\hat\phi(t,\vec x)+
\zeta(t,\vec x)\hat\pi(t,\vec x)\right)\d\vec{x}.
\label{pzk2}\eeq

\nowastrona

\subsubsection{Quantization in terms of $C^*$-algebras}

Let us mention yet another  equivalent
 approach to quantization,
using the language of {\em $C^*$-algebras}. 

Let $\CCR(\cY_{\KG})$ denote the {\em (Weyl) $C^*$-algebra of the CCR over $\cY_{\KG}$}. By definition, it is
generated by $W(\zeta)$, $\zeta\in\cY_{\KG}$, such that
\[W(\zeta_1)W(\zeta_2)=\e^{-\i\frac{\zeta_1\omega\zeta_2}{2}}W(\zeta_1+\zeta_2),\
\ \ W(\zeta)^*=W(-\zeta).\]
$\rr^{1,3}\rtimes  O^\uparrow(1,3)$ acts on $\CCR(\cY_{\KG})$ by $*$-automorphisms
defined by
\[\hat
r_{(y,\Lambda)}\left(W(\zeta)\right):=W\left(r_{(y,\Lambda)}(\zeta)\right).\] 
We are looking for a cyclic representation of this algebra with the
time evolution  generated by a positive Hamiltonian. 

\nowastrona

The solution is provided by the
 state on  $\CCR(\cY_{\KG})$ defined by
\begin{eqnarray*}
\psi\big(W (\zeta)\big)
&=&\exp\Big(-\frac12\langle\zeta|\zeta\rangle_{\cY}\Big).
\end{eqnarray*}
Let $(\cH_\psi,\pi_\psi,\Omega_\psi)$ be 
 the  GNS representation generated by the state $\psi$.
Then this representation has the required properties. 
$\cH_\psi$ can be identified with $\Gamma_\s(\cZ_\KG)$ and
the fields are related to the Weyl operators by
\[\pi_\psi(W(\zeta))=\e^{\i\hat\phi\lpar\zeta\rpar}.\]

\subsubsection{Two-point functions}

Note the identities
\begin{eqnarray}
(\Omega|\hat \phi(x)\hat\phi(y)\Omega)
&=&-\i D^{(+)}(x-y),\label{derive1}\\
(\Omega|\T(\hat\phi(x)\hat\phi(y))\Omega)
&=&-\i D^{\rm c}(x-y).
\end{eqnarray}

In fact,
\begin{eqnarray*}
(\Omega|\hat\phi(x)\hat\phi(y)\Omega)&=&
\int\int
\frac{\d\vec{k}\d\vec{k}'}{(2\pi)^3\sqrt{2\varepsilon }\sqrt{2\varepsilon '}}\e^{\i kx-\i k'y}(\Omega|\hat a(
k)\hat a^*( k')\Omega) \\&=&
\int\frac{\d\vec{k}}{(2\pi)^32\varepsilon(\vec k) }\e^{\i k(x-y)}\\&=&-\i D^{(+)}(x-y);\\
(\Omega|\T(\hat\phi(x)\hat\phi(y))\Omega)&=&
\theta(x^0-y^0)(\Omega|\hat\phi(x)\hat\phi(y)\Omega)+
\theta(y^0-x^0)(\Omega|\hat\phi(y)\hat\phi(x)\Omega)\\&=&
-\i
\theta(x^0-y^0)D^{(+)}(x-y)
-\i\theta(y^0-x^0)D^{(+)}(y-x)\\&=&-\i D^{\rm c}(x-y).
\end{eqnarray*}

\nowastrona

  (\ref{derive1}) implies the following identities for spacetime
 smeared fields and Weyl operators:
\begin{eqnarray}(\Omega|\hat\phi
[f]^2\Omega)&=&-\i\int\int f(x)D^{(+)}(x-y)f(y)\d x\d
  y,\label{derive4}\\
(\Omega|\e^{\i\hat\phi[f]}\Omega)&=&\exp\left(\frac{\i}{2}
\int\int f(x)D^{(+)}(x-y)f(y)\d x\d
  y\right).\label{derive5}\end{eqnarray}

Differentiating if needed (\ref{derive1}) with respect time we obtain the equal time correlation functions expressed as real symmetric kernels:
\begin{eqnarray}
(\Omega|\hat\phi(0,\vec x)\hat\phi(0,\vec y)\Omega)&=&-\i D^{(+)}(0,\vec x-\vec
  y),\label{derive2} \\
(\Omega|\hat\phi(0,\vec x)\hat\pi(0,\vec y)\Omega)&=&0,\label{derive7}\\
(\Omega|\hat\pi(0,\vec x)\hat\pi(0,\vec y)\Omega)&=&\i\partial_t^2
 D^{(+)}(0,\vec x-\vec
  y)\nonumber\\
&=&-\i(-\Delta_{\vec x}+m^2)
 D^{(+)}(0,\vec x-\vec
  y).
\label{derive3} 
\end{eqnarray}
This yields the identities for spatially smeared fields and Weyl operators, where the scalar product $\langle\cdot|\cdot\rangle_\cY$ on $\cY_\KG$ was introduced in 
(\ref{derive9}):
\begin{eqnarray}
(\Omega|\hat\phi\lpar\zeta\rpar^2\Omega)&=&
-\i\int\int\dot\zeta(0,\vec x)D^{(+)}(0,\vec x-\vec y)
\dot\zeta(0,\vec y)\d\vec x\d\vec y
\notag\\
&&-\i\int\int\zeta(0,\vec x)(-\Delta_{\vec x}+m^2)D^{(+)}(0,\vec
x-\vec y)
\zeta(0,\vec y)\d\vec x\d\vec y\notag\\
&=&\langle\zeta|\zeta\rangle_{\cY},\label{derive}\\
(\Omega|\e^{\i\hat\phi\lpar\zeta\rpar}\Omega)&=&
\exp\big(-\frac12\langle\zeta|\zeta\rangle_{\cY}\big).\label{derive6}
\end{eqnarray}
\nowastrona

\subsection{Neutral scalar bosons with a linear source}

\nowastrona

\subsubsection{Classical fields}

We go back to the classical theory.
 The  fields studied in
the previous subsection will be called {\em free  fields}. We change
slightly the notation:  free classical fields will be now 
denoted by $\phi_\fr(x)$, $\pi_\fr(x)$.
Clearly, they satisfy
\begin{eqnarray}
 (-\Box+m^2)\phi_\fr(x)&=&0,\label{sfa2a}\\
 \pi_\fr(x)&=&\dot\phi_\fr(x).\notag\end{eqnarray}

Fix a function
      \beq\rr^{1,3}\ni x\mapsto j(x)\in\rr,\label{source1}\eeq
which will be called the
      {\em (external) linear source}. 
In most of this subsection we will assume that 
(\ref{source1}) is Schwartz.
The {\em interacting fields}
      satisfy the equation
\begin{eqnarray} (-\Box+m^2)\phi(x)&=&-j(x),\label{sfa2}\\
 \pi(x)&=&\dot\phi(x).\label{poisson9}\end{eqnarray}
We also require  that the interacting fields  have the same equal-time Poisson brackets as the free fields:
\begin{eqnarray}
\{\phi(t,\vec x),\phi(t,\vec y)\}=\{\pi(t,\vec x),\pi(t,\vec y)\}&=&0,\nonumber
\\
\{\phi(t,\vec x),\pi(t,\vec y)\}&=&\delta(\vec x-\vec y).
\label{poisson2}\end{eqnarray}

\nowastrona
There are several, usually equivalent, ways to introduce interacting fields. 

One way is to treat them as functionals on the space of solutions to the free Klein-Gordon equation,  $\cY_\KG$. 
We can
 demand in addition that
 \begin{eqnarray}
\phi(\vec x)&:= \phi_\fr(0,\vec x)&=\phi(0,\vec x),\nonumber\\
\pi(\vec x)&:= \pi_\fr(0,\vec
x)&=\pi(0,\vec x).\label{pfp}\end{eqnarray}
This condition determines the field $\phi(x)$  uniquely:
\begin{eqnarray}\phi(x)&:=&\phi_\fr(x)\label{phii}\\
&&+\int\big( D^+(x-y)\theta(y^0)+ D^-(x-y)\theta(-y^0)\big)j(y)\d
  y.\nonumber\end{eqnarray}

Let us mention some alternative ways to define the interacting fields $\phi(x)$. First of all, there is nothing special about the time $t=0$ in (\ref{pfp}) -- we can replace it with any $t=t_0$. Alternatively, we can  demand
 \begin{eqnarray*} \lim_{t\to\infty}\left(\phi_\fr(t,\vec x)-\phi(t,\vec x)\right)=0,&& \lim_{t\to\infty}\left(\pi_\fr(t,\vec
x)-\pi(t,\vec x)\right)=0,\\\hbox{or }\ \ \ 
\lim_{t\to-\infty}\left(\phi_\fr(t,\vec x)-\phi(t,\vec x)\right)=0,&& \lim_{t\to-\infty}\left(\pi_\fr(t,\vec
x)-\pi(t,\vec x)\right)=0.\end{eqnarray*}
Another possibility is to introduce $\cY_\KG(j)$, the space of smooth real space-compact solutions of
\beq (-\Box+m^2)\zeta(x)=-j(x),\label{sfa2-}\eeq
and define $\phi(x)$ by
\[\langle \phi(x)|\zeta\rangle:=\zeta(x),\ \ \ 
\zeta\in\cY_\KG(j).\]

\nowastrona

\subsubsection{Lagrangian and Hamiltonian formalism}
\label{Lagrangian formalism and the stress-energy tensor}
We can obtain the equations (\ref{sfa2})
as the Euler-Lagrange equations for the {\em Lagrangian density} 
\begin{eqnarray}
\cL(x)=&
-\frac12\partial_\mu\phi(x)\partial^\mu\phi(x)-\frac12 m^2\phi(x)^2-j(x)\phi(x).
\label{lagra1-}\end{eqnarray}
The conjugate variable is
\begin{eqnarray*}
\pi(x)&:=&\frac{\partial
  \cL}{\partial\phi_{,0}( x)}=\partial_0\phi( x),\end{eqnarray*} 
just as in the free case.

The Legendre transformation
leads to the {\em  Hamiltonian density }
\begin{eqnarray*}
\cH(x)\ &:=&\frac12
\left(\pi( x)^2+\big(\vec\partial\phi( x)\big)^2 +m^2\phi(x)^2\right)+j(x)\phi(x).
\end{eqnarray*}
and the  (time-dependent) {\em Hamiltonian} 
\begin{eqnarray}
H(t)&=&\int \cH(t,\vec x)\d\vec x\label{phii1}\\
&=&\int\d\vec x\Big(\frac12
\pi(t,\vec x)^2+\frac12\big(\vec\partial\phi(t,\vec x)\big)^2 +\frac{m^2}{2}\phi(t,\vec x)^2+j(t,\vec x)\phi(t,\vec x)\Big).
\nonumber
\end{eqnarray}

The  Hamiltonian
 generates the dynamics:
\begin{eqnarray}\label{phii2}
\dot\phi(t,\vec x)=\{\phi(t,\vec x),H(t)\},\ \ \ 
\dot\pi(t,\vec x)=\{\pi(t,\vec x),H(t)\}.\end{eqnarray}

In (\ref{phii1}) and (\ref{phii2}), $H(t)$, $\phi(t,\vec x)$ and $\pi(t,\vec x)$ should be understood as the functions of the initial conditions at $t=0$. Therefore, we use the Lagrangian description  (see Subsubsect.
 \ref{Heisenberg picture}).

 We can also introduce the Hamiltonian in the Eulerian description, 
which  is convenient for quantization.
It uses the fields  $\phi(\vec x)$ and $\pi(\vec x)$ introduced in (\ref{pfp}):
\begin{eqnarray*}
H_\Eul(t)\label{phii3}
&=&\int\d\vec x\Big(\frac12
\pi(\vec x)^2+\frac12\big(\vec\partial\phi(\vec x)\big)^2 +\frac{m^2}{2}\phi(\vec x)^2+j(t,\vec x)\phi(\vec x)\Big).
\end{eqnarray*}

\nowastrona

\subsubsection{Quantization}
We will  use the notation $\hat\phi_\fr(x)$ for the {\em free quantum
  fields} studied in the previous subsection.
We are now looking for  {\em interacting quantum fields}
 $\hat \phi(x)$  satisfying
\beq (-\Box+m^2)\hat\phi(x)=-j(x).\label{sfa3c}\eeq
We also set \beq \hat\pi(x):=\dot{\hat\phi}(x)\label{poisson11}\eeq
  and require the equal time commutation relations
\begin{eqnarray}
[\hat\phi(t,\vec x),\hat\phi(t,\vec y)]=[\hat\pi(t,\vec x),\hat\pi(t,\vec y)]&=&0,\nonumber
\\{}
[\hat\phi(t,\vec x),\hat\pi(t,\vec y)]&=&\i\delta(\vec x-\vec y).
\label{poisson21}\end{eqnarray}

We would like to solve (\ref{sfa3c}) and (\ref{poisson21}) in terms of free fields.
That means, we are looking for  $\hat \phi(x)$  on
 the Hilbert space of the free Klein-Gordon fields, $\Gamma_\s(\cZ_\KG)$. 
 We will in addition demand that the interacting and free fields at time $t=0$ coincide:
\begin{eqnarray}\hat\phi(\vec x):&=&\hat\phi(0,\vec x)=\hat\phi_\fr(0,\vec x),\nonumber\\
\hat\pi(\vec x):&=&\hat\pi(0,\vec x)=\hat\pi_\fr(0,\vec x).
\label{time0}\end{eqnarray}
Clearly, the unique solution is obtained by decorating (\ref{phii}) with hats:
\begin{eqnarray}\hat\phi(x)&:=&\hat\phi_\fr(x)\nonumber\\
&&\hspace{-7ex}+\int \Big(D^+(x-y)\theta(y^0)+ D^-(x-y)\theta(-y^0)\Big)j(y)\d
  y.\label{sfa3c0}\end{eqnarray}
It can be written as
\beq\hat\phi(t,\vec x)=
\Texp\left(-\i\int_t^0 \hat H(s)\d s\right)\hat \phi(0,\vec
x)\Texp\left(-\i\int_0^t \hat H(s)\d s\right), \label{hamio1}\eeq
where the   the Schr\"odinger picture Hamiltonian is
\beq
\hat H(t):=\int\d \vec x\ {:}{\left(\frac12\hat\pi^2( \vec
  x)+\frac12\partial_i\hat\phi( 
\vec x)\partial_i\hat\phi( \vec x)
+\frac{m^2}{2}\hat \phi^2(\vec x)+j(t,\vec x)\hat\phi(\vec x)\right)}{:}.\label{hamio}\eeq
Note that $\hat H(t)$ is obtained from $H_\Eul(t)$ by the Wick quantization \ref{Weyl and Wick quantization}, which is expressed by decorating the fields with ``hats'' and putting the ``double dots''.

In principle, one could replace $\hat H(t)$ by $\hat H(t)+C(t)$ for any real function $t\mapsto C(t)$. The choice that we made  satisfies 
\beq (\Omega|\hat H(t)\Omega)=0,\ \ t\in\rr.\label{arbi}\eeq Condition (\ref{arbi}) is quite arbitrary -- the vector $\Omega$ is the ground state of of the free Hamiltonian at time zero -- in particular, it depends on the choice  $t=0$ in (\ref{time0}).


We also have the 
interaction picture Hamiltonian
\begin{eqnarray}
\hat H_\Int(t)&=&\int j(t,\vec x)\hat\phi_\fr(t,\vec x)\d \vec x.
\label{hamio4}\end{eqnarray}

\subsubsection{Operator valued source}

So far we assumed that $j(x)$ is a c-number. Most of the
 formalism works, at least formally, for {\em operator valued
   sources}.
 The main additional difficulty is the need to distinguish between the source in various pictures.

Let us start with the Schr\"odinger picture. Let $\rr^{1,3}\ni
x\mapsto \hat j(x)$ be an operator-valued function (or distribution) that commutes with time zero fields:
\[[\hat\phi(\vec x),\hat j(t,\vec y)]=[\hat\pi(\vec x),\hat j(t,\vec y)]=0,
\ \ \ \ \vec x,\vec y\in\rr^3,\ \ t\in\rr.\]
Define the {\em the Schr\"odinger picture Hamiltonian} $\hat H(t)$
by  (\ref{hamio}), where
$j(x)$ is replaced by $\hat j(x)$.

Then we  define the {\em  the Heisenberg picture fields} $\hat \phi(x)$,  $\hat \pi(x)$, as in (\ref{hamio1}).
We also have the {\em source in the  Heisenberg  picture}
\begin{eqnarray*}
\hat j_\Heis(t,\vec x)
&:=&
\Texp\left(-\i\int_t^0 \hat H(s)\d s\right)\hat j(t,\vec x
)\Texp\left(-\i\int_0^t \hat H(s)\d s\right)
\end{eqnarray*}
having the commutation relations
\begin{eqnarray*}
[\hat\phi(t,\vec x),\hat j_\Heis(t,\vec y)]=[\hat\pi(t,\vec x),\hat j_\Heis(t,\vec y)]&=&0.
\end{eqnarray*}
The Klein-Gordon equation (\ref{sfa3c})  and the relation  (\ref{sfa3c0}) generalize:
\begin{eqnarray}
(-\Box+m^2)\hat\phi(x)& =&-\hat j_\Heis(x),\label{sfa3c+}\\
\hat\phi(x)&:=& \hat\phi_\fr(x)\nonumber\\
&&\hspace{-12ex}+\int\Big( D^+(x-y)\theta(y^0)+ D^-(x-y)\theta(-y^0)\Big)\hat j_\Heis(y)\d
  y.\nonumber\end{eqnarray}

We can also introduce the {\em source in the interaction picture}
\begin{eqnarray*}
\hat j_\Int(t,\vec x)&:=&\e^{\i t\hat H_\fr}\hat j(t,\vec x)\e^{-\i t\hat H_\fr},
\end{eqnarray*}
satisfying the commutation relations
\begin{eqnarray*}
[\hat\phi_\fr(t,\vec x),\hat j_\Int(t,\vec y)]=[\hat\pi_\fr(t,\vec x),\hat j_\Int(t,\vec y)]&=&0,
\ \ \ \ \vec x,\vec y\in\rr^3,\ \ t\in\rr.
\end{eqnarray*}
The {\em interaction picture Hamiltonian} is
\begin{eqnarray*}
\hat H_\Int(t)&=&\int \hat j_\Int (t,\vec x)\hat\phi_\fr(t,\vec x)\d \vec x,\end{eqnarray*}
which is obtained from (\ref{hamio4}) by replacing
$j(t,\vec x)$  with $\hat j_\Int(t,\vec x)$.
\nowastrona
\subsubsection{Scattering operator}
We go back to a $c$-number source $j(x)$. The interaction picture Hamiltonian written in terms of creation and annihilation operators equals
\begin{eqnarray*}
\hat H_\Int(t)&=&
\int\frac{\d \vec k}{\sqrt{(2\pi)^3}\sqrt{2\varepsilon(\vec
    k)}}\Big(\e^{-\i t\varepsilon(\vec k)}\bar{j(t,\vec k)}\hat a(k)+\e^{\i t\varepsilon(\vec k)}
j(t,\vec k)\hat a^*(k)\Big).\end{eqnarray*} 
The
 {\em  scattering operator} (\ref{scatte1}) can be computed exactly. 
On the level of creation and annihilation operators it acts as
\begin{eqnarray}
\hat S\hat a^*(k)\hat S^*&=&\hat a^*(k)-\i\frac{ j\big(\varepsilon(\vec k),\vec k\big)}
{\sqrt{(2\pi)^3}\sqrt{2\varepsilon(\vec k)}},\label{implo1}\\
\hat S\hat a(k)\hat S^*&=&\hat a(k)+\i\frac{ \bar{j\big(\varepsilon(\vec k),\vec k\big)}}
{\sqrt{(2\pi)^3}\sqrt{2\varepsilon(\vec k)}}.\label{implo2}
\end{eqnarray}
We have an explicit formula:
\begin{eqnarray}\hat S
&=&\exp\left(\frac{\i}{2}\int\frac{|j(k)|^2}
{(k^2+m^2-\i0)}\frac{\d k}{(2\pi)^4}\right)\label{sca0}\\
&&\times\exp\left(-\i\int\frac{{j(\varepsilon(\vec k),\vec k)}}{\sqrt{2\varepsilon(\vec k)}}\hat
a^*( k)\frac{\d\vec k}{\sqrt{(2\pi)^3}}\right) 
\exp\left(-\i\int\frac{\bar{j(\varepsilon(\vec k),\vec k)}}{\sqrt{2\varepsilon(\vec k)}}\hat
a( k)\frac{\d\vec k}{\sqrt{(2\pi)^3}}\right). \notag
\end{eqnarray}
To see this, we insert $f(t,\vec k):=-\e^{\i t\varepsilon(\vec k)}\frac{j(t,\vec k)}{\sqrt{(2\pi)^3}\sqrt{2\varepsilon(\vec k)}}$ into
 (\ref{bch}). In particular, the exponent of (\ref{bch}) becomes:
\begin{eqnarray*}
&&-\int
\int\frac{\e^{\i(t_2-t_1)\varepsilon(\vec k)}\theta(t_1-t_2)\bar{j(t_1,\vec k)}j(t_2,\vec k)\d t_1\d t_2}{2\varepsilon(\vec k)}\frac{\d\vec k}{(2\pi)^3}
\\
&=&\i\int
\int\frac{|j(k^0,\vec k)|^2\d k^0\d \vec k}{2\varepsilon(\vec k)\big(\varepsilon(\vec k)-k^0-\i0\big)}\frac{\d k^0}{2\pi}\frac{\d\vec k}{(2\pi)^3}
\\
&=&\frac{\i}{2}\int
\int\frac{|j(k)|^2}{2\varepsilon(\vec k)}\Big(\frac{1}{(\varepsilon(\vec k)-k^0-\i0)}+\frac{1}{(\varepsilon(\vec k)+k^0-\i0)}\Big)\frac{\d k}{(2\pi)^4}
\\
&=&\frac{\i}{2}\int
\int\frac{|j(k)|^2}{\big(\varepsilon(\vec k)^2-(k^0)^2-\i0\big)}\frac{\d k}{(2\pi)^4}.
\end{eqnarray*}
Note that we used $\bar{j(k^0,\vec k)}=j(-k^0,-\vec k)$

For distinct $k_1,\dots,k_n$ on shell, set
\[|k_n,\dots,k_1):=\hat a^*(k_n)\cdots \hat a^*(k_1)\Omega.\]
Matrix elements of the scattering operator between such vectors are
 called {\em scattering amplitudes}:
\begin{eqnarray}\label{sca2}
&&\left(k_1^+,\dots,k_{n^+}^+|\,\hat S\, |k_{n^-}^-,\dots
,k_1^-\right)\\
&=&\exp\left(\frac{\i}{2}\int\frac{|j( k)|^2}
{(k^2+m^2-\i0)}\frac{\d k}{(2\pi)^4}\right)\frac{(-\i)^{n^++n^-}}{\sqrt{(2\pi)^{3(n^++n^-)}}}\notag
\\
&&\times
\frac{j(\varepsilon(\vec k_1^+),\vec k_1^+)}{\sqrt{2\varepsilon(\vec k_1^+)}}
\cdots
\frac{j(\varepsilon(\vec k_{n^+}^+),\vec k_{n^+}^+)}{\sqrt{2\varepsilon(\vec
    k_{n^+}^+)}}
\frac{\bar{j(\varepsilon(\vec k_{n^-}^-),\vec k_{n^-}^-)}}{\sqrt{2\varepsilon(\vec
    k_{n^-}^-)}}\cdots 
\frac{\bar{j(\varepsilon(\vec k_1^-),\vec
    k_1^-)}}{\sqrt{2\varepsilon(\vec k_1^-)}}.\notag 
\end{eqnarray}

 \nowastrona
\subsubsection{Green's functions}

Recall that the {\em $N$-point 
 Green's function} is defined for $x_{N},\dots,x_1$
as follows:
\begin{eqnarray}\notag
&&\langle\hat\phi(x_{N})\cdots\hat\phi(x_{1})\rangle\\
&:=&
\left(\Omega^+|\T\bigl(\hat\phi(x_N)\cdots\cdots\hat\phi(x_1
)\bigr)\Omega^-\right),\label{sca3} 
\end{eqnarray}
where
\begin{eqnarray*}
\Omega^\pm&:=&\lim_{t\to\pm\infty}
\Texp\left(-\i\int_0^t \hat H(s)\d s\right)\Omega\\
&=&
\Texp\left(-\i\int_0^{\pm\infty} \hat H_\Int(s)\d s\right)\Omega.
\end{eqnarray*}

One can organize Green's functions in terms of the {\em generating function}:
\begin{eqnarray}Z(f)\notag
&=&\sum_{N=0}^\infty\int\cdots\int 
\frac{(-\i)^N}{N!}\langle\hat\phi(x_{N})\cdots\hat\phi(x_{1})\rangle
f(x_N)\cdots f(x_1)\d x_N\cdots\d x_1\\
&=&
\left(\Omega^+\Big|\Texp\left(-\i\int_{-\infty}^{\infty} \Big(\hat H(t)+\int f(t,\vec x)\hat \phi(\vec x)\d \vec x\Big)\d t \right)\Omega^-\right)\notag\\
&=&
\left(\Omega\Big|\Texp\left(-\i\int_{-\infty}^{\infty} \hat H_\Int(t)\d t-\i\int f(x)\hat \phi_\fr(x)\d x\right)\Omega\right)\notag\\
&=&\exp\left(\frac{\i}{2}\int\frac{|j(k)+f(k)|^2}{(k^2+m^2-\i0)}\frac{\d
k}{(2\pi)^4}\right) \label{sca4}
.\end{eqnarray}
One can retrieve Green's functions from the generating function:
\beq
\langle\hat\phi(x_N)\cdots\hat\phi(x_1)\rangle=\i^N\frac{\partial^N}{\partial f(x_N)\cdots \partial f(x_1)}
Z(f)\Big|_{f=0}.\eeq
We introduce also {\em amputated Green's functions}:
\begin{eqnarray}
&&\langle\hat\phi(k_n)\cdots\hat\phi(k_1)\rangle_\amp\notag\\
&=&(k_n^2+m^2)\cdots(k_1^2+m^2)
\langle\hat\phi(k_n)\cdots\hat\phi(k_1)\rangle.\label{sca5}
\end{eqnarray}
Amputated Green's functions can be used to compute scattering amplitudes:
\begin{eqnarray}
&&\left(k_{n^+}^+,\dots,k_1^+|\,\hat S\, |k_{n^-}^-,\dots
,k_1^-\right)\label{sca6}\\
&=&\frac{
\langle\hat\phi(k_1^+)\cdots\hat\phi(k_{n^+}^+)\hat\phi(-k_{n^-}^-)\cdots
\hat\phi(-k_1^-)\rangle_\amp}{\sqrt{(2\pi)^{3(n^++n^-)}}
\sqrt{2\varepsilon(k_1^+)}
\cdots
\sqrt{2\varepsilon(k_{n^+}^+)}
\sqrt{2\varepsilon(k_{n^-}^-)}\cdots
\sqrt{2\varepsilon(k_1^-)}},\notag
\end{eqnarray}
where all $k_i^\pm$ are  on shell.
\nowastrona

\subsubsection{Path integral formulation}

Recall that the generating function equals
\begin{eqnarray}
Z(f)&=&\exp\left(\frac{\i}{2}\int\big(j(x)+f(x)\big)D^{\rm c}(x-y)\big(j(y)+f(y)\big)\d x\d y\right)\nonumber\\
&=&\exp\left(\frac{\i}{2}\int\frac{|j(k)+f(k)|^2}{(k^2+m^2-\i0)}
\frac{\d k}{(2\pi)^4}\right).\label{generi}\end{eqnarray}

We have the following expressions for the {\em action integral}:
\begin{eqnarray*}
\int \cL_\fr(x)\d x&=&-\int\frac12\phi(x)(-\Box+m^2)\phi(x)\d x,\\
\int \left(\cL(x)-\phi(x)f(x)\right)\d x&=&\int\cL_\fr(x)\d x-\int\phi(x)(j(x)+f(x))\d x.
\end{eqnarray*}
Consider heuristically the space of all (off-shell) configurations with the Lebesgue measure 
$\lpi_x\d\phi(x).$
Physicists like to rewrite (\ref{generi}) as
\begin{eqnarray}\label{causapro}
Z(f)&=&\frac{\int\lpi_x\d\phi(x)\exp\left(\i\int\big(\cL(x)-f(x)\phi(x)\big)\d x\right)}
{\int\lpi_x\d\phi(x)\exp\left(\i\int\cL_\fr(x)\d x\right)},
\end{eqnarray}
which follows by basic rules of Gaussian integrals.
Note that strictly speaking (\ref{causapro}) is ambiguous, since 
 $D^{\rm c}$, the causal propagator, is only one of many inverses (Green's functions) of
$-\Box+m^2$. The choice of the causal propagator
is an additional convention
that is  not explicitly contained in the expression
(\ref{causapro}). 
\subsubsection{Feynman rules}

Perturbative expansions can be organized in terms of Feynman
diagrams. The prescriptions how to draw Feynman diagrams and to
evaluate them are called {\em Feynman rules}.
We restrict ourselves to Feynman rules in the momentum space.

We have 1 kind of {\em lines} and 1 kind of {\em vertices}.
 At each vertex  just
one line ends. Vertices are denoted by solid dots. Lines have no distinguished orientation. However,
when we fix the orientation of a line, we can  associate to it a
 momentum $k$.

Diagrams for Green's functions, in addition to {\em internal lines} 
 have {\em external lines} ending with {\em insertion vertices}, which will be denoted by 
small circles. To compute Green's functions we do as follows:

\ben \item We draw all possible Feynman diagrams. More precisely, we put
$N$ dots  for insertion vertices, labelled $1,\dots,N$. We put $n$ dots, labelled $1,\dots,n$, for interaction vertices. Then we connect them
in all possible allowed ways. The expression for the diagram is then 
 divided by $n!$.
\item 
To each  vertex 
 we associate the factor  $-\i j(k)$, where $k$ is the momentum flowing towards this vertex.
\item To each line we associate the propagator
\[-\i D_\fr^{\rm c}(k)=\frac{-\i}{k^2+m^2-\i0}.\]
\item For internal lines we integrate over the variables with the measure
$\frac{\d^4 k}{(2\pi)^4}$.
\een

\begin{figure}[!h]
\centering
\includegraphics{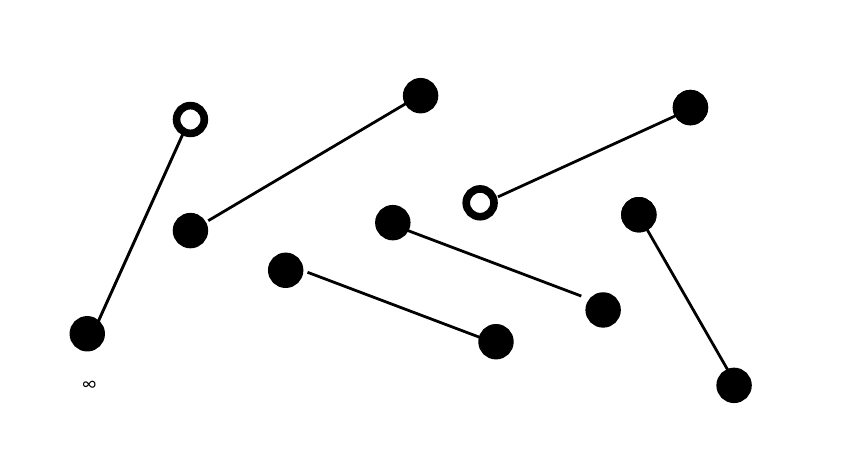}
\label{diag-1}
\caption{Diagram for Green's function.}
\end{figure}

\nowastrona

\begin{figure}[ht]
\centering
\includegraphics{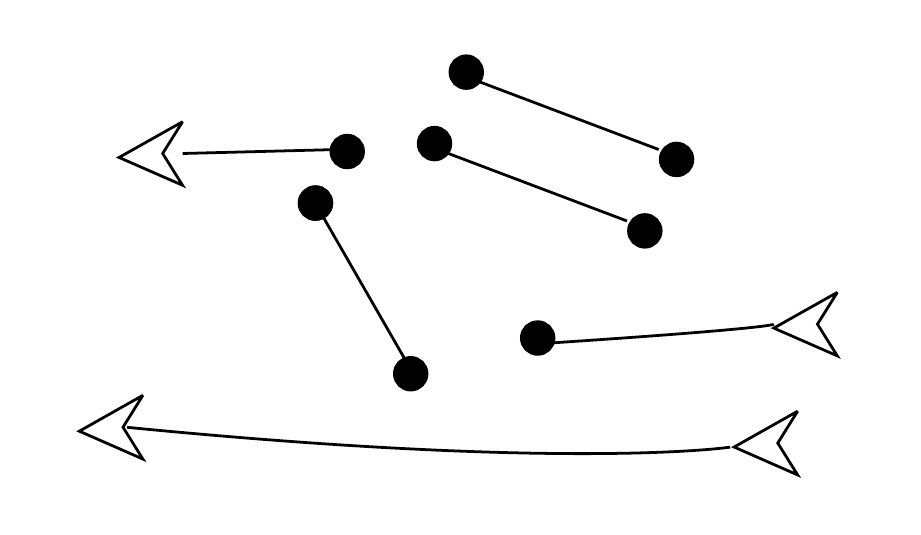}
\label{diag-2}
\caption{Diagram for a scattering amplitude.}
\end{figure}

Diagrams used to compute  scattering amplitudes with $N^-$ incoming and $N^+$
outgoing particles are similar to  diagrams 
for $N^-+N^+$-point Green's functions, except that instead of insertion vertices we have  {\em incoming and outgoing particles}.
For the   incoming lines, $-k$ are on-shell, for the
 outgoing lines, $k$ is on-shell.
The rules are changed only
concerning the external lines:
\begin{romanenumerate}\item
To each incoming external line we associate
 $\frac{1}{\sqrt{(2\pi)^32\varepsilon (\vec k)}}$.
\item  To each outgoing external line we associate
 $\frac{1}{\sqrt{(2\pi)^32\varepsilon (\vec k)}}$.
\end{romanenumerate}

\subsubsection{Vacuum energy}

Let $D$ denote the value of  the (unique) connected diagram with no external lines.
We have
\[\log(\Omega|\hat S\Omega)=\frac{\i}{2}\int\frac{|j(k)|^2}{(k^2+m^2-\i0)}\frac{\d k}{(2\pi)^4}
=\frac{D}{2}.\]

\begin{figure}[!h]
\centering
\includegraphics{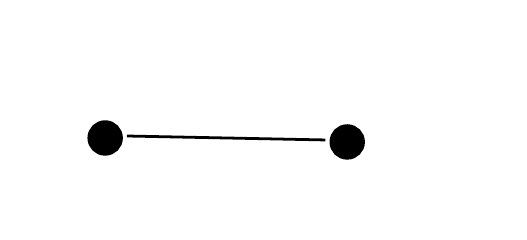}
\label{diag-3}
\caption{Diagram for vacuum energy.}
\end{figure}

We can derive it diagrammatically as follows. At the order $2m$ there are 
$\frac{(2m)!}{2^mm!}$ pairings. Hence
\[(\Omega|\hat S\Omega)=\sum_{m=0}^\infty
\frac{1}{(2m)!}\frac{(2m)!}{2^mm!}D^m=\exp(D/2).\]

\nowastrona

\subsubsection{Problems with the scattering operator}

$\hat S$ can be ill defined. 

First of all, \beq\Re\frac{1}{2}\int\frac{|j(k)|^2}
{(k^2+m^2-\i0)}\frac{\d k}{(2\pi)^4}=\frac{1}{2}\int\frac{|j(k)|^2}
{(k^2+m^2)}\frac{\d k}{(2\pi)^4}\label{pha}\eeq
 can be infinite. This is not a very serious problem. (\ref{pha}) is
 responsible only for the phase of the scattering amplitude and does
 not influence  scattering cross-sections.

We can try to remedy the problem by an apropriate
renormalization of the phase.
 In particular, in the case of a stationary source or, more generally,
 a source travelling with a constant velocity, we can use
the adiabatic switching and the Gell-Mann and Low construction to
obtain a meaningful  scattering operator.  
We will describe this construction in the next subsubsections.

The problem with $\hat S$  is
more serious if \[\Im \frac{1}{2}\int\frac{|j(k)|^2}
{(k^2+m^2-\i0)}\frac{\d k}{(2\pi)^4}=
\frac{1}{2}\int\frac{|j( \varepsilon(\vec k),\vec k)|^2}
{\varepsilon(\vec k)}\frac{\d \vec k}{(2\pi)^3}
\] is infinite. Then no unitary operator $\hat S$ satisfies the relations (\ref{implo1}) and (\ref{implo2}), see Thm \ref{transla}. The scattering
operator is ill defined.  However, as we describe in Subsubsect. 
\ref{Inclusive cross-section}, also in this situation there is a pragmatic solution --  we  can define  {\em inclusive cross-sections}.

Note that if $k\mapsto j(k)$ is Schwartz, then $\hat S$ is well-defined, even if $m=0$.

\subsubsection{Energy shift and scattering theory for a stationary source}

Suppose now that the source does not depend on time and is given by a
Schwartz function
$\rr^3\ni\vec x\mapsto j(\vec x)$. Then we have the time-independent Hamiltonian
\[\hat H=\int{:}\Big(\frac12\hat\pi(\vec x)^2+\frac12\big(\vec\partial\hat\phi(\vec
x)\big)^2+\frac{m^2}{2}\hat\phi(\vec x)^2+j(\vec x)\hat\phi(\vec x)\Big){:}\d\vec x.\]
By the method of completing the square (\ref{hov1}) we compute the infinum of $\hat H$:
\[E=-\frac{1}{2}\int j(\vec x)\frac{\e^{-m|\vec x-\vec y|}
}{4\pi|\vec
    x-\vec y|}j(\vec y)\d \vec x\d\vec y.\]

 Obviously, the standard  M{\o}ller operators $\hat S^\pm$ (\ref{gell1}) are ill defined and we need to use the Gell-Mann--Low construction.
 Let us replace $j$ with $j_\epsilon^\pm(t,\vec x):=\theta(\pm x)j(\vec x)\e^{-\epsilon|t|}$. Its Fourier transform is $j_\epsilon^\pm(k^0,\vec k)=\mp\i(k\mp\i\epsilon)^{-1}j(\vec k)$, Inserting this into
(\ref{sca0}) we obtain
\begin{eqnarray*}
\hat S_\epsilon^\pm
&=&\exp\Big(-\int\frac{j(\vec
  k)}{\sqrt{2\varepsilon(\vec k)}(\varepsilon(\vec k)\mp\i\epsilon)}a^*( k)\frac{\d\vec k}{\sqrt{(2\pi)^3}}\Big)\\ 
&&\times\exp\Big(
\int\frac{\bar{j(\vec k)}}{\sqrt{2\varepsilon(\vec k)}(\varepsilon(\vec k)\pm\i\epsilon)}a( k)\frac{\d\vec k}{\sqrt{(2\pi)^3}}\Big)\\&&\times
\exp\Big(-\frac{1}{2}\int\frac{|j(\vec
  k)|^2}{2\varepsilon(\vec k)(\varepsilon(\vec k)^2+\epsilon^2)
}\frac{\d\vec k}{(2\pi)^3}+
\frac{\i}{\epsilon2}\int\frac{|j(\vec
  k)|^2}{2(\varepsilon(\vec k)^2+\epsilon^2)
}\frac{\d\vec k}{(2\pi)^3}
\Big).
\end{eqnarray*}
Note that the phase of $\hat S_\epsilon^\pm$ behaves as $O(\epsilon^{-1})$. In the definition of $S_\GL^\pm$ we take this phase away and put $\epsilon\searrow0$,
 see  (\ref{qrowa}). We obtain
\begin{eqnarray*}
\hat S_\GL^\pm
&=&\exp\Big(-\int\frac{j(\vec
  k)}{\sqrt{2\varepsilon(\vec k)^3}}a^*( k)\frac{\d\vec k}{\sqrt{(2\pi)^3}}\Big)\\ 
&&\times\exp\Big(
\int\frac{\bar{j(\vec k)}}{\sqrt{2\varepsilon(\vec k)^3}}a( k)\frac{\d\vec k}{\sqrt{(2\pi)^3}}\Big)\\&&\times
\exp\Big(-\frac{1}{2}\int\frac{|j(\vec
  k)|^2}{2\varepsilon(\vec k)^3
}\frac{\d\vec k}{(2\pi)^3}\Big).
\end{eqnarray*}

If $m>0$ or if $\int j(\vec x)\d\vec x=0$, then $\hat H$ has a ground
state and the operators $\hat S_\GL^\pm$ are well defined. We have
\[\hat S_\GL^\pm \hat H_\fr=(\hat H-E)\hat S_\GL^\pm.\]

Note a somewhat disappointing feature: $\hat S_\GL^+=\hat S_\GL^-$, and hence the scattering operator $\hat S_\GL:=\hat S_\GL^{+*}\hat S_\GL^-=\one$ is trivial.

If $m=0$ and
 $\int j(\vec x)\d\vec x\neq 0$, then $\hat H$ has no ground state
(even though it is bounded from below) and 
the operators $\hat S_\GL^\pm$ are ill defined.

\subsubsection{Travelling source}
\label{Travelling source}

Consider now a  source of a profile given by a function $q\in C_{\rm
  c}^\infty(\rr^3)$ travelling with velocity $\vec v$. 
That means
\beq j(t,\vec x)=q(\vec x-t\vec v).\label{trav}\eeq
We note that the Fourier transform of (\ref{trav}) in the spatial variables 
equals
\[j(t,\vec k)=q(\vec k)\e^{-\i t\vec v\vec k}.\]
The interaction picture Hamiltonian becomes
\begin{eqnarray*}
\hat H_{\epsilon,\Int}(t)&=&
\int\frac{\d \vec k}{\sqrt{(2\pi)^3}\sqrt{2\varepsilon(\vec
    k)}}\Big(\e^{-\i t\big(\varepsilon(\vec k)-\vec v\vec k\big)}\bar{q(\vec k)}\hat a(k)+\e^{\i t\big(\varepsilon(\vec k)-\vec v\vec k\big)}
q(\vec k)\hat a^*(k)\Big).\end{eqnarray*} 
This is the interaction picture Hamiltonian for a time-independent perturbation where 
the 1-particle energy  $\varepsilon(\vec k)$ is replaced by $\varepsilon(\vec k)-\vec v\vec k$.

\nowastrona

We use the   
Gell-Mann--Low type adiabatic switching, so
that we replace $j$ by
\[j_\epsilon(t,\vec x):=\e^{-|t|
\epsilon}j(t,\vec x).\] 

We slightly generalize the Gell-Mann--Low M{\o}ller operators:
\begin{eqnarray*}
\hat S_\GL^\pm&=&\lim_{\epsilon\searrow0}\frac{|(\Omega|\hat S_\epsilon^\pm\Omega)|}{
(\Omega|\hat S_\epsilon^\pm\Omega)}\hat S_\epsilon^\pm\\
&=&\exp\Big(-\int\frac{q(\vec k)}{\sqrt{2\varepsilon(\vec k)}(\varepsilon(\vec
  k)-\vec v\vec k)}a^*( k)\frac{\d\vec k}{\sqrt{(2\pi)^3}}\Big)\\
&&\times\exp\Big(
\int\frac{\bar{q(\vec k)}}{\sqrt{2\varepsilon(\vec k)}(\varepsilon(\vec
  k)-\vec v\vec k)}a( k)\frac{\d\vec k}{\sqrt{(2\pi)^3}}\Big)\\&&\times
\exp\Big(-\frac{1}{2}\int\frac{|q(\vec k)|^2}{2\varepsilon(\vec k)(\varepsilon(\vec
  k)-\vec v\vec k)^2}\frac{\d\vec k}{(2\pi)^3}\Big).
\end{eqnarray*}

Note that if $|v|<1$ (if the source is  slower than light) and $m>0$, then
$\hat S_\GL^+=\hat S_\GL^-$ are well defined unitary operators.

 If $m=0$ and $\int q(\vec
x)\d\vec x\neq0$, then the infrared problem shows up: $\hat S_\GL^\pm$ are ill defined.
\nowastrona

It is interesting to  assume that the source has a different asymptotics in
the future and in the past. For simplicity, suppose that the change occurs
sharply at time $t=0$ and consider
\[j(t,\vec x)=\left\{\begin{array}{ll}q_-(\vec x-t\vec v_-),&t<0,\\
q_+(\vec x- t\vec v_+),& t>0.
\end{array}\right.\]
The following operator can be used as a scattering operator:
\begin{eqnarray}
&&\hat S_\GL^{+*}\hat S_\GL^-\label{gell-mann1}\\
&=&\exp\Bigg(\int\frac{1}{\sqrt{2\varepsilon(\vec k)}}\Big(\frac{q_+(\vec k)}{(\varepsilon(\vec
  k)-\vec v_+\vec k)}-
\frac{q_-(\vec k)}{(\varepsilon(\vec
  k)-\vec v_-\vec k)}\Big)
\hat a^*( k)\frac{\d\vec k}{\sqrt{(2\pi)^3}}\Bigg)\notag\\
&&\times
\exp\Bigg(\int\frac{1}{\sqrt{2\varepsilon(\vec k)}}\Big(-\frac{\bar q_+(\vec k)}{(\varepsilon(\vec
  k)-\vec v_+\vec k)}+
\frac{\bar q_-(\vec k)}{(\varepsilon(\vec
  k)-\vec v_-\vec k)}\Big)
\hat a( k)\frac{\d\vec k}{\sqrt{(2\pi)^3}}\Bigg)\notag\\
&&\times
\exp\Bigg(-\frac{1}{2}\int\frac{1}{2\varepsilon(\vec k)}\Big(\frac{|q_+(\vec k)|^2}{(\varepsilon(\vec
  k)-\vec v_+\vec k)^2}+
\frac{|q_-(\vec k)|^2}{(\varepsilon(\vec
  k)-\vec v_-\vec k)^2}\notag\\&&\hspace{5ex}-
\frac{2\bar{q_+(\vec k)}q_-(\vec k)}{(\varepsilon(\vec
  k)-\vec v_+\vec k)(\varepsilon(\vec
  k)-\vec v_-\vec k)}
\Big)\frac{\d\vec k}{(2\pi)^3}\Bigg).\notag
\end{eqnarray}
Let $m=0$. Then (\ref{gell-mann1}) is ill defined if\ben\item
 $\int q_+(\vec x)\d\vec x\neq \int
q_-(\vec x)\d\vec x$,\\
or
 \item 
 $\int q_+(\vec x)\d\vec x= \int
q_-(\vec x)\d\vec x\neq0$ and $v_+\neq v_-$. \een

\nowastrona
Alternatively, we can introduce
 first the scattering operator $\hat S_\epsilon$ with the adiabatically switched interaction. Then we can define another kind of scattering operator by taking $\epsilon\searrow0$ and renormalizing the phase:
\begin{eqnarray}
&&\lim_{\epsilon\searrow0}\frac{|(\Omega|\hat S_\epsilon\Omega)|}{
(\Omega|\hat S_\epsilon\Omega)}\hat S_\epsilon\label{gell-mann2}\\
&=&\exp\Bigg(\int\frac{1}{\sqrt{2\varepsilon(\vec k)}}\Big(\frac{q_+(\vec k)}{(\varepsilon(\vec
  k)-\vec v_+\vec k)}-
\frac{q_-(\vec k)}{(\varepsilon(\vec
  k)-\vec v_-\vec k)}\Big)
\hat a^*( k)\frac{\d\vec k}{\sqrt{(2\pi)^3}}\Bigg)\notag\\
&&\times
\exp\Bigg(\int\frac{1}{\sqrt{2\varepsilon(\vec k)}}\Big(-\frac{\bar q_+(\vec k)}{(\varepsilon(\vec
  k)-\vec v_+\vec k)}+
\frac{\bar q_-(\vec k)}{(\varepsilon(\vec
  k)-\vec v_-\vec k)}\Big)
\hat a( k)\frac{\d\vec k}{\sqrt{(2\pi)^3}}\Bigg)\notag\\
&&\times
\exp\Bigg(-\frac{1}{2}\int\frac{1}{2\varepsilon(\vec k)}\Big|\frac{q_+(\vec k)}{(\varepsilon(\vec
  k)-\vec v_+\vec k)}-
\frac{q_-(\vec k)}{(\varepsilon(\vec
  k)-\vec v_-\vec k)}\Big|^2\frac{\d\vec k}{(2\pi)^3}\Bigg).\notag
\end{eqnarray}
Note that (\ref{gell-mann1}) and  (\ref{gell-mann2}) differ only by a phase. 
 (\ref{gell-mann2}) is given by 
 (\ref{sca0})  where we replace
\[\int\frac{|j(k)|^2}
{(k^2+m^2-\i0)}\frac{\d k}{(2\pi)^4}\]
with
\begin{eqnarray*}
\Im\int\frac{|j(k)|^2}
{(k^2+m^2-\i0)}\frac{\d k}{(2\pi)^4}
&=&
\int\frac{|j(\varepsilon(\vec k),\vec k)|^2}{2\varepsilon(\vec
  k)}\frac{\d\vec k}{(2\pi)^3}.\end{eqnarray*}
Here, $j(k)$ is the Fourier transform of the source (\ref{trav}):
 \begin{eqnarray*}
j(k)&=&\int j(t,\vec x)\e^{-\i \vec k \vec x+\i k^0t}\d \vec x\d t\\
&=&-\frac{\i q_+(\vec k)}{\vec k\vec v_+-k^0-\i0}+
\frac{\i q_-(\vec k)}{\vec k \vec v_--k^0+\i0}.
\end{eqnarray*}

If  we do not like the adiabatic switching approach we can directly
define the 
M{\o}ller operators by removing the (possibly infinite) phase shift from
(\ref{sca0}).

\subsubsection{Scattering cross-sections}

We consider again an arbitrary source term $j$.
Given on-shell momenta of incoming particles $k_{n^-}^-,\dots
,k_1^-$ and outgoing particles
$k_1^+,\dots,k_{n^+}^+$ we can compute the scattering cross-section for the corresponding process, or actually its density w.r.t. the Lebesgue measure
$\d k_1^+\cdots\d k_{n^+}^+$:

\begin{eqnarray*}&&
\sigma\left(k_1^+,\dots,k_{n^+}^+;k_{n^-}^-,\dots
,k_1^-\right)\\&=&\left|\left(k_1^+,\dots,k_{n^+}^+|\,\hat S\, |k_{n^-}^-,\dots
,k_1^-\right)\right|^2\\
&=&\frac{1}{\sqrt{(2\pi)^{(n^++n^-)}}}\exp\Bigl(-\int\frac{|j( \varepsilon(\vec k),\vec k)|^2}
{2\varepsilon(\vec k)}\frac{\d\vec k}{(2\pi)^3}\Bigr)
\\
&&\times
\frac{|j(\varepsilon(\vec k_1^+),\vec k_1^+)|^2}{2\varepsilon(\vec k_1^+)}
\cdots
\frac{|j(\varepsilon(\vec k_{n^+}^+),\vec k_{n^+}^+)|^2}{2\varepsilon(\vec
    k_{n^+}^+)}
\frac{|j(\varepsilon(\vec k_{n^-}^-),\vec k_{n^-}^-)|^2}{2\varepsilon(\vec
    k_{n^-}^-)}\cdots 
\frac{|j(\varepsilon(\vec k_1^-),\vec k_1^-)|^2}{2\varepsilon(\vec k_1^-)}.
\end{eqnarray*}
Note that the crosssections are zero if $m=0$ and $\int j(x)\d x\neq0$. In the next subsubsection we describe how to cope with this problem.

\subsubsection{Inclusive cross-section}
\label{Inclusive cross-section}

Let $\delta>0$. The 1-particle Hilbert space can be split as
$\cZ=\cZ_{<\delta}\oplus\cZ_{>\delta}$ corresponding to the {\em soft
  momenta} $|\vec k|<\delta$ and {\em hard momenta} $|\vec k|>\delta$. Clearly,
\[\Gamma_\s(\cZ)\simeq\Gamma(\cZ_{<\delta})\otimes\Gamma(\cZ_{>\delta}),\ \ \ \ \ 
\Omega\simeq \Omega_{<\delta}\otimes \Omega_{>\delta}.
\]

Assume first that $m>0$ and the scattering operator is computed as above. Clearly, the scattering operator and scattering cross-sections factorize:
\[\hat S\simeq \hat S_{<\delta}\otimes \hat S_{>\delta},\ \ \ 
\sigma=\sigma_{<\delta}\ \sigma_{>\delta}.\]

 More precisely, let \beq |\vec q_1^+|,\dots,|\vec q_{m^+}^+|,
|\vec q_1^-|,\dots,|\vec q_{m^-}^-|<\delta.\label{inclo1}\eeq
Then we have the {\em soft scattering cross-sections}
\begin{eqnarray*}&&
\sigma_{<\delta}\left(q_1^+,\dots,q_{m^+}^+;q_{m^-}^-,\dots
,q_1^-\right)\\&=&\left|\left(q_1^+,\dots,q_{m^+}^+|\,\hat S_{<\delta}\, |q_{m^-}^-,\dots
,q_1^-\right)\right|^2\\
&=&\frac{1}{\sqrt{(2\pi)^{(m^++m^-)}}}\exp\Bigl(-\int_{|\vec q|<\delta}\frac{|j( \varepsilon(\vec q),\vec q)|^2}
{2\varepsilon(\vec q)}\frac{\d \vec q}{(2\pi)^3}\Bigr)
\\
&&\times
\frac{|j(\varepsilon(\vec q_1^+),\vec q_1^+)|^2}{2\varepsilon(\vec q_1^+)}
\cdots
\frac{|j(\varepsilon(\vec q_{m^+}^+),\vec q_{m^+}^+)|^2}{2\varepsilon(\vec
    q_{m^+}^+)}
\frac{|j(\varepsilon(\vec q_{m^-}^-),\vec q_{m^-}^-)|^2}{2\varepsilon(\vec
    q_{m^-}^-)}\cdots 
\frac{|j(\varepsilon(\vec q_1^-),\vec q_1^-)|^2}{2\varepsilon(\vec q_1^-)}.
\end{eqnarray*}

Likewise, let \beq |\vec k_1^+|,\dots,|\vec k_{n^+}^+|,
|\vec k_1^-|,\dots,|\vec k_{n^-}^-|>\delta.\label{inclo2}\eeq
The corresponding {\em hard scattering cross-section} are
\begin{eqnarray*}&&
\sigma_{>\delta}\left(k_1^+,\dots,k_{n^+}^+;k_{n^-}^-,\dots
,k_1^-\right)\\&=&\left|\left(k_1^+,\dots,k_{n^+}^+|\,\hat S_{>\delta}\, |k_{n^-}^-,\dots
,k_1^-\right)\right|^2\\
&=&\frac{1}{\sqrt{(2\pi)^{(n^++n^-)}}}\exp\Bigl(-\int_{|\vec k|>\delta}\frac{|j( \varepsilon(\vec k),\vec k)|^2}
{2\varepsilon(\vec k)}\frac{\d \vec k}{(2\pi)^3}\Bigr)
\\
&&\times
\frac{|j(\varepsilon(\vec k_1^+),\vec k_1^+)|^2}{2\varepsilon(\vec k_1^+)}
\cdots
\frac{|j(\varepsilon(\vec k_{n^+}^+),\vec k_{n^+}^+)|^2}{2\varepsilon(\vec
    k_{n^+}^+)}
\frac{|j(\varepsilon(\vec k_{n^-}^-),\vec k_{n^-}^-)|^2}{2\varepsilon(\vec
    k_{n^-}^-)}\cdots 
\frac{|j(\varepsilon(\vec k_1^-),\vec k_1^-)|^2}{2\varepsilon(\vec k_1^-)}.
\end{eqnarray*}

We have
\begin{eqnarray}
&&\sigma_{>\delta}\left(k_1^+,\dots,k_{n^+}^+;k_{n^-}^-,\cdots
,k_1^-\right)\label{infro}\\
& =& 
\sigma\left(k_1^+,\dots,k_{n^+}^+;k_{n^-}^-,\cdots
,k_1^-\right)
\nonumber
\\
&&\nonumber+
\sum_{j=1}^\infty\int_{|\vec q_1|<\delta}\cdots\int_{|\vec q_j|<\delta}
\sigma\left(k_1^+,\dots,k_{n^+}^+,q_1,\dots,q_j;k_{n^-}^-,\cdots
,k_1^-\right)\d\vec q_1\cdots\d\vec q_j.
\end{eqnarray}
$\sigma_{>\delta}$ describes the
 experiment that does not measure
outgoing particles of momentum less than $\delta$ and in the incoming
state 
 there are no particles of momentum less than $\delta$.
Actually, we would have obtained 
the same scattering cross-section if the part of the incoming state
below the momentum $\delta$ was arbitrary. This is an example of an {\em inclusive cross-section} -- a cross-section which involves summing over many unobserved final states.

If $m\searrow0$, the {\em soft scattering operator} $\hat S_{<\delta}$ has no limit. All  $\sigma_{<\delta}$ go to zero. In fact, they are proportional to
\[\sigma_{<\delta}=\exp\Bigl(-\int_{|\vec q|<\delta}\frac{|j( \varepsilon(\vec q),\vec q)|^2}{2\varepsilon(\vec q)}\frac{\d \vec q}{(2\pi)^3}\Bigr)
.\]
The {\em hard scattering operator} $\hat S_{>\delta}$ and
$ \sigma_{>\delta}$
 have well defined limits and can have a physical meaning.

One can imagine various experimental scenarios that lead to different inclusive cross-sections. 
For example, imagine
 that our apparatus  does not detect  the details of an outgoing state if the total energy of soft particles is less than $\delta$. 
This leads to the following inclusive cross-section:
\begin{eqnarray}\notag
&&\sigma_{>\delta}^\app\left(k_1^+,\dots,k_{n^+}^+;k_{n^-}^-,\cdots
,k_1^-\right)\label{infro1}\ :=\ 
\sigma\left(k_1^+,\dots,k_{n^+}^+;k_{n^-}^-,\cdots
,k_1^-\right)
\\
&&\nonumber
+\sum_{j=1}^\infty\int_{\varepsilon(\vec q_1)+\cdots+
\varepsilon(\vec
q_j)<\delta}
\sigma\left(k_1^+,\dots,k_{n^+}^+,q_1,\dots,q_j;k_{n^-}^-,\cdots
,k_1^-\right)\d\vec q_1\cdots\d\vec q_j.
\end{eqnarray}
Note that both $\sigma_{>\delta}$ and $\sigma_{>\delta}^\app$ are proportional to one another:
\begin{eqnarray}\notag
&&\frac{
\sigma_{>\delta}^\app\left(k_1^+,\dots,k_{n^+}^+;k_{n^-}^-,\cdots
,k_1^-\right)}
{\sigma_{>\delta}\left(k_1^+,\dots,k_{n^+}^+;k_{n^-}^-,\cdots
,k_1^-\right)}\\
&:=&(\Omega_{<\delta}|\hat S_{<\delta}^*\one_{[0,\delta]}(\hat H_\fr)\hat S_{<\delta}\Omega_{<\delta})\ =\ 
\sigma_{<\delta}(;)\\
\notag&&\nonumber
+\sum_{j=1}^\infty\int_{\varepsilon(\vec q_1)+\cdots+
\varepsilon(\vec
q_j)<\delta}
\sigma_{<\delta}\left(q_1,\dots,q_j;\right)\d\vec q_1\cdots\d\vec q_j\nonumber.
\end{eqnarray}
This ratio is in practice not very interesting
 -- it contributes a common numerical factor to all scattering cross-sections for hard particles.

\subsection{Neutral scalar bosons with a mass-like perturbation}

\subsubsection{Classical fields}

A scalar field can be also perturbed by a mass-like perturbation. Classically, this is expressed by the equation
\beq\label{poisson8}
(-\Box +m^2)\phi(x)=-\kappa(x)\phi(x),\eeq
where $\rr^{1,3}\ni x\mapsto \kappa(x)$ is a given function. In most of this subsection we will assume that $\kappa$  is  Schwartz and $m>0$. We introduce also $\pi(x):=\dot\phi(x)$.
 
Let us define the corresponding retarded and advanced propagators as
the unique distributional solutions of
\beq
\big(-\Box_x +m^2+\kappa(x)\big)D^\pm(x,y)=\delta(x-y),\eeq
satisfying
\[\supp D^\pm\subset\{x,y\ :\ x\in J^\pm(y)\}.\]
We also generalize the Pauli-Jordan function:
\[D(x,y):=D^+(x,y)-D^-(x,y).\]
Note that 
\[\supp D\subset\{x,y\ :\ x\in J(y)\}.\]

The function $D$ can be used to solve the initial value problem of (\ref{poisson8}):
\begin{eqnarray}
\phi(t,\vec x)&=&-\int\partial_sD(t,\vec x,s,\vec y)\Big|_{s=0}\phi(0,\vec
y)\d \vec y\notag \\&&\ +\int  D(t,\vec x,0,\vec y)\pi(0,\vec
y)\d \vec y.\label{poio1+}
\end{eqnarray}

We would like to interpret the classical  field $\phi(x)$ satisfying 
(\ref{poisson8}) as a functional on the space $\cY_\KG$ coinciding with the free field at time $0$, as in (\ref{pfp}).
By (\ref{poio1+}), this  allows us to express uniquely the field $\phi$ in terms of the free field.

\subsubsection{Lagrangian and Hamiltonian formalism}
The  Lagrangian density
is
\[\cL(x)=-\frac{1}{2}\partial_\mu\phi(x)\partial^\mu\phi(x)-\frac12(m^2+\kappa(x))\phi(x)^2.\]
As in
Subsubsect.
\ref{Lagrangian formalism and the stress-energy tensor}, the variable conjugate to $\phi(x)$ is  $\pi(x)$.
We easily obtain the Hamiltonian density 
\[\cH(x)=\frac12\pi^2(x)+\frac12\big(\vec\partial\phi(x)\big)^2+\frac12(m^2+\kappa(x))\phi^2(x),\]so that 
the full Hamiltonian generating the dynamics is
\[H(t)=\int\cH(t,\vec x)\d\vec x.\]

\subsubsection{Dynamics in the interaction picture}
The classical interaction picture
Hamiltonian can be expressed in terms
of plane wave functionals:
\begin{eqnarray}\label{areleft}
H_\Int(t)&=&\frac12\int\kappa(t,\vec x)\phi_\fr^2(t,\vec x)\d \vec x\\
&=&\frac{1}{2}\int\frac{\d\vec k_1\d \vec k_2\kappa(t,\vec k_1+\vec k_2)}
{(2\pi)^3\sqrt{2\varepsilon(\vec k_1)}\sqrt{2\varepsilon(\vec k_2)}}
\Big(\e^{-\i t\varepsilon(\vec k_1)-\i t\varepsilon(\vec k_2)} a(-k_1) a(-k_2)\nonumber\\
&&\ \ \ +2\e^{\i t\varepsilon(\vec k_1)-\i t\varepsilon(\vec k_2)} a^*(k_1) a(-k_2)
+\e^{\i t\varepsilon(\vec k_1)+\i t\varepsilon(\vec k_2)}a^*(k_1) a^*(k_2)
\Big).\nonumber\end{eqnarray}
Consider the equations of motion 
in the interaction picture:
\begin{eqnarray*}
\dot a_{t}^*(k)
&=&
\{ a_{t}^*(k),H_\Int(t)\}\\
& =&
\i\int
\frac{\d\vec k_1\kappa(t,-\vec k+\vec k_1)}
{(2\pi)^3\sqrt{2\varepsilon(\vec k)}\sqrt{2\varepsilon(\vec k_1)}}\\&&\ \ \ \ 
\times\Big(\e^{-\i t\varepsilon(\vec k)-\i t\varepsilon(\vec k_1)}a_t(-k_1)+
\e^{-\i t\varepsilon(\vec k)+\i t\varepsilon(\vec k_1)}a_t^*(k_1)\Big),
\\
a_{0}^*(k)&=&
a^*(k).
\end{eqnarray*}
The solution of these equations at two times are related by a matrix
of the form \beq\left[\begin{array}{cc}p_{t_+,t_-}&q_{t_+,t_-}\\[1.1ex]
\bar{q_{t_+,t_-}}&\bar{p_{t_+,t_-}}\end{array}\right]\label{pada51}\eeq
 in the following way:
\[\left[\begin{array}{c}a_{t_+}^*(k)\\[1.8ex]a_{t_+}(k)\end{array}\right]
=\int\d\vec k_1
\left[\begin{array}{cc}p_{t_+,t_-}(k,k_1)&q_{t_+,t_-}(k,k_1)\\[1.8ex]
\bar{q_{t_+,t_-}(k,k_1)}&\bar{p_{t_+,t_-}(k,k_1)}\end{array}\right]
\left[\begin{array}{c}a_{t_-}^*(k_1)\\[1.8ex]a_{t_-}(k_1)\end{array}\right].
\]
 (\ref{pada51}) 
has a limit as $t_+,-t_-\to\infty$,
which can be called the {\em classical scattering operator}.

One can try to solve the equations of motion by iterations. The first iteration is often (at least in the quantum context) called the {\em Born approximation}, and it gives the following formula for the elements of (\ref{pada51}):
\begin{eqnarray*}
p_{t_+,t_-}^\Born(k,k_1)&=&\delta(\vec k-\vec k_1)+
\i\int_{t_-}^{t_+}
\d s\frac{\kappa(s,-\vec k+\vec k_1)}
{(2\pi)^3\sqrt{2\varepsilon(\vec k)}\sqrt{2\varepsilon(\vec k_1)}}
\e^{-\i s\varepsilon(\vec k)+\i s\varepsilon(\vec k_1)},\\
q_{t_+,t_-}^\Born(k,k_1)&=&\i\int_{t_-}^{t_+}\d s\frac{
\kappa(s,-\vec k+\vec k_1)}
{(2\pi)^3\sqrt{2\varepsilon(\vec k)}\sqrt{2\varepsilon(\vec k_1)}}
\e^{-\i s\varepsilon(\vec k)-\i s\varepsilon(\vec k_1)}.
\end{eqnarray*}

\subsubsection{Quantization}\label{shale2}
Again, we are looking for quantum fields $\rr^{1,3}\mapsto\hat
\phi(x)$ satisfying
\beq\label{poisson81}
(-\Box +m^2)\hat\phi(x)=-\kappa(x)\hat\phi(x),\eeq
 with the conjugate field $\hat\pi(x):=\dot{\hat\phi}(x)$ having the
equal time commutators (\ref{poisson21}), and coinciding with the free field at time $0$, as in (\ref{time0}).
The solution is
given by putting ``hats'' onto
 (\ref{poio1+}).

We would like to check whether the classical scattering operator and the
classical dynamics are implementable
in the Fock space for nonzero $\kappa$. By Thm \ref{shale}, we need to check the
{\em Shale condition}, that is, whether the off-diagonal elements of  (\ref{pada51}) are square
integrable. For simplicity, we will restrict ourselves to the Born
approximation; the higher order terms do not change the conclusion.

The verification of the Shale condition is easier for the scattering operator.
Consider
\beq
q_{\infty,-\infty}^\Born(k,k_1)=\i\int_{-\infty}^\infty\d s\frac{
\kappa(s,-\vec k+\vec k_1)}
{(2\pi)^3\sqrt{2\varepsilon(\vec k)}\sqrt{2\varepsilon(\vec k_1)}}
\e^{-\i s\varepsilon(\vec k)-\i s\varepsilon(\vec k_1)}.\label{squa}\eeq
Recall that  $\kappa$ is a Schwartz function. Therefore, we can
integrate  by parts as many times as we want:
\beq
q_{\infty,-\infty}^\Born(k,k_1)=\i^{n+1}\int_{-\infty}^\infty\d s\frac{
\partial_s^n\kappa(s,-\vec k+\vec k_1)}
{(2\pi)^3\sqrt{2\varepsilon(\vec k)}\sqrt{2\varepsilon(\vec k_1)}}
\frac{\e^{-\i s\varepsilon(\vec k)-\i s\varepsilon(\vec k_1)}}{\big(\varepsilon(\vec k)+\varepsilon(\vec k_1)\big)^n}.\label{squa1}\eeq
This decays in $\vec k$ and $\vec k_1$  as any inverse power, and
hence 
is square integrable on $\rr^3\times\rr^3$. Therefore the classical
scattering operator is implementable.

Next let us check the implementability of the dynamics, believing
again
that it is sufficient to check the Born approximation.
We integrate by parts once:
\begin{eqnarray}\nonumber
&&q_{t_+,t_-}^\Born(k,k_1)\\
\nonumber
&=&\frac{
-\kappa(t_+,-\vec k+\vec k_1)\e^{-\i t_+\varepsilon(\vec k)-\i t_+\varepsilon(\vec k_1)}+\kappa(t_-,-\vec k+\vec k_1)\e^{-\i t_-\varepsilon(\vec k)-\i t_-\varepsilon(\vec k_1)}}
{(2\pi)^3\sqrt{2\varepsilon(\vec k)}\sqrt{2\varepsilon(\vec k_1)}
\big(\varepsilon(\vec k)+\varepsilon(\vec k_1)\big)}
\\
&&+\int_{t_-}^{t_+}\d s\frac{
\partial_s\kappa(s,-\vec k+\vec k_1)\e^{-\i s\varepsilon(\vec k)-\i s\varepsilon(\vec k_1)}}
{(2\pi)^3\sqrt{2\varepsilon(\vec k)}\sqrt{2\varepsilon(\vec k_1)}\big(\varepsilon(\vec k)+\varepsilon(\vec k_1)\big)}.
\label{squa4}\end{eqnarray}
Using that $\kappa(s,\vec k+\vec k_1)$ decays fast in the second variable, we see that (\ref{squa4}) can be estimated by
\[\frac{C}{(\varepsilon(\vec k)+\varepsilon(\vec k_1))^2},\]
which is square integrable. 
Therefore, the dynamics is implementable for any $t_-,t_+$.

By a similar computation we check that if we freeze  $t_0\in\rr$,
the dynamics generated by the momentary Hamiltonian $ H_\Int(t_0)$ is
implementable.

\subsubsection{Quantum Hamiltonian}

We may try to write the quantum Hamiltonian as
\beq\hat H(t):=
\int{:}\Big(\frac12\hat\pi^2(\vec x)+\frac12\big(\vec\partial\hat\phi(\vec
x)\big)^2+\frac12(m^2+\kappa(t,\vec x))\hat\phi^2(x)\Big){:}\d\vec x.\label{hamiu}\eeq
We will see later on that the Wick-ordered expression (\ref{hamiu}) does not define an operator. However we will use it to derive the Feynman rules, which unfortunately will lead to divergent diagrams. 

Formally  (\ref{hamio1}) remains true if we add a time dependent constant $C(t)$ to (\ref{hamiu}). We will see that in order to 
define correct Hamiltonians
 $\hat H(t)$  this constant has to be infinite. We will obtain bounded from below Hamiltonians $\hat H_\ren(t)$, however the vacuum will not be
 contained in their form domain. Therefore, the condition
 $(\Omega|\hat H_\ren(t)\Omega)=0$ for all $t$, which is equivalent to the
 Wick ordering, cannot be imposed. 

The  interaction picture Hamiltonian is
\begin{eqnarray}
\hat H_\Int(t)&=&\frac12\int\kappa(t,\vec x){:}\hat\phi_\fr^2(t,\vec x){:}\d \vec x\\
&=&\frac{1}{2}\int\frac{\d\vec k_1\d \vec k_2\kappa(t,\vec k_1+\vec k_2)}
{(2\pi)^3\sqrt{2\varepsilon(\vec k_1)}\sqrt{2\varepsilon(\vec k_2)}}
\Big(\e^{-\i t\varepsilon(\vec k_1)-\i t\varepsilon(\vec k_2)}\hat a(-k_1)\hat a(-k_2)\nonumber\\
&&\ \ \ +2\e^{\i t\varepsilon(\vec k_1)-\i t\varepsilon(\vec k_2)}\hat a^*(k_1)\hat a(-k_2)
+\e^{\i t\varepsilon(\vec k_1)+\i t\varepsilon(\vec k_2)}\hat a^*(k_1)\hat a^*(k_2)
\Big).\nonumber\end{eqnarray}

As in the case of linear sources, we define the scattering operator,
scattering amplitudes,
Green's functions, amputated Green's functions and  the generating
function, see (\ref{sca2})--(\ref{sca6}).

\subsubsection{Path integral formulation}

The generating function (and hence all the other quantities introduced
above) can be computed exactly. It is
\begin{eqnarray}\nonumber
Z(f)&=&\left(\det\Big(\big(-\Box+m^2\big)\big(-\Box+m^2+\kappa-\i0\big)^{-1}\exp\Bigl(\kappa\frac{1}{-\Box+m^2-\i0}\Bigr)\Big)\right)^{\frac12}\\\nonumber
&&\times\exp \left(\frac{\i}{2} f(-\Box+m^2+\kappa-\i0)^{-1}f\right)\\\nonumber
&=&\left(\det\bigl(\one+\kappa D_\fr^{\rm c}\bigr)^{-1}\exp\bigl(\kappa D_\fr^{\rm c}\bigr)\right)^{\frac12}\\
&&\times\exp\left(\frac{\i}{2} f D_\fr^{\rm c}\left(\one+\kappa D_\fr^{\rm c}\right)^{-1} f\right)
.\label{causapro2}
\end{eqnarray}
Here, the determinant is understood (at least formally) as the Fredholm determinant on the space $L^2(\rr^{1,3})$. The term $\exp\bigl(\kappa D_\fr^{\rm c}\bigr)^{\frac12}$ is responsible for the Wick ordering.

Similarly as in the case of
(\ref{causapro}), (\ref{causapro2}) is often expressed 
 in terms of path integrals as
\begin{eqnarray}\label{causapro1}
C\int\lpi_x\d\phi(x)\exp\left(\i\int\big(\cL(x)-f(x)\phi(x)\big)\d x\right).
 &&\end{eqnarray}
Here, $C$ is a normalization constant, which  does not depend on $f$.
Again, the formula (\ref{causapro1}) is only symbolic, the full information is contained  in
(\ref{causapro2}).

\subsubsection{Feynman rules}

\begin{figure}[!h]
\centering
\includegraphics{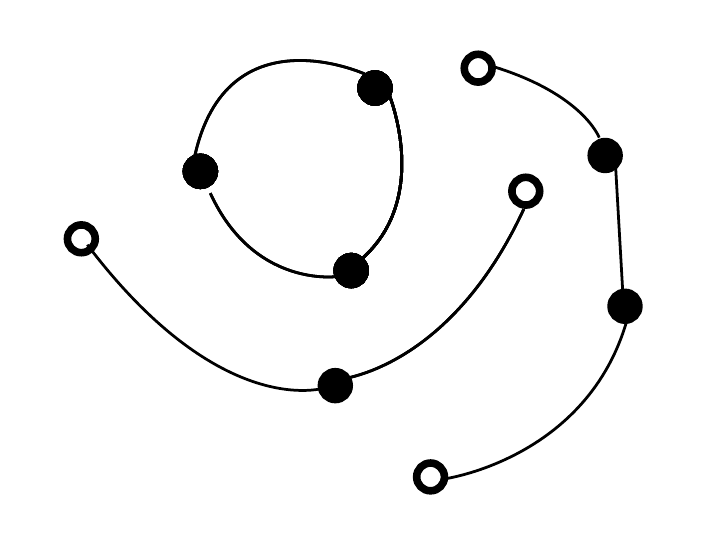}
\label{diag-4}
\caption{Diagram for Green's function.}
\end{figure}

\begin{figure}[!h]
\centering
\includegraphics{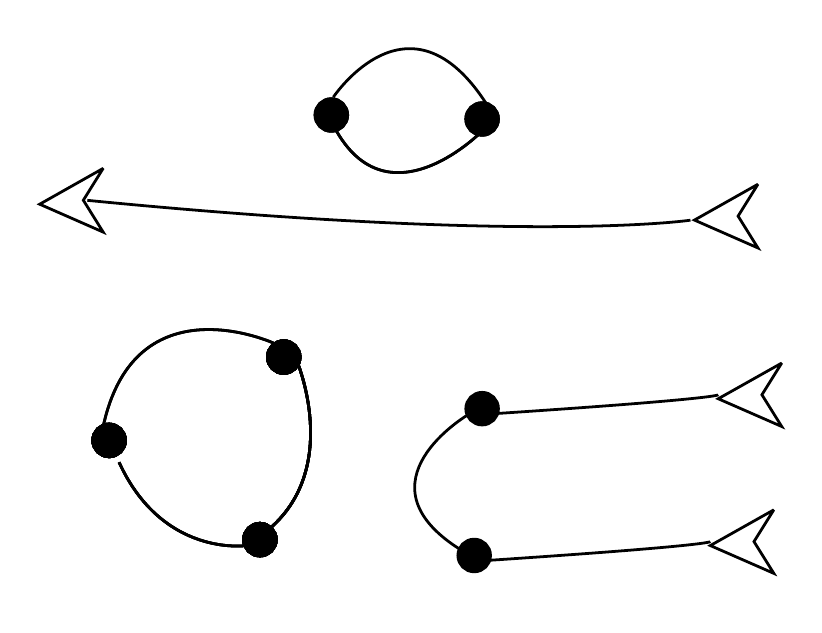}
\label{diag-5}
\caption{Diagram for scattering amplitude.}
\end{figure}

Feynman rules are similar as in the case of a linear source.
 The difference is that now  vertices have 2-legs.
  The rule (2) for calculating Green's functions changes: for each vertex with incoming momenta $k_1, k_2$ we insert the number $-\i\kappa(k_1+k_2)$, where $k_1$ and $k_2$ are the momenta of lines entering the vertex.
Another difference is that we do not allow a line to begin and end at the same vertex -- this is because we use the Wick ordered $\hat H(t)$.

Diagrams  can be decomposed into connected components of two kinds:
\begin{enumerate}\item
lines ending at insertion vertices (for Green's functions) or on-shell particles (for scattering amplitudes) with $0,1,2,\dots$ interaction vertices;
\item loops with $2,3,\dots$ interaction vertices.
\end{enumerate}
Note that  loops
 with 1 interaction vertex do not appear because of the Wick ordering.

Diagrams without loops (both for Green's functions and
scattering amplitudes) are finite, because the external momenta are
fixed and on 
 interaction vertices we have the fast decaying function $\kappa$.

Consider a loop with 4-momenta $k_1,\dots,k_n$  flowing
around it. On vertices we have the function $\kappa$, which essentially
 identifies $k_i$ with $k_{i+1}$.
 The propagators give the power $|k_i|^{-2}$. Thus we are left with 4
 degrees of freedom and the integrand that behaves as
 $|k|^{-2n}$. This is integrable if $n>2$, but divergent for 
 the 2-vertex loop. We will see that only the imaginary part of this
 diagram is divergent.

\subsubsection{Vacuum energy}

The classical scattering operator is well defined. 
The quantum scattering operator, if computed naively (that is, using
the Wick ordered Hamiltonian) is ill defined. Its problem comes from
the overall phase, which is not fixed by the classical transformation.

One can say that this phase has no physical meaning, since it does not
appear in scattering cross-sections. 
  However, it may be relevant for a more complete theory. We will see
  that there is a natural choice of this phase, which   leads to a
  renormalized scattering operator $\hat S_\ren(\kappa)$. We will also see
  that
 there is a natural renormalized  Hamiltonian $\hat H_\ren(t)$.

\begin{figure}[!h]
\centering
\includegraphics{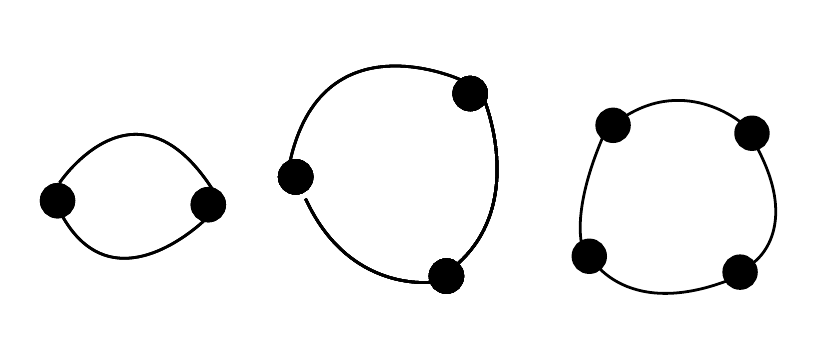}
\label{diag-6}
\caption{Vacuum energy}
\end{figure}

The logarithm of the {\em vacuum-to-vacuum scattering amplitude} times the imaginary unit
will be called the {\em vacuum energy}. It can be
computed exactly:
\begin{eqnarray}\nonumber\cE&:=&
\i\log(\Omega|\hat S\Omega)\,=\,\i\log Z(0)\\\nonumber
&\hspace{-3ex}=&\hspace{-2ex}\frac{\i}{2}\Tr\Big(\log({-}\Box{+}m^2{-}\i0)-
\log({-}\Box{+}m^2{+}\kappa{-}\i0)+\kappa({-}\Box{+}m^2{-}\i0)^{-1}\Big)
\\\nonumber
&=&\frac{\i}{2}\Tr\Big(-\log(1+\kappa D_\fr^{\rm c})+\kappa D_\fr^{\rm c}\Big)\\
&=&\i\sum_{n=2}^\infty\frac{(-1)^{n}}{2n}\Tr(\kappa D_\fr^{\rm c})^n
=:\sum_{n=2}^\infty \cE_n.\label{paks}\end{eqnarray}
Here, $\Tr$ is understood (at least formally) as the usual trace of operators on $L^2(\rr^{1,3})$. $\cE_n$ is
the $n$th order contribution to the vacuum energy. Note that for $n=1$ there is no contribution because of the Wick ordering and for $n=2$  it is 
divergent. 

We have
$\cE_n=\i\frac{D_n}{2n}$, where
$D_n=(-1)^{n}\Tr(\kappa D_\fr^{\rm c})^n$ is the value of  the loop with $n$ vertices.
This is a special case of a more general rule saying that to compute
$\log(\Omega|\hat S\Omega)$ we need to sum over all connected diagrams with
no external lines divided by the symmetry factor (the order of the
group of the symmetries of the diagram). In the case of a loop with
$n$ vertices its group of symmetries is the $n$th dihedral group,
hence the symmetry factor is $2n$.

\subsubsection{Pauli-Villars renormalization}
\label{sec-PV}

The lowest contribution to the vacuum energy is of the second order and comes from the  loop with two vertices.
Formally, it can be written as
\[\cE_2=
\int\kappa(-k)\kappa(k)\pi(k)\frac{\d k}{(2\pi)^4}=\int|\kappa(k)|^2\pi(k)
\frac{\d k}{(2\pi)^4},\]
where the right hand side defines
 the {\em vacuum energy function}
 $\pi(k)$. 
Unfortunately, computed naively, $\pi(k)$ 
 is logarithmically divergent.

The renormalization of a mass-like perturbation is not very difficult and can be done in many ways. We will describe 3 methods of renormalization. All of them will lead to the same {\em renormalized vacuum energy function} $\pi^\ren(k)$.

 We start with the {\em Pauli-Villars method}. In the context of a mass-like perturbation, the Pauli-Villars regularization consists in introducing an additional fictitious field that has a (large) mass $M$ and appears only in loops. (Thus we ignore diagrams involving external lines of the fictitious particle). In addition, each loop of the fictitios field has a (nonphysical) coefficient  $-1$. We organize our computations by setting
 $m_0=m$, $C_0=1$,
$m_1=M$,  and $C_1=-1$.
The {\em Pauli-Villars regularized vacuum energy function} is the sum of the  loop of the physical particle and of the fictitious one:

\nowastrona

\begin{eqnarray*}
&&4\pi_M(k^2)
\,=\,\i\int\frac{\d^4
  q}{(2\pi)^4}\sum_iC_i
\frac{1}
{((q+\frac12k)^2+m_i^2-\i0)((q-\frac12k)^2+m_i^2-\i0)}
\\
&=&-\i\int\frac{\d^4
  q}{(2\pi)^4} \int_0^\infty\d\alpha_1\int_0^\infty\d\alpha_2
\sum_iC_i\exp\left(-\i(\alpha_1+\alpha_2)\Big(q^2+\frac14k^2+m_i^2\Big)
-\i(\alpha_1-\alpha_2)qk\right) 
\\
&=&
-\frac{1}{(4\pi)^2}\int_0^\infty\d\alpha_1\int_0^\infty\d\alpha_2\sum_iC_i\frac{1}{(\alpha_1+\alpha_2)^2}
\exp\left(-\i(\alpha_1+\alpha_2)m_i^2
-\i\frac{\alpha_1\alpha_2}{\alpha_1+\alpha_2}k^2 \right)\\
&=&
-\frac{1}{(4\pi)^2}
\int_0^1\d v\int_0^\infty\frac{\d\rho}{\rho}
\sum_iC_i
\exp\left(-\i\rho\left( m_i^2
+\frac{(1-v^2)
  k^2}{4}\right)\right)\\
&=&\frac{1}{(4\pi)^2}
\int_0^1\d v
\sum_iC_i\log\Big(m_i^2
+\frac{ k^2(1-v^2)}{4}-\i0\Big)\\
&=&\frac{1}{(4\pi)^2}
\int_0^1\d v\sum_iC_i\left(
\log\left(1+\frac{(1-v^2) k^2}{4m_i^2}-\i0\right)+\log m_i^2\right).
\end{eqnarray*}
\nowastrona
We used the identities (\ref{ide1}) and (\ref{ide2}).
 We  inserted
\beq 1=\int_0^\infty\d\rho\delta(\rho-\alpha_1-\alpha_2),\label{inser}\eeq
and then changed the variables as $\alpha_1=\rho\frac{(1-v)}{2}$,
$\alpha_2=\rho\frac{(1+v)}{2}$, so that
$\d\alpha_1\d\alpha_2=\frac{1}{2}\rho\d v\d\rho$.
We also used the
symmetry $v\mapsto -v$ to restrict the integration from $[-1,1]$ to
$[0,1]$.
At the end we use the identity (\ref{ide3}).

We define the {\em renormalized vacuum energy function} as
\begin{eqnarray}\notag
\pi^\ren(k^2)
&:=&
\lim_{M\to\infty}\big(\pi_M(k^2)-\pi_M(0)\big)\\\notag
&=&
\lim_{M\to\infty}\Big(\pi_M(k^2)+\frac{1}{4(4\pi)^2}\log\frac{M^2}{m^2}\Big)
\\
&=&
\frac{1}{4(4\pi)^2}\int_0^1\log\Big(1+\frac{k^2(1-v^2)}{4m^2}-\i0\Big)\d
v.\label{piren1+}
\end{eqnarray}
Note that
$\pi^\ren(0)=0$. Using (\ref{ide4}) 
we obtain 
\begin{eqnarray*}&&\pi^\ren(k^2)\\&=&\frac{1}{4(4\pi)^2}\Bigg(
\frac{\sqrt{k^2+4m^2}}{\sqrt{k^2}}\log\frac{\sqrt{k^2+4m^2}+\sqrt{k^2}}
{\sqrt{k^2+4m^2}-\sqrt{k^2}}-2\Bigg),\ \ \ 0<k^2.\end{eqnarray*}
Using the analyticity and $\log\frac{x+\i y}{x-\i y}=2\i\arctan\frac{y}{x}$ we can extend this formula for $k^2<0$:
\begin{eqnarray*}&&
\pi^\ren(k^2)\\
&=&\frac{1}{4(4\pi)^2}
\Bigg(\frac{\sqrt{k^2+4m^2}}{\sqrt{-k^2}}2\arctan\frac{\sqrt{-k^2}}{\sqrt{k^2+4m^2}}-2\Bigg), \ \ \ \ \ \ \ \ \ -4m^2<k^2<0;\\
&=&\frac{1}{4(4\pi)^2}
\Bigg(\frac{\sqrt{-k^2-4m^2}}{\sqrt{-k^2}}\Big(
\log\frac{\sqrt{-k^2-4m^2}+\sqrt{-k^2}}
{\sqrt{-k^2-4m^2}-\sqrt{-k^2}}-\i\pi\Big)-2\Bigg), \ \ k^2<-4m^2.
\end{eqnarray*}


\nowastrona

\subsubsection{Renormalization of the vacuum energy}

The renormalized 2nd order vacuum energy is
\begin{eqnarray*}\cE_2^\ren&=&\int\pi^\ren(k)|\kappa(k)|^2\frac{\d k}{(2\pi)^4}\\ &=&\lim_{M\to\infty}
\int\big(\pi_M(k)-\pi_M(0)\big)|\kappa(k)|^2\frac{\d k}{(2\pi)^4}\\
&=&\lim_{M\to\infty}\Bigg(
\frac{1}{(2\pi)^4}\int\pi_M(k)|\kappa(k)|^2\frac{\d
k}{(2\pi)^4}-\pi_M(0)\int\kappa(x)^2\d x\Bigg).
\end{eqnarray*}
The full renormalized vacuum energy has a compact formula:
\begin{eqnarray}\nonumber
\cE^\ren&=&\cE_2^\ren+\sum_{n=3}^\infty\cE_n\\
&=&-\frac{\i}{2}\Tr\Big(\log(1+\kappa D_\fr^{\rm c})-\kappa D_\fr^{\rm
  c}+\frac{(\kappa D_\fr^{\rm c})^2}{2}\Big)\nonumber
\\
&&+\int|\kappa(k)|^2\pi^\ren(k)\frac{\d k}{(2\pi)^4}.
\label{paks1}\end{eqnarray}

We can formally write $\pi_\infty(k):=\lim\limits_{M\to\infty}\pi_M(k)$
(which is typically  infinite). 
Note that the renormalized scattering operator $\hat S_\ren$ is a well defined unitary operator and the renormalized Hamiltonian $\hat H_\ren(t)$ is a well defined self-adjoint operator:
\begin{eqnarray}
\hat S_\ren&=&\e^{\i\pi_\infty(0)\int \kappa(x)^2\d x} \hat S,\\
\hat H_\ren(t)&=&\hat H(t)-\pi_\infty(0)\int\kappa(t,\vec x)^2\d\vec
x.\label{formo1}\end{eqnarray}


The counterterm 
has an infinite
coefficient $\pi_\infty(0)$. Otherwise, it is quite well behaved -- it
 depends locally on the interaction, and therefore the renormalization preserves the Einstein causality.
 This manifests itself in the identity
\[\hat S_\ren(\kappa_2)\hat S_\ren(\kappa_1)=\hat S_\ren(\kappa_2+\kappa_1),\]
whenever $\supp \kappa_2$ is later than  $\supp \kappa_1$.

Formally, the correct Lagrangian density is
\[\cL_\ren(x)=\cL(x)+\pi_\infty(0)\kappa(x)^2.\]

\subsubsection{Method of dispersion relations}

There exists an alternative method to renormalize and compute the
vacuum energy. We start with computing  the imaginary part
of $\pi(k)$ without a regularization, which gives a finite result:
\begin{eqnarray*}
\Im\pi^\ren(k^2)
&=&\Im \,\frac{\i}{4}\int\frac{\d^4
  q}{(2\pi)^4}
\frac{1}
{((q+\frac12k)^2+m^2-\i0)((q-\frac12k)^2+m^2-\i0)}\\
&=&\Im\frac{1}{4(4\pi)^2}
\int_0^1\d v\left(
\log\left(1+\frac{(1-v^2)
  k^2}{4m^2}-\i0\right)+\log m^2\right).
\end{eqnarray*}
Using $\log(t-\i0)=\log|t|-\i\pi\theta(-t)$, we see that the imaginary part of the logarithm is very simple. Hence
\begin{eqnarray*}
\Im \pi^{\ren}(k^2)
&=&
-\frac{\pi}{4(4\pi)^2}\int_0^1\theta\Big(-1-\frac{(1-v^2)k^2}{4m^2}\Big)\d v\\
&=&-\frac{\pi}{4(4\pi)^2\sqrt{-k^2}}
\sqrt{\Big|-k^2-4m^2\Big|_+}.
\end{eqnarray*}
We can obtain the real part by using the fact that $\pi^\ren(0)=0$ and the once subtracted dispersion relations for the lower complex halfplane, see Thm \ref{dysp1}:
\begin{eqnarray}
\Re\pi^\ren(k^2)&=&
-\frac{1}{\pi}\cP\int_{-\infty}^{-4m^2}\d s\Im\pi^\ren(s)
\left(\frac{1}{s-k^2}-\frac{1}{s}\right).
\label{pdf}\end{eqnarray}
\nowastrona

\subsubsection{Wick rotation}

Recall that the causal propagator is defined as
\begin{eqnarray*}
\rr^{1,3}\ni p\mapsto
D^{\rm c}( p)&=&
\frac{1}{ p^2+m^2-\i0}.\end{eqnarray*}
It can be interpreted as a boundary value of  a holomorphic function
\begin{eqnarray}
\Big(\cc\backslash\big(]-\infty,-m]\cup[m,\infty[\big)\Big)
\times\rr^3&\ni&(p^0,\vec p)\label{holom1}\\
\mapsto\, D^{\rm c}(p)&=&\frac1{-(p^0)^2+\vec p^2+m^2}\,=\,\frac1{ p^2+m^2}.
\notag
\end{eqnarray}
The {\em physical region} $\rr^{1,3}$ of (\ref{holom1}) lies at the boundary--on $]0,\infty[\times\rr^3$ from above and on $\hbox{$]-\infty,0[\times\rr^3$}$ from below:
\[ D^{\rm c}(p)=\lim_{\phi\searrow0}D^{\rm c}(\e^{\i\phi}p^0,\vec p).\]

Define the {\em Euclidean scalar product}
as 
\[\langle p|q\rangle_\E:=p^0q^0+\vec p\vec q,\]
and the {\em Euclidean propagator}
\begin{eqnarray}\label{holom2}
\Big(\cc\backslash\big(]-\i\infty,-\i m]\cup[\i m,\i\infty[\big)\Big)
\times\rr^3&\ni&
(p^0,\vec p)\\
\mapsto \, D^{\E}(p^0,\vec p)\ :=\ D(\i p^0,\vec p)\,=\, \frac1{(p^0)^2+\vec p^2+m^2}
&=&\frac1{\langle p|p\rangle_\E^2+m^2}.\notag\end{eqnarray}
Clearly, we can express the causal propagator in terms of the Euclidean propagator with help of the {\em Wick rotation}:
\[D^{\rm c}(p^0,\vec p)=\lim_{\phi\nearrow\pi/2} D^\E(\e^{-\i\phi}p^0,\vec p).\]

Suppose now that a physical quantity is given by an integral
\beq \rr^{1,3}\ni p\mapsto F(p):=\int\frac{\d^4 q}{(2\pi)^4}\frac{G(p^2,pq,q^2)}{\big(a p^2+2bpq+c q^2+m^2-\i0\big)^n},\label{phy}\eeq
where $G$ is holomorphic and the matrix $\left[\begin{array}{cc}a&b\\b&c\end{array}\right]$ is positive definite.
Then instead of $F$ we can consider
the holomorphic function
\begin{eqnarray}\label{boundi}
\Big(\cc\backslash\big(]-\infty,-m]\cup[m,\infty[\big)\Big)
\times\rr^3&\ni&(p^0,\vec p)\\
\mapsto F(p)&:=&\int\frac{\d^4 q}{(2\pi)^4}\frac{G(p^2,pq,q^2)}{\big(a p^2+2bpq+c q^2+m^2\big)^n},\notag\end{eqnarray}
where there is no need to put $\i0$, because the denominator is automatically invertible. The physical function (\ref{phy}) is the boundary value of (\ref{boundi}):
\[\lim_{\phi\searrow0}F(\e^{\i\phi}p^0,\vec p).\]
 We can also introduce the Euclidean version of $F$ given by
\begin{eqnarray*}F^\E(p)&=&F^\E(p^0,\vec p)\ :=\ F(\i p^0,\vec p)\\
&=&\int\frac{\i\d^4 q}{(2\pi)^4}\frac{G(\langle p|p\rangle_\E^2,\langle p|q\rangle_\E,\langle q|q\rangle_\E^2)}{\big(a \langle p|p\rangle_\E^2+2b\langle p|q\rangle_\E+c\langle q|q\rangle_\E^2+m^2\big)^n},\end{eqnarray*}
where  in the integral we substituted $(\i q^0,\vec q)$ for $(q^0,\vec q)$. This substitution can be reached from the original variables inside the holomorphy domain by the Wick rotation, hence it does not affect the integral.
$F^\E$ is holomorphic on the domain of (\ref{holom2}). We can retrieve the physical values of $F$ from $F^\E$ by
\[F(p^0,\vec p)=\lim_{\phi\nearrow\pi/2} F^\E(\e^{-\i\phi}p^0,\vec p).\]

In what follows, whenever we use  Euclidean functions such as $F^\E$, we will use the Euclidean scalar product $\langle p|q\rangle_\E$. We will denote this scalar product simply by $pq$, since its use will be obvious from the context.

\subsubsection{Dimensional renormalization}

Let us renormalize the vacuum energy by yet another method -- the method of dimensional regularization. We will use the Euclidean quantities. 

Let us first compute formally the 2-vertex loop:
\begin{eqnarray}\notag
4\pi^\E(k^2)
&=&-\int\frac{\d^4
  q}{(2\pi)^4}
\frac{1}
{((q+\frac12k)^2+m^2)((q-\frac12k)^2+m^2\big)}
\\\notag
&=&-\frac12\int_{-1}^1\d v
\int\frac{\d^4
  q}{(2\pi)^4}
\frac1{\big(q^2+\frac{k^2}{4}+m^2+vqk\big)^2}
\\&=&-\int_{0}^1\d v
\int\frac{\d^4
  q}{(2\pi)^4}
\frac1{\big(q^2+\frac{k^2}{4}(1-v^2)+m^2\big)^2},
\label{pdf2}\end{eqnarray}
where we used the Feynman identity (\ref{dim-feyn}), replaced $q+\frac{vp}{2}$ with $q$, used the symmetry $v\to-v$ to  replace $\frac12\int_{-1}^1\d v$ with  $\int_0^1\d v$.
 After this preparation, we use the dimensional regularization:
\beq
\int\frac{\d q^4}{(2\pi)^4}\ \hbox{ is replaced by }\
\frac{\mu^{4-d}\Omega_d}{(2\pi)^d}\int_0^\infty|q|^{d-1}\d|q|,\eeq
where $\Omega_d$ is the ``area of the unit sphere in $d$ dimension'', see (\ref{dim1}). Thus instead of (\ref{pdf2}) we consider its dimensionally regularized version:
\begin{eqnarray}
4\pi^{\E,d}(k^2)&=&-\frac{\mu^{4-d}\Omega_d}{(2\pi)^d}\int_0^1\d v\int_0^\infty\frac{|q|^{d-1}}
{\big(q^2+\frac{k^2}{4}(1-v^2)+m^2\big)^2}\d|q|\notag\\
&\simeq&-\frac{1}{(4\pi)^2}\int_0^1\d v
\Big(-\gamma +\log(\mu^24\pi)-\log\Big(
\frac{k^2}{4}(1-v^2)+m^2\Big)\Big)\notag\\&&
-\frac{1}{(4\pi)^2(2-d/2)}.
\label{pdf2+}\end{eqnarray}
To renormalize we demand that $\pi^{\E,\ren}(0)=0$. Thus
\begin{eqnarray*}
\pi^{\E,\ren}(k^2)&=&\lim_{d\to4}\Big(\pi^{\E,d}(k^2)
-\pi^{\E,d}(0)\Big)\\
&=&\frac{1}{4(4\pi)^2}\int_0^1\d v
\log\Big(1+
\frac{k^2}{4m^2}(1-v^2)\Big),
\end{eqnarray*}
which coincides  with the Wick rotated 
result obtained by the Pauli-Villars method.
Thus the renormalization of (\ref{pdf2+}) amounts to choosing
\beq \log\frac{\mu^2}{m^2}=\gamma-\log4\pi,\label{dim9}\eeq
dropping the pole term  and setting $d=4$.

\subsubsection{Energy shift}

Suppose that the perturbation does not depend on time and is given by
a Schwartz function $\rr^3\ni \vec x\mapsto \kappa(\vec x)$. The naive
(Wick ordered) 
 Hamiltonian is
\[\hat H:=
\int{:}\Big(\frac12\hat\pi^2(\vec x)+\frac12\big(\vec\partial\hat\phi(\vec
x)\big)^2+\frac12\big(m^2+\kappa(\vec x)\big)\hat\phi^2(x)\Big){:}\d\vec x\]
The infimum of a  quadratic Wick ordered Hamiltonian can be computed exactly (\ref{bogol1a}):
\begin{eqnarray*}
E&=&
\Tr\Big(\frac12(-\Delta+m^2+\kappa)^{1/2}-\frac12(-\Delta+m^2)^{1/2}
-\frac14(-\Delta+m^2)^{-1/2}\kappa\Big)
\\&=&\int\Tr\Bigg(\frac{-\Delta+m^2+\kappa}{(-\Delta+m^2+\kappa+\tau^2)}
-
\frac{-\Delta+m^2}{(-\Delta+m^2+\tau^2)} -\frac{\tau^2}{(-\Delta+m^2+\tau^2)^2}\kappa\Bigg)\frac{\d\tau}{2\pi}\\
&=&\int\Tr\Bigg(
\frac{\tau^2}{(-\Delta+m^2+\tau^2)}
-\frac{\tau^2}{(-\Delta+m^2+\kappa+\tau^2)}
-\frac{\tau^2}{(-\Delta+m^2+\tau^2)^{2}}\kappa\Bigg)\frac{\d\tau}{2\pi}\\
&=&-\int\Tr
\frac{1}{(-\Delta+m^2+\tau^2)^2}\kappa
\frac{1}{(-\Delta+m^2+\kappa+\tau^2)}\kappa\tau^2
\frac{\d\tau}{2\pi}
\\
&=&\sum_{n=2}^\infty(-1)^{n-1}\int\Tr
\frac{1}{(-\Delta+m^2+\tau^2)^2}\kappa\Big(
\frac{1}{(-\Delta+m^2+\tau^2)}\kappa\Big)^{n-1}\tau^2
\frac{\d\tau}{2\pi}\\
&=&\sum_{n=2}^\infty\frac{(-1)^{n}}{2n}\int\Tr
\Big(
\frac{1}{(-\Delta+m^2+\tau^2)}\kappa\Big)^n
\frac{\d\tau}{2\pi}
.
\end{eqnarray*}
Above, we rewrote the square root by using the identities
(\ref{sqrt}) and (\ref{sqrt1}), expanded the denominator in the Neumann series and at the end we used the identity (\ref{sqrt2}).
Note that the $n$th term of the above expansion corresponds to the loop with $n$ vertices. They are all well defined except for $n=2$, which needs renormalization. We can guess that the  renormalized energy shift is
\begin{eqnarray}\label{guess}
&&E^\ren\ =\ \int\pi^\ren(0,\vec k)|\kappa(\vec k)|^2\frac{\d\vec k}{(2\pi)^3}
+\int\Tr
\frac{1}{(-\Delta+m^2+\tau^2)}\kappa\\
&&\times\frac{1}{(-\Delta+m^2+\tau^2)}\kappa
\frac{1}{(-\Delta+m^2+\kappa+\tau^2)}\kappa\frac{1}{(-\Delta+m^2+\tau^2)}\tau^2
\frac{\d\tau}{2\pi},\notag
\end{eqnarray}
where we rewrote the sum of terms  with $n\geq3$ in a compact form, and
 $\pi^\ren$ was introduced in (\ref{piren1+}).


Another way to derive the expression for $E^\ren$ is to use Sucher's formula. We introduce the adiabatically switched perturbation $\e^{-\epsilon|t|}\kappa(\vec x)$ multiplied by a coupling constant $\lambda$, which will be put to $1$ at the end. The Fourier transform of 
the switching factor $\e^{-\epsilon|t|}$ is $\frac{2\epsilon}{\epsilon^2+\tau^2}$
Therefore,
\begin{eqnarray*}\cE_\epsilon^\ren&=&\i\log(\Omega|\hat S_\epsilon^\ren\Omega)
\\
&=&\lambda^2\int\pi^\ren(\tau,\vec k)\frac{4\epsilon^2}{(\epsilon^2+\tau^2)^2}|\kappa(\vec k)|^2\frac{\d\tau\d \vec k}{(2\pi)^4}+
O(\lambda^3).
\end{eqnarray*}
By Sucher's formula,
\begin{eqnarray*}
E^\ren&=&
\lim_{\epsilon\searrow0}\frac{\i\epsilon\lambda}{2}\partial_\lambda
\log(\Omega|\hat S_\epsilon^\ren\Omega) \\
&=&\lim_{\epsilon\searrow0}\lambda^2
\int\pi^\ren(\tau,\vec k)\frac{4\epsilon^3}{(\epsilon^2+\tau^2)^2}|\kappa(\vec k)|^2\frac{\d\tau\d \vec k}{(2\pi)^4}+O(\lambda^3)\\
&=&\lambda^2
\int\pi^\ren(0,\vec k)|\kappa(\vec k)|^2\frac{\d \vec k}{(2\pi)^3}+O(\lambda^3),
\end{eqnarray*}
where we used $\int\frac{4\epsilon^3}{(\epsilon^2+\tau^2)^2}\d\tau=2\pi$.
Eventually, we put $\lambda=1$ and we obtain (\ref{guess}).

\init\section{Massive photons}



 Let $m>0$. In this section we discuss the quantization of the {\em Proca equation}
\begin{eqnarray}
-\partial_\mu F^{\mu\nu}(x)+m^2 A^\nu(x)&=&0,
\label{proc-}
\end{eqnarray}
where
\beq F^{\mu\nu}:=\partial^\mu A^\nu-\partial^\nu A^\mu.\label{ff-}\eeq

\nowastrona

Beside the free equation, 
we will also consider the Proca equation interacting with a given vector function $J^\mu$,
called an {\em external 
4-current}:
\begin{eqnarray}
-\partial_\mu F^{\mu\nu}(x)+m^2 A^\nu(x)&=&-J^\nu(x).
\label{ff1}\end{eqnarray}
We will assume that
  the 4-current is {\em conserved}, that is
\beq \partial _\nu J^\nu(x)=0.
\label{current-}\eeq


There are several possible approaches to the Proca equation on the
classical and, especially, quantum level. In particular, one can use from
the beginning  the reduced phase space,
both for the classical
description and quantization. This is the approach that we will treat
as the standard one. Alternative approaches will be discussed later.

\nowastrona
\subsection{Free massive photons}
\label{free1}
\subsubsection{Space of solutions}

Let $\cY_{\Pr}$, resp.  $\cc\cY_{\Pr}$
 denote the set of 
real, resp.  complex 
smooth space-compact solutions of the Proca equation
\beq-\partial^\mu (\partial_\mu  \zeta_\nu-\partial_\nu \zeta_\mu)
+m^2 \zeta_\nu(x)=0.\label{proc0}\eeq

\nowastrona

It is easy to see that for $\zeta_1,\zeta_2\in\cc\cY_\Pr$ the following expression defines a {\em conserved 4-current}:
\begin{eqnarray}
&&j_{\Pr}^\mu(\zeta_1,\zeta_2,x)\label{proco1}\\
&:=&\big(\partial^\mu \zeta_{1\nu}(x)-\partial_\nu\zeta_{1}^\mu(x)\big)\zeta_2^\nu(x)
-\zeta_{1\nu}(x)\big(\partial^\mu \zeta_2^\nu(x)-\partial^\nu
\zeta_{2}^\mu(x)\big).\notag \end{eqnarray}

$\cY_{\Pr}$ is a symplectic space with the {\em symplectic form}
\begin{eqnarray}
\zeta_1\omega_\Pr \zeta_2
&=&\int_\cS j_{\Pr}^\mu(\zeta_{1},\zeta_2,x)\d s_\mu(x)\label{pro1}\\
&=&\int\left(-\left(\dot{\vec\zeta}_1(t,\vec x)
-\vec\partial\zeta_{10}(t,\vec x)\right)\vec\zeta_2(t,\vec x)+\vec\zeta_1(t,\vec x)
\left(\dot{\vec\zeta}_2(t,\vec x)-\vec\partial\zeta_{10}(t,\vec x)\right)\right)\d\vec x,\notag
\end{eqnarray}
\nowastrona
where $\cS$ is any Cauchy surface.

The Poincar\'{e}{} group $\rr^{1,3}\rtimes O(1,3)$ 
acts on $\cY_{\Pr}$ by
\[r_{(y,\Lambda)}\zeta_\mu(x):=\Lambda_\mu^\nu\zeta_\nu\left((y,\Lambda)^{-1}x\right).\]
$r_{(y,\Lambda)}$ are symplectic for $\Lambda\in O^\uparrow(1,3)$, otherwise they are antisymplectic. 


\subsubsection{Classical 4-potentials}
We introduce the functionals $A_\mu(x)$ called {\em  4-potentials}.
 They act on $\zeta\in\cY_\Pr$ giving
\begin{eqnarray*}\langle A_\mu(x)|\zeta\rangle:=\zeta_\mu(x).
\end{eqnarray*}
On $\cY_{\Pr}^\t$ we have the action of the Poincar\'{e}{} group
$(y,\Lambda)\mapsto r_{(y,\Lambda)}^{\t-1}$. Note
that
\[r_{(y,\Lambda)}^{\t-1}A_\mu(x)=(\Lambda^{-1})_\mu^\nu A_\nu( \Lambda x+y).\]

We also introduce the {\em field tensor} and the {\em electric field vector}:
\begin{eqnarray*}
F_{\mu\nu}(x)&:=&
\partial_\mu A_\nu(x)-\partial_\nu A_\mu(x),\\
 \ E_i(x)&:=&F_{0i}(x)\ =\ \dot A_i-\partial_i A_0.\end{eqnarray*}

Clearly, the free Proca equation (\ref{proc-}) is satisfied. Equivalently, we have
\begin{eqnarray}
(-\Box+m^2)A_\mu(x)&=&0,\label{proc1v}\\
\partial^\nu A_\nu(x)&=&0.
\label{lorenz1v}\end{eqnarray}

Yet another equivalent system of equations convenient for further analysis is
\begin{eqnarray}
(-\Delta+m^2)A_0+\div \dot{\vec A}&=&0,\label{ph0}\\
(-\Box+m^2)\vec A&=&0.\label{ph4}\end{eqnarray}
(\ref{lorenz1v}) can be rewritten as
\begin{eqnarray}\dot A_0&=&\div \vec A.\label{ph3}
\end{eqnarray}
Thus only $\vec A$ is dynamical: $A_0$ can be computed from $\vec A$.
Taking the divergence of the definition of the electric field
$
\vec E=\dot{\vec A}-\vec\partial A_0$,
 then using (\ref{ph3}) and (\ref{ph4}), we can express $A_0$ in
 terms of  $\vec E$:
\begin{eqnarray}
m^2A_0&=&-\div \vec E.\label{ph2}
\end{eqnarray}
Finally, we have the following version of the evolution equations in
terms of $\vec E$, $\vec A$ with only first order derivatives:
\begin{eqnarray}
\dot{\vec A}&=&\vec E-\frac{1}{m^2}\vec\partial\div\vec E,\label{ph1}\\
\dot{\vec E}&=&-(-\Delta+m^2)\vec A-\vec\partial\div\vec A.\label{ph1a}
\end{eqnarray}
\nowastrona

\subsubsection{Poisson brackets}

The symplectic form on $\cY_\Pr$  (\ref{pro1}) can be written as
\[\omega_\Pr=\int  \vec A(t,\vec x)\wedge \vec E(t,\vec x)\d\vec x.\]
It leads to 
a {\em Poisson bracket} on
 functions on $\cY_{\Pr}$:
\begin{eqnarray}
\{A_i(t,\vec x),A_j(t,\vec y)\}=\{ E_i(t,\vec x), E_j(t,\vec y)\}&=&0,
\nonumber
\\
\{A_i(t,\vec x),E_j(t,\vec y)\}&=&\delta_{ij}\delta(\vec x-\vec y).
\label{poisson3}\end{eqnarray}
We have
\[
\{A_\mu(x),A_\nu(y)\}=\left(g_{\mu\nu}-\frac{\partial_\mu\partial_\nu}{m^2}\right)
D(x-y),\]
where $D(x-y)$ is the Pauli-Jordan function.

Indeed, this follows after we insert (\ref{ph1}), (\ref{ph2}) and (\ref{ph3})
into 
\begin{eqnarray*}
\vec A(t,\vec x)&=&\int \left(D(t,\vec x-\vec y)\dot{\vec A}(0,\vec
y)
+\dot D(t,\vec x-\vec y)\vec A(0,\vec y)
\right)\d \vec y,\\
A_0(t,\vec x)&=&\int \left(D(t,\vec x-\vec y)\dot A_0(0,\vec
y)
+\dot D(t,\vec x-\vec y)A_0(0,\vec y)
\right)\d \vec y,
\end{eqnarray*}
and then we commute them with $A_0(0,\vec x)$ and $\vec A(0,\vec x)$.



\nowastrona

\subsubsection{Smeared 4-potentials}

We can use the symplectic form to pair distributions and solutions.
For $\zeta\in\cY_{\Pr}$, the corresonding
{\em spatially smeared 4-potential} is the functional on
$\cY_{\Pr}$
 given by
\[\langle A\lpar\zeta\rpar|\rho\rangle:= \bar\zeta\omega\rho.\]
Note that
\[\{A\lpar\zeta_1\rpar,A\lpar\zeta_2\rpar\}=\bar\zeta_1\omega\bar \zeta_2.\]
\beq
A\lpar\zeta\rpar =\int\left(-\bar{(\dot{\vec{\zeta}}(t,\vec x)-\partial\zeta^0(t,\vec
x))}\vec A(t,\vec x)+
\bar{\vec\zeta(t,\vec x)}\vec E(t,\vec x)\right)\d\vec{x}.
\label{pzk3}\eeq

\nowastrona

Another way of smearing the 4-potentials is also useful.
For a space-time vector valued functions
$f\in C_{\rm c}^\infty(\rr^{1,3},\rr^{1,3})$
the corresponding {\em space-time smeared 4-potential } is
\beq A[f]:=\int f_\mu(x)A^\mu(x)\d x.\label{space-time2}\eeq
Note that $A[f]=A\lpar \zeta\rpar$, where
\[\zeta_\mu=-D* f_\mu+\frac{\partial_\mu\partial^\nu}{m^2}D* f_\nu.\]
Adding
 to $f^\mu$ a derivative $\partial^\mu\chi$ for $\chi\in C_{\rm
   c}^\infty(\rr^{1,3})$ does not change (\ref{space-time2}).

\nowastrona
\subsubsection{Lagrangian formalism and stress-energy tensor}
Consider the Lagrangian density in the off-shell formalism
\[
\cL:=-\frac14 F_{\mu\nu}F^{\mu\nu}-\frac{m^2}{2}A_\mu A^\mu.
\]
The resulting Euler-Lagrange equations  
\[\frac{\partial\cL}{\partial
  A_\alpha}=\partial_\mu\left(\frac{\partial\cL}{\partial
  A_{\alpha,\mu}}\right)\] 
coincide with the Proca equation.

The  {\em canonical stress-energy tensor}, which follows directly from the Noether Theorem, equals
\begin{eqnarray*}
\cT_\can^{\mu\nu}&=&g^{\mu\nu}\cL-\frac{\partial\cL}{\partial
  A_{\alpha ,\mu} }A_\alpha^{\ ,\nu} \\
&=&-g^{\mu\nu}\Big(\frac14F_{\alpha\beta}F^{\alpha\beta}+\frac{m^2}{2}A_\alpha
A^\alpha\Big)+F^{\mu\alpha} A_\alpha^{,\nu}.
\end{eqnarray*}
One usually prefers to replace it with the {\em Belifante-Rosenfeld
  stress-energy tensor}. It is defined as
\begin{eqnarray*}
\cT^{\mu\nu}&=&
\cT_\can^{\mu\nu}+\partial_\alpha\Sigma^{\mu\nu\alpha}\\
&=&-g^{\mu\nu}\Big(\frac14F_{\alpha\beta}F^{\alpha\beta}+\frac{m^2}{2}A_\alpha
A^\alpha\Big)+m^2A^\mu A^\nu+F^{\mu\alpha}F^{\nu}_{\ \alpha},
\end{eqnarray*}
where
\beq \Sigma^{\mu\nu\alpha}=-\Sigma^{\alpha\nu\mu}
:=F^{\mu\alpha}A^\nu.\label{sigma}\eeq
On solutions of the Euler-Lagrange equations we have
\[\partial_\mu\cT_\can^{\mu\nu}=\partial_\mu\cT^{\mu\nu}=0.\]
In addition, $\cT^{\mu\nu}$ is symmetric.

As discussed before, the variables $A_0(x)$ are not dynamical.
To pass to the Hamiltonian formalism, we introduce the variable conjugate to $A^i(x)$ 
\[\partial_{\dot A^i(x)}\cL(x)=E_i(x).\]

The Hamiltonian and the momentum density obtained from the Belifante-Rosen  tensor are
\begin{eqnarray*}\cH(x):=\cT^{00}(x)&=&
\frac12\vec E^2(x)+\frac{1}{2m^2}(\div \vec E)^2(x)+(\rot \vec
A)^2(x)+
\frac{m^2}{2}\vec A^2(x),
\\
\cP^j(x):=\cT^{0j}(x)&=&m^2A^0(x)A^j(x)+E^i(x) F^{ji}(x).
\end{eqnarray*}

The total  Hamiltonian and  momentum obtained from both Belifante-Rosen and canonical stress-energy tensor coincide:
\begin{eqnarray}\notag
H&:=&\int \cH(t,\vec x)\d\vec x=\int \cT_\can^{00}(t,\vec x)\d\vec
x,\\\notag
  P^j&:=&\int \cP^j(t,\vec x)\d\vec x=\int \cT_\can^{0j}(t,\vec x)\d\vec
x.\label{pyko}\end{eqnarray}
Using (\ref{ph1}) and (\ref{ph1a}) we check that $H$ generates the equations
of motion and $\vec P$ the translations.

It is also natural to introduce 
\beq\cS(x):= E_i(x)\epsilon^{ijk}\partial_k A_j(x),\label{pola1}\eeq
and its spatial integral
\beq
S:=\int \cS(t,\vec x)\d\vec x.\label{pola2}\eeq
We are  not aware of an established name of these quantities. We will call 
(\ref{pola1}) the {\em polarization density} 
 and  (\ref{pola2}) the {\em polarization}.

The observables $H,\vec P, S$ are in involution.

\subsubsection{Diagonalization of the equations of motion}

For $\vec k\in\rr^3$, $\vec k\neq \vec0$  fix two  spatial vectors $\vec e_1(\vec k),\vec e_2(\vec k)$ that form an oriented 
orthonormal
basis of the plane orthogonal to $\vec k$. Define
\[\vec e(\vec k,\pm1):=\frac{1}{\sqrt2}\left(\vec e_1(\vec k)\pm\i \vec e_2(\vec k)\right).\]
Note that 
\begin{eqnarray*}
\vec k\times \vec e(\vec k,\pm1)&=&\pm \i|\vec k|\vec e_\pm(\vec k),\\
\vec e(\vec k,\sigma)\cdot\vec k&=&0,\\
\bar{e_i(\vec k,\sigma)}e_i(\vec k,\sigma')&=&\delta_{\sigma,\sigma'},\\
\sum_{\sigma=\pm1}
\bar{e_i(\vec k,\sigma)}e_j(\vec k,\sigma)&=&\delta_{ij}-\frac{k_i
  k_j}{\vec{ k}^2}. \end{eqnarray*}

Let $k\in\rr^{1,3}$ with $k^0=\varepsilon(\vec k)=\sqrt{\vec k^2+m^2}$.
Introduce
\begin{eqnarray}\label{ede1}
u(k,0)&:=&\Big(\frac{|\vec
  k|}{m},\frac{\varepsilon(\vec k)\vec k}{m|\vec k|}\Big),\\
u(k,
\pm1)&:=&\left(0,\vec e(\vec k,\pm1)\right).\label{ede}
\end{eqnarray}

\nowastrona

Note that
\begin{eqnarray*}u_\mu(k,\sigma)k^\mu&=&0,\\
\bar{u_\mu(k,\sigma)}u^\mu(k,\sigma')&=&\delta_{\sigma,\sigma'},\\
\sum_{\sigma=0,\pm1}
\bar{u_\mu(k,\sigma)}u_\nu(k,\sigma)&=&g_{\mu\nu}+\frac{k_\mu
  k_\nu}{m^2}. \end{eqnarray*}

Set
\begin{eqnarray*}
\vec A_t(\vec k)&=&\int \vec A(t,\vec x)\e^{-\i \vec k\vec x}\frac{\d\vec x}{\sqrt{(2\pi)^3}},\\
\vec E_t(\vec k)&=&\int \vec E(t,\vec x)\e^{-\i \vec k\vec x}\frac{\d\vec x}{\sqrt{(2\pi)^3}}.\end{eqnarray*}
We have the equations of motion
\begin{eqnarray*}
\dot{\vec A}_t(\vec k)&=&\vec E_t(\vec k)+\frac{\vec k}{m^2}\vec k{\cdot} \vec E_t(\vec k),\\ 
\dot{\vec E}_t(\vec k)&=&-(\vec k^2+m^2)\vec A_t(\vec k)+\vec k\ \vec k{\cdot} \vec A_t(\vec k),
 \end{eqnarray*}
the relations
\[ A_i^*(\vec k)= A_i(-\vec k),\ \  E_i^*(\vec k)= E_i(-\vec k),\]
and the Poisson brackets
\begin{eqnarray*}
\{A_{ti}^*(\vec k),A_{tj}(\vec k')\}=\{ E_{ti}^*(\vec k), E_{tj}(\vec k')\}&=&0,
\nonumber
\\
\{ A_{ti}^*(\vec k),E_{tj}(\vec k')\}&=&\delta_{ij}\delta(\vec k-\vec k').
\notag\end{eqnarray*}

Set
\begin{eqnarray*}
A_t(\vec k, \pm 1)&:=&\bar{\vec  e(\vec k,\pm1)}{\cdot} \vec A_t(\vec k),\\
E_t(\vec k,\pm 1)&:=& \bar{\vec e(\vec k,\pm1)}{\cdot} \vec E_t(\vec k),\\
A_t(\vec k, 0)&:=& \frac{m}{\varepsilon(\vec k)}\frac{\vec k}{|\vec k|}{\cdot}\vec A_t(\vec k),\\
E_t(\vec k,0)&:=&  \frac{\varepsilon(\vec k)}{m}\frac{\vec k}{|\vec k|}{\cdot}\vec E_t(\vec k).
\end{eqnarray*}

We have the equations of motion
\begin{eqnarray*}
\dot{ A}_t(\vec k,\sigma)&=&E_t(\vec k,\sigma),\\ 
\dot{ E}_t(\vec k,\sigma)&=&-\varepsilon(\vec k)^2 A_t(\vec k,\sigma).
 \end{eqnarray*}
the relations
\[ A_t^*(\vec k,\sigma)= A_t(-\vec k,-\sigma),\ \ E_t^*(\vec k,\sigma)= E_t(-\vec k,-\sigma),\]
and the Poisson brackets
\begin{eqnarray}
\{A_{t}^*(\vec k,\sigma),A_{t}(\vec k',\sigma')\}=\{ E_{t}^*(\vec k,\sigma), E_{t}(\vec k',\sigma')\}&=&0,
\label{pois1}
\\
\{ A_{t}^*(\vec k,\sigma),E_{t}(\vec k',\sigma')\}&=&\delta_{\sigma\sigma'}\delta(\vec k-\vec k').
\notag\end{eqnarray}

We set
\begin{eqnarray*}
a_t(k,\sigma)&:=&
\sqrt{\frac{\varepsilon (\vec k)}{2}}A_t(\vec k,\sigma)
+\frac{\i}{\sqrt{2\varepsilon (\vec k)}}E_t(\vec k,\sigma),\\
 a_t^*(k,\sigma)&:=&
\sqrt{\frac{\varepsilon (\vec k)}{2}} A_t^*(\vec k,\sigma)
-\frac{\i}{\sqrt{2\varepsilon (\vec k)}} E_t^*(\vec k,\sigma).
 \end{eqnarray*}

We have the equations of motion
\begin{eqnarray*}
\dot a_t( k,\sigma)&=&-\i \varepsilon(\vec k)a_t( k,\sigma),\\
\dot{ a^*_t}( k,\sigma)&=&\i \varepsilon(\vec k) a^*_t( k,\sigma).
\end{eqnarray*}
We will usually write $ a( k,\sigma)$, $ a^*( k,\sigma)$ for 
$ a_0( k,\sigma)$, $ a_0^*( k,\sigma)$, so that
\begin{eqnarray*}
 a_t( k,\sigma)&=&\e^{-\i t\varepsilon(\vec k)}a( k,\sigma),\\
 a^*_t( k,\sigma)&=&\e^{\i t\varepsilon(\vec k)} a^*( k,\sigma).
\end{eqnarray*}
The direct definitions of  $ a( k,\sigma)$, $ a^*( k,\sigma)$ are
\begin{eqnarray*}
a(k,\pm1)&=&
\int\frac{\d\vec x}{\sqrt{(2\pi)^3}}\e^{-\i\vec k\vec x}\Bigg(
\sqrt{\frac{\varepsilon (\vec k)}{2}}\bar{\vec e(\vec k,\pm1)}\vec A(0,\vec x)
+\frac{\i}{\sqrt{2\varepsilon (\vec k)}}\bar{\vec e(\vec k,\pm1)}\vec E(0,\vec x)\Bigg),
\\
a^*(k,\pm1)&=&
\int\frac{\d\vec x}{\sqrt{(2\pi)^3}}\e^{\i\vec k\vec x}\Bigg(
\sqrt{\frac{\varepsilon (\vec k)}{2}}\vec e(\vec k,\pm1)\vec A(0,\vec x)
-\frac{\i}{\sqrt{2\varepsilon (\vec k)}}\vec e(\vec k,\pm1)\vec E(0,\vec x)\Bigg),
\end{eqnarray*}
\begin{eqnarray*}
a(k,0)&=&
\int\frac{\d\vec x}{\sqrt{(2\pi)^3}}\e^{-\i\vec k\vec x}\Bigg(
\frac{m}{\sqrt{2\varepsilon (\vec k)}}\frac{\vec k}{|\vec k|}\vec A(0,\vec x)
+\frac{\i}{m}\sqrt{\frac{\varepsilon (\vec k)}{2}}\frac{\vec k}{|\vec k|}
\vec E(0,\vec x)\Bigg),\\
a^*(k,0)&=&
\int\frac{\d\vec x}{\sqrt{(2\pi)^3}}\e^{\i\vec k\vec x}\Bigg(
\frac{m}{\sqrt{2\varepsilon (\vec k)}}\frac{\vec k}{|\vec k|}\vec A(0,\vec x)
-\frac{\i}{m}\sqrt{\frac{\varepsilon (\vec k)}{2}}\frac{\vec k}{|\vec k|}
\vec E(0,\vec x)\Bigg). \end{eqnarray*}
Their Poisson brackets are
\begin{eqnarray*}
\{a(\vec k,\sigma), a(\vec k',\sigma')\}=\{ a^*(\vec k,\sigma), a^*(\vec k',\sigma')\}&=&0,\\
\{a(\vec k,\sigma), a^*(\vec k',\sigma')\}&=&-\i\delta(\vec k-\vec
k')\delta_{\sigma,\sigma'}. \end{eqnarray*}
The 4-potentials can be written as
%
\begin{eqnarray}
A_\mu(x)&=& \sum_{\sigma=0,\pm1}
\int\frac{\d\vec{k}}{\sqrt{(2\pi)^3}\sqrt{2\varepsilon (\vec k)}}
 \left(u_\mu(k,\sigma)\e^{\i kx}a( k,\sigma)+
\bar{u_\mu(k,\sigma)}\e^{-\i kx} a^*( k,\sigma)\right).\notag
\end{eqnarray}

We have accomplished the 
diagonalization of the Hamiltonian, momentum, polarization and symplectic form: 

\begin{eqnarray*}
H&=& \sum_{\sigma=0,\pm1}\int\d\vec k\varepsilon (\vec k) a^*(k,\sigma) a( k,\sigma),\\
\vec P&=& \sum_{\sigma=0,\pm1}\int\d\vec k  \vec k a^*(k,\sigma) a(
k,\sigma),\\
S&=& \sum_{\sigma=0,\pm1}\int\d\vec k \sigma
|\vec k|a^*(k,\sigma) a( k,\sigma),\\
\i\omega&=&
 \sum_{\sigma=0,\pm1}\int a^*(k,\sigma)\wedge
  a(k,\sigma)
\d\vec{k}.
\end{eqnarray*}

\subsubsection{Plane waves}

A {\em plane wave} is defined as
\beq |k,\sigma)^\mu=\frac{1}{\sqrt{(2\pi)^3}\sqrt{2\varepsilon(\vec k)}}
u^\mu(k,\sigma)\e^{\i kx},\label{planewave1a}\eeq
with $k^0=\varepsilon(\vec k)=\sqrt{\vec k^2+m^2}$. We have
\begin{eqnarray*}
\i(\bar{k,\sigma}|\omega|k',\sigma')=\i(k,\sigma|\omega|\bar{k',\sigma'})&=&0,\\
\i(\bar{k,\sigma}|\omega|\bar{k',\sigma'})=-\i(k,\sigma|\omega|k',\sigma')&=&\delta(\vec k-\vec k')
\delta_{\sigma,\sigma'}.\end{eqnarray*}
$a(k,\sigma)$ can be called {\em plane wave functionals}:
\begin{eqnarray*}
a(k,\sigma)&=&-\i A\lpar |k,\sigma)\rpar\\
&=&
-\i\int
\Bigl(\bigl(\partial_t( k,\sigma|x)_i\  \ -\partial_i ( k,\sigma|x)_0\bigr)
 A_i(0,\vec x)-
( k,\sigma|x)_i E_i(0,\vec
x)\Bigr)\d \vec x,\\
a^*(k,\sigma)&=&\i A\lpar |-k,\sigma)\rpar\\
&=&
\i\int
\Bigl(\bigl(\partial_t\bar{( k,\sigma|x)}_i\  \ -\partial_i \bar{( k,\sigma|x)}_0\bigr)
 A_i(0,\vec x)-
\bar{( k,\sigma|x)}_i E_i(0,\vec
x)\Bigr)\d \vec x
. \end{eqnarray*}


\nowastrona


\nowastrona


\subsubsection{Positive frequency space}

 $\cW_\Pr^{(+)}$ will denote the subspace of $\cc
\cY_\Pr$
 consisting of {\em positive frequency
 solutions}:
\begin{eqnarray*}
\cW_\Pr^{(+)}&:=&\{g\in\cc\cY_\Pr\ :\ 
 \bar{(k,\sigma|}\omega g=0,\ \sigma=\pm,0\}.
\end{eqnarray*}

\nowastrona

Every $g\in
\cW_\Pr^{(+)}$ can be written as
\[g_\mu(x)= \sum_{\sigma=0,\pm1}
\int\frac{\d\vec{k}}{\sqrt{(2\pi)^3}\sqrt{2\varepsilon (\vec k)}}
\e^{\i kx}
u_\mu(k,\sigma)\langle
a(k,\sigma)|g\rangle.
\]
For $g_1,g_2\in\cW_\Pr^{(+)}$ we define the  scalar product
\beq(g_1|g_2):=\i\bar g_1\omega g_2=
 \sum_{\sigma=0,\pm1}\int\bar{\langle a(k,\sigma)|g_1\rangle}\langle
 a(k,\sigma )|g_2\rangle\d\vec{k}.\label{tyt1}\eeq

We set $\cZ_\Pr$ to be the completion of $\cW_\Pr^{(+)}$ in this scalar
product.
$\rr^{1,3}\rtimes  O^\uparrow(1,3)$ leaves $\cZ_\Pr$ invariant.

We have 
\[\langle
 a(k,\sigma )|g  \rangle=(k,\sigma |g).\]
We can identify
$\cZ_\Pr$ with $L^2(\rr^3,\cc^3)$ and rewrite (\ref{tyt1}) as
\[(g_1|g_2)=
 \sum_{\sigma=0,\pm1}\int\bar{(k,\sigma|g_1)}(k,\sigma |g_2)\d\vec{k}.\]

\nowastrona

\subsubsection{Spin averaging}
For a given $k\in\rr^{1,3}$ with $k^2=m^2$, let $M,N$ be vectors with
\[M^\mu k_\mu= N^\nu k_\nu=0.\]
The following identity allows us to {\em average over spin}
 and is useful in
computations of scattering cross-sections:
\beq
\sum_{\sigma=0,\pm1} \bar{M^\mu u_\mu(k,\sigma)}u_\nu(k,\sigma) N^\nu=
\bar{M^\mu }N_\nu.\label{spino+}\eeq

\nowastrona

In fact,
\[\sum_{\sigma=0,\pm1} \bar{ u_\mu(k,\sigma)}u_\nu(k,\sigma) =g_{\mu\nu}
+\frac{ k_\mu k_\nu}{m^2}.\]
Therefore,
the left hand side of (\ref{spino+}) equals
\begin{eqnarray*}
\bar{M^\mu} g_{\mu\nu} N^\nu+\frac{( \bar 
M\cdot k)(N\cdot  k)}{m^2}.\end{eqnarray*}
But 
\[ k\cdot M= k\cdot N=0.\]

\nowastrona

\nowastrona


\nowastrona

\subsubsection{Quantization}

We want to construct  $(\cH,\hat H,\Omega)$ satisfying the standard requirements of QM (1)-(3)
and 
a self-adjoint operator-valued 
distribution $\rr^{1,3}\ni x\mapsto \hat A_\mu(x)$ 
such that, setting $
\vec{\hat E}=\dot{\vec{\hat A}}-\vec\partial \hat A_0$, we have
\ben\item 
$
-\partial^\mu (\partial_\mu \hat A_\nu-\partial_\nu \hat A_\mu)
+m^2 \hat A_\nu(x)=0$;
\item
$[\hat A_i(0,\vec x),\hat A_j(0,\vec y)]=[ \hat E_i(0,\vec x), \hat 
E_j(0,\vec y)]=0$,
\\
$[\hat A_i(0,\vec x),\hat E_j(0,\vec y)]=\i\delta_{ij}\delta(\vec x-\vec y)$;
\item 
$\e^{\i t\hat H}\hat A_\mu(x^0,\vec x)\e^{-\i t\hat H}=\hat A_\mu(x^0+t,\vec x)$;
\item 
$\Omega$ is cyclic for $\hat A_\mu(x)$.\een

\nowastrona




\nowastrona

The above problem has an essentially unique  solution, which we describe below.
 
For the Hilbert space we should  take the bosonic Fock space
 $\cH=\Gamma_\s(\cZ_\Pr)$
and for $\Omega$  the Fock
vacuum. With $\cZ_\Pr\simeq L^2(\rr^3,\cc^3)$ and $k$ on shell we have creation operators
\[\hat a^*(k,\sigma)=\hat a^*\big(|k,\sigma)\big),\]
written in both ``physicist's'' and
  ``mathematician's notation'' satisfying
\begin{eqnarray*}
[\hat a( k,\sigma), \hat a( k',\sigma')]=[\hat a^*( k,\sigma),\hat a^*( k',\sigma')]&=&0,\\{}
[\hat a( k,\sigma),\hat a^*( k',\sigma')]&=&\delta(\vec k-\vec
k')\delta_{\sigma,\sigma'}. \end{eqnarray*}
Therefore the {\em quantum 4-potentials}
\begin{eqnarray}&&\label{+++}\hat A_\mu(x)\\&=&
\sum_{\sigma=0,\pm1}
\left(u_\mu(k,\sigma)\e^{\i kx}\hat a( k,\sigma)+ 
\bar{u_\mu(k,\sigma)}\e^{-\i kx}\hat a^*( k,\sigma)\right)\notag\end{eqnarray}
satisfy the required commutation relations.
\nowastrona
The {\em quantum Hamiltonian,  momentum and polarization} are
\begin{eqnarray*}
\hat H&=&\sum_{\sigma=0,\pm1}\int \varepsilon(\vec k)
\hat a^*(k,\sigma)\hat a( k,\sigma)\d\vec{k},\\
\vec{\hat P}&=&\sum_{\sigma=0,\pm1}
\int \vec k \hat a^*( k,\sigma)\hat a( k,\sigma)\d\vec{k},\\
\hat S&=&\sum_{\sigma=0,\pm1}\int \sigma
|\vec k|\hat a^*(k,\sigma) \hat a( k,\sigma) \d\vec k.
\end{eqnarray*}


\nowastrona

The group $\rr^{1,3}\rtimes  O^\uparrow(1,3)$ is  unitarily implemented on $\cH$
by
 $U(y,\Lambda):=\Gamma\Bigl(r_{(y,\Lambda)}\Big|_{\cZ_\Pr}\Bigr)$
We have
\[U(y,\Lambda)\hat A_\mu(x)U(y,\Lambda)^*=\Lambda_\mu^\nu\hat
A_\nu\bigl((y,\Lambda)x\bigr).\] 
Moreover,
\begin{eqnarray*} [\hat A_\mu(x),\hat A_\nu(y)]&=&-\i
 \left(g_{\mu\nu}-\frac{\partial_\mu\partial_\nu}{m^2}\right)D(x-y).
\end{eqnarray*}

Note the identities
\begin{eqnarray}\notag
(\Omega|\hat  A_\mu(x)\hat A_\mu(y)\Omega)
&=&-\i\left(g_{\mu\nu}-\frac{\partial_\mu\partial_\nu}{m^2}\right)
D^{(+)}(x-y),\\
(\Omega|\T(\hat A_\mu(x)\hat A_\nu(y))\Omega)
&=&-\i\left(g_{\mu\nu}-\frac{\partial_\mu\partial_\nu}{m^2}\right)D^{\rm
  c}(x-y).\label{causa}
\end{eqnarray}

\nowastrona

For  $f\in C_{\rm c}^\infty(\rr^{1,3},\rr^{1,3})$ set
\[\hat  A[f]:=\int f^\mu(x) \hat A_\mu(x)\d x.\]
We obtain a family  that  satisfies
 the  Wightman axioms with
$\cD:=\Gamma_\s^\fin(\cZ_\Pr)$.

For an open  set $\cO\subset \rr^d$ we set
\[\fA(\cO):=\left\{\exp(\i \hat A[f])\ :\ f\in C_{\rm
  c}^\infty(\cO,\rr^{1,3})\right\} 
.\]
The algebras $\fA(\cO)$ satisfy the  Haag-Kastler axioms.

\nowastrona

\subsection{Massive photons with an external 4-current}

\subsubsection{Classical 4-potentials}

We return to the classical Proca equation.
We assume that \beq\rr^{1,3}\ni x\mapsto J(x)=[J^\mu(x)]\in \rr^{1,3}\label{source2}\eeq
 is a given  function  called an {\em external 4-current}, which satisfies
\begin{eqnarray} \partial _\nu J^\nu(x)&=&0.
\label{current}
\end{eqnarray} 
In most of this subsection we will assume that 
(\ref{source2}) is Schwartz.

In its presence  the Proca equation takes
the form
\begin{eqnarray}
-\partial^\mu (\partial_\mu A^\nu-\partial^\nu A_\mu)
+m^2 A^\nu(x)&=&-J^\nu(x).
\label{proc}
\end{eqnarray}

Note that (\ref{proc}) and (\ref{current}) imply the {\em Lorentz condition}
\begin{eqnarray}
\partial_\nu A^\nu(x)&=&0.\label{lorenz}\end{eqnarray}
We have therefore
\beq (-\Box+m^2)A^\mu(x)=-J^\mu(x).\eeq

The temporal component of (\ref{proc}) has no time derivative:
\beq
-\Delta A_0+\partial_0\div \vec A+m^2A_0=-J_0.\eeq
Therefore, we can compute $A_0$ in terms of $\vec A$ at the same time:
\beq A_0=-(-\Delta+m^2)^{-1}(\partial_0\div \vec A+J_0).
\eeq
The only dynamical variables are the spatial components, satisfying the
equation
\beq
(\partial_0^2-\Delta+m^2)\vec A=-\vec J.\eeq

\nowastrona

\subsubsection{Lagrangian and Hamiltonian formalism}

The  Lagrangian density is
\begin{eqnarray*}
\cL&:=&-\frac14 F_{\mu\nu}F^{\mu\nu}-\frac{m^2}{2}A_\mu A^\mu-J_\mu A^\mu\\
&=&-\frac12 \partial_\mu A_\nu\partial^\mu A^\nu-\frac12\partial_\mu
A_\nu\partial^\nu A^\mu-
\frac{m^2}{2}A_\mu A^\mu-J_\mu A^\mu\\
&=&-\frac12(\rot \vec A)^2+\frac12(\vec\partial A_0)^2+\frac12\bigl(\dot{\vec
  A}\bigr)^2- \dot{\vec A}\vec\partial
A_0+\frac{m^2}{2}A_0^2-\frac{m^2}{2}\vec A^2-\vec J\vec A+J_0A_0.
\end{eqnarray*}

\nowastrona


As noted before, only spatial components $\vec A(x)$ are dynamical and the
conjugate variable is $\vec E(x)=\dot{\vec A}(x)-\vec\partial A_0(x)$.
Thus  we have the standard
 Poisson brackets:
\begin{eqnarray}
\{A_i(t,\vec x),A_j(t,\vec y)\}=\{ E_i(t,\vec x), E_j(t,\vec y)\}&=&0,
\nonumber
\\
\{A_i(t,\vec x),E_j(t,\vec y)\}&=&\delta_{ij}\delta(\vec x-\vec y).
\label{poisson3+}\end{eqnarray}
We  can compute $A_0$ in terms of $\vec E$:
\beq A_0=-\frac{1}{m^2}(J_0+\div\vec E).\label{lus}\eeq

\nowastrona

The  {\em canonical Hamiltonian density} is
\begin{eqnarray*}
\cH^\can(x)&=&-\cL(x)+\frac{\partial\cL(x)}{\partial
  \dot A_i(x)} \dot A_i(x) \\
&=&
\frac12(\rot \vec A)^2(x)-\frac12(\vec\partial A_0)^2(x)-\frac12\bigl(\dot{\vec
  A}\bigr)^2(x)
\\&&-\frac{m^2}{2}A_0^2(x)+\frac{m^2}{2}\vec A^2(x)+\vec J(x)\vec A(x)-J_0(x)A_0(x).
\end{eqnarray*}
We add to it a  spatial divergence
$\div\big(\vec E(x) A^0(x)\big)$ and express it in terms of $\vec A$,
$\vec E$, obtaining the usual Hamiltonian density
\begin{eqnarray*}\cH(x)&:=&
\frac12\vec E^2(x)+\frac12(\rot \vec A)^2(x)\\&&+\frac{m^2}{2}A_0^2(x)+ 
\frac{m^2}{2}\vec A^2(x)
+\vec J(x)\vec A(x).\end{eqnarray*}
 The 
 Hamiltonian
\begin{eqnarray}
H(t)&=&\int \cH(t,\vec x)\d\vec x=\int \cH^\can(t,\vec x)\d\vec
x\label{tutu}\end{eqnarray}
generates the equations
of motion. Using
the  splitting of $\vec A$ and $\vec E$ into the transversal and
longitudinal part, as in (\ref{nice}), we can rewrite $H(t)$ as
\begin{eqnarray}\notag H(t)&=&
\int\d\vec x\Big(\frac12\vec{  E}_\tr^2(t,\vec x)
+\frac12 \vec { A}_\tr(t,\vec x)(-\Delta+m^2) \vec { A}_\tr(t,\vec x)
+\vec J( x) \vec { A}_\tr( x)\Big)
\\
&&+
 \int\d\vec x\Big(
\frac12\big((-\Delta)^{-1/2} \div \vec E(t,\vec x)\big)^2+\frac{1}{2m^2}\big(J^0(t,\vec x)-\div
\vec E( t,\vec x)\big)^2\notag \\
&&
+\frac{m^2}{2}\big((-\Delta)^{-1/2}\div \vec A(t,\vec x)\big)^2\Big).
\label{tra}
\end{eqnarray}



\nowastrona
We can interpret interacting fields as functionals on
$\cY_\Pr$ satisfying
\begin{eqnarray}\vec A(0,\vec x)=\vec A_\fr(0,\vec x),&&\vec E(0,\vec x)=\vec E_\fr(0,\vec x).
\notag\end{eqnarray}

\nowastrona

\subsubsection{Quantization}
\label{quant-proca}

We are looking for  operator valued distributions 
 $\rr^{1,3}\ni x\mapsto \hat A_\mu(x)$ 
satisfying
\begin{eqnarray*}
-\partial_\mu (\partial^\mu \hat A^\nu(x)-\partial^\nu \hat A^\mu(x))
+m^2 \hat A^\nu(x)&=&-J^\nu(x),
\end{eqnarray*}
 having the standard equal time commutation  relations with $\hat E^i:=\dot {\hat A}^i-\partial_i\hat A_0$
\begin{eqnarray*}
[\hat A_i(0,\vec x),\hat A_j(0,\vec y)]=[ \hat E_i(0,\vec x), \hat 
E_j(0,\vec y)]&=&0,
\\{}
[\hat A_i(0,\vec x),\hat E_j(0,\vec y)]&=&\i\delta_{ij}\delta(\vec x-\vec y).
\end{eqnarray*}
We will assume that $\vec{\hat A}$, $\vec{\hat E}$ coincide with free fields at $t=0$:
\begin{eqnarray*}
\hat A^i(\vec x)&:=&\hat A^i(0,\vec x)=\hat A_\fr^i(0,\vec x),\\
\hat E^i(\vec x)&:=&\hat E^i(0,\vec x)=\hat E_\fr^i(0,\vec x).
\end{eqnarray*}

We have
\[\hat A_\mu(t,\vec x):=\Texp\left(-\i\int_t^0 \hat H(s)\d s\right)\hat
A_\mu(\vec
x)\Texp\left(-\i\int_0^t \hat H(s)\d s\right), \]
where the Hamiltonian $\hat H(t)$, and the corresponding Hamiltonian in the
interaction picture are
\begin{eqnarray*}
\hat H(t)&=&\int\d \vec x\ {:}\Bigl(
\frac12\vec{ \hat E}^2(\vec x)+\frac{1}{2m^2}(J^0(t,\vec x)-\div\vec{\hat
  E}(\vec x))^2\\
&&\ \ \ +\frac12(\rot \vec {\hat A})^2(\vec x)
+
\frac{m^2}{2}\vec {\hat A}^2(\vec x)
+\vec J(t,\vec x)\vec{\hat A}(\vec x)\Bigr){:}\\&&
\\
\hat H_\Int(t)&=&
\int\d \vec x\Bigl(
-\frac{1}{m^2} J^0(t,\vec x)\div\vec{\hat
  E}_\fr(t,\vec x)+\vec J(t,\vec x)\vec{\hat A}_\fr(t,\vec x)
+\frac{1}{2m^2}J^0(t,\vec x)^2\Bigr)
.\end{eqnarray*}
Using (\ref{+++}) and
$\div\vec{\hat E}_\fr=-(-\Delta+m^2)^{-1}\div\dot{\vec{\hat A}}_\fr$, we express
the interaction picture Hamiltonian in terms of creation/annihilation operators:
\begin{eqnarray*}
\hat H_\Int(t)&=&
\int\frac{\d \vec k}{\sqrt{(2\pi)^3}\sqrt{2\varepsilon(\vec
    k)}}\Big(\e^{\i t\varepsilon(\vec k)}J_\mu(t,\vec
k)\bar{u^\mu(\vec k,\sigma)}\hat a^*(k,\sigma)\\&&\ \ +
\e^{-\i t\varepsilon(\vec k)}
\bar{J_\mu(t,\vec k)}u^\mu(\vec k,\sigma)\hat a(k,\sigma)\Big)
+\int\frac{\d\vec k}{(2\pi)^{3}}\frac{|J^0(t,\vec k)|^2}{2m^2}
.\end{eqnarray*}

We can compute the  scattering operator
\nowastrona
\begin{eqnarray*}
\hat S&=&\exp\left(-\frac{\i}{2}\int\frac{ \d k}{(2\pi)^4} \bar{J^\mu(k)}D_{\mu\nu}^{0}(k) J^\nu(k)\right)\\
&&\times\exp\left(-\i\sum_{\sigma=0,\pm1}\int\frac{\d \vec k}{\sqrt{(2\pi)^3}}
      a^*(k,\sigma)\frac{\bar{u_\mu(k,\sigma)}}{\sqrt{2\varepsilon(\vec k)}}J^\mu(\varepsilon(\vec k),\vec k)\right)\\&&
\times\exp\left(-\i\sum_{\sigma=0,\pm1}\int\frac{\d \vec k}{\sqrt{(2\pi)^3}}
      a(k,\sigma)\frac{u_\mu(k,\sigma)}{\sqrt{2\varepsilon(\vec k)}}\bar{J^\mu(\varepsilon(\vec k),\vec k)}\right),
\end{eqnarray*}
where
\beq D_{\mu\nu}^{0}(k)=
\frac{1}{m^2+k^2-\i0}\left(g_{\mu\nu}+\frac{k_\mu
    k_\nu}{m^2}\right).\label{qhw+}\eeq
\nowastrona
(The superscript $0$ over $D_{\mu\nu}^0$ will be explained later on).

For $x_{N},\dots,x_1$, the {\em $N$-point 
 Green's function} is defined
as follows:
\begin{eqnarray*}
&&\langle\hat A_{\mu_N}(x_{N})\dots\hat A_{\mu_1}(x_1)\rangle\\&:=&
\left(\Omega^+|\T\big(\hat A_{\mu_N}(x_{N})\cdots\hat A_{\mu_1}(x_1)\big)\Omega^-\right).
\end{eqnarray*}
Green functions can be organized into the {\em generating function}
\begin{eqnarray*}
&&\sum_{n=0}^\infty\int\cdots\int 
\langle\hat A_{\mu_N}(x_{N})\cdots\hat A_{\mu_1}(x_{1})\rangle
(-\i)^Nf^{\mu_N}(x_N)\cdots f^{\mu_1}(x_1)\d x_N\cdots\d x_1\\
&=&
\left(\Omega\Big|\Texp\left(-\i\int_{-\infty}^{\infty} \hat
H_\Int(s)\d s-\i\int f^\mu(x)\hat A_\mu(x)\d
x\right)\Omega\right)\ =:\ Z(f).\end{eqnarray*}
The {\em amputated $N$-point 
 Green's functions} are
\begin{eqnarray*}
&&\langle\hat A_{\mu_N}(k_{N})\cdots\hat A_{\mu_1}(k_1)\rangle_\amp\\&:=&
(k_N^2+m^2)\cdots(k_1^2+m^2)
\langle\hat A_{\mu_N}(k_{N})\cdots\hat A_{\mu_1}(k_1)\rangle
.\end{eqnarray*}

For $k_1,\dots,k_N$ on shell, set
\[|k_N,\sigma_N,\dots,k_1,\sigma_1):=\hat a^*(k_N,\sigma_N)\cdots
\hat a^*(k_1,\sigma_1)\Omega.\] 
Amputated Green's functions can be used to compute {\em scattering amplitudes}:
\begin{eqnarray*}
&&\left(k_{n^+}^+,\sigma_{n^+}^+,\dots,k_1^+,\sigma_1^+|\,\hat S\, |k_{n^-}^-,\sigma_{n^-}^-,\dots
,k_1^-,\sigma_1^-\right)\\[3ex]
&=&\frac{
\bar{ u^{\mu_1^+}(k_1^+,\sigma_1^+)}\cdots \bar{ u^{\mu_{n^+}^+}(k_{n^+}^+,\sigma_{n^+}^+)}
 u^{\mu_{n^-}^-}(k_{n^-}^-,\sigma_{n^-}^-)\cdots
u^{\mu_1^-}(k_1^-,\sigma_1^-)}{(2\pi)^{\frac{n^++n^-}{2}}
\sqrt{2\varepsilon(k_1^+)}\cdots
\sqrt{2\varepsilon(k_{n^+}^+)}
\sqrt{2\varepsilon(k_{n^-}^-)}\cdots\sqrt{
2\varepsilon(k_1^-)}}\\
&&\times \langle\hat A_{\mu_1^+}(k_1^+)\cdots
\hat A_{\mu_{n^+}^+}(k_{n^+}^+)\hat A_{\mu_{n^-}^-}(-k_{n^-}^-)\cdots
\hat A_{\mu_1^-}(-k_1^-)\rangle_\amp.
\end{eqnarray*}

\nowastrona

\subsubsection{Causal propagators}
\label{Propagators for massive QED}

The causal propagator used to compute Green's functions and
 scattering amplitudes that follows
directly from the interaction picture Hamiltonian is $\left(g_{\mu\nu}-\frac{\partial_\mu\partial_\nu}{m^2}\right)D^{\rm
  c}$, see
(\ref{causa}).
If we compute  scattering amplitudes,
we can pass from this  propagator  to another by adding
$k_\mu f_\nu(k)+f_\mu(k)k_\nu$
for an arbitrary function $f_\mu(k)$.

\nowastrona

To see this note  that after adding
$k_\mu f_\nu(k)+f_\mu(k)k_\nu$ the contribution of each line 
changes by
\[J^\mu(k)\left(k_\mu f_\nu(k)+f_\mu(k)k_\nu\right)J^\nu(k),\]
which is zero, because $k_\mu J^\mu(k)=0$.
For scattering amplitudes, external lines do not involve the
propagator. Therefore, scattering amplitudes do not change. 

Below we will list a number of useful
causal propagators. (In principle, they should be decorated by the superscript ${\rm c}$, for {\em  causal}, which we however suppress).

For any $\alpha\in\rr$, we can pass to the following
 propagators
\begin{eqnarray*}
D_{\mu\nu}^{\alpha}&=&\frac{1}{m^2+k^2-\i0}\left(g_{\mu\nu}+(1-\alpha)\frac{k_\mu
    k_\nu}{\alpha k^2+m^2}\right).
\end{eqnarray*}
The above propagator for $\alpha=0$ was obtained in the Hamiltonian approach, see (\ref{causa}).
For $\alpha=1$
we obtain the so-called propagator in the {\em  Feynman gauge}
\[D_{\mu\nu}^{\rm Feyn}(k)=
\frac{1}{m^2+k^2-\i0}.\]
 $\alpha=\infty$ corresponds to the propagator in the {\em Landau} or {\em Lorentz gauge}:
\begin{eqnarray*}
D_{\mu\nu}^{\rm Lan}&=&\frac{1}{m^2+k^2-\i0}\left(g_{\mu\nu}-\frac{k_\mu
    k_\nu}{ k^2}\right).
\end{eqnarray*}

\nowastrona
We can introduce the propagator in the
{\em Yukawa gauge}:
\begin{eqnarray*}
D_{00}^{\rm Yuk}=-\frac{1}{m^2 +\vec k^2},\ \ \ D_{0j}^{\rm Yuk}
=0,\ \ \ 
D_{ij}^{\rm Yuk}=\frac{1}{m^2+k^2-\i0}\left(\delta_{ij}-\frac{k_ik_j}{m^2+\vec k^2}\right).
\end{eqnarray*}
We have $D_{\mu\nu}^{\rm Yuk}=D_{\mu\nu}^{\rm Feyn}+k_\mu f_\nu^{\rm
  Yuk}(k)+f_\mu^{\rm Yuk}(k)k_\nu$,
where
\[f_0^{\rm Yuk}(k)=\frac{k_0}{(k^2+m^2-\i0)2(m^2+\vec k^2)},\
 \ f_i^{\rm Yuk}(k)=-\frac{k_i}{(k^2+m^2-\i0)2(m^2+\vec
  k^2)}. \]
(The  propagator in the Yukawa gauge is the massive analog of the propagator in the Coulomb gauge.)

\nowastrona

The propagator in the {\em temporal gauge} is
\begin{eqnarray*}
D_{00}^{\rm tem}=0,\ \ \ D_{0j}^{\rm tem}=0,\ \ \ 
D_{ij}^{\rm tem}=\frac{1}{k^2+m^2-\i0}\left(\delta_{ij}-\frac{k_ik_j}{k_0^2}\right).
\end{eqnarray*}
We have $D_{\mu\nu}^{\rm tem}=D_{\mu\nu}^{\rm Feyn}+k_\mu f_\nu^{\rm
  tem}(k)+f_\mu^{\rm tem}(k)k_\nu$, where
\[f_0^{\rm tem}(k)=\frac{1}{(m^2+k^2-\i0)2 k_0},\ \ f_i^{\rm tem}
(k)=-\frac{k_i}{(m^2+k^2-\i0)2
  k_0^2}. \]

\nowastrona
\subsubsection{Feynman rules}

Perturbation expansion can be organized with help of Feynman diagrams,
which are very similar to diagrams for neutral fields interacting with
a linear source.
We have 1 kind of lines and 1 kind of vertices.
 At each vertex  just
one line ends.

To compute 
 Green's functions we do as follows:

\ben \item In the $n$th order we draw all possible
Feynman diagrams with $n$ vertices and external lines.
\item 
To each  vertex 
 we associate the factor  $-\i J^\mu(k)$.
\item To each line we associate the propagator
$-\i D_{\mu\nu}^0(k)=\frac{-\i}{m^2+k^2-\i0}\left(g_{\mu\nu}+\frac{k_\mu
    k_\nu}{m^2}\right)$.
\item For internal lines we integrate over the variables with the measure
$\frac{\d^4 k}{(2\pi)^4}$.
\een

\nowastrona


To compute scattering amplitudes with $N^-$ incoming and $N^+$
outgoing particles we draw the same diagrams as
for $N^-+N^+$-point Green's functions.
The rules are changed only
concerning the external lines.
\begin{romanenumerate}\item
 With each incoming external line we associate
 $\frac{1}{\sqrt{(2\pi)^32\varepsilon(\vec k)}}u(k,\sigma)$.
\item  With each outgoing external line we associate
 $\frac{1}{\sqrt{(2\pi)^32\varepsilon(\vec k)}}\bar{u(k,\sigma)}$.
\end{romanenumerate}

If we prefer, we can use a different causal propagator instead of $D_{\mu\nu}^0$. Green's
functions change, because of external lines, however scattering amplitudes do not.

\subsubsection{Path integral formulation}
\label{Path integral formulation}

We can compute exactly the generating function:
\begin{eqnarray}\label{genor}
&&Z(f)\\
&=&\exp\left(\frac{\i}{2}\int\bar{J^\mu(k)+f^\mu(k)}\frac{(g_{\mu\nu}+m^{-2}k_\mu k_\nu)}{(k^2+m^2-\i0)}
(J^\nu(k)+f^\nu(k))\d k\right).\notag
\end{eqnarray}

This can be rewritten in the path integral formalism.
Recall that
\begin{eqnarray*}
\int \cL_\fr(x)\d x&=&-\int\frac12\Big(\partial_\mu
A_\nu(x)\partial^\mu A^\nu(x)-\partial_\mu A_\nu(x)\partial^\nu
A^\mu(x)\\
&&+m^2 A_\mu (x)A^\mu(x)\Big)\d x\\
&=&-\int\frac12 A_\mu(x)\left(g^{\mu\nu}(-\Box+m^2)+\partial^\mu\partial^\nu\right) A_\nu(x)\d x,\\
\int (\cL(x)-f^\mu(x)A_\mu)(x)\d x&=&\int\cL_\fr(x)\d x-\int( J^\mu+f^\mu(x))(x)A_\mu(x)\d x.
\end{eqnarray*}

Note that $D_{\mu\nu}^0(k)=\frac{g_{\mu\nu}+m^{-2}k_\mu k_\nu}{k^2+m^2-\i0}$, or in the position representation
$D_{\mu\nu}^0=(g_{\mu\nu}-m^{-2}\partial_\mu\partial_\nu)D^{\rm c}$ is one of the inverses of $g^{\mu\nu}(-\Box+m^2)+\partial^\mu\partial^\nu$.
Therefore, (\ref{genor}) is often formally rewritten as
\begin{eqnarray}\label{causapro3}
Z(f)&=&
\frac{\int\lpi_\mu\lpi_x\d A_\mu(x)\exp\left(\i\int\big(\cL(x)-(J^\mu(x)+f^\mu(x)) A_\mu(x)\big)\d x\right)}
{\int\lpi_\mu\lpi_x\d A_\mu(x)\exp\left(\i\int\cL_\fr(x)\d x\right)}.
\nonumber\end{eqnarray}

 Let $D^\bullet_{\mu\nu}$ be one of the propagators considered in 
 Subsubect. \ref{Propagators for massive QED}. Let $B_\bullet^{\mu\nu}$ be its inverse.
We have the corresponding ``free action''
\begin{eqnarray*}
 T_{\bullet\fr}
&=&-\frac12\int A_\mu(x) B_\bullet^{\mu\nu}(x-y) A_\nu(y)\d x\d y.
\end{eqnarray*}
We define the corresponding generating function as
\begin{eqnarray}\label{causapro4}&& Z_\bullet(f)
\\
&:=&\exp\left(\frac{\i}{2}\int\bar{(J^\mu(k)+f^\mu(k)} D_{\bullet\mu\nu}(k)
(J^\nu(k)+f^\nu(k))\d k\right)\nonumber
\\
&=&\exp\left(\frac{\i}{2}\int (J^\mu(x)+f^\mu(x))D_{\bullet\mu\nu}(x-y)( J^\nu(y)+f^\nu(y))\d x\d y\right)\nonumber\\
&=&\frac{\int\lpi_\mu\lpi_x  d A_\mu(x)\exp\left(\i T_{\bullet\fr}+\i\int (J^\mu(x) +f^\mu(x))A_\mu(x)\d x\right)}
{\int\lpi_\mu\lpi_x\d A_\mu(x)\exp\left(\i T_{\bullet\fr}\right)}
.\nonumber\end{eqnarray}
In general, $Z_\bullet(f)$ differs for various propagators $D^\bullet_{\mu\nu}$, unless $f$ satisfies the Lorentz condition. 
 However, all  $Z_\bullet(f)$  can be used to compute the same scattering operator.

Likewise, the Euler-Lagrange equations obtained from those various action integrals differ from the Proca equation. However, $\cY_\Pr$ belong always to their solutions.

If we take the Lagrangian
\begin{eqnarray}&&-\frac12\Big(\partial_\mu A^\nu(x) \partial^\mu A_\nu(x) +m^2A^\nu(x)A_\nu(x)\notag\\&&+(\alpha-1)\partial_\mu A^\mu(x)\partial_\nu A^\nu(x)\Big),\label{lagran}\end{eqnarray}
then we obtain the propagator $D_{\mu\nu}^\alpha$.
Indeed, 
\[g^{\mu\nu}(k^2+m^2)+(\alpha-1)k^\mu k^\nu\]
is the inverse of $D_{\mu\nu}^\alpha(k)$.

If we  restrict the integration by the Lorentz condition 
\beq \partial_\mu A^\mu(x)=0.\label{lorioma}\eeq
and take the Lagrangian (\ref{lagran}) (they now coincide for all $\alpha$),  then we obtain the propagator in the Landau/Lorentz gauge.

If we take the Lagrangian
\begin{eqnarray*}&&-\frac12\Big(\partial_\mu A_i(x) \partial^\mu A_i(x)+m^2A_i(x)A_i(x)\\&&+\frac{1}{m^2}\partial_\mu\partial_i A_i(x)\partial^\mu\partial_j A_j(x)+\partial_i A_i(x)\partial_j A_j(x)\\
&&-
\partial_iA_0(x)\partial_iA_0(x)-m^2A_0(x)^2\Big),\end{eqnarray*}
we obtain $D_{\mu\nu}^{\rm Yuk}$.
Indeed,
\[(k^2+m^2)\left(\delta_{ij}+\frac{k_ik_j}{m^2}\right)-
\delta_{\mu0}\delta_{0\nu}(\vec k^2+m^2)\]
is the inverse of $D_{\mu\nu}^{\rm Yuk}(k)$.

If we take the action
\begin{eqnarray*}&&-\frac12\int\left(\partial_\mu A_i(x) \partial^\mu A_i(x)+m^2A_i(x)A_i(x)\right)\d x\\
&&-\frac12\int\Big(\partial_\mu\partial_i
  A_i(x)(-\Box)^{-1}(x-y)\partial^\mu\partial_jA_j(y)\\&&\ \ \ +
\partial_i A_i(x)(-\Box)^{-1}(x-y)\partial_jA_j(y)\Big)\d x\d y
,\end{eqnarray*}(which is nonlocal and does not involve $A_0$),  we obtain $D_{\mu\nu}^{\rm tem}$.
Indeed, 
\[(k^2+m^2)\left(\delta_{ij}-\frac{k_ik_j}{k^2}\right)\]
is the inverse of $D_{ij}^{\rm tem}(k)$.

\subsubsection{Energy shift}

Suppose that the 4-current is stationary and is given by a Schwartz
function $\rr^3\ni\vec x\mapsto J^\mu(\vec x)$. Note that $\div\vec
J(\vec x)=0$.

Using the quantum version of (\ref{tra}), we can write the Hamiltonian as \begin{eqnarray*}
\hat H
&=&
\int\d \vec x\ {:}\Bigl(
\frac12\vec{ \hat E}_\tr^2(\vec x)
+\frac12 \vec {\hat A}_\tr(\vec x)(-\Delta+m^2) \vec {\hat A}_\tr(\vec x)
+\vec J(\vec x) \vec {\hat A}_\tr(\vec x)\Bigr){:}\\
&&\hspace{-8ex}+
\int\d \vec x\ {:}\Bigl(\frac12 \big((-\Delta)^{-1/2}\div\vec{\hat E}(\vec x)\big)^2
+\frac{1}{2m^2}\big(J^0(\vec x)-\div\vec{\hat
  E}(\vec x)\big)^2\\&&
+\frac{m^2}{2}\big((-\Delta^{-1/2}\div\vec{\hat{A}}(\vec x)\big)^2\Bigr){:}
.\end{eqnarray*}
By (\ref{hov1}), the
infimum of $\hat H$ is
\begin{eqnarray*}
E&=&-\frac12\int\int\d\vec x\d\vec y \vec J(\vec x)\frac{\e^{-m|\vec
    x-\vec y|}}{4\pi |\vec x-\vec y|}\vec J(\vec y)\\
&&+
\frac12\int\int\d\vec x\d\vec y J^0(\vec x)\frac{\e^{-m|\vec
    x-\vec y|}}{4\pi |\vec x-\vec y|}J^0(\vec y).
\end{eqnarray*}


\subsection{Alternative approaches}
\nowastrona 
 \subsubsection{Classical 4-potentials without the Lorentz condition}\label{sec-lor}

So far our treatment of massive photons was based on the Proca equation (\ref{proc-}). 
As we remember, the Proca equation is equivalent to the Klein-Gordon equation for vector fields (\ref{proc1v}) together with the Lorentz condition (\ref{lorenz1v}). This suggests an alternative approach to the massive photons.

In this approach one considers  first the Klein-Gordon equation on
functions with values in $\rr^{1,3}$:
\beq (-\Box+m^2)\zeta_\mu(x)=0.\label{maxi1m}\eeq
The space of 
 smooth real space-compact solutions of 
(\ref{maxi1m}) will be denoted by $\cY_\rv$.
The following 4-current
\begin{eqnarray*}
j_{\rv}^\mu(\zeta_1,\zeta_2,x)
&:=&\partial^\mu \zeta_{1,\nu}(x)\zeta_2^\nu(x)
-\zeta_{1,\nu}(x)\partial^\mu \zeta_2^\nu(x) \end{eqnarray*}
is conserved, that is
\[\partial_\mu j_\rv^\mu(x)=0.\]
It defines in the usual way a symplectic form on $\cY_\rv$
\begin{eqnarray*}
\zeta_1\omega_\rv \zeta_2&=&
\int_\cS  j_{\rv}^\mu(\zeta_{1},\zeta_2,x)\d s_\mu(x)\\
&=&\int\left(-\dot \zeta_{1\nu}(t,\vec x)\zeta_2^\nu(t,\vec x)+
\zeta_{1\nu}(t,\vec x)\dot\zeta_2^\nu(t,\vec x)\right)\d\vec{x},
\end{eqnarray*}
where $\cS$ is any Cauchy surface.
\nowastrona

One introduces the 4-potentials $A^\mu(x)$ as the
functionals on $\cY_\rv$ defined by
\[\langle A^\mu(x)|\zeta\rangle:=\zeta^\mu(x).\]
We clearly have
\beq (-\Box +m^2)A_\mu(x)=0.\label{maxi2m}\eeq 
\nowastrona

We can use the Lagrangian
\[\cL(x):=-\frac12 A_{\mu,\nu}(x)A^{\mu,\nu}(x)-\frac{m^2}{2}A_\mu (x)A^\mu(x).\]
The conjugate variables are
\[\Pi_\mu(x):=\frac{\partial}{\partial \dot A^\mu(x)}\cL(x)=\dot A_\mu(x).\]
The Poisson structure is
given by the equal time brackets
\begin{eqnarray*}
\{A_\mu(t,\vec x), A_\nu(t,\vec y)\}\ =\ 
\{\Pi_\mu(t,\vec x),\Pi_\nu(t,\vec y)\} &=&0,\\
\{A_\mu(t,\vec x),\Pi_\nu(t,\vec y)\}&=&g_{\mu\nu}\delta(\vec x-\vec y).
\end{eqnarray*}

The stress-energy tensor is
\begin{eqnarray*}
\cT^{\mu\nu}&=&-\frac{\partial\cL}{\partial A_{\alpha ,\mu}}A_{\alpha}^{\ ,\nu}+g^{\mu\nu}\cL\\
&=& A^{\ ,\mu}_\alpha A^{\alpha,\nu}-\frac{1}{2} g^{\mu\nu}\big(A_{\alpha,\beta}
A^{\alpha,\beta}+m^2A_\alpha A^\alpha\big).\end{eqnarray*}
The Hamiltonian and momentum density are
\begin{eqnarray*}
\cH(x)\, =\, \cT^{0 0}(x)&=&
\frac12\Pi_\mu(x)\Pi^\mu(x)+
\frac12 A_{\mu,i}(x)A^{\mu,i}(x)+\frac{m^2}{2}A_\mu(x) A^\mu(x),\\
\cP^i(x)\,=\,\cT^{0i}(x)&=&-\Pi_\mu(x)A^{\mu,i}(x).
\end{eqnarray*}
As usual, we can define the Hamiltonian and momentum
\begin{eqnarray}
H&=&\int\cH(t,\vec x)\d\vec x,\label{paksa}\\
P^j&=&\int\cP^j(t,\vec x)\d\vec x.\notag\end{eqnarray}
The Hamiltonian  (\ref{paksa})
is unbounded from below.

\subsubsection{The Lorentz condition}

Introduce two subspaces  of $\cY_\rv$ 
\begin{eqnarray*}
\cY_\Lor&:=&\{\zeta\in\cY_\rv\ : \ \partial_\mu\zeta^\mu=0\},\\
\cY_\sc&:=&\{\zeta\in\cY_\rv\ :
\ \zeta^\mu=\partial^\mu\chi,\ \ \chi\in\cY_\KG\}.
\end{eqnarray*}
Note that $\cY_\rv=\cY_\Lor\oplus\cY_\sc$ is a decomposition into
symplectically orthogonal subspaces each preserved by the
Poincar\'{e}{} group. If $\zeta\in\cY_\rv$, then its projection onto
$\cY_\sc$ is
\beq \zeta_\sc^\mu:=\frac{1}{m^2}\partial^\mu\partial_\nu\zeta^\nu.
\label{projec}\eeq

Elements of  $\cY_\Lor$ satisfy the Proca equation, so that we can
make the identification 
\[\cY_\Lor=\cY_\Pr.\]
On $\cY_\Lor$ the forms  $\omega_\rv$ and $\omega_\Pr$
coincide.

Clearly, we are back with the theory that was used in most of this
section. 
 In particular, the Hamiltonian (\ref{paksa}) restricted to 
$\cY_\Lor$ is now positive.

\subsubsection{Diagonalization of the equations of motion}

In order to diagonalize the Hamiltonian, besides the vectors
$u(k,\sigma)$ with $\sigma=0,\pm1$ introduced in
(\ref{ede1}),
  we will need the  vectors for the scalar plane waves
\[u(k,\sc):=\frac{1}{m}(\varepsilon(\vec k),\vec k).\]
Note that
\begin{eqnarray*}
\bar{u_\mu(k,\sigma)}u^\mu(k,\sigma')&=&\delta_{\sigma,\sigma'},\\
\sum_{\sigma}
\bar{u_\mu(k,\sigma)}u_\nu(k,\sigma)&=&g_{\mu\nu}. \end{eqnarray*}

Set
\begin{eqnarray*}
\vec A_t(\vec k)&=&\int \vec A(t,\vec x)\e^{-\i \vec k\vec x}\frac{\d \vec x}{\sqrt{(2\pi)^3}},
\\
\vec \Pi_t(\vec k)&=&\int \vec \Pi(t,\vec x)\e^{-\i \vec k\vec x}\frac{\d \vec x}{\sqrt{(2\pi)^3}}
.\end{eqnarray*}
We have the equations of motion
\begin{eqnarray*}
\dot{ A}_t(\vec k)&=&\Pi_t(\vec k),\\
\dot{ \Pi}_t(\vec k)&=&-\varepsilon(\vec k)^2A_t(\vec k),
 \end{eqnarray*}
the relations
\[ A_t^*(\vec k)= A_t(-\vec k),\ \  \Pi_t^*(\vec k)= \Pi_t(-\vec k),\]
and the Poisson brackets
\begin{eqnarray}
\{ A_{t\mu}^*(\vec k),A_{t\nu}(\vec k')\}=\{ \Pi_{t\mu}^*(\vec k), \Pi_{t\nu}(\vec k')\}&=&0,
\nonumber
\\
\{A_{t\nu}^*(\vec k') ,\Pi_{t\mu}(\vec k)\}&=&g_{\mu\nu}\delta(\vec k-\vec k').
\label{poisson31-}\end{eqnarray}

Set
\begin{eqnarray*}
A_t(\vec k, \sigma)&:=&\bar{ u_\mu(\vec k,\sigma)}  A_t^\mu(\vec k),\\
\Pi_t(\vec k,\sigma)&:=& \bar{ u_\mu(\vec k,\sigma)} \Pi_t^\mu(\vec k).\\
\end{eqnarray*}

We have the equations of motion
\begin{eqnarray*}
\dot{ A}_t(\vec k,\sigma)&=&\Pi_t(\vec k,\sigma),\\ 
\dot{ \Pi}_t(\vec k,\sigma)&=&-\varepsilon(\vec k)^2 A_t(\vec k,\sigma),\\
 \end{eqnarray*}
the relations
\[ A_t^*(\vec k,\sigma)= A_t(-\vec k,-\sigma),\ \  \Pi_t^*(\vec k,\sigma)= \Pi_t(-\vec k,-\sigma),\]
and the Poisson brackets
\begin{eqnarray}
\{ A_{t}^*(\vec k,\sigma),A_{t}(\vec k',\sigma')\}=\{  \Pi_{t}^*(\vec k,\sigma), \Pi_{t}(\vec k',\sigma')\}&=&0,
\nonumber
\\
\{A_{t}^*(\vec k,\sigma),\Pi_{t}(\vec k',\sigma')\}&=&\kappa_{\sigma\sigma'}\delta(\vec k-\vec k').
\label{poisson32-}\end{eqnarray}
where $\kappa_{\sigma,\sigma'}=1$  for $\sigma=\sigma'=\pm 1,0$ and $\kappa_{\sc,\sc}=-1$.
We set
\begin{eqnarray*}
a_t(k,\sigma)&:=&
\sqrt{\frac{\varepsilon (\vec k)}{2}}A_t(\vec k,\sigma)
-\frac{\i}{\sqrt{2\varepsilon (\vec k)}}\Pi_t(\vec k,\sigma),\\
 a_t^*(k,\sigma)&:=&
\sqrt{\frac{\varepsilon (\vec k)}{2}} A_t^*(\vec k,\sigma)
+\frac{\i}{\sqrt{2\varepsilon (\vec k)}} \Pi_t^*(\vec k,\sigma).
 \end{eqnarray*}

We have the equations of motion
\begin{eqnarray*}
\dot a_t( k,\sigma)&=&-\i \varepsilon(\vec k)a_t( k,\sigma),\\
\dot{ a_t^*}( k,\sigma)&=&\i \varepsilon(\vec k) a_t^*( k,\sigma).
\end{eqnarray*}
and the Poisson brackets
\begin{eqnarray*}
\{a( k,\sigma), a( k',\sigma')\}=\{a^*( k,\sigma), a^*( k',\sigma')\}&=&0,\\
\{a( k,\sigma), a^*( k',\sigma')\}&=&-\i\kappa_{\sigma,\sigma'}\delta(\vec k-\vec
k'). \end{eqnarray*}

We diagonalize the Hamiltonian and momentum:
\begin{eqnarray*}
H&=& \sum_{\sigma=0,\pm1}\int\d\vec k\varepsilon (\vec k)a^*(k,\sigma) a( k,\sigma)
-\int\d\vec k\varepsilon (\vec k) a^*(k,\sc) a( k,\sc),\\
\vec P&=& \sum_{\sigma=0,\pm1}\int\d\vec k  \vec k a^*(k,\sigma) a(
k,\sigma)-
\int\d\vec k  \vec ka^*(k,\sc) a(
k,\sc).
\end{eqnarray*}
 The 4-potentials can be decomposed as
\begin{eqnarray*}
 A_\mu(x)&=& \sum_{\sigma}
\int\frac{\d\vec{k}}{\sqrt{(2\pi)^3}\sqrt{2\varepsilon (\vec k)}}
 \left(u_\mu(k,\sigma)\e^{\i kx} a( k,\sigma)+
\bar{u_\mu(k,\sigma)}\e^{-\i kx}a^*( k,\sigma)\right).
\end{eqnarray*}

Clearly, the restriction to $\cY_\Lor$ amounts to
 dropping all  scalar
components.

\subsubsection{Positive frequency space}

 $\cW_\rv^{(+)}$ will denote the subspace of $\cc
\cY_\rv$
 consisting of {\em positive frequency
 solutions}:
\begin{eqnarray*}
\cW_\rv^{(+)}&:=&\{ g\in\cc\cY_\Pr\ :\ \bar{(k,\sigma|}\omega_\rv g
=0,\ \ \sigma=\pm,0,\sc\}.
\end{eqnarray*}

\nowastrona

Every $g\in
\cW_\rv^{(+)}$ can be written as
\[g_\mu(x)= \sum_{\sigma=0,\pm1,\sc}
\int\frac{\d\vec{k}}{\sqrt{(2\pi)^3}\sqrt{2\varepsilon (\vec k)}}
\e^{\i kx}
u_\mu(k,\sigma)\langle
a(k,\sigma)|g\rangle.
\]
For $g_1,g_2\in\cW_\rv^{(+)}$ we have a natural scalar product
\begin{eqnarray}\notag
( g_1| g_2):=\i\bar g_1\omega_\rv g_2&=&
 \sum_{\sigma=0,\pm1}\int\bar{\langle a(k,\sigma)| g_1\rangle}\langle
 a(k,\sigma )| g_2\rangle\d\vec{k}\\&&\hskip 2ex
-\int\bar{\langle a(k,\sc)| g_1\rangle}\langle
 a(k,\sc )| g_2\rangle\d\vec{k}\notag\\
&=&
 \int g^{\mu\nu}\bar{\langle a_\mu(k)|g_1\rangle}\langle
a_\nu(k)|g_2\rangle\d\vec{k}
.\label{scal}\end{eqnarray}
Unfortunately, 
 the
above definition gives an indefinite scalar product. We can also
introduce a positive definite scalar product, which unfortunately is
not covariant: 
\[(g_1|g_2)_+:=
\sum_\mu \int \bar{\langle a_\mu(k)|g_1\rangle}\langle
a_\mu(k)|g_2\rangle\d\vec{k}.\] 
The positive frequency space  $\cW_\rv^{(+)}$ equipped with the scalar product (\ref{scal}) can be completed in the norm given by $(\cdot|\cdot)_+$. It will  be called
$\cZ_\rv$. 
It is an example of the so-called {\em Krein space}, which
is a space with an indefinite scalar product and has a topology given by
a
positive scalar product.

 Using the projection (\ref{projec}),
$\cW_\rv^{(+)}$ can be  decomposed into the direct
sum of orthogonal subspaces  $\cW_\Lor^{(+)}$ 
and $\cW_\sc^{(+)}$. On  $\cW_\Lor^{(+)}$ the scalar product
(\ref{scal}) is positive definite, on $\cW_\sc^{(+)}$ it is negative
definite. Their completions will be denoted
$\cZ_\Lor$ and $\cZ_\sc$. 

Every $\zeta\in\cY_\rv$ can be uniquely written as $\zeta=\zeta^{(+)}+\bar{\zeta^{(+)}}$, where $\zeta^{(+)}\in\cW_\rv^{(+)}$. This allows us to define a real  scalar product on $\cY_\rv$:
\begin{eqnarray}\label{derive9=}\langle\zeta_1|\zeta_2\rangle_\cY&:=&\Re(\zeta_1^{(+)}|\zeta_2^{(+)})\\
&=&\int\int\dot\zeta_{1\mu}(0,\vec x)(-\i)D^{(+)}(0,\vec x-\vec y)
\dot\zeta_2^\mu(0,\vec y)\d\vec x\d\vec y\nonumber\\
&&+
\int\int\zeta_{1\mu}(0,\vec x)(-\Delta_{\vec x}+m^2)(-\i)D^{(+)}(0,\vec x-\vec y)
\zeta_2^\mu(0,\vec y)\d\vec x\d\vec y\nonumber.\end{eqnarray}
Again, (\ref{derive9=}) is positive definite on $\cY_\Lor$ and negative definite on $\cY_\sc$.

\nowastrona
\subsubsection{``First quantize, then reduce''}

The quantization described in Subsect. \ref{free1} will be called
{\em ``first reduce, then quantize''}. There exist alternative methods of
quantization, which  use the
 symplectic space $\cY_\rv$ introduced in (\ref{maxi1m}) as
the basis. There are two basic ways  to implement this idea.

The first
insists on using
 only {\em positive definite Hilbert
spaces}. Unfortunately,   the Hamiltonian turns out  to be  unbounded from below.

In  the {\em Gupta-Bleuler approach} the 4-potentials $\hat A^\mu(x)$ evolve with positive frequencies. Unfortunately, it
uses an indefinite scalar
product.

\nowastrona
\subsubsection{Quantization without reduction
 on a positive definite Hilbert space}
\label{pada21}

In this approach we use the Hilbert space
\beq
\Gamma_\s(\cZ_\Lor\oplus\bar\cZ_\sc)\label{pada31}\eeq
 equipped with a positive
definite scalar product. More explicitly,
we replace $a(k,\sigma)$ with  $\hat a(k,\sigma)$ for $\sigma=0,\pm1$. We replace  $a(k,\sc)$ with $\hat b^*(k,\sc)$.
They satisfy the standard
commutation relations 
\begin{eqnarray*}
[\hat  a(k,\sigma),\hat  a^*(k',\sigma')]&=&\delta_{\sigma,\sigma'}\delta(\vec k-\vec k'),\\{}
[\hat  b(k,\sc),\hat  b^*(k',\sc)]&=&\delta(\vec k-\vec k').\end{eqnarray*}
 $\hat a(k,\sigma)$, $\hat b(k,\sc)$ kill the vacuum:
\[\hat a(k,\sigma)\Omega=\hat b(k,\sc)\Omega=0.\]

 The quantized 4-potentials, Hamiltonian and momentum  become
\begin{eqnarray*}
\hat A_\mu(x)&=& \sum_{\sigma=0,\pm1}
\int\frac{\d\vec{k}}{\sqrt{(2\pi)^3}\sqrt{2\varepsilon (\vec k)}}
 \left(u_\mu(k,\sigma)\e^{\i kx}\hat a( k,\sigma)+
\bar{u_\mu(k,\sigma)}\e^{-\i kx}\hat a^*( k,\sigma)\right)\\
&&+
\int\frac{\d\vec{k}}{\sqrt{(2\pi)^3}\sqrt{2\varepsilon (\vec k)}}
 \left(u_\mu(k,\sc)\e^{\i kx}\hat b^*( k,\sc)+
\bar{u_\mu(k,\sc)}\e^{-\i kx}\hat b( k,\sc)\right),
\end{eqnarray*}

\begin{eqnarray*}
\hat H&=& \sum_{\sigma=0,\pm1}\int\d\vec k\varepsilon (\vec k)\hat a^*(k,\sigma)\hat a( k,\sigma)-
\int\d\vec k\varepsilon (\vec k)\hat b^*(k,\sc) \hat b( k,\sc)
,\\
\hat P&=& \sum_{\sigma=0,\pm1}\int\d\vec k  \vec k
\hat a^*(k,\sigma)\hat a( k,\sigma)-
\int\d\vec k \vec k\hat b^*(k,\sc) \hat b( k,\sc)
.\end{eqnarray*}
The propagator  in the position representation is given by
\[\Bigl(\Omega|\T\bigl(\hat A_\mu(x)\hat A_\nu(y)\bigr)\Omega\Bigr)=
-\i\Bigl(g_{\mu\nu}-\frac{2}{m^2}\partial_\mu\partial_\nu\Bigr)D^{\rm c}(x-y),\]
and in the momentum representation
\[\frac{-\i}{k^2+m^2-\i0}\left(g_{\mu\nu}+2\frac{k_\mu k_\nu}{m^2}\right).\]
\nowastrona
It is an example of a propagator from the class considered in Subsubsect. 
\ref{Propagators for massive QED}.
Note also that
\beq(\Omega|\hat A\lpar\zeta\rpar^2\Omega)=
\langle\zeta|\zeta\rangle_{\cY}+\frac{2}{m^2}\langle
\partial_\mu\zeta^\mu|\partial_\nu\zeta^\nu\rangle_\cY,\label{pada}\eeq
which is the scalar product
(\ref{derive9=}) corrected by a term  given by the scalar product (\ref{derive9}). Note that (\ref{pada}) is positive definite.

Vectors built by applying  fields
satisfying the Lorentz condition to the vacuum 
will be called {\em physical}.
 Equivalently, physical vectors are elements of the
algebraic Fock space built on $\cW_\Lor^{(+)}$. After the completion
 the physical space coincides with $\Gamma_\s(\cZ_\Lor)$. Thus we obtain the
 same space as in the method ``first reduce, then quantize''.

It will be convenient to describe this method in the $C^*$-algebraic language. 
Let $\CCR(\cY_{\rv})$ denote the {\em (Weyl) $C^*$-algebra of the CCR over $\cY_{\rv}$}, that is, the $C^*$-algebra
generated by $W(\zeta)$, $\zeta\in\cY_{\rv}$, such that
\beq W(\zeta_1)W(\zeta_2)=\e^{-\i\frac{\zeta_1\omega_\rv\zeta_2}{2}}W(\zeta_1+\zeta_2),\
\ \ W(\zeta)^*=W(-\zeta).\label{cstar}\eeq
We have the obvious action of $\rr^{1,3}\rtimes O^\uparrow(1,3)$
 on $\CCR(\cY_{\rv})$ by $*$-automorphisms:
\[\hat
r_{(y,\Lambda)}\left(W(\zeta)\right):=W\left(r_{(y,\Lambda)}(\zeta)\right).\] 
Choose the state   on $\CCR(\cY_{\rv})$ defined by
\begin{eqnarray}&&
\psi\big(W (\zeta)\big)\label{pada41}
\\
&=&\exp\Big(-\frac12\langle\zeta|\zeta\rangle_{\cY}-\frac{1}{m^2}\langle
\partial_\mu\zeta^\mu|\partial_\nu\zeta^\nu\rangle_\cY\Big)\notag
\end{eqnarray}
Let $(\cH_\psi,\pi_\psi,\Omega_\psi)$ be 
 the GNS representation generated by the state $\psi$.
Using (\ref{pada}) we see that
$\cH_\psi$ can be identified with $\Gamma_\s(\cZ_\Lor\oplus\bar\cZ_\sc)$ and
the fields are related to the Weyl operators by
\[\pi_\psi(W(\zeta))=\e^{\i\hat A\lpar\zeta\rpar}.\]

\subsubsection{The Gupta-Bleuler approach}

This approach also uses the symplectic space
$\cY_\rv$ as the basic input. It follows almost verbatim the usual steps
of quantization of the Klein-Gordon equation.
We  introduce the bosonic Fock space $\Gamma_\s(\cZ_\rv)$,
which  has an indefinite scalar product and can be viewed as a Krein space.

We replace $a(k,\sigma)$ by $\hat a(k,\sigma)$.
The commutation relations have a wrong sign for the scalar component:
\begin{eqnarray*}
[\hat  a(k,\sigma),\hat  a^*(k',\sigma')]&=&\kappa_{\sigma,\sigma'}\delta(\vec k-\vec k').\end{eqnarray*}
The annihilation operators kill the vacuum:
\[\hat a(k,\sigma)\Omega=0.\]
The expressions for the Hamiltonian, momentum and 4-potentials are the same as in the classical case:
\begin{eqnarray*}
\hat H&=& \sum_{\sigma=0,\pm1}\int\d\vec k\varepsilon (\vec k)\hat a^*(k,\sigma) \hat a( k,\sigma)
-\int\d\vec k\varepsilon (\vec k)\hat a^*(k,\sc)\hat a( k,\sc),\\
\vec{\hat P}&=& \sum_{\sigma=0,\pm1}\int\d\vec k  \vec k \hat a^*(k,\sigma)\hat a(
k,\sigma)-
\int\d\vec k  \vec k\hat a^*(k,\sc)\hat a(
k,\sc).
\end{eqnarray*}
\begin{eqnarray*}
\hat  A_\mu(x)&=&\sum_{\sigma}
\int\frac{\d\vec{k}}{\sqrt{(2\pi)^3}\sqrt{2\varepsilon (\vec k)}}
 \left(u_\mu(k,\sigma)\e^{\i kx}\hat a( k,\sigma)+
\bar{u_\mu(k,\sigma)}\e^{-\i kx}\hat a^*( k,\sigma)\right).
\end{eqnarray*}
Note that all eigenvalues of $\hat H $ are positive,  however its
expectation values (wrt the indefinite scalar product) can be negative.
We have
\begin{eqnarray*}
(\Omega|\hat A_\mu(x)\hat A_\nu(y)\Omega)
&=&-\i g_{\mu\nu}
D^{(+)}(x-y),\\
(\Omega|\T(\hat A_\mu(x)\hat A_\nu(y))\Omega)
&=&-\i g_{\mu\nu}D^{\rm
  c}(x-y). 
\end{eqnarray*}
In particular, the  2-point Green's function is the propagator
in the Feynman gauge.
Smeared 4-potentials $\hat A\lpar g\rpar$ are well
defined operators. 
\nowastrona

Similarly as in the previous method, 
vectors created by applying  fields
satisfying the Lorentz condition to the vacuum  will be called {\em physical}.
Again we obtain the
algebraic Fock space built on $\cW_\Lor^{(+)}$. This space is positive definite
and after the completion coincides with $\Gamma_\s(\cZ_\Lor)$. Thus
 the physical space is the same as before.

\nowastrona

\init\section{Massless photons}


In this section we discuss the quantization of the {\em Maxwell equation}
\begin{eqnarray}
-\partial_\mu F^{\mu\nu}(x)&=&0,
\label{maxwell}\end{eqnarray}
where, as in the previous section,
\[ F^{\mu\nu}:=\partial^\mu A^\nu-\partial^\nu A^\mu.\]

\nowastrona 

We will also consider an external
{\em conserved 4-current}, that is
  a  vector function  $J^\nu(x)$
satisfying
\beq \partial _\nu J^\nu(x)=0.\label{current3}\eeq
The Maxwell equation in the presence of the current $J$ reads
\begin{eqnarray}
\partial_\mu F^{\mu\nu}(x)&=&J^\nu(x).
\end{eqnarray}

Similarly as in the massive
case, there are several possible approaches to the Maxwell equation on the
classical and, especially, quantum level.
 The approach based from
the beginning on the reduced phase space,  both for the classical
description and quantization, will be treated
as the standard one. 
The situation is however somewhat more complicated than in the massive
case, since the Lorentz condition is not enough to fully reduce the
phase space.
Alternative approaches will be discussed later.

We try to make the discussion of massive and massless photons as parallel as possible. This is not entirely straightforward. In particular,  the massless limit is quite subtle -- to describe it one needs to fix the time coordinate. The covariant  4-potential  converges as $m\searrow0$ in an appropriate sense to the  noncovariant 4-potential in the Coulomb gauge.

\subsection{Free massless photons}

\subsubsection{Space of solutions and the gauge invariance}

It is well known that  the Maxwell  equation 
\begin{eqnarray}
-\partial_\mu \big(\partial^\mu\zeta^\nu(x)-\partial^\nu\zeta^\mu(x)\big)&=&0
\label{maxwell2}\end{eqnarray}
is invariant w.r.t. the replacement of
$\zeta_\mu$ with $\zeta_\mu+\partial_\mu\chi$, where $\chi$ is an
arbitrary smooth function on the space-time. 
In particular, there is no uniqueness of the Cauchy problem for
(\ref{maxwell2}).


This property is called 
{\em gauge invariance}.
It poses problems both for the classical and quantum theory.
One could avoid the problem of  gauge invariance by considering
fields and not 4-potentials as basic objects. However,  
when one quantizes the  Maxwell equation with a
  4-current, it is more convenient to
use 4-potentials. Therefore, we will stick to 4-potentials.

\nowastrona

 There
exist several  ways to cope with  gauge invariance.
The  approach that we will use as the standard one can be called {\em first reduce, then quantize}. In this approach we start
with the  Maxwell equation in the form
(\ref{maxwell2}). Note that it coincides with the Proca equation with
 $m=0$.
We will use objects defined in the context of the Proca equation,
where we replace $\Pr$ with $\widetilde{\Max}$ to indicate that
the mass is zero.

Thus the space of smooth space compact solutions of 
(\ref{maxwell2}) is denoted
$\cY_{\widetilde{\Max}}$ and (\ref{proco1}) defines a conserved
4-current, which we now call $j_{\widetilde\Max}$,
that leads to the form defined as in
(\ref{pro1}):
\begin{eqnarray}
&&\zeta_1\omega_{\widetilde\Max} \zeta_2\label{maxw}\\
&=&\int\left(-\left(\dot{\vec\zeta}_1(t,\vec x)-\vec\partial
\zeta_{10}(t,\vec x)\right)\vec\zeta_2(t,\vec x)+\vec\zeta_1(t,\vec x)
\left(\dot{\vec\zeta}_2(t,\vec x)
-\vec\partial\zeta_{10}(t,\vec x)\right)\right)\d\vec x.\notag
\end{eqnarray}
Unfortunately, this form is  only {\em presymplectic} not symplectic (it is degenerate).

(\ref{maxw}) does not depend on the gauge. To see this it is enough
to note that if $\zeta_2=\partial\chi$, and $\zeta_1$ is a solution of
the Maxwell equation, then the integrand of (\ref{maxw}) is a spatial
divergence, so (\ref{maxw}) is then zero.


\nowastrona

We say that a solution $\zeta$ of the Maxwell equation is in the {\em
  Coulomb gauge}  
if \[\zeta_0=0,\ \ \ \div\vec\zeta=0.\]
A functon in $C^\infty(\rr^3,\rr^3)$ will be called {\em transversal} if its divergence vanishes.

 Note that  every $\zeta\in \cY_{\widetilde{\Max}}$
 is gauge-equivalent to a unique solution of the Maxwell equation
 in the Coulomb
 gauge, denoted
 by $\zeta^\Coul$, where
\beq\chi(t,\vec x)=-(-\Delta)^{-1}\div\vec\zeta(t,\vec x),\ \ 
\zeta_\mu^\Coul+\partial_\mu\chi=\zeta_\mu.\label{meyer}\eeq
Neither $\chi$ nor $\zeta^\Coul$  have to be space-compact. The Stokes theorem yields however that
$\int\div\vec\zeta(t,\vec x)\d\vec x=0$, therefore
 $\chi$ and $\zeta^\Coul$ behave like $O(|\vec x|^{-2})$ because of
(\ref{beha}). 

The presymplectic form can be written as
\begin{eqnarray}
\zeta_1 \omega_{\widetilde{\Max}}\zeta_2&=&\zeta_1^\Coul
\omega_{\widetilde{\Max}}\zeta_2^\Coul\label{gaugo+} \\
&=&\int\left(-\dot{\vec\zeta}_1^\Coul(t,\vec x){\vec\zeta}_2^\Coul(t,\vec x)+
{\vec\zeta}_1^\Coul(t,\vec x)\dot{\vec\zeta}_2^\Coul(t,\vec x)\right)\d\vec x.
\notag\end{eqnarray}
Note that the integrand of (\ref{gaugo+}) behaves as $O(|\vec x|^{-4})$, hence is integrable.
\nowastrona
\bep
Let $\zeta\in \cY_{\widetilde{\Max}}$. We have the following equivalence:
\ben\item $\zeta\in\Ker \omega_{\widetilde{\Max}}$.
\item $\zeta^\Coul=0$.
\item $\zeta=\partial\chi$.
\een\label{pro3}\eep
\proof (2)$\Rightarrow$(3) follows from (\ref{meyer}).

The implication (3)$\Rightarrow$(1) follows from the gauge invariance
of the form $\omega_{\widetilde\Max}$.

Let us prove (1)$\Rightarrow$(2).
Let   $\zeta^\Coul\neq0$. Then one of the transversal functions
$\rr^3\ni \vec x\mapsto\vec \zeta(0,\vec x),\dot{\vec \zeta}(0,\vec x)$ is nonzero. Therefore we can find transversal functions $\vec u$, $\vec v$ in $C_{\rm c}^\infty(\rr^3,\rr^3)$ such that
\beq
\int\left(-\vec u(\vec x)\vec\zeta^\Coul(0,\vec x)+
v(\vec x)\dot{\vec\zeta}^\Coul(0,\vec x)\right)\d\vec x\neq0.\label{meyer1}\eeq
There exists a unique $\xi\in C_\sc^\infty(\rr^4,\rr^4)$ such that
\[\dot\xi(0,\vec x)=\bigl(0,\vec u(\vec x)\bigr),\ \ \ 
\xi(0,\vec x)=\bigl(0,\vec v(\vec x)\bigr),\ \ \Box\xi=0.\]
$\xi$ clearly belongs to $ \cY_{\widetilde{\Max}}$ and is in the Coulomb gauge. We have
\begin{eqnarray*}\xi \omega_{\widetilde{\Max}}\zeta&=&\xi \omega_{\widetilde{\Max}}\zeta^\Coul\\
&=&\int\left(-\dot{\vec\xi}(0,\vec x){\vec\zeta}^\Coul(0,\vec x)+
{\vec\xi}(0,\vec x)\dot{\vec\zeta}^\Coul(0,\vec x)\right)\d\vec x,
\end{eqnarray*}
which equals (\ref{meyer1}) and is nonzero. Hence $\zeta\not\in
\Ker \omega_{\widetilde{\Max}}$. \qed

Define
  \[\cY_{\Max}
 := \cY_{{\widetilde{\Max}}}/\Ker \omega_{\widetilde{\Max}}.\]
In other words, $\cY_{\Max}$ is obtained by the {\em symplectic reduction} of the presymplectic space  $ \cY_{\widetilde{\Max}}$.
Clearly,  $\cY_{\Max}$ is equipped with a
natural  {\em symplectic form} $\omega_\Max$. $\rr^{1,3}\rtimes  O^\uparrow(1,3)$
acts on $\cY_{\Max}$ by symplectic transformations.

By Prop. \ref{pro3}, $\cY_\Max$ consists of gauge equivalence classes of  $ \cY_{\widetilde{\Max}}$.

Analogously we define the  space $\cc\cY_\Max$ of  gauge classes
of complex smooth space-compact solutions of (\ref{maxwell2}).

\nowastrona

\subsubsection{Classical 4-potentials}

$A_\mu(x)$ denotes the functional on $ \cY_{\widetilde\Max}$ given by
\beq \langle A_\mu(x)|\zeta\rangle:=\zeta^\mu(x).\label{ambi}\eeq
Obviously, $A_\mu(x)$ is not defined on 
 $\cY_{\Max}$.

We introduce also the functional $A_\mu^\Coul(x)$ on $ \cY_{\widetilde\Max}$,
called
the 
the {\em classical 4-potential in the Coulomb gauge},
\[A_0^\Coul(x):=0,\ \ \vec A^\Coul(x):=\vec A_\tr(x)=\vec A(x)-\vec\partial\Delta^{-1}\div\vec A(x).\]
Note that
\[\langle A_\mu^\Coul(x)|\zeta\rangle=\langle A_\mu(x)|\zeta^\Coul\rangle=\zeta_\mu^{\Coul}(x),\]
where $\zeta^\Coul$ on the right hand side is 
 the representative of the class $\zeta$ in the Coulomb gauge.
$A^\Coul(x)$ does not depend on the gauge, hence can be interpreted as a functional on
 $\cY_{\Max}$. It is not, however, Lorentz covariant.

Moreover, we introduce the functionals $F_{\mu\nu}(x)$ on $ \cY_{\widetilde\Max}$,
 called the
{\em fields}:
\[\langle  F_{\mu\nu}(x)|\zeta\rangle:=\partial_\mu\zeta_\nu(x)-
\partial_\nu\zeta_\mu(x).\]
They also do not depend on the gauge, hence can be interpreted as functionals on  $\cY_{\Max}$. They are moreover Lorentz covariant.

We will write $E_i(x)=F_{0i}(x)$.
Clearly, $\vec E=\partial_t\vec {A}^\Coul$ and \beq \div\vec A^\Coul(x)=0,\ \ 
\div\vec E(x)=0.\label{constraint}\eeq

In what follows we will usually drop the subscript $\Coul$ from
$A^\Coul(x)$. This introduces a possible ambiguity with $A(x)$
defined in (\ref{ambi}). However, when we speak about $\cY_\Max$, then
(\ref{ambi}) is ill defined,  only $A^\Coul(x)$ is well defined, so we
think that the risk of confusion is small.



\nowastrona

The symplectic structure on the space $\cY_{\Max}$
\[\omega_{\Max}=\int  A^i_{}(t,\vec x)\wedge E_i(t,\vec x)\d\vec x\]
together with the constraint (\ref{constraint}) leads to 
a {\em Poisson bracket} on
the level of functions on $\cY_{\Max}$:
\begin{eqnarray}
\{A_{{}\,i}(t,\vec x),A_{{}\,j}(t,\vec y)\}=\{ E_i(t,\vec x), E_j(t,\vec y)\}&=&0,
\nonumber
\\
\{A_{{}\,i}(t,\vec x),E_j(t,\vec
y)\}&=&\left(\delta_{ij}-\frac{\partial_i\partial_j}{\Delta}\right)
\delta(\vec x-\vec y) .
\nonumber\end{eqnarray}
From the above relations we deduce 
\[
\{A_{{}\,i}(x),A_{{}\,j}(y)\}=
\left(\delta_{ij}-\frac{\partial_i\partial_j}{\Delta}\right) 
D(x-y).\]

\nowastrona


\subsubsection{Smeared 4-potentials}

We can use the symplectic form to pair distributions and solutions.
For $\zeta\in\cY_{\Max}$  
we introduce  the corresponding
{\em spatially smeared 4-potentials}, which is the
 functional on $\cY_{\Max}$ given by
\[\langle A_{}\lpar\zeta\rpar|\rho\rangle:=\rho\omega_\Max \bar\zeta,\ \ \ \rho\in\cc\cY_\Max.\]
Note that
\[\{A_{}\lpar\zeta_1\rpar,A_{}\lpar\zeta_2\rpar\}=\bar\zeta_1\omega_\Max \bar\zeta_2.\]
\beq
A_{}\lpar\zeta\rpar =\int\left(-\bar{\dot\zeta_\mu(t,\vec x)}A_{}^\mu(t,\vec x)+
\bar{\zeta_\mu(t,\vec x)}E^\mu(t,\vec x)\right)\d\vec{x}.
\label{pzk4}\eeq
Let us stress that $A_{}\lpar\zeta\rpar$ depends on $\zeta$ only modulo gauge
transformations and is Lorentz covariant.

\nowastrona

We can also introduce {\em space-time smeared 4-potentials in the Coulomb gauge}, which are the
functionals on  $\cY_{\Max}$, for
$f\in C_{\rm c}^\infty(\rr^{1,3},\rr^{1,3})$ given by
\beq A_{}[f]:=\int f^\mu(x)A_\mu(x)\d x.\label{space-time}\eeq
Note that $A_{}[f]=A_{}\lpar \zeta\rpar$, where
\[\zeta_i=-D*\left( f_i-\frac{\partial_i\partial^j}{\Delta} f_j\right),\
 \ \zeta_0=0.\]
(\ref{space-time}) is not Lorentz covariant. To see this it is enough to note that it does not depend on  $f^0$.
 Replacing $[f^\mu]$ with $[f^\mu+\partial^\mu\chi]$ for $\chi\in C_{\rm
   c}^\infty(\rr^{1,3})$ does not change (\ref{space-time}), because 
 $\partial_\mu A^\mu(x)=0$. 

\nowastrona
\subsubsection{Lagrangian formalism and the stress-energy tensor}


The Euler-Lagrange equations for the Lagrangian density
\[
\cL:=-\frac14 F_{\mu\nu}F^{\mu\nu}
\]
  coincide with the Maxwell equation.


The  {\em canonical stress-energy tensor} is
\begin{eqnarray*}
\cT_\can^{\mu\nu}&=&g^{\mu\nu}\cL-\frac{\partial\cL}{\partial
  A_{\alpha,\mu}} A_{\alpha}^{\ ,\nu} \\
&=&-g^{\mu\nu}\frac14F_{\alpha\beta}F^{\alpha\beta}-F^{\mu\alpha} A^{,\nu}_{\ \alpha}.
\end{eqnarray*}
One usually replaces it with the {\em Belifante-Rosenfeld
  stress-energy tensor}. It is defined as
\begin{eqnarray*}
\cT^{\mu\nu}&=&
\cT_\can^{\mu\nu}+\partial_\alpha\Sigma^{\mu\nu\alpha}\\
&=&-g^{\mu\nu}\frac14F_{\alpha\beta}F^{\alpha\beta}+F^{\mu\alpha}F^{\nu}_{\ \alpha},
\end{eqnarray*}
where
\beq \Sigma_{\mu\nu\alpha}=-\Sigma_{\alpha\nu\mu}
:=F_{\mu\alpha}A_\nu.\label{sigma0}\eeq
On solutions of the Euler-Lagrange equations we have
\[\partial^\mu\cT_{\mu\nu}^\can=\partial^\mu\cT_{\mu\nu}=0.\]
In addition, $\cT_{\mu\nu}$ is symmetric.

To pass to the Hamiltonian formalism we use the Coulomb gauge, writing $A_\mu$ for $A_\mu^\Coul$.
Recall that in this gauge $A_0=0$ and $\div \vec A=0$.
The variable conjugate to $A_i$ is
$\partial_{\dot A_i}\cL=E^i$, which also satisfies $\div \vec E=0$. We express $\cT_{\mu\nu}^\can$ and $\cT_{\mu\nu}$ in terms of $\vec A$ and $\vec E$.
We introduce the Hamiltonian, momentum and {\em polarization
  density} 
\begin{eqnarray*}\cH(x):=\cT^{00}(x)&=&
\frac12\left(\vec E^2(x)+(\rot \vec A)^2(x)\right),
\\
\cP^j(x):=\cT^{0j}(x)&=&E^i(x)F^{ji}(x),\\
\cS(x)&=& E_i(x)\epsilon^{ijk}\partial_k A_j(x)
.\end{eqnarray*}

 They yield the 
 Hamiltonian, momentum and  {\em polarization} as in (\ref{pyko})
 satisfying analogous properties. 


\subsubsection{Diagonalization of the equations of motion}

As in the massive case, we would like  to diagonalize simultaneously the Hamiltonian, momentum,
polarizaton and symplectic form. 

For $\vec k\in\rr^{3}$ we set $k=(\varepsilon,\vec k)$, $\varepsilon(\vec k):=\sqrt{\vec k^2}$.
The vectors $u(k,
\pm1)$ are defined as in (\ref{ede}). $u(k,0)$ are not defined at all.
\nowastrona


\nowastrona

For $\sigma=\pm1$, define the following functionals on
$\cY_{\Max}$, called {\em plane wave functionals}:
\begin{eqnarray*}&&a(k,\sigma)\\
&=&
\int\Bigl(\sqrt{\frac{\varepsilon (\vec k)}{2}}\e^{-\i \vec
  x\vec k}\bar{u_j(k,\sigma)}A_{}^j(0,\vec x)-\frac{\i}{\sqrt{2\varepsilon (\vec k)}}\e^{-\i\vec k\vec x}\bar{u_j(k,\sigma)}
E^j(0,\vec
x)\Bigr)\frac{\d \vec x}{\sqrt{(2\pi)^3}}
. \end{eqnarray*}

\nowastrona
We have accomplished the promised  diagonalization
\begin{eqnarray*}
H&=&\sum_{\sigma=\pm1}
\int\d\vec k \varepsilon (\vec k)a^*(k,\sigma) a( k,\sigma),\\
\vec P&=&\sum_{\sigma=\pm1}\int\d\vec k \vec k a^*(k,\sigma) a(
k,\sigma),\\
S&=&\sum_{\sigma=\pm1}\int\d\vec k \sigma
|\vec k|a^*(k,\sigma) a( k,\sigma),\\
\i\omega_\Max&=&
 \sum_{\sigma=\pm1}\int a^*(k,\sigma)\wedge
  a(k,\sigma)\d\vec{k}.
\end{eqnarray*}


\nowastrona

The 4-potentials can be written as
%
\begin{eqnarray*}
A_{{} \mu}(x)&=& \sum_{\sigma=\pm1}
\int\frac{\d\vec{k}}{\sqrt{(2\pi)^3}\sqrt{2\varepsilon (\vec k)}}
 \left(u_\mu(x,\sigma)\e^{\i kx}a( k,\sigma)+
\bar{u_\mu(x,\sigma)}\e^{-\i kx}a^*( k,\sigma)\right).
\end{eqnarray*}

Plane waves are defined as in the massive case, with
$\sigma=\pm1$. We have
\[a(k,\sigma)=\i A_{}\lpar |k,\sigma)\rpar\]
and
\begin{eqnarray*}
A_{{} \mu}(x)
&=& \sum_{\sigma=\pm1}
\int\left((x|k,\sigma)a(k,\sigma)+\bar{(x|k,\sigma)}
a^*(k,\sigma)\right)\d\vec k.
\end{eqnarray*}

\nowastrona

\subsubsection{Positive frequency space}
\label{posi1}

 $\cW_\Max^{(\pm)}$ will denote the subspace of $\cc\cY_\Max$
 consisting of classes of solutions that in the Coulomb gauge have 
 positive,
resp.  negative frequencies.

Every $g\in\cW_\Max^{(+)}$ can be written as
\[g(x)= \sum_{\sigma=\pm1}
\int\frac{\d\vec{k}}{\sqrt{(2\pi)^3}\sqrt{2\varepsilon (\vec k)}}
\e^{\i kx}u(k,\sigma)\langle
a(k,\sigma)|g\rangle.
\]
For $g_1,g_2\in\cW_\Max^{(+)}$ we define the scalar product
\beq(g_1|g_2):=\i\bar g_1\omega_\Max g_2=
 \sum_{\sigma=\pm1}\int\bar{\langle a(k,\sigma)|g_1\rangle}\langle
 a(k,\sigma )|g_2\rangle\d\vec{k}.\label{tyt2}\eeq

The definition of $\cW_\Max^{(+)}$ depends on the choice of coordinates. It is however easy to see that the space  $\cW_\Max^{(+)}$ is invariant w.r.t. $\rr^{1,3}\rtimes  O^\uparrow(1,3)$.

We set $\cZ_\Max$ to be the completion of $\cW_\Max^{(+)}$ in this scalar
product.

We have 
\[\langle
 a(k,\sigma )|g  \rangle=(k,\sigma |g).\]
We can identify
$\cZ_\Max$ with $L^2(\rr^3,\cc^2)$  and rewrite (\ref{tyt2}) as
\[(g_1|g_2)=
 \sum_{\sigma=\pm1}\int\bar{(k,\sigma|g_1)}(k,\sigma |g_2)\d\vec{k}.\]

We can identify $\cY_\Max$  with 
$\cW_\Max^{(+)}$ and transport the  scalar product onto $\cY_\Max$,
which for $\zeta_1,\zeta_2$ is given by
\begin{eqnarray*}
\langle\zeta_1|\zeta_2\rangle_\cY&:=&\Re(\zeta_1^{(+)}|\zeta_2^{(+)})\\
 &=&
\int\int \bigl(\dot\zeta_{1i}^\Coul(
0,    \vec x)(-\i)D^{(+)}(0,\vec x-\vec y)\dot\zeta_{2i}^\Coul(0,\vec y)
\d \vec x\d
 \vec  y
\\
&&\hspace{-7ex} +\int\int\zeta_{1i}^\Coul(0,
\vec    x)(-\Delta_{\vec x})(-\i)D^{(+)}(0,\vec x-\vec
y)\zeta_{2i}^\Coul
(0,\vec y)
\d \vec x\d
 \vec  y.\end{eqnarray*}

\nowastrona

\subsubsection{Spin averaging}

Let us describe
the {\em spin averaging identities} useful in computations of scattering
cross-sections. 
For a given $k\in\rr^{1,3}$ with $k^2=0$, let $M,N$ be vectors with
\[M^\mu k_\mu= N^\nu k_\nu=0.\]
Then we have
\beq
\sum_{\sigma=\pm1} \bar{M^\mu u_\mu(k,\sigma)}u_\nu(k,\sigma) N^\nu=
\bar{M^\mu }N_\nu.\label{spino}\eeq

\nowastrona

To see (\ref{spino}), note that
\[\sum_{\sigma=\pm1} \bar{ u_\mu(k,\sigma)}u_\nu(k,\sigma) =
g_{\mu\nu}+\delta_{\mu 0}\delta_{\nu 0}
-\frac{\vec k_\mu \vec k_\nu}{|\vec k|^2}.\]
Therefore,
the left hand side of (\ref{spino}) equals
\begin{eqnarray*}
\bar{M^\mu} g_{\mu\nu} N^\nu+\bar{M^0}N^0
-\frac{(\bar{\vec M}\vec k)(\vec N\vec k)}{|\vec k|^2}.\end{eqnarray*}
But 
\[M^0=\frac{\vec k\vec M}{|\vec k|},\ N^0=\frac{\vec k\vec N}{|\vec k|}.\]


\nowastrona

\subsubsection{Quantization}


We would like to quantize the Maxwell equation starting from the
symplectic space
 $\cY_{\Max}$. We will use the 4-potentials in the
Coulomb gauge (where, as usual, we drop the superscript $\Coul$).
The quantization is similar to the Proca equation based on
$\cY_{\Pr}$ described in  Subsubsect. \ref{quant-proca}, with  Condition 
(1)   replaced by
\begin{eqnarray*}
-\Box\hat A_{i}(x)=0, &&
 \partial_i \hat A_{i}(x)=0,\ \ \ 
\hat A_{0}(x)=0,\end{eqnarray*} 
and Condition (2)  replaced by
\begin{eqnarray}
[\hat A_{i}(0,\vec x),A_{j}(0,\vec y)]=[\hat E_i(0,\vec x),
  \hat E_j(0,\vec y)]&=&0,
\nonumber
\\{}
[\hat A_{i}(0,\vec x),\hat E_j(0,\vec
y)]&=&\i\left(\delta_{ij}-\frac{\partial_i\partial_j}{\Delta}\right)
\delta(\vec x-\vec y) .
\nonumber\end{eqnarray}

The above problem has a  solution unique up to a unitary equivalence. 
We set $\cH:=\Gamma_\s(\cZ_\Max)$.
 The creation
operators will be denoted by 
 \begin{eqnarray*}
\hat a^*(k,\sigma)&=&\hat a^*( |k,\sigma)).\end{eqnarray*}
 $\Omega$ will be the Fock vacuum. We set
\[\hat A_{i}(x):=\int\frac{\d\vec{k}}{\sqrt{(2\pi)^3}\sqrt{2\varepsilon(\vec k)}}
\sum_{\sigma=\pm1}
\left(u_i(k,\sigma)\e^{\i kx}\hat a( k,\sigma)+ 
\bar{u_i(k,\sigma)}\e^{-\i kx}\hat a^*( k,\sigma)\right),\]
\nowastrona

The  quantum
Hamiltonian,  momentum and  polarization are
\begin{eqnarray*}
\hat H&:=&\sum_{\sigma=\pm1}\int \hat a^*(k,\sigma)\hat a(k,\sigma)\varepsilon(\vec k)
\d\vec{k},\\
\vec {\hat P}&:=&\sum_{\sigma=\pm1}
\int \hat a^*( k,\sigma)\hat a( k,\sigma)\vec k\d\vec{k},\\
\hat S&:=&\sum_{\sigma=\pm1}\int\d\vec k \sigma
|\vec k|\hat a^*(k,\sigma) \hat a( k,\sigma)
.
\end{eqnarray*}

\nowastrona

The whole group
$\rr^{1,3}\rtimes  O^\uparrow(1,3)$ is  unitarily implemented on $\cH$
by
 $U(y,\Lambda):=\Gamma\Bigl(r_{(y,\Lambda)}\Big|_{\cZ_\Max}\Bigr)$
We have
\[U(y,\Lambda)\hat F_{\mu\nu}(x)U(y,\Lambda)^*=\Lambda_\mu^{\mu'}
\Lambda_\nu^{\nu'}\hat
F_{\mu'\nu'}\bigl((y,\Lambda)x\bigr).\] 
Moreover,
\begin{eqnarray*} [\hat A_{j}(x),\hat A_{i}(y)]&=&-\i
\left(\delta_{ij}-\frac{\partial_i\partial_j}{\Delta}\right)D(x-y).
\end{eqnarray*}

Note the identities
\begin{eqnarray*}
(\Omega|\hat A_{{}i}(x)\hat A_{{}j}(y)\Omega)
&=&-\i\left(\delta_{ij}-\frac{\partial_i\partial_j}{\Delta}\right)
D^{(+)}(x-y),\\
(\Omega|\T(\hat A_{{}i}(x)\hat A_{{}j}(y))\Omega)
&=&-\i\left(\delta_{ij}-\frac{\partial_i\partial_j}{\Delta}\right)D^{\rm
  c}(x-y). 
\end{eqnarray*}

\nowastrona

The family \[C_{\rm c}^\infty(\rr^{1,3},\rr^{1,3})
\ni f\mapsto \hat A_{}[f]:=\int f^\mu(x)\hat A_{\mu}(x)\d x\]
 with
$\cD:=\Gamma_\s^\fin(\cZ_\Max)$
does not satisfy  the  Wightman axioms because of two problems:
the noncausality of the commutator and the absence of the Poincar\'{e}{}
covariance. 

If we replace  $\hat A_\mu$ with $\hat F_{\mu\nu}$, we restore the
causality and the Poincar\'{e}{}
covariance.

For an open  set $\cO\subset \rr^{1,3}$ we set
\[\fA(\cO):=\{\exp(\i \hat F[f] ) :\ f\in C_{\rm c}^\infty(\cO,\otimes_\a^2\rr^{1,3})\}
.\]
The algebras $\fA(\cO)$ satisfy the  Haag-Kastler axioms.


\subsubsection{Quantization in terms of $C^*$-algebras}
Let $\CCR(\cY_{\widetilde\Max})$ denote the {\em (Weyl) $C^*$-algebra of canonical
commutation relations} over $\cY_{\widetilde\Max}$. By definition, it is
generated by $W(\zeta)$, $\zeta\in\cY_{\widetilde\Max}$, such that
\[W(\zeta_1)W(\zeta_2)=\e^{-\i\frac{\zeta_1\omega_{\widetilde\Max}\zeta_2}{2}}W(\zeta_1+\zeta_2),\
\ \ W(\zeta)^*=W(-\zeta).\]
$\rr^{1,3}\rtimes  O^\uparrow(1,3)$ acts on $\CCR(\cY_{\widetilde\Max})$ by $*$-automorphisms
defined by
\[\hat
r_{(y,\Lambda)}\left(W(\zeta)\right):=W\left(r_{(y,\Lambda)}(\zeta)\right).\] 
We are looking for a cyclic representation of this algebra with the
time evolution generated by a positive Hamiltonian.

\nowastrona

Consider the state  on $\CCR(\cY_{\widetilde\Max})$ defined for $\zeta\in\cY_{\widetilde\Max}$ by
\begin{eqnarray*}
\psi\big(W (\zeta)\big)
&=&\exp\big(-\frac12\langle\zeta|\zeta\rangle_{\cY}\big).
 \end{eqnarray*}

\nowastrona
Note that the state is gauge and Poincare invariant.
Let  $(\cH_\psi,\pi_\psi,\Omega_\psi)$ be the GNS representation.
$\cH_\psi$ is naturally isomorphic to  $\Gamma_\s(\cZ_\Max)$.
$\Omega_\psi$ can be identified with the vector $\Omega$. 
$\pi_\psi(W(\zeta))$ can be identified with $\e^{\i\hat A_{}\lpar\zeta\rpar}$. In particular, if $\zeta_1$ and $\zeta_2$ are gauge equivalent, then $\hat A\lpar\zeta_1\rpar=\hat A\lpar\zeta_2\rpar$.
However, $\hat A(x)$ in the sense of  (\ref{ambi}) is not well defined. 

\nowastrona

\subsection{Massless photons with  an external 4-current}

\subsubsection{Classical fields}

We return to the classical Maxwell equation. 
We consider an  external 
4-current given by  function
 $\rr^{1,3}\ni x\mapsto J(x)=[J^\mu(x)]\in \rr^{1,3}$ satisfying
\begin{eqnarray} \partial _\nu J^\nu(x)&=&0.
\label{current1}
\end{eqnarray} In most of this subsection we assume that $J$ is Schwartz.
The Maxwell equation reads
\beq
-\partial_\mu\partial^\mu A_\nu+\partial_\nu\partial_\mu A^\mu=-J_\nu.
\eeq

Let $\zeta$ be a solution of
\beq
-\partial_\mu\partial^\mu \zeta_\nu+\partial_\nu\partial_\mu \zeta^\mu=-J_\nu.
\label{maxww}\eeq
We write separately the temporal and spatial equations:
\begin{eqnarray*}
-\Delta \zeta_0+\div\dot{\vec \zeta}&=&-J_0,\\
\left(\partial_0^2-\Delta\right) \vec \zeta-\vec\partial \dot \zeta_0
+\vec\partial\div\vec \zeta&=&-\vec J.
\end{eqnarray*}
\nowastrona
We can compute $\zeta_0$ in terms of $\vec \zeta$ at the same time:
\beq \zeta_0(x)=\Delta^{-1}(J_0+\partial_0\div \vec 
\zeta)(x).
\eeq
We can insert this into spatial equations, using $\dot J_0=\div \vec J$, obtaining
\beq
\Box\vec \zeta_\tr=\vec J_\tr
,\label{trans}\eeq
where
\begin{eqnarray*}
\vec \zeta_\tr&:=&
\vec \zeta-\vec\partial\Delta^{-1}\div\vec
\zeta,\\
\vec J_\tr&:=&\vec J-\vec\partial\Delta^{-1}\div\vec
J.\end{eqnarray*}
Thus   $\vec\zeta_\tr$ can be treated as the only dynamical variables.  $\div
\vec \zeta=:\Theta$ is an arbitrary space-time function, which can be used to determine $\zeta_0$.

The simplest choice is $\Theta=0$, which corresponds to the {\em Coulomb gauge}:
\begin{eqnarray}
 \zeta_0&=&\Delta^{-1}J_0,\notag\\
\Box \vec \zeta&=&\vec J_\tr,\notag\\
\div \vec \zeta&=&0.\label{coulo}
\end{eqnarray}
The Coulomb gauge seems to be the most natural gauge for the Hamiltonian approach.

Let $\zeta$ be a space compact solution of (\ref{maxww}).
 Setting
\[\zeta_\mu^\Coul:=\zeta_\mu+\partial_\mu\chi,\]
where $\chi(t,\vec x):=(-\Delta)^{-1}\div\vec \zeta(t,\vec x)$, we obtain a solution of (\ref{coulo}).
 $ \zeta^\Coul$ is the unique solution of  (\ref{coulo})  gauge equivalent to $\zeta$. It  does not have to be space compact.

Thus the {\em classical 4-potential in the Coulomb gauge} $A^\Coul(x)$ 
satisfies
\begin{eqnarray*}
A_0^\Coul&=&-(-\Delta)^{-1}J_0,\\
\Box \vec A^\Coul&=&\vec J_\tr,\\
\div \vec A^\Coul&=&0.
\end{eqnarray*}



The electric field is $\vec E=\dot{\vec A}-\vec\partial A_0$. It is easy to see that if we use the Coulomb gauge, then $\vec E_\tr=\dot{\vec A}^\Coul$.

 Similarly as in the previous subsection,  we will drop the superscript $\Coul$ in what follows.

\nowastrona

\subsubsection{Lagrangian and Hamiltonian formalism}
The  Lagrangian density is
\begin{eqnarray*}
\cL&:=&-\frac14 F_{\mu\nu}F^{\mu\nu}-J_\mu A^\mu\\
&=&-\frac12(\rot \vec A)^2+\frac12(\vec\partial A_0)^2+\frac12\bigl(\dot{\vec
  A}\bigr)^2- \dot{\vec A}\vec\partial
A_0-\vec J\vec A+J_0A_0.
\end{eqnarray*}



We will  use the Coulomb gauge. Thus we  assume that
$A_0=\Delta^{-1}J_0$ and the only dynamical variable is $\vec A$ satisfying
 $\div \vec A=0$.
Using the transversality of $\vec A$ we can rewrite the Lagrangian density as
\begin{eqnarray*}
\cL&=&-\frac12(\vec\partial\vec A)^2+\frac12(\dot{\vec A})^2-\vec J\vec A\\
&&+\frac12(\vec\partial A_0)^2+J_0A_0.
\end{eqnarray*}
The conjugate variable is $\vec E_\tr(x)=\dot{\vec A}(x)$. 
Thus we have the Poisson brackets
\begin{eqnarray}
\{A_{{}\,i}(t,\vec x),A_{{}\,j}(t,\vec y)\}=\{ E_{\tr\,i}(t,\vec x), E_{\tr\,j}(t,\vec y)\}&=&0,
\nonumber
\\
\{A_{{}\,i}(t,\vec x),E_{\tr\,j}(t,\vec
y)\}&=&\left(\delta_{ij}-\frac{\partial_i\partial_j}{\Delta}\right)
\delta(\vec x-\vec y) .
\nonumber\end{eqnarray}
Note that $\vec E$ differs from $\vec E_\tr$ by a c-number function $-\vec\partial A_0$. Therefore, $\vec E$ satisfies the same commutation relations as $\vec E_\tr$.

The canonical Hamiltonian density is
\begin{eqnarray*}
\cH^\can(x)&=&-\cL(x)+\vec E_\tr \dot A_i(x) \\
&=&
\frac12(\vec\partial \vec A)^2(x)+\frac12\bigl(\vec E_\tr\bigr)^2(x)
+\vec J(x)\vec A(x)\\
&&
-\frac12(\vec\partial A_0)^2(x)
-J_0(x)A_0(x).
\end{eqnarray*}
We add to it a  spatial divergence
$\div\big(A_0(x)\vec\partial A_0(x)\big)$ and express $A_0$ in terms of $J_0$.
 We obtain the usual Hamiltonian density
\begin{eqnarray*}
\cH(x)
&=&
\frac12\vec E_\tr^2(x)+\frac12(\vec\partial \vec A)^2(x)
+\vec J(x)\vec A(x)\\
&&+\frac12J_0(-\Delta)^{-1}J_0(x).
\end{eqnarray*}

Similarly as in the massive case, the 
 Hamiltonian
\begin{eqnarray}
H(t)&=&\int \cH(t,\vec x)\d\vec x=\int \cH^\can(t,\vec x)\d\vec
x\label{tutum}\end{eqnarray}
generates the equations
of motion and we can interpret interacting fields as functionals on
$\cY_\Max$ satisfying
\begin{eqnarray}\vec A(0,\vec x)=\vec A_\fr(0,\vec x),&&\vec E(0,\vec x)=\vec E_\fr(0,\vec x).
\notag\end{eqnarray}




\nowastrona

\subsubsection{Quantization}

To quantize the Maxwell equation in the
presence of an external 4-current we will use the Coulomb gauge, dropping  as usual the subscript $\Coul$. 

\nowastrona

We are looking for  quantum 4-potentials
 $\rr^{1,3}\ni x\mapsto \hat A_\mu(x)$
 satisfying
\begin{eqnarray*}
\hat A_0^\Coul&=&-(-\Delta)^{-1}J_0,\\
\Box \vec {\hat A}^\Coul&=&\vec J_\tr,\\
\div \vec{\hat A}^\Coul&=&0,
\end{eqnarray*}
having the following commutation relations 
 with  $\vec E(x)=\dot{\vec{\hat
  A}}(x)-\vec\partial\hat A_0(x)$.
\begin{eqnarray}
[\hat A_{{}\,i}(t,\vec x),\hat A_{{}\,j}(t,\vec y)]=[\hat E_i(t,\vec x), E_j(t,\vec y)]&=&0,
\nonumber
\\{}
[A_{{}\,i}(t,\vec x),E_j(t,\vec
y)]&=&\i\left(\delta_{ij}-\frac{\partial_i\partial_j}{\Delta}\right)
\delta(\vec x-\vec y) .
\nonumber\end{eqnarray}
The above conditions determine  $\hat A_0$.
To fix $\vec{\hat A}$ and $\vec{\hat E}$ we assume that they coincide  with
their free quantum counterparts at $t=0$: 
\begin{eqnarray*}
\vec {\hat A}(0,\vec x)&=&\vec {\hat A}_\fr(0,\vec x)=:\vec{\hat A}(\vec x),\\
\vec {\hat E}(0,\vec x)&=&\vec {\hat E}_\fr(0,\vec x)=:\vec{\hat E}(\vec x).
\end{eqnarray*}

\nowastrona
The  Schr\"odinger picture Hamiltonian
 and the corresponding interaction picture Hamiltonian are
\begin{eqnarray*}
\hat H(t)&=&\int\d \vec x\ {:}\Bigl(
\frac12\vec{ \hat E}^2(\vec x)
+\frac12(\vec\partial \vec {\hat A})^2(\vec x)
+\vec J(t,\vec x)\vec{\hat A}(\vec x)\Bigr){:}\\&&
+\frac12\int\int\d\vec x\d\vec y
 J^0(t,\vec x)\frac{1}{4\pi|\vec x-\vec y|}J^0(t,\vec y),
\\
\hat H_\Int(t)&=&
+\int\d \vec x
\vec J(t,\vec x)\vec{\hat A}_\fr(t,\vec x)\\
&&+ \frac12\int\int\d\vec x\d\vec y
J^0(t,\vec x)\frac{1}{4\pi|\vec x-\vec y|}J^0(t,\vec y)
.\end{eqnarray*}

\nowastrona

The {\em scattering operator} 
can be computed exactly:
\begin{eqnarray}
\hat S&=&
\exp\left(\frac{\i}{2}\int \frac{\d k}{(2\pi)^4} \bar J^\mu(k)D_{\mu\nu}^\Coul(k) J^\nu(k)
\right)\notag\\
&&\times\exp\left(-\i\sum_{\sigma=\pm1}\int\frac{\d \vec k}{\sqrt{(2\pi)^3}}
      \hat a^*(k,\sigma)\frac{\bar{u_\mu(k,\sigma)}}{\sqrt{2\varepsilon(\vec k)}}J^\mu(\varepsilon(\vec k),\vec k)\right)\notag\\&&
\times\exp\left(-\i\sum_{\sigma=\pm1}\int\frac{\d \vec k}{\sqrt{(2\pi)^3}}
      \hat a(k,\sigma)\frac{u_\mu(k,\sigma)}{\sqrt{2\varepsilon(\vec k)}}\bar{J^\mu(\varepsilon(\vec k),\vec k)}\right),\label{exa1}
\end{eqnarray}
where the  propagator in the {\em Coulomb gauge} is defined as
\begin{eqnarray*}
D_{00}^\Coul=-\frac{1}{\vec k^2},\ \ \ D_{0j}^\Coul=0,\ \ \ 
D_{ij}^\Coul=\frac{1}{k^2-\i0}\left(\delta_{ij}-\frac{k_ik_j}{\vec k^2}\right).
\end{eqnarray*}
We did not use the fact that $J^\mu$ is conserved.

\nowastrona

\subsubsection{Causal propagators}

If we compute  scattering amplitudes,
we can pass from the   propagator in the Coulomb gauge to another by adding
$k_\mu f_\nu(k)+f_\mu(k)k_\nu$
for an arbitrary function $f_\mu(k)$.
 Let us list a number of useful
propagators in other gauges.

We distinguish the family of propagators
\[\frac{1}{k^2-\i0}\left(g_{\mu\nu}+\left(\frac1\alpha-1\right)\frac{k_\mu k_\nu}{k^2}\right).\]
Some of them have special names:
\begin{eqnarray*}
D_{\mu\nu}^{\rm Lan}&:=
 \frac{1}{k^2-\i0}\left(g_{\mu\nu}-\frac{k_\mu k_\nu}{k^2}\right)&\hbox{\em Landau or Lorentz gauge},\\
D_{\mu\nu}^{\rm Feyn}&:=\frac{1}{k^2-\i0}g_{\mu\nu}\ \ \ \ \ \ \ \ \ &\hbox{\em Feynman gauge},\\
D_{\mu\nu}^{\rm FY}&:=\frac{1}{k^2-\i0}\left(g_{\mu\nu}+2\frac{k_\mu k_\nu}{k^2}\right)&\hbox{\em Fried and Yennie gauge}.
\end{eqnarray*}

\nowastrona

We have $D_{\mu\nu}^\Coul=D_{\mu\nu}^{\rm Feyn}+k_\mu
f_\nu^\Coul(k)+f_\mu^\Coul(k)k_\nu$, 
where
\[f_0^\Coul(k)=\frac{k_0}{(k^2-\i0)2\vec k^2},\ \ f_i^\Coul
(k)=-\frac{k_i}{(k^2-\i0)2\vec
  k^2}. \]

The propagator in the {\em temporal gauge}
\begin{eqnarray*}
D_{00}^{\rm tem}=0,\ \ \ D_{0j}^{\rm tem}=0,\ \ \ 
D_{ij}^{\rm tem}=\frac{1}{k^2-\i0}\left(\delta_{ij}-\frac{k_ik_j}{k_0^2}\right).
\end{eqnarray*}
We have $D_{\mu\nu}^{\rm tem}=D_{\mu\nu}^{\rm Feyn}+k_\mu f_\nu^{\rm
  tem}(k)+f_\mu^{\rm tem}(k)k_\nu$, where
\[f_0^{\rm tem}(k)=\frac{1}{(k^2-\i0)2 k_0},\ \ f_i^{\rm tem}
(k)=-\frac{k_i}{(k^2-\i0)2
  k_0^2}. \]

\subsubsection{Path integral formulation}

 Let $D^\bullet_{\mu\nu}$ be one of the propagators considered in 
 Sect. \ref{Propagators for massive QED}. Let $B_\bullet^{\mu\nu}$ be its inverse.
Then we can use the corresponding action to express the generating function by path integrals, as described in Sect.
 \ref{Path integral formulation}, where this approach
 for massive vector fields was considered.

The discussion of the propagators $D_{\mu\nu}^\alpha$ and $D_{\mu\nu}^{\rm tem}$ is an obvious generalization of the massive case.

To obtain the propagator in the Coulomb gauge $D_{\mu\nu}^{\Coul}$,  we take the Lagrangian
\begin{eqnarray*}&&-\frac12\Big(\partial_\mu A_i(x) \partial^\mu A_i(x)
-
\partial_iA_0(x)\partial_iA_0(x)\Big),\end{eqnarray*}
and restrict the integration by the condition
\[\div\vec A(x)=0.\]

\nowastrona

\subsubsection{The $m\to0$ limit}

Assume that $J^\mu$ is a conserved 4-current. Using the propagator in the Yukawa gauge we can write the scattering
operator for a positive mass as
\begin{eqnarray*}
\hat S&=&\exp\Bigg(\frac{\i}{2}\int \frac{\d k}{(2\pi)^4} \bar{J^i(k)}\frac{1}{m^2+k^2-\i0}
\bigg(g_{ij}-\frac{k_i
      k_j}{m^2+\vec k^2}\bigg) J^i(k)\\&&\ \ -\frac{\i}{2}
\int \frac{\d k}{(2\pi)^4} \frac{1}{\vec k^2+m^2}
|J^0(k)|^2\Bigg)\\
&&\times\exp\left(-\i\sum_{\sigma=0,\pm1}\int\frac{\d \vec k}{\sqrt{(2\pi)^3}}
     \hat a^*(k,\sigma)\frac{\bar{u_\mu(k,\sigma)}}{\sqrt{2\varepsilon(\vec k)}}J^\mu(k)\right)\\&&
\times\exp\left(-\i\sum_{\sigma=0,\pm1}\int\frac{\d \vec k}{\sqrt{(2\pi)^3}}
     \hat a(k,\sigma)\frac{u_\mu(k,\sigma)}{\sqrt{2\varepsilon(\vec
      k)}}\bar{J^\mu(k)}\right)
\\&=&\hat S_\tr\otimes \hat S_\lg,
\end{eqnarray*}
\nowastrona
(In the expressions where we use the 3-dimensional integration $\d\vec k$, the 4-momenta are on shell, that is, $k=(\varepsilon(\vec k),\vec k)$).
Here, the {\em transversal scattering operator}
is
\begin{eqnarray*}
\hat S_\tr&=&\exp\Bigg(\frac{\i}{2}\int \frac{\d k}{(2\pi)^4}\bar{ J^i(k)}\frac{1}{m^2+k^2-\i0}\bigg(g_{ij}-\frac{k_i
      k_j}{k^2}\bigg) J^j(k)\\&&\ \ -\frac{\i}{2}\int \frac{\d k}{(2\pi)^4}
\frac{1}{\vec k^{2}+m^2}|J^0(k)|^2\Bigg)\\
&&\times\exp\left(-\i\sum_{\sigma=\pm1}\int\frac{\d \vec k}{\sqrt{(2\pi)^3}}
     \hat a^*(k,\sigma)\frac{\bar{u_\mu(k,\sigma)}}{\sqrt{2\varepsilon(\vec k)}}J^\mu(k)\right)\\&&
\times\exp\left(-\i\sum_{\sigma=\pm1}\int\frac{\d \vec k}{\sqrt{(2\pi)^3}}
     \hat a(k,\sigma)\frac{u_\mu(k,\sigma)}{\sqrt{2\varepsilon(\vec
      k)}}\bar{J^\mu(k)}\right)
\end{eqnarray*}
and converges to the massless scattering operator in the Coulomb gauge
as $m\searrow0$.
\nowastrona
The {\em longitudinal scattering operator}
is
\begin{eqnarray*}
\hat S_\lg&=&\exp\left(\frac{\i}{2} m^2\int \frac{\d k}{(2\pi)^4}  \bar{J^i(k)}\frac{1}{m^2+k^2-\i0}\frac{k_i
      k_j}{(m^2+\vec k^2)\vec k^2} J^j(k)\right)\\
&&\times\exp\left(-\i\int\frac{\d \vec k}{\sqrt{(2\pi)^3}}
      \hat a^*(k,0)\frac{\bar{u_\mu(k,0)}}{\sqrt{2\varepsilon(\vec k)}}J^\mu(k)\right)\\&&
\times\exp\left(-\i\int\frac{\d \vec k}{\sqrt{(2\pi)^3}}
      \hat a(k,0)\frac{u_\mu(k,0)}{\sqrt{2\varepsilon(\vec
      k)}}\bar{J^\mu(k)}\right).
\end{eqnarray*}
\nowastrona
This  can be rewritten  as
\begin{eqnarray*}
\hat S_\lg&=&\exp\left(\frac{\i}{2} m^2\int \frac{\d k}{(2\pi)^4} 
\frac{|\vec J\cdot\vec k|^2}{(m^2+k^2)(m^2+\vec k^2)\vec k^2}\right)\\
&&\times
\exp\left(- \frac{1}{2}\int \frac{\d k}{(2\pi)^4}
\frac{m^2|J^0(k)|^2}{2\varepsilon(\vec k)\vec k^2}\right)\\
&&\times\exp\left(\i\int\frac{\d \vec k}{\sqrt{(2\pi)^3}}
      \hat a^*(k,0)\frac{m J^0(k)}{|\vec k|\sqrt{2\varepsilon(\vec k)}}\right)
\\&&
\times\exp\left(\i\int\frac{\d \vec k}{\sqrt{(2\pi)^3}}
     \hat  a(k,0)\frac{m\bar{J^0(k)}}{|\vec k|\sqrt{2\varepsilon(\vec      k)}}\right),
\end{eqnarray*}
where the integral on the first line should be understood as the principal
value.
Thus $\hat S_\lg$, under rather general circumstances, converges to $\one$.

\subsubsection{Current produced by a travelling particle}

\nowastrona
Consider a
 classical particle travelling along the trajectory 
$t\mapsto \vec y(t)$ with a constant
profile $q(\vec x)$. Then its 4-current equals
\[J(t,\vec x)=q(\vec x-\vec y(t))\Big(1,\frac{\d \vec y(t)}{\d t}\Big).\]
Assume that $\vec y(t)=t\vec v^\pm$ for 
$\pm t>0$.
 Then \begin{eqnarray*}
J^\mu(k)&=&\int J^\mu(t,\vec x)\e^{-\i \vec k \vec x+\i k^0t}\d x\d t\\
&=&\left(-\frac{\i(1,\vec v_+)^\mu}{\vec k\vec v_+-k^0-\i0}+
\frac{\i(1,\vec v_-)^\mu}{\vec k \vec v_--k^0+\i0}\right) q(\vec k)\\&=&
\left(-\frac{\i p_+^\mu}{kp_+-\i0}+
\frac{\i p_-^\mu}{kp_-+\i0}\right) q(\vec k),\end{eqnarray*}
where $p^\pm=\frac{m}{\sqrt{1-(\vec v^\pm)^2}}(1,\vec v^\pm)$.

\nowastrona

Consider photons of mass $m\geq0$ coupled to the 4-current
$J^\mu$. 
Similarly as in Subsubsect. \ref{Travelling source}, we
define the scattering operator $\hat S_\GL$ by  replacing
\[\int \frac{\d k}{(2\pi)^4} \bar {J^\mu(k)}D_{\mu\nu}(k) J^\nu(k)\]
in  (\ref{exa1}) 
with
\beq \Im\int \frac{\d k}{(2\pi)^4} \bar{ J^\mu(k)}D_{\mu\nu}(k) J^\nu(k) .\label{poos}\eeq
(\ref{poos}) is infrared divergent if $m=0$, $\int q(\vec x)\d\vec
x\neq0$  and $\vec v_+\neq \vec v_-$.

We could try to justify the use of $\hat S_\GL$ similarly as in 
Subsubsect. \ref{Travelling source}, by introducing the Gell-Mann--Low
adiabatic switching. This justification is adopted by many physicists,
eg.
\cite{Lab}. One could criticize this approach, since after  multiplying 
by the switching function $\e^{-\epsilon |t|}$ the 4-current is no
longer conserved. Therefore, as indicated above,
we prefer to define the scattering
operator $\hat S_\GL$  simply by removing the (typically infinite)
phase shift.

\subsubsection{Energy shift}

Suppose that the 4-current is stationary and is given by a Schwartz
function $\rr^3\ni\vec x\mapsto J^\mu(\vec x)$ with $\div\vec
J(\vec x)=0$.

The Hamiltonian is given by
\begin{eqnarray*}
\hat H
&=&
\int\d \vec x\ {:}\Bigl(
\frac12\vec{ \hat E}_\tr^2(\vec x)
+\frac12 \big( \vec\partial\vec {\hat A}(\vec x)\big)^2
+\vec J(\vec x) \vec {\hat A}(\vec x)\Bigr){:}\\
&&+
\frac12\int\int\d\vec x\d\vec y J^0(\vec x)\frac{1}{4\pi |\vec x-\vec y|}J^0(\vec y).
\end{eqnarray*}
By (\ref{hov1}),  the
infimum of $\hat H$ is
\begin{eqnarray*}
E&=&-\frac12\int\int\d\vec x\d\vec y \vec J(\vec x)\frac{1}{4\pi |\vec x-\vec y|}\vec J(\vec y)\\
&&+
\frac12\int\int\d\vec x\d\vec y J^0(\vec x)\frac{1}{4\pi |\vec x-\vec y|}J^0(\vec y).
\end{eqnarray*}

\subsection{Alternative approaches}
\nowastrona 

\subsubsection{Manifestly Lorentz covariant formalism}

So far, our  treatment of the Maxwell equation
 was based on the Coulomb gauge, which
depends on the choice of the temporal coordinate. One can ask
whether massless vector fields can be studied in a manifestly
covariant fashion.


\nowastrona

Let $\Xi$ be an arbitrary space-time function. The Maxwell
equation allow us to impose
 a {\em generalized  Lorentz condition}
\beq \partial_\mu A^\mu=\Xi.\label{maxo0}\eeq
The Maxwell equation together with (\ref{maxo0}) imply
\beq-\Box A^\mu=-J^\mu+\partial^\mu\Xi.\label{maxo}\eeq

The function $\Xi$ has no physical meaning. Therefore it is natural to
adopt the simplest choice $\Xi=0$, that is the usual {\em Lorentz condition},
 for which (\ref{maxo}) reads
$-\Box A^\mu=-J^\mu$. We will discuss this approach in what
follows. For simplicity, we will limit ourselves to free fields.

\subsubsection{The Lorentz condition}

Recall that the Proca equation is equivalent to the Klein-Gordon
equation for vector fields together with the Lorentz
condition. Therefore, one can first develop its theory on the symplectic space
$\cY_\rv$, and then reduce it to the subpace $\cY_\Lor$, as described
before.

One can follow a similar route for the Maxwell equation. However,
there is a difference: the reduction by the Lorentz condition is
insufficient, one has to make an additional reduction.

Anyway, let us start as described in Subsubsect. \ref{sec-lor} by introducing
the space $\cY_\rv$, the form $\omega_\rv$, the subspace 
$\cY_\Lor$, the 4-potentials $A_\mu(x)$, $\Pi_\mu(x):=\dot A_\mu(x)$, where now
$m=0$. 

In the massive case $\cY_\Lor$ was symplectic (that means, the form $\omega_\rv$ restricted to $\cY_\Lor$ was nondegenerate). This is no longer true
in the massless case. Instead, the following is true.

\bep
 $\cY_\Lor$ is coisotropic. That means, if $\zeta$ is symplectically orthogonal to $\cY_\Lor$, then $\zeta\in\cY_\Lor$. \eep

\proof
Using 
 $-\Box \partial_\mu A^\mu(x)=0$ we see that, for any fixed $t$, we can replace
\beq \partial_\mu A^\mu(x)=0\label{loriom}\eeq
  with 
\begin{eqnarray}\label{maxo1}
0&=& \partial_\mu A^\mu(t,\vec x)=(-\Pi^0+\partial_i
A^i)(t,\vec x),\\ 
0&=&\partial_\mu\Pi^\mu(t,\vec x)=(-\Delta
A^0+\partial_i\dot A^i)(t,\vec x) 
\label{maxo2}
\end{eqnarray} 
as the defining conditions for $\cY_\Lor$. 
$\cY_\Lor$ is coisotropic iff
\begin{eqnarray}
\{ \partial_\mu A^\mu(t,\vec x),\partial_\mu A^\mu(t,\vec y)\}&=&0,\label{coiso1}\\
\{ \partial_\mu \Pi^\mu(t,\vec x),\partial_\mu\Pi^\mu(t,\vec y)\}&=&0,\label{coiso2}\\
\{ \partial_\mu A^\mu(t,\vec x),\partial_\mu\Pi^\mu(t,\vec y)\}&=&0.
\label{coiso3}
\end{eqnarray}
It is clear that 
(\ref{coiso1}) and (\ref{coiso2} are true.
To see (\ref{coiso3}) we compute:
\begin{eqnarray*} 
 &&\{\partial_\mu A^\mu      
(t,\vec x),\partial_\nu \Pi^\nu(t,\vec y)\}\\
&=&\Delta_{\vec y}\delta(\vec x-\vec y)+\partial_{\vec
  x_i}\partial_{\vec y_i}\delta(\vec x-\vec y)\ =\ 0. 
\end{eqnarray*}
\qed

$\cY_\Lor$ is a subspace of $\cY_{\widetilde\Max}$ and on
$\cY_\Lor$
the forms $\omega_{\widetilde\Max}$ and $\omega_\rv$
coincide.
\bep
Any $\zeta\in\cY_{\widetilde\Max}$  is gauge equivalent to an element
of $\cY_\Lor$.
\eep

\proof We can find smooth functions $\xi_+$ and $\xi_-$ such that
\[\partial_\mu\zeta^\mu =\xi_++\xi_-,\]
$\xi_-$ is past space compact and
$\xi_+$ is future space compact. By using the advanced and retarded
Green's functions  we can 
solve
\[-\Box\chi_-=\xi_-,\ \ \ -\Box\chi_+=\xi_+,\]
where $\chi_-$ is past space compact and $\chi_+$ is future space
compact. 
Then 
$\zeta_\mu+\partial_\mu\chi$
belongs to $\cY_\Lor$. \qed

 Therefore, the symplectically reduced $\cY_\Lor$
coincides with the symplectically reduced $\cY_{\widetilde\Max}$,
that is, with $\cY_\Max$.
This shows that both approaches to the Maxwell equation are equivalent
on the classical level.

\subsubsection{Positive frequency space}

 $\cW_\Lor^{(\pm)}$ will denote the subspace of $\cc\cY_\Lor$
 consisting of solutions that  have 
 positive,
resp.  negative frequencies.

For $g_1,g_2\in\cW_\Lor^{(+)}$ we define the scalar product
\begin{eqnarray} (g_1|g_2)&:=&\i\bar g_1\omega_\rv g_2\notag\\
&=&\i\bar g_1^\Coul\omega_\rv g_2^\Coul.
\label{scal+}\end{eqnarray}
Note that the definition  (\ref{scal+}) does not
depend on the choice of coordinates and is invariant wrt. the group
$\rr^{1,3}\rtimes  O^\uparrow(1,3)$.

The scalar product is positive semidefinite, but not strictly positive
definite.
Let $\cW_{\Lor,0}^{(+)}$ be the subspace
of elements $\cW_\Lor^{(+)}$ with a zero norm. Using 
 Prop. \ref{pro3} we see that   $\cW_{\Lor,0}^{(+)}$ 
 consists of
 pure gauges. The factor space  $\cW_\Lor^{(+)}/\cW_{\Lor,0}^{(+)}$ has a
 nondegenerate scalar product. Its completion is naturally isomorphic to
the space $\cZ_\Max$, which we constructed in Subsubsect. \ref{posi1}.

We have a natural identification of $\cY_\Lor$ with
$\cW_\Lor^{(+)}$ given by the obvious projection. For
$\zeta\in\cY_\Lor$ we will denote by $\zeta^{(+)}$ the corresponding
element of
$\cW_\Lor^{(+)}$.
This identification allows us to define a positive semidefinite
 scalar product on $\cY_\Lor$:
\begin{eqnarray}
\nonumber\langle\zeta_1|\zeta_2\rangle_{\cY}&:=&\Re(\zeta_1^{(+)}|\zeta_2^{(+)})\\
&=&\int\int\dot\zeta_{1i}^{\Coul}(0,\vec x)(-\i)D^{(+)}(0,\vec x-\vec y)
\dot\zeta_{2i}^\Coul(0,\vec y)\d\vec x\d\vec y\nonumber\\
&&\hspace{-7ex}+
\int\int\zeta_{1i}^\Coul(0,\vec x)(-\Delta_{\vec x})(-\i)D^{(+)}(0,\vec x-\vec y)
\zeta_{2i}^\Coul(0,\vec y)\d\vec x\d\vec y\nonumber.
\label{derive90}\end{eqnarray}

\nowastrona
\subsubsection{``First quantize, then reduce''}

One can  try to use the
 symplectic space $\cY_\rv$ of real vector valued
solutions of the Klein-Gordon equation
 as
the basis for quantization. In the literature, this starting point is
employed by two approaches. 

The first, which we call the {\em
  approach with a subsidiary condition}
has the advantage that it  uses only positive definite Hilbert
spaces. Unfortunately, in this approach
there are problems with the 4-potential $\hat A^\mu(x)$. Besides, the full
Hilbert space turns out to be non-separable.

In  the {\em Gupta-Bleuler approach} the 4-potentials $\hat A^\mu(x)$ are
well defined and covariant. Unfortunately it
uses indefinite scalar
product spaces.

\nowastrona
\subsubsection{Quantization with a subsidiary condition}

The quantization of the Proca equation described
in Subsubsec. \ref{pada21} is problematic in the zero mass limit. If $m=0$,
we cannot use the Hilbert space (\ref{pada31}) for the quantization, since it is not well defined.

However,  the $C^*$-algebraic formulation  survives the $m\searrow0$ limit.
In particular, 
 the {\em (Weyl) $C^*$-algebra of canonical
commutation relations} over $\cY_\rv$, introduced in (\ref{cstar}) and denoted $\CCR(\cY_\rv)$,
 is well defined also for $m=0$ 
and is invariant wrt the Poincar\'{e}{} group.

Strictly speaking, the spaces $\cY_\rv$ and hence the algebras $\CCR(\cY_\rv)$ are different for various $m$. If we fix a Cauchy subspace we can identify them by using the initial conditions.

Recall that in the massive case
\beq(\Omega|\hat A\lpar\zeta\rpar^2\Omega)=
\langle\zeta|\zeta\rangle_{\cY}+\frac{2}{m^2}\langle
\partial_\mu\zeta^\mu|\partial_\nu\zeta^\nu\rangle_\cY.\label{pada11}\eeq 
Recall that   $\zeta\in\cY_\Lor$ iff $\partial_\mu\zeta^\mu=0$. Therefore,
in the limit  $m\searrow0$,
\begin{eqnarray*}
(\Omega|\hat A\lpar\zeta\rpar^2\Omega) &=&\left\{\begin{array}{ll}
\langle\zeta|\zeta\rangle_{\cY},&
\zeta\in\cY_\Lor,\\
+\infty,&
\zeta\not\in\cY_\Lor.
\end{array}\right.
\end{eqnarray*}
 So,  the following  state  on $\CCR(\cY_\rv)$ is
 the limit of the state (\ref{pada41}) for $m\searrow0$:
\begin{eqnarray*}
\psi\big(W (\zeta)\big) &=&\left\{\begin{array}{ll}
\exp\Big(-\frac12\langle\zeta|\zeta\rangle_{\cY}\Big),&
\zeta\in\cY_\Lor,\\
0,&
\zeta\not\in\cY_\Lor.
\end{array}\right.
\end{eqnarray*}

\nowastrona

Let $(\cH_\psi,\pi_\psi,\Omega_\psi)$
 denote the {\em GNS representation}
for
this state. 
We can identify
\beq
J:\cH_\psi\to
l^2\left(\cY_\rv/\cY_\Lor,\Gamma_\s(\cZ_\Max)\right).\label{secto}\eeq
To describe this identification, first note that 
 $\cY_\rv/\cY_\Lor$ can be parametrized by 
smooth space-compact functions
\[\Xi=\partial_\mu\zeta^\mu,\]
which can be called
the {\em values of the Lorentz condition}. For each $\Xi$ choose $\zeta_\Xi\in\cY_\rv$ such that $\partial_\mu\zeta_\Xi^\mu=\Xi$. We demand that
\[\Big(J\pi_\psi\big(W(\zeta_\Xi)\big)\Omega_\psi\Big)(\Xi)=\begin{cases}
\Omega
,&\partial_\mu\zeta^\mu=\Xi,\\
0,&\partial_\mu\zeta^\mu\neq\Xi.\end{cases}
\]
Then $J$  is given by
\[\left(J\pi_\psi\big(W(\zeta)\big)\Omega_\psi\right)(\Xi)
=\begin{cases}
\e^{\frac{\i}{2}\zeta\omega_\rv\zeta_\Xi}
\e^{\i\hat A\lpar\zeta-\zeta_\Xi\rpar}\Omega
,&\partial_\mu\zeta^\mu=\Xi,\\
0,&\partial_\mu\zeta^\mu\neq\Xi.\end{cases}
 \]
Note that $\cH_\psi$ is {\em non-separable} -- it is an uncountable
 direct sum of {\em
  superselection sectors} corresponding to various values of the
 Lorentz condition. All these superselection sectors are separable. 

Special role is played by the (separable) 
subspace (superselection sector) corresponding to the
Lorentz condition $\Xi=0$. We can choose $\zeta_{\Xi=0}=0$ and thus this subspace is naturally isomorphic to $\Gamma_\s(\cZ_\Max)$ with the fields  obtained by the usual quantization obtained by the method  ``first reduce, then quantize''.

\nowastrona

Note that $\pi_\psi(W(\zeta))$ maps between various sectors of
(\ref{secto}) if $\zeta\not\in\cY_\Lor$. The unitary group
$\rr\ni t\mapsto \pi_\psi\left(W(t\zeta)\right)$ is strongly
continuous if and only if $\zeta\in\cY_\Lor$. If this is the case,
we can write
$\pi_\psi(W(\zeta))=\e^{\i \hat A\lpar \zeta\rpar}$.
We have $\hat A\lpar \zeta_1\rpar=\hat A\lpar
\zeta_2\rpar$ if in addition
 $\zeta_1$ differs from $\zeta_2$ by a pure gauge.
$\hat A\lpar\zeta\rpar$ is ill defined if $\zeta\not\in\cY_\Lor$.

To my knowledge, the approach that we described above, restricted to the
$0$th sector, was essentially one of
 the first approaches to the quantization of
Maxwell equation. It is typical for older
presentations, eg. \cite{JR}.
However, without the language of $C^*$-algebras it is
somewhat awkward to describe.
 One usually says that the Lorentz 
condition $\partial_\mu \hat A^\mu(x)=0$ is enforced on the Hilbert space
of states and constitutes a {\em subsidiary condition}.

\nowastrona

\subsubsection{ The Gupta-Bleuler approach}

The Gupta-Bleuler approach follows the same lines as
in the massive case until we arrive at the algebraic 
Fock space built on
$\cW_\Lor^{(+)}$. As we know, the scalar product on $\cW_\Lor^{(+)}$
is only semidefinite. We factor  $\cW_\Lor^{(+)}$ by the null space of
its scalar product, obtaining
 $\cW_{\Max}^{(+)}$. We complete it, obtaining $\cZ_{\Max}$ and
we take the  corresponding Fock space $\Gamma_\s(\cZ_{\Max})$ -- this
coincides with
 the usual
quantization.

Equivalently, we can take the (algebraic) Fock space over
 $\cW_\Lor^{(+)}$. It has a natural semidefinite product. We divide by
its null space and take the completion. Again, the resulting Hilbert
space can be naturally identified with $\Gamma_\s(\cZ_{\Max})$.
\nowastrona

\init\section{Charged scalar bosons}
\label{Charged scalar bosons}

In this section we consider again the {\em Klein-Gordon equation}
\beq(-\Box+m^2)\psi(x)=0.\label{comkg}\eeq
This time we will quantize the space of its {\em complex solutions}.

\nowastrona

The formalism used in physics to describe complex fields, and
especially to quantize them, is
different from the real case, therefore we devote to it a separate section.

The advantage of complex fields, as compared with real fields, is the 
possibility to
include an {\em external electromagnetic 4-potential} $A(x)=[A^\mu(x)]$ and to consider the equation
\[\left(-\left(\partial_\mu+\i A_\mu(x)\right)\left(\partial^\mu+\i A^\mu(x)\right)
+m^2\right)\psi(x)=0.\]

\nowastrona

\subsection{Free charged scalar bosons}

\subsubsection{Classical fields}

 $\cW_\KG$ will denote the space of
{\em smooth space-compact complex solutions of the Klein-Gordon equation}
\beq(- \Box+m^2) \zeta=0.\label{KG1ze}\eeq
(In the context of neutral fields, it was denoted $\cc\cY_\KG$, because it was
an auxiliary object, the {\em complexification of the phase space}
$\cY_\KG$. Now it is the basic object, the {\em phase space} itself). 

Clearly, the space  $\cW_\KG$ is equipped with a {\em  complex
  conjugation} 
$\zeta\mapsto\bar\zeta$ and a {\em $U(1)$ symmetry}
$\zeta\mapsto\e^{\i\theta}\zeta$, $\theta\in\rr/2\pi\zz=U(1)$.

If $T$ is a real linear functional on $\cW$,  then
we have  two kinds of natural complex conjugations of 
$T$:
\begin{eqnarray}
\langle \bar T|\zeta\rangle:= \bar{ \langle  T|\bar\zeta\rangle},&&
\langle T^*|\zeta\rangle:= \bar{ \langle  T|\zeta}\rangle.
\label{sense}\end{eqnarray}
 Both
maps $T\mapsto \bar T$ and  $T\mapsto T^*$ are antilinear. 
When restricted to the real subspace $\cY_\KG\subset\cW_\KG$, the functionals $\bar T$ and $T^*$ coincide.

A special role is played by {\em complex linear} functionals on $\cW$. The space of such functionals will be denoted $\cW^\#$. If $T\in\cW^\t$, then $\bar T\in\cW^\t$, unlike  $T^*$, which is antilinear.


\nowastrona

In the neutral case a crucial role was played by the conserved 4-current  $j_\mu(\zeta_1,\zeta_2)$,
where $\zeta_1,\zeta_2\in\cY_\KG$; see (\ref{curr}).
In the charged case we will use its sesquilinear version defined on
$\cW_\KG$:
\begin{eqnarray}
\cj^\mu(\bar\zeta_1,\zeta_2,x)&:=&\bar{\partial^\mu\zeta_1(x)}\zeta_2(x)
-\bar{\zeta_1(x)}\partial^\mu \zeta_2(x).\label{versio}\end{eqnarray}
If
we decompose elements of $\cW_\KG$ into their real and imaginary part
$\zeta=\zeta_\R+\i\zeta_\I$, then the real part of the 4-current splits into a part depending on $\zeta_\R$ and on $\zeta_\I$:
\begin{eqnarray*}
&&\Re j^\mu(\bar \zeta_1,\zeta_2,x)\\
&=&\partial^\mu\zeta_{\R,1}(x)\zeta_{\R,2}(x)-\zeta_{\R,1}(x)\partial^\mu
\zeta_{\R,2}(x)\\&&
 +  \partial^\mu\zeta_{\I,1}(x)\zeta_{\I,2}(x)-\zeta_{\I,1}(x)\partial^\mu
\zeta_{\I,2}(x).
\end{eqnarray*}
Thus $\cW_\KG$ can be viewed as the direct sum of two symplectic spaces
with the form
\[\Re\bar
\zeta_1\omega\zeta_2=\zeta_{\R,1}\omega\zeta_{\R,2}+
\zeta_{\I,1}\omega\zeta_{\I,2} .\]

\nowastrona

For $x\in\rr^{1,3}$, one can introduce
the fields $\phi_\R(x)$,  $\phi_\I(x)$,  $\pi_\R(x)$, 
 $\pi_\I(x)$ as the real linear
 functionals on $\cW_\KG$ given by
\begin{eqnarray}
 \langle \phi_\R(x)|\zeta\rangle:=\Re\zeta(x),&& \langle
 \phi_\I(x)|\zeta\rangle:=\Im\zeta(x),\label{real1}\\
\langle\pi_\R(x)|\zeta\rangle:=\Re\dot\zeta(x),&&
\langle\pi_\I(x)|\zeta\rangle:=\Im\dot\zeta(x).\label{real2}
\end{eqnarray}
Clearly, we have the usual equal time Poisson brackets (we write only the
non-vanishing ones):
\begin{eqnarray}
\{\phi_\R(t,\vec x),\pi_\R(t,\vec y)\}
=\{\phi_\I(t,\vec x),\pi_\I(t,\vec y)\}&=&\delta(\vec x-\vec y).
\label{poisson7}\end{eqnarray}

\nowastrona

In practice instead of (\ref{real1}) and (\ref{real2}) one prefers to use {\em complex  fields}  $\psi(x),\eta(x)\in\cW^\#$ defined by
\begin{eqnarray*}
 \langle \psi(x)|\zeta\rangle:=\frac{1}{\sqrt2}\zeta(x),&& \langle
 \psi^*(x)|\zeta\rangle:=\frac{1}{\sqrt2}\bar{\zeta(x)},\\
\langle\eta(x)|\zeta\rangle:=\frac{1}{\sqrt2}\dot\zeta(x),&&
\langle\eta^*(x)|\zeta\rangle:=\frac{1}{\sqrt2}\bar{\dot\zeta(x)}.
\end{eqnarray*}
Clearly, 
\begin{eqnarray*}
&&\psi(x)=\frac{1}{\sqrt2}\bigl(\phi_\R(x)+\i\phi_\I(x)\bigr),\ \ \ 
\psi^*(x)=\frac{1}{\sqrt2}\bigl(\phi_\R(x)-\i\phi_\I(x)\bigr),\\
&&\eta(x)=\frac{1}{\sqrt2}\bigl(\pi_\R(x)+\i\pi_\I(x)\bigr),\ \ \ 
\eta^*(x)=\frac{1}{\sqrt2}\bigl(\pi_\R(x)-\i\pi_\I(x)\bigr).
\end{eqnarray*}


\nowastrona

Note that
\beq\psi(t,\vec x)=\int \dot D(t,\vec x-\vec y)\psi(0,\vec y)\d\vec{y}+
\int  D(t,\vec x-\vec y)\eta(0,\vec y)\d\vec{y}.\label{poisson1a}\eeq

The only non-vanishing equal-time
 Poisson brackets are
\begin{eqnarray}
\{\psi(t,\vec x),\eta^*(t,\vec y)\}
=\{\psi^*(t,\vec x),\eta(t,\vec y)\}&=&\delta(\vec x-\vec y).
\label{poisson1}\end{eqnarray}
Using (\ref{poisson1a}) we obtain
\begin{eqnarray*}\{\psi(x),\psi(y)\}=\{\psi^*(x),\psi^*(y)\}&=&0,\\
 \{\psi(x),\psi^*(y)\}&=&D(x-y).\end{eqnarray*}

\nowastrona


\nowastrona

\subsubsection{Smeared fields}

We can use the symplectic form to pair distributions and solutions.
For $\zeta\in\cW_\KG$ the corresponding
{\em spatially smeared fields} are the functionals on $\cW_\KG$ given by
\begin{eqnarray*}
\langle \psi\lpar\zeta\rpar|\rho\rangle&:=&\frac{1}{\sqrt2}\bar\zeta\omega\rho,\\
\langle\psi^*\lpar\zeta\rpar|\rho\rangle&:=&\frac{1}{\sqrt2}
\zeta\omega\bar\rho,\ \ \
\rho\in\cW_\KG. 
\end{eqnarray*}
Equivalently,
\begin{eqnarray*}
\psi\lpar\zeta\rpar&=&
\int\left(-\bar{\dot\zeta(t,\vec x)}\psi(t,\vec x)+
\bar{\zeta(t,\vec x)}\eta(t,\vec x)\right)\d\vec x,\\
\psi^*\lpar\zeta\rpar&=&
\int\left(-\dot\zeta(t,\vec x)\psi^*(t,\vec x)+
\zeta(t,\vec x)\eta^*(t,\vec x)\right)\d\vec x.\end{eqnarray*}
Note that
\begin{eqnarray*}\{\psi\lpar\zeta_1\rpar,\psi\lpar\zeta_2\rpar\}=
\{\psi^*\lpar\zeta_1\rpar,\psi^*\lpar\zeta_2\rpar\}&=&0,\\
\{\psi\lpar\zeta_1\rpar,\psi^*\lpar\zeta_2\rpar\}
&=&\bar\zeta_1\omega \zeta_2.
\end{eqnarray*}
\nowastrona

We can also introduce {\em space-time smeared fields}.
To a space-time function $f\in C_{\rm   c}^\infty(\rr^{1,3},\cc)$ we associate
\begin{eqnarray*}
\psi[f]&:=&\int \bar{f(x)}\psi(x)\d x,\\
\psi^*[f]&:=&\int f(x)\psi^*(x)\d x.
\end{eqnarray*}
Clearly,
\begin{eqnarray*}\{\psi[f_1],\psi[f_2]\}=
\{\psi^*[f_1],\psi^*[f_2]\}&=&0,\\
\{\psi[f_1],\psi^*[f_2]\}&=&\int\int \bar{f_1(x)}D(x-y)f_2(y)\d x\d y,
\end{eqnarray*}
\[ \psi[f]=-\psi\lpar D*f\rpar,\ \ \ 
\psi^*[f]=-\psi^*\lpar D*f\rpar.\]
\subsubsection{Lagrangian formalism}

In the Lagrangian formalism we use the complex off-shell fields $\psi(x)$ and $\psi^*(x)$ as the basic variables. We introduce the Lagrangian density
\begin{eqnarray*}
\cL(x)=&
-\partial_\mu\psi^*(x)
\partial^\mu\psi(x)- m^2\psi^*(x)\psi(x).
\end{eqnarray*}
The Euler-Lagrange equations 
\beq
\partial_{\psi^*}\cL-
\partial_\mu\frac{\partial \cL}{\partial\psi_{,\mu}^*}=0,\ \ \    \ 
\partial_{\psi}\cL-
\partial_\mu\frac{\partial \cL}{\partial\psi_{,\mu}}=0
\label{euler-lagrange1a}\eeq
yield (\ref{comkg}).
The {\em variables conjugate} to $\psi(x)$ and $\psi^*(x)$ are
\begin{eqnarray*}
\eta^*(x)&:=&\frac{\partial \cL}{\partial\psi_{,0}(x)}=
\partial_0\psi^*(x),\\
\eta(x)&:=&\frac{\partial \cL}{\partial\psi_{,0}^*(x)}=
\partial_0\psi(x).
\end{eqnarray*}

\subsubsection{Classical 4-current}

The Lagrangian is invariant w.r.t. the $U(1)$ symmetry $\psi\mapsto\e^{-\i\theta}\psi$. The Noether 4-current associated to this symmetry is called simply
the {\em  4-current}. It is
\begin{eqnarray*}\cJ^\mu(x)&:=&
\i\Big(\psi^*(x)\frac{\partial\cL(x)}{\partial\psi_{,\mu}^*}
-\frac{\partial\cL(x)}{\partial\psi_{,\mu}}\psi(x)\Big)\\&=&
\i\bigl(\partial^\mu\psi^*(x)\psi(x)-\psi^*(x)\partial^\mu\psi(x)\bigr).
\end{eqnarray*} 
It is conserved on shell and real:
\begin{eqnarray*}\partial_\mu \cJ^\mu(x)&=&0,\\
\cJ^\mu(x)^*&=&\cJ^\mu(x).
\end{eqnarray*}
Up to a coefficient, it coincides with (\ref{versio}) viewed as a
quadratic form: 
\begin{eqnarray*}
\langle \cJ^\mu(x)|\zeta\rangle&=& \frac{\i}{2} j^\mu(\bar\zeta,\zeta,x)\\
&=&\frac{\i}{2}\bigl(\bar{\partial^\mu\zeta(x)}\zeta(x)
-\bar{\zeta(x)}\partial^\mu \zeta(x)\bigr).\end{eqnarray*}

The 0th component of the 4-current is called the {\em charge density}
\[\cQ(x):=\cJ^0(x)=
\i\bigl(-\eta^*(x)\psi(x)+\psi^*(x)\eta(x)\bigr).\] 
We have the relations
\begin{eqnarray}
\nonumber
\{  \cQ(t,\vec x), \psi(t,\vec y)\}&=&\i\psi(t,\vec y)\delta(\vec
x-\vec y),\\ {}\nonumber
\{  \cQ(t,\vec x), \eta(t,\vec y)\}&=&\i\eta(t,\vec y)\delta(\vec
x- \vec y),\\{} \label{westill0}
\{  \cQ(t,\vec x),  \cQ(t,\vec y)\}&=&0
.\end{eqnarray}

The {\em (total) charge}
\[Q:=\int \cQ(t,\vec x)\d\vec x\]
 is conserved (does not depend on time).

For $\chi\in C_{\rm c}^\infty(\rr^3,\rr)$, let $\alpha_\chi$ denote the
 $*$-automorphism of the algebra of functions on $\cW_\KG$
 defined by
\begin{eqnarray}\notag\alpha_\chi(\psi(0,\vec x))&:=&\e^{-\i\chi(\vec
    x)}\psi(0,\vec x),\\
\alpha_\chi(\eta(0,\vec x))&:=&\e^{-\i\chi(\vec x)}\eta(0,\vec x).
\label{gaugo}\end{eqnarray}
Obviously, 
\begin{eqnarray}\notag\alpha_\chi(\psi^*(0,\vec x))&=&\e^{\i\chi(\vec
    x)}\psi^*(0,\vec x),\\
\alpha_\chi(\eta^*(0,\vec x))&=&\e^{\i\chi(\vec x)}\eta^*(0,\vec x).
\label{gaugo1}\end{eqnarray}
(\ref{gaugo}) is called
the {\em gauge transformation} at time $t=0$ corresponding to $\chi$.
Set
\beq
 Q(\chi)=\int\chi(\vec x)\cQ(0,\vec x) \d\vec
x.\label{illo-}\eeq $Q(\chi)$ generates
the one-parameter group of gauge 
transformations $\rr\ni s\mapsto\alpha_{s\chi}$ (\ref{gaugo}). In other words,  
for any classical observable $B$ (a function on $\cW_\KG)$)
 \begin{eqnarray*}
\partial_s\alpha_{s\chi}(B)&=&\big\{\alpha_{s\chi}(B), Q(\chi)\big\},\\
\alpha_{0\chi}(B)&=&B.
\end{eqnarray*}

\nowastrona

\subsubsection{Stress-energy tensor}

\nowastrona
The Lagrangian is invariant w.r.t. space-time translations. This leads to the {\em stress-energy tensor}
\begin{eqnarray*}
\cT^{\mu\nu}(x)&:=&-\frac{\partial \cL(x)}{\partial\psi_{,\mu}(x)}\partial^\nu\psi(x)-
\partial^\nu\psi^*(x)\frac{\partial \cL(x)}{\partial\psi^{*}_{,\mu}(x)}
+g^{\mu\nu}\cL(x)\\
&=&
\partial^\mu\psi^*(x)\partial^\nu\psi(x)+
\partial^\nu\psi^*(x)\partial^\mu\psi(x)\\&&
-
g^{\mu\nu}\left(\partial_\alpha\psi^*(x)\partial^\alpha\psi(x)+
m^2\psi^*(x)\psi(x)\right).  
\end{eqnarray*}
It is conserved on shell
\[\partial_\mu \cT^{\mu\nu}(x)=0.\]
The components of the stress-energy tensor with the first temporal coordinate
are called the  Hamiltonian density and  momentum
  density.
We express them on-shell in terms of $\psi(x)$, $\psi^*(x)$, $\eta(x)$
and $\eta^*(x)$:
\begin{eqnarray*}
\cH(x)\ :=\ \cT^{00}(x)&=&
\eta^*( x)\eta(x)+\vec\partial\psi^*( x)\vec\partial\psi(x) +m^2\psi^*(x)\psi(x),\\
\cP^i(x)\ :=\ \cT^{0i}(x)&=&-\eta^*(x)\partial^i
 \psi(x)-\partial^i
 \psi^*(x)\eta(x).\end{eqnarray*}
 $\cH(x)$ and $\vec\cP(x)$ acting on $\zeta\in\cW_\KG$ yield
\begin{eqnarray*}
\langle\cH(x)|\zeta\rangle&=&\frac12|\dot\zeta( x)|^2+\frac12|\vec\partial\zeta( x)|^2
+\frac{m^2}{2}|\zeta(x)|^2,\\
\langle\vec\cP(x)|\zeta\rangle&=&-\frac12\bar{\dot\zeta(x)}\vec\partial\zeta( x)-\frac12\bar{\vec\partial\zeta( x)}\dot\zeta(x).\end{eqnarray*}

We can define the 
   Hamiltonian and  momentum
\[H=\int \cH(t,\vec x)\d\vec x,
\ \ \ \vec P=\int \vec\cP(t,\vec x)\d\vec x.
\]
$H$ and $\vec P$
are the generators of the time and space translations. The observables
 $H$,
$P_1$, $P_2$, $P_3$ and $Q$ are in involution.
\nowastrona

\subsubsection{Diagonalization of the equations of motion}


Recall that in the neutral case the generic notation for the energy-momentum was $k$. The on-shell condition was $k^2+m^2=0$, $k^0>0$. In other words, $k^0=\varepsilon(\vec k):=\sqrt{\vec k^2+m^2}$.

In the charged case, following \cite{GR}, it will be convenient
 to use different letters for the generic notation of the
energy-momentum. In the charged case, the
energy-momentum will be denoted generically by $p$ with the on-shell condition
 $p^2+m^2=0$, $p^0>0$.
We will also use a different letter for the energy: $E(\vec
p):=\sqrt{\vec p^2+m^2}$. In other words,  
$p=(E(\vec p),\vec p)$.

Define
\begin{eqnarray*}
\psi_t(\vec p)&:=&\int\psi(t,\vec x)\e^{-\i \vec p \vec x}\d\vec x,\\
\eta_t(\vec p)&:=&\int\eta(t,\vec x)\e^{-\i \vec p \vec x}\d\vec x.\end{eqnarray*}
Clearly, the only nonvanishing Poisson brackets are
\begin{eqnarray*}
\{\psi_t(\vec p),\eta_t^*(\vec p')\}=\{\psi_t^*(\vec p),\eta_t(\vec p')\}&=&
(2\pi)^3\delta(\vec p-\vec p').
\end{eqnarray*}
The equations of motion are
\begin{eqnarray*}
\dot\psi_t(\vec p)&=&\eta_t(\vec p),\\
\dot\eta_t(\vec p)&=&-E^2(\vec p)\psi_t(\vec p).
\end{eqnarray*}

\nowastrona
For on-shell $p\in\rr^{1,3}$ define
\begin{eqnarray*}
a_t(p)&=&
\frac{1}{\sqrt{(2\pi)^3}}\Big(\sqrt{\frac{E (\vec p)}{2}}\psi(t,\vec p)+\frac{\i
}{\sqrt{2E (\vec p)}}\eta(t,\vec p)\Big),\\
a_t^*(p)&=&\frac{1}{\sqrt{(2\pi)^3}}
\Big(\sqrt{\frac{E (\vec p)}{2}}\psi^*(t,\vec p)-\frac{\i}{\sqrt{2E (\vec p)}}\eta^*(t,\vec p)\Big),\\
b_t(p)&=&
\frac{1}{\sqrt{(2\pi)^3}}\Big(\sqrt{\frac{E (\vec p)}{2}}\psi^*(t,\vec p)+\frac{\i
}{\sqrt{2E (\vec p)}}\eta^*(t,\vec p)\Big),\\
b_t^*(p)&=&
\frac{1}{\sqrt{(2\pi)^3}}\Big(\sqrt{\frac{E (\vec p)}{2}}\psi(t,\vec p)-\frac{\i}{\sqrt{2E (\vec p)}}\eta(t,\vec p)\Big).
\end{eqnarray*}

We have the equations of motion
\begin{eqnarray*}
\dot a_t(p)=-\i  E(\vec p)a_t( p),&&
\dot b_t(p)=-\i  E(\vec p)b_t( p),\\
\dot{ a}_t^*( p)=\i E(\vec p)a_t^*( p),&& \dot{b}_t^*( p)=\i E(\vec p)b_t^*( p).
\end{eqnarray*}
We will usually write $a(p)$, $a^*(p)$, $b(p)$, $b^*(p)$ instead of $a_0(p)$, $ a_0^*(p)$, $b_0(p)$, $b_0^*(p)$, so that
\begin{eqnarray*}
a(p)&=&
\int\Big(\sqrt{\frac{E (\vec p)}{2}}\psi(0,\vec x)+\frac{\i
}{\sqrt{2E (\vec p)}}\eta(0,\vec x)\Big)\e^{-\i \vec p \vec x}\frac{\d\vec x}{\sqrt{(2\pi)^3}},\\
a^*(p)&=&
\int\Big(\sqrt{\frac{E (\vec p)}{2}}\psi^*(0,\vec x)-\frac{\i}{\sqrt{2E (\vec p)}}\eta^*(0,\vec x)\Big)\e^{\i \vec p \vec x}\frac{\d\vec x}{\sqrt{(2\pi)^3}},\\
b(p)&=&
\int\Big(\sqrt{\frac{E (\vec p)}{2}}\psi^*(0,\vec x)+\frac{\i
}{\sqrt{2E (\vec p)}}\eta^*(0,\vec x)\Big)\e^{-\i \vec p \vec x}\frac{\d\vec x}{\sqrt{(2\pi)^3}},\\
b^*(p)&=&
\int\Big(\sqrt{\frac{E (\vec p)}{2}}\psi(0,\vec x)-\frac{\i}{\sqrt{2E (\vec p)}}\eta(0,\vec x)\Big)\e^{\i \vec p \vec x}\frac{\d\vec x}{\sqrt{(2\pi)^3}},
\end{eqnarray*}
\begin{eqnarray*}
a_t(p)=\e^{-\i tE(\vec p)}a( p),&&b_t(p)=\e^{-\i tE(\vec p)}b( p),\\
 a_t^*( p)=\e^{\i tE(\vec p)}a^*( p),&& b_t^*( p)=\e^{\i tE(\vec p)}b^*( p).
\end{eqnarray*}
The only non-vanishing 
 Poisson bracket are
\begin{eqnarray*}
\{a(p), a^*(p')\}&=&\{b(p), b^*(p')\}=-\i
\delta(\vec p-\vec p').\end{eqnarray*}

We have  the following expressions for the
fields:
\begin{eqnarray*}
\psi(x)&=&\int\frac{\d\vec{p}}{\sqrt{(2\pi)^3}\sqrt{2E(\vec p)}} \left(\e^{\i
  px} a(p)+
\e^{-\i px}b^*(p)\right),\\
\eta(x)&=&\int\frac{\d\vec{p}\sqrt{E(\vec p)}}{\i\sqrt{(2\pi)^3}\sqrt2} \left(\e^{\i
  px} a(p)-
\e^{-\i px}b^*(p)\right).
\end{eqnarray*}
We have accomplished the diagonalization of the basic observables:
\begin{eqnarray*}
H&=&\int\d\vec p
 E(\vec p)\big( a^*( p)a(p)+b^*(p)b(p)\big),\\
\vec P&=&\int\d\vec p
 \vec p \big(a^*( p)a(p)+b^*(p)b(p)\big),\\
Q&=&\int\d\vec p
 \big( a^*( p)a(p)-b^*(p)b(p)\big).
\end{eqnarray*}

Every 
 $\zeta\in\cW_{\KG}$ can be written as
\beq\zeta(x)=\int\frac{\d\vec{p}}{\sqrt{(2\pi)^3}\sqrt{E(\vec p)}}
\Big( \e^{\i px}\langle a( p)|\zeta\rangle +
\e^{-\i px}\langle b^*( p)|\zeta\rangle\Big)
.\label{pafa.1}\eeq
Note that the plane wave functional $a(k)$ of the neutral case is slightly different from its counterpart $a(p)$ of the charged case. The former acts on the real space $\cY_\KG$ and the latter on the complex space $\cW_\KG$, but the latter is not simply the complexification of the former -- compare (\ref{pafa}) and (\ref{pafa.1}) and notice the absence of $\sqrt{2}$.

\subsubsection{Plane waves}

In the charged case we  use  almost the same plane waves  as those introduced in the neutral
case in (\ref{planewave1aa}).  There are three differences: the
generic notation for the energy-momentum is now $p$,   plane waves
with a negative frequency $p^0$ are now on the equal footing as those
with a positive frequency, and there is a factor $\sqrt2$ is missing in the denominator. Thus for $p\in\rr^{1,3}$ with $p^2+m^2=0$ we define
\beq (x|p)=\frac{1}{\sqrt{(2\pi)^3}\sqrt{E(\vec p)}}
\e^{\i px}.\eeq

Let $p^0,p^{0'}>0$.  We have
\begin{eqnarray*}
\i(-p|\omega|p')=\i(-p|\omega|p')&=&0,\\
-\i(-p|\omega|-p')=\i(p|\omega|p')&=&2\delta(\vec p-\vec p')
.\end{eqnarray*}

\nowastrona

$a(p)$ and $b(p)$ can be called {\em plane wave functionals}:
\begin{eqnarray*}
a(p)&=&\frac{\i}{\sqrt2}\psi\lpar|p)\rpar,\\
b^*(p)&=&-\frac{\i}{\sqrt2}\psi\lpar|-p)\rpar.\\
\end{eqnarray*}
Thus for every  $\zeta\in\cW_{\KG}$ we have
\begin{eqnarray}
\langle a(p)|\zeta\rangle&=&\frac{\i}{2}(p|\omega\zeta,\label{pqd1}\\
\langle b^*(p)|\zeta\rangle&=&-\frac{\i}{2}(-p|\omega\zeta.\label{pqd2}
\end{eqnarray}


\nowastrona

\nowastrona

\subsubsection{Positive and negative frequency subspace}
\label{positi}

\nowastrona
When we discussed neutral scalar fields we introduced 
 {\em positive/negative frequency spaces}, which in the notation used in the charged case can be defined by
\begin{eqnarray*}
\cW_\KG^{(+)}&:=&\{g\in\cc\cY_\KG\ :\ 
 (p|\omega g=0,\ \ \ p^0<0\},\\
\cW_\KG^{(-)}:=\bar{\cW_\KG^{(+)}}&=&\{g\in\cc\cY_\KG\ :\ 
 (p|\omega g=0,\ \ \ p^0>0\}.
\end{eqnarray*}
Every $\zeta\in\cW_\KG$ can be uniquely decomposed as
$\zeta=\zeta^{(+)}+\zeta^{(-)}$ with $\zeta^{(\pm)}\in
\cW_\KG^{(\pm)}$.

We equip  $\cW_\KG^{(+)}$
with
 the scalar product 
\begin{eqnarray}
(\zeta_1^{(+)}|\zeta_2^{(+)})&:=&
\frac{\i}{2}\bar{\zeta_1^{(+)}}\omega\zeta_2^{(+)}=
\int\bar{\langle a(p)|\zeta_1^{(+)}}\rangle\langle a(p)|\zeta_2^{(+)}
\rangle\d\vec{p}.
\label{tyt3}\end{eqnarray}
We set $\cZ_\KG^{(+)}$ to be the completion of 
$\cW_\KG^{(+)}$ in this scalar product. By (\ref{pqd1}),
\[\langle a(p)|\zeta^{(+)}\rangle=(p|\zeta^{(+)}).\]
 $\cZ_\KG^{(+)}$ can be identified with $L^2(\rr^3)$ and (\ref{tyt3}) rewritten as
\begin{eqnarray*}
(\zeta_1^{(+)}|\zeta_2^{(+)})&=&
\int\bar{(p|\zeta_1^{(+)})}(p|\zeta_2^{(+)})
\d\vec{p}.
\end{eqnarray*}

Instead of $\cW_\KG^{(-)}$ for quantization we will use the
corresponding
 complex
conjugate space denoted $\bar{\cW_\KG^{(-)}}$ and equipped
with the scalar product
\begin{eqnarray}
(\bar\zeta_1^{(-)}|\bar\zeta_2^{(-)})&:=&
\frac{\i}{2}\zeta_1^{(-)}\omega\bar{\zeta_2^{(-)}}=
\int\bar{\langle b(p)|\zeta_1^{(-)}\rangle}\langle
  b(p)|\zeta_2^{(-)}
\rangle\d\vec{p}.\label{tyt4}
\end{eqnarray}
We set $\cZ_\KG^{(-)}$ to be the completion of 
$\bar{\cW_\KG^{(-)}}$ in this scalar product. 
By (\ref{pqd2}),
\[\langle b(p)|\zeta^{(-)}\rangle=(\bar{-p}|\bar\zeta^{(-)}).\]
 $\cZ_\KG^{(-)}$ can be identified with $L^2(\rr^3)$ and (\ref{tyt4}) rewritten as
\begin{eqnarray*}
(\bar\zeta_1^{(-)}|\bar\zeta_2^{(-)})&=&
\int\bar{(\bar{- p}|\bar\zeta_1^{(-)})}(\bar{-p}|\bar\zeta_2^{(-)})
\d\vec{p}.
\end{eqnarray*}

Note that $\bar{\cW_\KG^{(-)}}=\cW_\KG^{(+)}$, where we use the usual
(internal) complex
conjugation in $\cW_\KG$.  Therefore in principle we could identify
  $\cZ_\KG^{(-)}$ and
$\cZ_\KG^{(+)}$. This identification will be important for the
definition of the charge conjugation. Normally, however, we treat 
 $\cZ_\KG^{(-)}$ and
$\cZ_\KG^{(+)}$ as two separate Hilbert spaces.

$\rr^{1,3}\rtimes  O^\uparrow(1,3)$  acts on  $\cZ_\KG^{(+)}$
and $\cZ_\KG^{(-)}$ in a natural way. 

\nowastrona

\subsubsection{Quantization}

In principle, we could quantize  the complex Klein-Gordon
equation  as a pair of real Klein-Gordon fields.
However, we will use 
 the formalism of quantization of charged bosonic systems \cite{DeGe}. 

\nowastrona

We want to construct  $(\cH,\hat H,\Omega)$
 satisfying the usual requirements of QM (1)-(3) and
an operator valued
distribution \beq
\rr^{1,3}\ni x\mapsto \hat\psi(x)\label{distro1}\eeq
satisfying, with $\eta(x):=\dot\psi(x)$,
 \ben\item 
$(-\Box+m^2)\hat\psi(x)=0$;
\item the only non-vanishing $0$-time commutators are 
 \beq
[\hat\psi(0,\vec x),\hat\eta^*(0,\vec y)]=\i \delta(\vec x-\vec y),\ \ 
[\hat\psi^*(0,\vec x),\hat\eta(0,\vec y)]=\i \delta(\vec x-\vec y);\label{nonva}\eeq
\item 
$\e^{\i t\hat H}\hat\psi(x^0,\vec x)\e^{-\i t\hat H}=\hat\psi(x^0+t,\vec x);$
\item 
$\Omega$ is cyclic for $\hat\psi(x)$,  $\hat\psi^*(x)$.\een

\nowastrona

\nowastrona

The above problem has an essentially unique  solution,
which we describe below. 

We set \[\cH:=\Gamma_\s(\cZ_\KG^{(+)}\oplus\cZ_\KG^{(-)}).\]
Creation/annihilation operators  for the particle space $\cZ_\KG^{(+)}\simeq L^2(\rr^3)$ are denoted with the letter $a$ and for the antiparticle space $\cZ_\KG^{(-)}\simeq L^2(\rr^3)$ with the letter $b$. Thus, for 
 $p$ on the
mass shell,  using physicist's notation on the left and mathematician's on the right, creation operators for particles/antiparticles are written as
\begin{eqnarray}
 \hat a^*(p)&=&\hat a^*\big(|p)\big),\label{qrt2=}\\
\hat b^*(p)&=&\hat b^*\big(\bar{|-p)}\big).\label{qrt2+}\end{eqnarray}
 $\Omega$ is the Fock
vacuum. 
\nowastrona
The quantum field is
\begin{eqnarray*}
\hat\psi(x)&:=&\int\frac{\d\vec{p}}{\sqrt{(2\pi)^3}\sqrt{2E(\vec p)}} \left(\e^{\i
  px}\hat a(p)+
\e^{-\i px}\hat b^*(p)\right),
\\
\hat\eta(x)
&:=&\int\frac{\d\vec{p}\sqrt{E(\vec p)}}{\i\sqrt{(2\pi)^3}\sqrt2} \left(\e^{\i
  px} \hat a(p)-
\e^{-\i px}\hat b^*(p)\right).
\end{eqnarray*}
The  quantum Hamiltonian,  momentum and   charge are
\begin{eqnarray}
\hat H&:=&\int \left(\hat a^*( p)\hat a(p)+\hat b^*( p)\hat b(p)\right)E(\vec
p)
\d\vec{p},\label{equi1}\\
\vec {\hat P}&:=&\int \left(\hat a^*( p)\hat a(p)+\hat b^*( p)\hat
  b(p)\right)
\vec p\d\vec{p},\nonumber\\
\hat Q&:=&\int \left(\hat a^*(p)\hat a(p)
- \hat b^*( p)\hat b( p)\right)\d\vec{p}.\nonumber
\end{eqnarray}

Equivalently, for any $t$
\begin{eqnarray*}
\hat H&=&
\int
{:}\Bigl(\hat\eta^*(t,\vec x)\hat\eta(t,\vec x)+\vec\partial\hat\psi^*(t,\vec x)\vec\partial\hat\psi(t,\vec x) +m^2\hat\psi^*(t,\vec x)\hat\psi(t,\vec x)\Bigr){:}\d\vec x,\\
\vec{\hat P}&=&\int{:}\Bigl(-\hat\eta^*(t,\vec x)\vec\partial
 \hat\psi(t,\vec x)-\vec\partial
 \hat\psi^*(t,\vec x)\hat\eta(t,\vec x)\Bigr){:}\d\vec x,\\
\hat Q&=&
\i\int{:}\Bigl(-\hat\eta^*(t,\vec x)\hat\psi(t,\vec x)
+\hat\psi^*(t,\vec x)\hat\eta(t,\vec x)\Bigr){:}\d\vec x.
\end{eqnarray*}
\nowastrona
Thus all these operators are expressed in terms of the Wick
 quantization of their classical expressions.

Note that the whole group
$\rr^{1,3}\rtimes  O^\uparrow(1,3)$  acts unitarily on $\cH$
by $U(y,\Lambda):=\Gamma\Bigl(r_{(y,\Lambda)}\Big|_{\cZ_\KG^{(+)}}
\Bigr) \otimes\Gamma\Bigl(\bar r_{(y,\Lambda)}\Big|_{\cZ_\KG^{(-)}}\Bigr)$, with
\[U(y,\Lambda)\hat\psi(x)U(y,\Lambda)^*=\hat\psi\bigl((y,\Lambda)x\bigr).\]
Moreover,
\[[\hat\psi(x),\hat\psi^*(y)]=-\i D(x-y),\ \ 
 [\hat\psi(x),\hat\psi(y)]=0.\]

Note the identities
\begin{eqnarray*}
(\Omega|\hat\psi(x)\hat\psi^*(y)\Omega)
&=&-\i D^{(+)}(x-y),\\
(\Omega|\T(\hat\psi(x)\hat\psi^*(y))\Omega)
&=&-\i D^{\rm c}(x-y).
\end{eqnarray*}

For  $f\in C_{\rm c}^\infty(\rr^{1,3},\cc)$ we set
\begin{eqnarray*}\hat\psi[f]&:=&\int\bar{f(x)}\hat\psi(x)
\d x,\\
\hat \psi^*[f]&:=&\int{f(x)}\hat\psi^*(x)
\d x.\end{eqnarray*}
We obtain an operator valued distribution satisfying
 the  Wightman axioms with
$\cD:=\Gamma_\s^\fin(\cZ_\KG^{(+)}\oplus\cZ_\KG^{(-)})$.

For an open  set $\cO\subset \rr^{1,3}$ the field algebra  is defined as
\[\fF(\cO):=\left\{\exp\left(\i\hat\psi^*[f]+\i\hat\psi[f]\right)\ :\ f\in C_{\rm
  c}^\infty(\cO,\cc)\right\}''.\] The observable algebra $\fA(\cO)$ is the subalgebra of
$\fF(\cO)$ fixed by the automorphism 
\[B\mapsto\e^{\i\theta\hat
  Q}B\e^{-\i\theta\hat
 Q}.\]
The algebras $\fF(\cO)$ and $\fA(\cO)$ satisfy the {\em Haag-Kastler axioms}.

\nowastrona

\subsubsection{Quantum 4-current}
\label{Quantized current}

Let us try to introduce the {\em (quantum) 4-current density}  by
\begin{eqnarray}\notag
\hat \cJ^\mu(x)
&=&\frac{\i}{2}\Big(\partial^\mu \hat\psi^*(x)\hat\psi(x)+
\hat\psi(x)\partial^\mu \hat\psi^*(x)\\
&&\hskip 1ex-\hat\psi^*(x)\partial^\mu \hat\psi(x)
-\partial^\mu\hat\psi(x)\hat\psi^*(x) \Big).\label{deffi}\end{eqnarray}
In Subsubsect. 
\ref{Quantized discrete symmetries} later on  we will introduce a certain unitary operator $C$ called the charge conjugation satisfying $C\Omega=\Omega$, $C\hat\cJ^\mu(x) C^{-1}=-\hat\cJ^\mu(x)$. The existence of such $C$ implies
\[(\Omega|\hat\cJ^\mu(x)\Omega)=0.\]
Therefore, (\ref{deffi}) can be replaced with the following equivalent definition:
\begin{eqnarray}
\hat \cJ^\mu(x)
&=&\i{:}\left(\partial^\mu \hat\psi^*(x)\hat\psi(x)
-\hat\psi^*(x)\partial^\mu \hat\psi(x)\right){:}.\label{deffi1}\end{eqnarray}
Thus $\hat\cJ^\mu(x)$ can be defined both as the Weyl quantization (\ref{deffi}) and the Wick quantization  (\ref{deffi1}) of the corresponding quadratic classical expression.

Formally, we can check the relations
\begin{eqnarray*}\partial^\mu \hat\cJ_\mu(x)&=&0,\\
\hat\cJ^\mu(x)^*&=&\hat\cJ^\mu(x).
\end{eqnarray*}

In particular, we have the {\em (quantum) charge density}
\[\hat\cQ(x):=\hat\cJ^0(x)=
\i{:}\bigl(-\hat\eta^*(x)\hat\psi(x)+\hat\psi^*(x)\hat\eta(x)\bigr){:}\] 
with the relations
\begin{eqnarray}
\nonumber
[\hat\cQ(t,\vec x), \hat\psi(t,\vec y)]&=&-\hat\psi(t,\vec y)\delta(\vec
x-\vec y),\\ {}\nonumber
[\hat\cQ(t,\vec x), \hat\eta(t,\vec y)]&=& -\hat\eta(t,\vec y)\delta(\vec
x- \vec y),\\{} \label{westill-q}
[\hat \cQ(t,\vec x),  \hat\cQ(t,\vec y)]&=&0
.\end{eqnarray}


Similarly, as in the classical case,
for $\chi\in C_{\rm c}^\infty(\rr^3,\rr)$, let $\alpha_\chi$ denote the corresponding {\em gauge transformation at time $t=0$}
defined as the
 $*$-automorphism of the algebra generated by the fields operators
 satisfying
\begin{eqnarray}\notag\alpha_\chi(\hat\psi(0,\vec x))&:=&\e^{-\i\chi(\vec
    x)}\hat\psi(0,\vec x),\\
\alpha_\chi(\hat\eta(0,\vec x))&:=&\e^{-\i\chi(\vec x)}\hat\eta(0,\vec x).
\label{gaugo-q}\end{eqnarray} 
Obviously, 
\begin{eqnarray}\notag\alpha_\chi(\hat\psi^*(0,\vec x))&=&\e^{\i\chi(\vec
    x)}\hat\psi^*(0,\vec x),\\
\alpha_\chi(\hat\eta^*(0,\vec x))&=&\e^{\i\chi(\vec x)}\hat\eta^*(0,\vec x).
\label{gaugo1-q}\end{eqnarray}

Assume that $\chi\neq0$. Let us check whether $\alpha_\chi$ is unitarily implementable.
On the level of annihilation operators
we have
\begin{eqnarray*}
\alpha_\chi(\hat a(p))&=&
\int\int\left(\sqrt{\frac{E(\vec p_1)}{E(\vec
    p)}}+\sqrt{\frac{E(\vec p)}{E(\vec p_1)}}\right)\frac{\d\vec
  x\d\vec p_1}{2(2\pi)^{3}}\e^{\i(\vec p_1-\vec p)\vec x-\i e\chi(\vec x)}\hat a(p_1)\\
&&\!\!\!\!\!+
\int\int\left(\sqrt{\frac{E(\vec p_1)}{E(\vec
    p)}}-\sqrt{\frac{E(\vec p)}{E(\vec p_1)}}\right)\frac{\d\vec
  x\d\vec p_1}{2(2\pi)^{3}}\e^{-\i(\vec p_1+\vec p)\vec x-\i e\chi(\vec x)}\hat b^*(p_1).
\end{eqnarray*}
Let $q_\chi(\vec p,\vec p_1)$ denote the integral kernel on the second line above.
By the Shale criterion (Thm \ref{shale}), we need to check whether it is square integrable.
Now
\begin{eqnarray}\label{qwacz2}
&&\left(\sqrt{\frac{E(\vec p_1)}{E(\vec
    p)}}-\sqrt{\frac{E(\vec p)}{E(\vec p_1)}}\right)\\
&=&\frac{(|\vec p_1|-|\vec    p|)(|\vec p_1|+|\vec    p|)}
{\big(E(\vec p)+E(\vec p_1)\big)\sqrt{E(\vec p)E(\vec p_1)}}.
\notag\end{eqnarray}
 After 
integrating in $\vec x$ we obtain fast decay of $q_\chi$ in $\vec p+\vec p_1$, which in particular allows us to control the term $|\vec p_1|-|\vec    p|$.
We obtain
\[\int|q(\vec p,\vec p_1)|^2\d\vec p\sim \frac{C}{E(\vec p_1)^{2}},\]
which is not  integrable. Thus 
$\alpha_\chi$ is
 not implementable.

Formally, if we set
\beq
\hat Q(\chi):=\int  \chi(\vec x)\hat \cQ(0,\vec x)\d\vec
x,\label{illo-q}\eeq then $\e^{\i e\hat Q(\chi)}$ implements the gauge
transformation:
 \[\alpha_\chi(\hat B)=\e^{\i e\hat Q(\chi)}\hat B\e^{-\i e\hat Q(\chi)}.\]
But we know that  $\alpha_\chi$ is not implementable.
Thus for nonzero $\chi$ (\ref{illo-q}) cannot be defined as a closable operator.


However, the {\em (quantum) charge}  
\[\hat Q=\int \hat 
\cQ(t,\vec x)\d\vec x,\]
as we have already seen, is a well defined operator.



For further reference let us express the charge and 
current density
in terms of creation and annihilation operators:
\begin{eqnarray*}\hat \cQ(x)&=&
\int\int \frac{\d \vec p_1\d \vec p_2}{2(2\pi)^{3}}\left(\sqrt{\frac{E(\vec p_1)}{E(\vec p_2)}}+\sqrt{\frac{E(\vec p_2)}{E(\vec p_1)}}\right)\\&&\times\left(
\e^{-\i x p_1+\i x p_2}\hat a^*(p_1)\hat a(p_2)
-\e^{\i x p_1-\i x p_2}\hat b^*(p_2)\hat b(p_1)\right)\\
&&+
\int\int \frac{\d \vec p_1\d \vec p_2}{2(2\pi)^{3}}\left(\sqrt{\frac{E(\vec p_1)}{E(\vec p_2)}}-\sqrt{\frac{E(\vec p_2)}{E(\vec p_1)}}\right)\\&&\times\left(
-\e^{-\i xp_1-\i xp_2}\hat a^*(p_1)\hat b^*(p_2)
+\e^{\i xp_1+\i xp_2}\hat b(p_1)\hat a(p_2)\right),\\
\vec{\hat \cJ}(x)&=&
\int\int \frac{\d \vec p_1\d \vec p_2}{2(2\pi)^{3}\sqrt{E(\vec p_1)E(\vec p_2)}}(\vec p_1+\vec p_2)\\&&\times\left(
-\e^{-\i xp_1+\i xp_2}\hat a^*(p_1)\hat a(p_2)
+\e^{\i xp_1-\i xp_2}\hat b^*(p_2)\hat b(p_1)\right)\\
&&+
\int\int 
 \frac{\d \vec p_1\d \vec p_2}{2(2\pi)^{3}\sqrt{E(\vec p_1)E(\vec p_2)}}(\vec p_1-\vec p_2)
\\&&\times\left(
-\e^{-\i xp_1-\i xp_2}\hat a^*(p_1)\hat b^*(p_2)
+\e^{\i xp_1+\i xp_2}\hat b(p_1)\hat a(p_2)\right).
\end{eqnarray*}

\nowastrona

\subsubsection{Quantization in terms of smeared fields}

An alternative equivalent formulation of the quantization
program uses  smeared fields instead of point fields.
Instead of
(\ref{distro}) we look for an antilinear function
\[\cW_\KG\ni\zeta\mapsto\hat\psi\lpar\zeta\rpar\]
with values in closed operators such that
\ben\item
$[\hat\psi\lpar\zeta_1\rpar,\hat\psi^*\lpar\zeta_2\rpar]
=\i\bar\zeta_1\omega \zeta_2,$\hspace{6ex}
$[\hat\psi\lpar\zeta_1\rpar,\hat\psi\lpar\zeta_2\rpar]
=0.$
\item
$\hat\psi\lpar r_{(t,\vec0)}\zeta\rpar
=\e^{\i t\hat H}\hat\psi\lpar\zeta\rpar\e^{-\i t\hat H}.$
\item 
$\Omega$ is cyclic for the algebra generated by
$ \psi\lpar\zeta\rpar$, $ \psi^*\lpar\zeta\rpar$.\een

\nowastrona

One can pass between these two versions of the quantization  by
\beq
\hat\psi\lpar\zeta\rpar =\int\left(-\bar{\dot\zeta(t,\vec x)}\hat\psi(t,\vec x)+ 
\bar{\zeta(t,\vec x)}\hat\eta(t,\vec x)\right)\d\vec{x}.
\eeq

\nowastrona

\subsection{Charged scalar bosons in an external 4-potential}

\subsubsection{Classical fields}

Let us go back to the classical theory.
Let \beq\rr^{1,3}\ni x\mapsto
A(x)=[A^\mu(x)]\in\rr^{1,3}\label{extera}\eeq
 be a given function
called the {\em (external electromagnetic) 4-potential}. In most of this
subsection we will assume that (\ref{extera}) is Schwartz. The (complex)
  Klein-Gordon equation
in the external 4-potential $A$ is
\beq\left(-(\partial_\mu+\i eA_\mu(x))(\partial^\mu+\i
eA^\mu(x))+m^2\right)\psi(x)=0.\label{sfa}\eeq 
If $\psi$ satisfies (\ref{sfa}) and $\rr^{1,3}\ni x\mapsto
\chi(x)\in\rr$ is smooth,
 then $\e^{-\i e\chi}\psi$ satisfies (\ref{sfa}) with $A$
replaced with $A+\partial\chi$.

In this subsection, the field satisfying the Klein-Gordon equation with $A=0$ will be denoted $\psi_\fr$.

\nowastrona
The retarded/advanced Green's function is defined as the unique
solution of
\beq\left(-(\partial_\mu+\i eA_\mu(x))(\partial^\mu+\i
e A^\mu(x))+m^2\right)D^\pm(x,y)=\delta(x-y)\label{sfa-c}\eeq 
satisfying
\[\supp D^\pm\subset\{x,y\ :\ x\in J^\pm(y)\}.\]
We generalize the Pauli-Jordan function:
\[D(x,y):=D^+(x,y)-D^-(x,y).\]
Clearly,\[
 \supp D\subset\{x,y\ :\ x\in J(y)\}.\]

\nowastrona

The {\em Cauchy problem} of (\ref{sfa}) can be expressed
 with help of the 
function $D$:
\begin{eqnarray}\label{decora+}
\psi(t,\vec x)
&=&-\int_{\rr^3}\partial_s D(t,\vec x;s,\vec y)\big|_{s=0}\psi(0,\vec y)
\d\vec{y}\\&&+\int_{\rr^3}
D(t,\vec x;0,\vec y)\dot\psi(0,\vec y)
\d\vec{y}.\notag
\end{eqnarray}


\nowastrona



\nowastrona

We would like to introduce a field $\rr^{1,3}\ni
x\mapsto \psi(x)$ satisfying (\ref{sfa}).
As we will see shortly, the conjugate field is
\[\eta(x):=\partial_0\psi(x)+\i eA_0(x)\psi(x).\]
For definiteness, we will assume that $\psi(x)$, $\eta(x)$ act on $\cW_\KG$ and at time $t=0$ coincide with free fields:
\begin{eqnarray*}
\psi(0,\vec x)&=&\psi_\fr(0,\vec x),\\
\eta(0,\vec x)&=&\eta_\fr(0,\vec x).
\end{eqnarray*}
 This determines the field $\psi$ uniquely:
\begin{eqnarray}\label{decora}
\psi(t,\vec x)
&=&-\int_{\rr^3}\partial_s D(t,\vec x;s,\vec y)\big|_{s=0}\psi_\fr(0,\vec y)
\d\vec{y}\\&&+\int_{\rr^3}
D(t,\vec x;0,\vec y)\big(\eta_\fr(0,\vec y)-\i eA_0(0,\vec y)\psi_\fr(0,\vec
y)\big)
\d\vec{y}.\notag
\end{eqnarray}

\subsubsection{Lagrangian and Hamiltonian formalism}

Consider the   Lagrangian density
\begin{eqnarray*}
\cL(x)=&
-\big(\partial_\mu-\i eA_\mu(x)\big)\psi^*(x)
\big(\partial^\mu+\i eA^\mu(x)\big)\psi(x)- m^2\psi^*(x)\psi(x).
\end{eqnarray*}
The Euler-Lagrange equations (\ref{euler-lagrange1a})
yield (\ref{sfa}).

\nowastrona

Let us introduce the {\em variable conjugate} to $\psi^*(x)$ and $\psi(x)$:
\begin{eqnarray*}
\eta(x)&:=&\frac{\partial \cL}{\partial_0\psi^*(x)}=
\partial_0\psi(x)+\i eA_0(x)\psi(x),\\
\eta^*(x)&=&\frac{\partial \cL}{\partial_0\psi(x)}=
\partial_0\psi^*(x)-\i eA_0(x)\psi^*(x).
\end{eqnarray*}

We introduce the   Hamiltonian density
\begin{eqnarray*}
\cH( x)&=&\frac{\partial\cL(x)}{\partial\dot\psi(x)}\dot\psi(x)+
\frac{\partial\cL(x)}{\partial\dot\psi^*(x)}\dot\psi^*(x)-\cL(x)\\
&=&\eta^*(x)\eta(x)+\i eA_0(x)\left(\psi^*(x)\eta(x)-
\eta^*(x)\psi(x)\right)\\
&&+(\partial_i-\i eA_i(x))\psi^*(x)(\partial_i+\i eA_i(x))\psi(x)
+m^2\psi^*( x)\psi(x)\\
&=&\eta^*(x)\eta(x)+
\partial_i\psi^*(x)\partial_i\psi(x)\\
&&+\i eA_0(x)\left(\psi^*(x)\eta(x)-
\eta^*(x)\psi(x)\right)-\i eA_i(x)\left(\psi^*(x)\partial_i\psi(x)
-\partial_i\psi^*(x)\psi(x)\right)
\\
&& +e^2\vec A(x)^2\psi^*(x)\psi(x)+m^2\psi^*( x)\psi(x).\end{eqnarray*}
 The   Hamiltonian
\[H(t)=\int \cH(t,\vec x)\d\vec x
\]
can be used to generate the dynamics
\begin{eqnarray*}
\dot\psi(t,\vec x)=\{\psi(t,\vec x),H(t)\},&&
\dot\eta(t,\vec x)=\{\eta(t,\vec x),H(t)\}.
\end{eqnarray*}




The interaction picture Hamiltonian can be 
partially expressed in terms of the free 4-current density:
\begin{eqnarray}\notag
 H_\Int(t)
&=&\int  \d\vec x\Bigl( e  A_\mu(t,\vec x)  \cJ_\fr^\mu(t,\vec x)
+e^2\vec A(t,\vec x)^2\psi_\fr^*(t,\vec x)\psi_\fr(t,\vec
x)\Bigr)\\\label{qwacz}
&=&\int\d\vec x \Bigl(e  A_0(t,\vec x) \cQ_\fr(t,\vec x)
\\
&&+e \vec A(t,\vec x) \vec \cJ_\fr(t,\vec x)
+e^2\vec A(t,\vec x)^2\psi_\fr^*(t,\vec x)\psi_\fr(t,\vec x)\Bigr)\notag
\end{eqnarray}
\begin{eqnarray*}&=&
\frac{e}{2}\int\int \frac{\d \vec p_1\d \vec p_2}{(2\pi)^{3}}\left(\sqrt{\frac{E(\vec p_1)}{E(\vec p_2)}}+\sqrt{\frac{E(\vec p_2)}{E(\vec p_1)}}\right)\\&&\times\left(
A_0(t,\vec p_1-\vec p_2)\e^{\i tE(\vec p_1)-\i t E(\vec p_2)}a^*(p_1)a(p_2)
-A_0(t,-\vec p_1+\vec p_2)\e^{-\i tE(\vec p_1)+\i t E(\vec p_2)}b(p_1)b^*(p_2)\right)\\
&&+\frac{e}{2}\int\int \frac{\d \vec p_1\d \vec p_2}{(2\pi)^{3}}\left(\sqrt{\frac{E(\vec p_1)}{E(\vec p_2)}}-\sqrt{\frac{E(\vec p_2)}{E(\vec p_1)}}\right)\\&&\times\left(
A_0(t,\vec p_1+\vec p_2)\e^{\i tE(\vec p_1)+\i t E(\vec p_2)}a^*(p_1)b^*(p_2)
-A_0(t,-\vec p_1-\vec p_2)\e^{-\i tE(\vec p_1)-\i t E(\vec p_2)}b(p_1)a(p_2)\right)\\
&&+\frac{e}{2}\int\int \frac{\d \vec p_1\d \vec p_2}{(2\pi)^{3}\sqrt{E(\vec p_1)E(\vec p_2)}}(\vec p_1+\vec p_2)\\&&\times\left(-
\vec A(t,\vec p_1-\vec p_2)\e^{\i tE(\vec p_1)-\i t E(\vec p_2)}a^*(p_1)a(p_2)
+\vec A(t,-\vec p_1+\vec p_2)\e^{-\i tE(\vec p_1)+\i t E(\vec p_2)}b(p_1)b^*(p_2)\right)\\
&&+\frac{e}{2}\int\int 
 \frac{\d \vec p_1\d \vec p_2}{(2\pi)^{3}\sqrt{E(\vec p_1)E(\vec p_2)}}(\vec p_1-\vec p_2)
\\&&\times\left(
-\vec A(t,\vec p_1+\vec p_2)\e^{\i tE(\vec p_1)+\i t E(\vec p_2)}a^*(p_1)b^*(p_2)
+\vec A(t,-\vec p_1-\vec p_2)\e^{-\i tE(\vec p_1)-\i t E(\vec p_2)}b(p_1)a(p_2)\right)\\
&&+\frac{e^2}{2}\int\int\frac{\d\vec p_1\d \vec p_2}
{(2\pi)^3\sqrt{E(\vec p_1)}\sqrt{E(\vec p_2)}}\\
&&\times\Big(\vec A^2(\vec p_1-\vec p_2)
\e^{\i tE(\vec p_1)-\i t E(\vec p_2)} a^*(p_1) a(p_2)
+\vec A^2(-\vec p_1+\vec p_2)
\e^{-\i tE(\vec p_1)+\i t E(\vec p_2)} b(p_1) b^*(p_2)\\
&&+\vec A^2(\vec p_1+\vec p_2)
\e^{\i tE(\vec p_1)+\i t E(\vec p_2)} a^*(p_1) b^*(p_2)
+\vec A^2(-\vec p_1-\vec p_2)
\e^{-\i tE(\vec p_1)-\i t E(\vec p_2)} b(p_1) a(p_2)\Big).\end{eqnarray*}

\subsubsection{Classical discrete symmetries}
\label{Classical discrete symmetries}

Choose $\xi_C\in\cc$, $|\xi_C|=1$.
If $\zeta$ solves the Klein-Gordon equation with the 4-potential $A$, then so does
$\xi_C\bar\zeta$ with the 4-potential $-A$. Thus replacing
\begin{eqnarray*}
&&\psi(x),\psi^*(x),A(x)\\
&\hbox{with }&\bar\xi_C\psi^*(x),\xi_C\psi(x),-A(x)\end{eqnarray*}
is a symmetry of the complex Klein-Gordon equation with an external
4-potential
 (\ref{sfa}). It is called {\em charge conjugation} and denoted $\cC$.

Choose $\xi_P\in\{1,-1\}$.
Recall that ${\rm P}(x^0,\vec x):=(x^0,-\vec x)$ denotes the space
inversion. Replacing \begin{eqnarray*}
&&\psi(x),\psi^*(x),\big(A_0(x),\vec A(x)\big)\\
&\hbox{with }&
\xi_P\psi({\rm P}x),\xi_P\psi^*({\rm P}x),\big(A_0({\rm P}x),-\vec A({\rm
  P}x)\big)\end{eqnarray*} is a symmetry of  (\ref{sfa})  called {\em parity} and denoted $\cP$.

Choose $\xi_T\in\cc$, $|\xi_T|=1$.
Recall that ${\rm T}(x^0,\vec x):=(-x^0,\vec x)$ denotes the time
reflection. Replacing \begin{eqnarray*}
&&\psi(x),\psi^*(x),\big(A_0(x),\vec A(x)\big)\\
&\hbox{with }&\bar\xi_T\psi^*({\rm T}x),\xi_T\psi({\rm T}x),\big(A_0({\rm T}x),-\vec A({\rm
  T}x)\big)\end{eqnarray*} is a symmetry of  (\ref{sfa})  called {\em time reversal} and denoted $\cT$.

 The composition of $\cC$, $\cP$ and $\cT$ has an especially simple form if
$\bar\xi_C\xi_P\xi_T=1$. It consists in
replacing 
 \begin{eqnarray*}
&&\psi(x),\psi^*(x), A(x)\\
&\hbox{with }&
\psi(-x),\psi^*(-x),-A(-x).\end{eqnarray*}
It is called the {\em CPT
  symetry} and is denoted $\cX$.

$\cC$, $\cP$, $\cT$ and  $\cX$ commute with one another
and we have the relations
\[\cC^2=\cP^2=\cT^2=\cX=\id.\]

\subsubsection{Quantization}
\label{Quantization-bosons}
We are looking for a quantum field satisfying
\beq\left(-(\partial_\mu+\i eA_\mu(x))(\partial^\mu+\i
eA^\mu(x))+m^2\right)\hat\psi(x)=0.\label{sfa1a}\eeq
We set
\[\hat\eta(x):=\partial_0\hat\psi(x)+\i eA_0(x)\hat\psi(x).\]
We will assume that $\hat\psi$, $\hat\eta$ act on the Hilbert space of free fields
\[\Gamma_\s(\cZ_\KG^{(+)}\oplus\cZ_\KG^{(-)}),\]
and at time $t=0$ we have
\begin{eqnarray*}
\hat\psi(\vec x)\,:=\,\hat\psi(0,\vec x)&=&\hat\psi_\fr(0,\vec x),\\
\hat\eta(\vec x)\,:=\,\hat\eta(0,\vec x)&=&\hat\eta_\fr(0,\vec x).\end{eqnarray*}

The solution is unique and is
obtained by
decorating (\ref{decora}) with ``hats''.

We would like to ask whether the quantum fields are implemented by a unitary dynamics. Equivalently, we want to check if the classical dynamics generated by
$H_\Int(t)$ satisfies the Shale criterion (Thm \ref{shale}).

By following the discussion of Subsubsect. \ref{shale2} we check that the classical scattering operator is unitarily implementable. 

The Shale criterion is
satisfied for the dynamics from $t_-$ to $t_+$ 
iff the spatial part of the 4-potential is the same at the initial and final time:
\beq\vec A(t_+,\vec x)=\vec A(t_-,\vec x),\ \ \vec x\in\rr^3.\label{pqiq}\eeq

To see this note that
$H_\Int(t)$   consists of three terms described in  (\ref{qwacz}).

The term $e^2\vec A(t,\vec x)^2\psi_\fr^*(t,\vec x)\psi_\fr(t,\vec x)$ is very similar to the mass-like perturbation considered already in Subsubsect. \ref{shale2}, which did not cause problems with the Shale criterion for the dynamics for any $t_+,t_-$.

The same is true for the term $e  A_0(t,\vec x) \cQ_\fr(t,\vec x)$. Indeed, a similar term was  discussed before in the context of gauge
transformations, see in particular (\ref{qwacz2}). Then there was a
problem with the square integrability. But now we can integrate by
parts, which improves the decay.

The term $e \vec A(t,\vec x) \vec \cJ_\fr(t,\vec x)$ is problematic -- it has worse behavior for large momenta then the previous two terms. The integration by parts creates
 a boundary term that is not square integrable unless (\ref{pqiq}) holds, when it vanishes.


\subsubsection{Quantum Hamiltonian}

Formally, the  fields undergo a unitary dynamics given by
\[\hat\psi(t,\vec x):=\Texp\left(-\i\int_t^0 \hat H(s)\d s\right)\hat \psi(0,\vec
x)\Texp\left(-\i\int_0^t \hat H(s)\d s\right), \]\nowastrona
\noindent
where the  Schr\"odinger picture Hamiltonian is
\begin{eqnarray}\nonumber
\hat H(t)&=&\int\d\vec x
\Bigl(\hat\eta^*(\vec x)\hat\eta(\vec x)
+\i eA_0(t,\vec x){:}\bigl(\hat \psi^*(\vec
x)\hat \eta(\vec x)-
\hat\eta^*(\vec x)\hat\psi(\vec x)\bigr){:}\\
\nonumber
&&+(\partial_i-\i eA_i(t,\vec x))\hat \psi^*(\vec x)(\partial_i+\i eA_i(t,\vec x))
\hat \psi(\vec x)
\\
&&+m^2\hat\psi^*(\vec  x)\hat\psi(\vec x)
\Bigr).
\label{nmun3}\end{eqnarray}
Note that the above Hamiltonian  is formally the Weyl quantization of its corresponding classical expressions. This is perhaps not obvious the way it is written. To see this we should note that equal time $\hat\psi$ and $\hat\psi^*$ commute, the same is true for equal time $\hat\eta$ and $\hat\eta^*$, finally the mixed term can be expressed by the 4-current where the Wick and Weyl quantizations coincide, see Subsubsect.
\ref{Quantized current}.

In any case, the analysis of the previous subsubsection shows that  the above Hamiltonian is often ill defined and should be understood as a formal
expression, even when we try renormalize by adding a constant $C(t)$. We will need it to develop perturbation expansion for the quantum scattering operator and to compute the energy shift. 

(\ref{nmun3}) can be compared with the free Hamiltonian without the Wick ordering, which differs from (\ref{equi1}) by an (infinite) constant:
\begin{eqnarray}
\hat H_\fr&=&\int\d\vec x
\Bigl(\hat\eta^*(\vec x)\hat\eta(\vec x)
+\partial_i\hat \psi^*(\vec x)\partial_i
\hat \psi(\vec x)
+m^2\hat\psi^*(\vec  x)\hat\psi(\vec x)
\Bigr).
\label{nmun3free}\end{eqnarray}
This leads to the following interaction picture Hamiltonian:
\begin{eqnarray}\nonumber
\hat H_\Int(t)
&=&\int\d\vec x\Bigl(e A_\mu(t,\vec x)\hat \cJ^\mu_\fr(t,\vec x)
+e^2\vec A(t,\vec x)^2\hat\psi_\fr^*(t,\vec x)\hat\psi_\fr(t,\vec x)
\Bigr)
\\\nonumber
&=&\int\d\vec x\Bigl(e A_\mu(t,\vec x)\hat \cJ_\fr^\mu(t,\vec x)
+e^2A(t,\vec x)^2\hat\psi_\fr^*(t,\vec x)\hat\psi_\fr(t,\vec x)
\\&&
+e^2A_0(t,\vec x)^2\hat\psi_\fr^*(t,\vec x)\hat \psi_\fr(t,\vec x)
\Bigr).\label{nmun4}
\end{eqnarray}

\subsubsection{Quantized discrete symmetries}
\label{Quantized discrete symmetries}

The discrete symmetries considered in
Subsubsect. \ref{Classical discrete symmetries} remain true when we
decorate
 the fields
with ``hats''. Thus on the level of quantum
observables
the discrete symmetries are the same as in the classical
case.

A separate discussion is needed concerning the implementation of these
symmetries by unitary or antiunitary operators on the Hilbert
space $\Gamma_\s(\cZ_\KG^{(+)}\oplus\cZ_\KG^{(-)})$.
 We will discuss this for free fields, that is, for $A=0$.  Free fields are used to compute 
 the scattering operator for the
4-potential $A$, denoted  by $\hat S(A)$. Therefore, our analysis will lead to some identities for $\hat S(A)$.

First consider the charge conjugation. As we have already pointed out in Subsubsect. \ref{positi},
 the spaces
$\cZ_\KG^{(+)}$ and $\cZ_\KG^{(-)}$ can be naturally
identified. Therefore, we can define a unitary operator on
$\cZ_\KG^{(+)}\oplus\cZ_\KG^{(-)}$
\[\chi(g_1,\bar g_2):=(\xi_C\bar g_2,\bar\xi_Cg_1).\]
Clearly,
\[\chi|p)=\xi_C\bar{|-p)},\ \ \ \chi\bar{|-p)}=\bar\xi_C|p).\]
We set
$C:=\Gamma(\chi)$. We have $C^2=\one$, $C\Omega=\Omega$,
\[ C\hat\psi_\fr(x)C^{-1}=\xi_C\hat\psi_\fr^*(x),\ \ C\hat\psi_\fr^*(x)C^{-1}=\bar\xi_C\hat\psi_\fr(x),\]
\[C\hat\cQ_\fr(x)C^{-1}=-\hat\cQ_\fr(x),\ \ \ 
C\vec{\hat\cJ}_\fr(x)C^{-1}=-\vec{\hat\cJ}_\fr(x),\]
\[C\hat S(A)C^{-1}=\hat S(-A).\]

Define a unitary operator on
$\cZ_\KG^{(+)}\oplus\cZ_\KG^{(-)}$
\[\pi(g_1,\bar g_2):=\big(\xi_P g_1\circ{\rm P},\xi_P\bar{g_2\circ{\rm P}}\big).\]
(The circle denotes the composition of two functions).
Clearly,
\[\pi|E,\vec p)=\xi_P|E,-\vec p),\ \ \pi|{-}E,{-}\vec p)=\xi_P|{-}E,\vec p).\]
We have a natural implementation of the parity $P:=\Gamma(\pi)$.
It satisfies
 $P^2=\one$, $P\Omega=\Omega$,
\[ P\hat\psi_\fr(x)P^{-1}=\xi_P\hat\psi_\fr({\rm P}x),\ \ P\hat\psi_\fr^*(x)P^{-1}=\xi_P\hat\psi_\fr^*({\rm P}x),\]
\[P\hat\cQ_\fr(x)P^{-1}=\hat\cQ_\fr({\rm P}x),\ \ \ 
P\vec{\hat\cJ}_\fr(x)P^{-1}=-\vec{\hat\cJ}_\fr({\rm P}x),\]
\[P\hat S(A^0,\vec A)P^{-1}=\hat S(A^0\circ{\rm P},-\vec A\circ{\rm P}).\]

Define
 the following  antiunitary operator on
$\cZ_\KG^{(+)}\oplus\cZ_\KG^{(-)}$:
\[\tau(g_1,\bar g_2):=\big(\xi_T\bar {g_1\circ {\rm T}}, \bar\xi_T g_2\circ{\rm T}\big).\]
Clearly,
\[\tau|E,\vec p)=\xi_T |E,-\vec p),\ \ \tau|{-}E,{-}\vec p)=\bar\xi_T|{-}E,\vec p).\]
Set $T:=\Gamma(\tau)$.
 We have $T^2=\one$, $T\Omega=\Omega$,
\[ T\hat\psi_\fr(x)T^{-1}=\bar\xi_T\hat\psi_\fr^*({\rm T}x),\ \ T\hat\psi_\fr^*(x)T^{-1}=\xi_T\hat\psi_\fr({\rm T}x),\]
\[T\hat\cQ_\fr(x)T^{-1}=\hat\cQ_\fr({\rm T}x), \ \ 
T\vec{\hat\cJ}_\fr(x)T^{-1}=-\vec{\hat\cJ}_\fr({\rm T}x),\]
\[T\hat S(A^0,\vec A)T^{-1}=\bar{\hat S(A^0\circ{\rm T},-\vec A\circ{\rm T})}.\]

\subsubsection{$2N$-point Green's functions}

For $y_{N},\dots y_1,x_{N},\dots,x_1$, the {\em $2N$ point 
 Green's function} are defined
as follows:
\begin{eqnarray*}
&&\big\langle\hat\psi^*(y_1)\cdots\hat\psi^*(y_N)\hat\psi(x_{N})\cdots\hat\psi(x_{1})\big\rangle\\&:=&
\left(\Omega^+|\T\left(
\hat \psi^*(y_1)\cdots\hat \psi^*(y_N)\hat \psi(x_{N})\cdots\hat
\psi(x_{1})\right)
\Omega^-\right).
\end{eqnarray*}

One can organize Green's functions in terms of the {\em generating function}:
\begin{eqnarray*}&&Z(g,\bar g)\\
&:=&\sum_{n=0}^\infty\int\cdots\int 
\frac{(-1)^N}{(N!)^2}\big\langle\hat\psi^*(y_1)\cdots\hat\psi^*(y_N)\hat\psi(x_{N})\cdots\hat\psi(x_{1})\big\rangle\\
&&\times g(y_1)\cdots g(y_N)\bar{ g(x_N)}\cdots\bar{ g(x_1)}\d y_1\cdots\d y_N\d x_N\cdots\d x_1\\
&=&
\left(\Omega\Big|\Texp\left(-\i\int_{-\infty}^{\infty} \hat H_\Int(t)\d t-\i\int g(x)\hat \psi_\fr^*(x)\d x-\i\int \bar{ g(x)}\hat \psi_\fr(x)\d x\right)\Omega\right)
.\end{eqnarray*}

One can retrieve Green's functions from the generating function:
\begin{eqnarray*}&&
\big\langle\hat\psi^*(y_1)\cdots\hat\psi^*(y_N)\hat\psi(x_N)\cdots\hat\psi(x_1)\big\rangle\\&=&
(-1)^{N}\frac{\partial^{2N}}{\partial g(y_1)\cdots\partial g(y_N)\partial \bar {g(x_N)}\cdots \partial \bar{ g(x_1)}}
Z(g,\bar g)\Big|_{g=\bar g=0}.\end{eqnarray*}

We introduce also the {\em amputated Green's function}
\begin{eqnarray*}&&
\big\langle\hat\psi^*(p_1')\cdots\hat\psi^*(p_N')\hat\psi(p_N)\cdots\hat\psi(p_1)\big\rangle_\amp\\
&:=&
\bigl((p_1')^2+m^2\bigr)\cdots\bigl((p_{N}')^2+m^2\bigr)
\bigl((p_{N})^2+m^2\bigr)\cdots
\bigl((p_1)^2+m^2\bigr)\\&&
\times \big\langle\hat\psi^*(p_1')\cdots\hat\psi^*(p_{N}')\hat\psi(p_N)\cdots\hat\psi(p_1)\big\rangle.\end{eqnarray*}


Set
\[|\bar{-p_{n'}'},\dots,\bar{-p_1'},p_n,\dots,p_1):=\hat b^*(p_{n'}')\cdots
\hat b^*(p_1')\hat a^*(p_n)\cdots\hat  a^*(p_1)\Omega.\]
One can compute {\em scattering amplitudes} from the amputated Green's functions:
\begin{eqnarray*}&&
\left(
\bar{-p_{n^{+\prime}}^{+\prime}},\dots,p_{n^+}^+,\dots
|\,\hat S\, 
|\bar{-p_{n^{-\prime}}^{-\prime}},\cdots,
p_{n^-}^-,\cdots\right)\\&=&\frac{ \big\langle\cdots \hat\psi(p_{n^+}^+)\cdots\hat\psi^*(p_{n^{+\prime}}^{+\prime})
\hat\psi(-p_{n^{-\prime}}^{-\prime})\cdots
\hat\psi^*(-p_{n^-}^-)\cdots
\big\rangle_\amp}{\sqrt{(2\pi)^{3(n^++n^{+\prime}+n^{-\prime}+n^-)}}\cdots
\sqrt{2 E (p_{n^+}^+)}\cdots
\sqrt{2 E (p_{n^{+\prime}}^{+\prime})}
\sqrt{2 E (p_{n^{-\prime}}^{-\prime})}\cdots
\sqrt{2 E (p_{n^-}^-)}\cdots}
,\end{eqnarray*}
where all $p_i^\pm$, $p_i^{\pm\prime}$ are  on shell.

\subsubsection{Path integral formulation}

Since the Hamiltonian that we consider is quadratic, we can compute exactly 
 the generating function in terms of the Fredholm determinant on $L^2(\rr^{1,3})$:
\begin{eqnarray}&&Z(g,\bar g)\label{patho}\\
&=&\det\big(-\Box+m^2\big)
\big(-(\partial_\mu+\i e A_\mu(x))(\partial^\mu+\i e A^\mu(x))+m^2-\i0\big)^{-1}
\notag\\
&&\times\exp\left(\i \bar g
\left((\partial_\mu+\i e A_\mu(x))(\partial^\mu+\i e A^\mu(x))+m^2-\i0\right)^{-1}g\right)\notag\\
&=&\det\left(\one+\Bigl(-
\i e A_\mu(x)\partial^\mu-\i e \partial^\mu A_\mu(x)+e^2 A_\mu(x) A^\mu(x)\Bigr)D_\fr^{\rm c}\right)^{-1}\notag\\
&&\times\exp\left(\i \bar g D_\fr^{\rm c}
\Bigl(\one+\bigl(
-\i e A_\mu(x)\partial^\mu-\i e \partial^\mu A_\mu(x)+e^2 A_\mu(x) A^\mu(x)\bigr)
D_\fr^{\rm c}\Bigr)^{-1}g\right)
.\notag\end{eqnarray}
Let us stress that the above formulas are based on the  formal expression for the Hamiltonian
 (\ref{nmun4}) where we used the Weyl quantization, in contrast to the analogous formula (\ref{paks}) for the mass-like perturbation, which were Wick ordered.
 The expression is to a large degree ill-defined.

Formally, (\ref{patho}) can be rewritten in terms of path integrals  as
\begin{eqnarray*}
&&\frac{\int\lpi_y\d\psi^*(x)\lpi_{y'}\d\psi(y)
\exp\left(\i\int\big(\cL(x)-g(x)\psi^*(x)-\bar{g(x)}\psi(x)\big)\d x\right)}
{\int\lpi_y\d\psi^*(y)\lpi_{y'}\d\psi(y')
\exp\left(\i\int\cL_\fr(x)\d x\right)}.
\end{eqnarray*}

\nowastrona
\subsubsection{Feynman rules}

Let us describe the Feynman rules for the charged scalar field in an external  4-potential.
We have 1 kind of lines and 2 kinds of vertices. Each line has an arrow. At each
vertex two lines meet, one with an arrow pointing towards, one with an arrow
pointing away from the vertex. The 1-photon vertex is denoted by an attached
``photon line'' ending with a small cross. The 2-photon vertex has  two ``photon lines'',
each ending with a cross. Note that the ``photon lines'' are in this context only decorations of the vertices -- there are no photons in this theory. They are usually denoted by wavy, sometimes dashed  lines. For typographical reasons we use dashed lines.

To compute Green's functions we do as follows:
\ben \item We draw all possible Feynman diagrams.
\item 
\begin{romanenumerate}\item 
To each 1-photon vertex 
 we associate the factor
\[\i e (p^+_\nu+p_\nu^-)  A^\nu(p^+-p^-).\]
\item To each 2-photon  vertex 
 we associate the factor
\[-\i e^2 
(A^\nu A_\nu)(p^+-p^-).\]
\end{romanenumerate}
\item To each line we associate the propagator
\[-\i D_\fr^{\rm c}(p)=\frac{-\i}{p^2+m^2-\i0}.\]
\item We integrate over the variables of internal lines
 with the measure
$\frac{\d^4 p}{(2\pi)^4}$.
\een

\nowastrona

It is immediate to derive the Feynman rules for charged scalar bosons from the
path integral formula (\ref{patho}).

The derivation of the Feynman rules within 
the Hamiltonian formalism using the Dyson expansion of the scattering
operator
is relatively
complicated, since one has to use not only the two-point functions of ``configuration space fields'' $\psi,\psi^*$, but also of conjugate fields $\eta,\eta^*$ \cite{IZ}:
\begin{eqnarray*}
(\Omega|\T(\hat\psi_\fr(x)\hat\psi_\fr^*(y))\Omega)
&=&-\i D_\fr^{\rm c}(x-y),\\
(\Omega|\T(\hat\eta_\fr(x)\hat\psi_\fr^*(y))\Omega)
&=&-\i\partial_{x^0} D_\fr^{\rm c}(x-y),\\
(\Omega|\T(\hat\psi_\fr(x)\hat\eta_\fr^*(y))\Omega)
&=&-\i \partial_{y^0}D_\fr^{\rm c}(x-y),\\
(\Omega|\T(\hat\eta_\fr(x)\hat\eta_\fr^*(y))\Omega)
&=&-\i \partial_{x^0}\partial_{y^0}D_\fr^{\rm c}(x-y)-\i\delta(x-y).
\end{eqnarray*}

\begin{figure}[!h]
\centering
\includegraphics{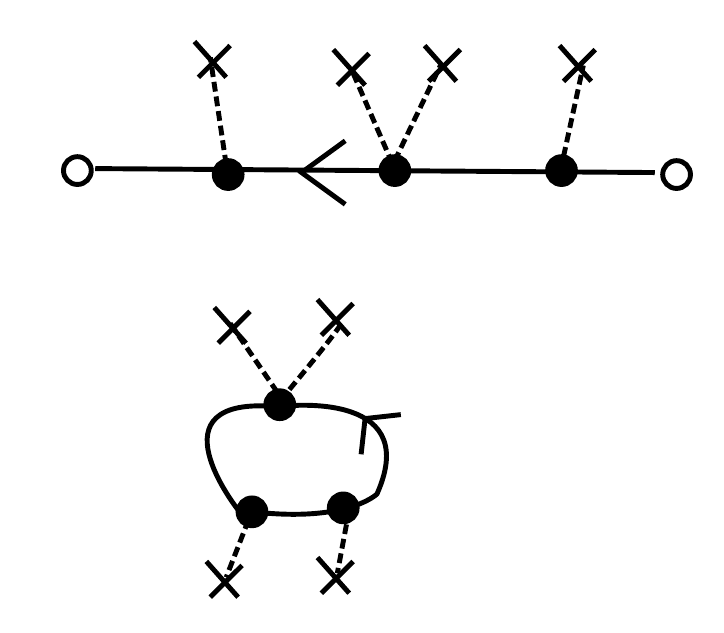}
\caption{Diagram for Green's function.\label{diag-7}}
\end{figure}

\begin{figure}[!h]
\centering
\includegraphics{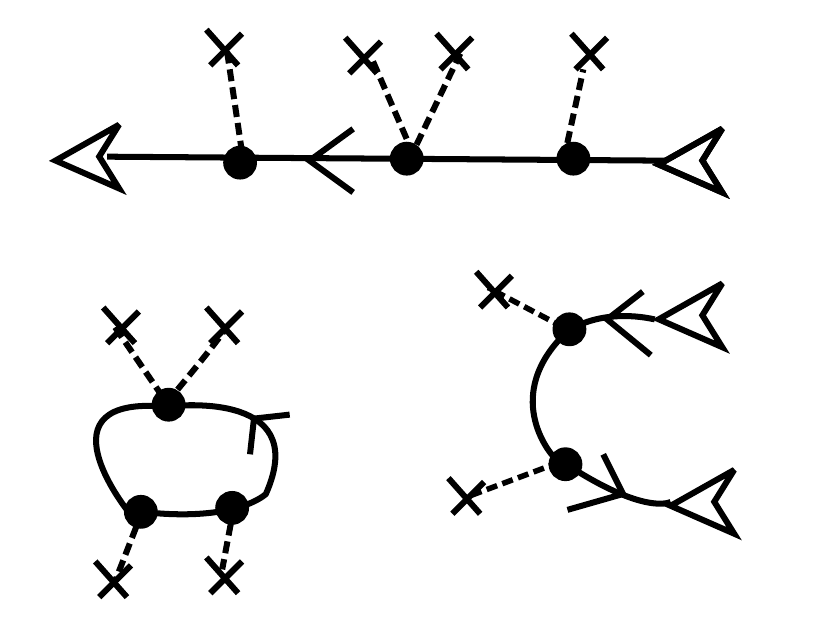}
\caption{Diagram for scattering amplitudes.\label{diag-8}
}
\end{figure}

\nowastrona

To compute  scattering amplitudes with $N^-$ incoming and $N^+$
outgoing particles we draw similar diagrams as
for $N^-+N^+$-point Green's functions, where as usual 
the incoming lines are drawn on the right and outgoing lines on the left.
The rules are changed only
concerning the external lines.
\begin{romanenumerate}\item
 With each incoming external line we associate
\begin{itemize}  \item charged boson: $\frac{1}{\sqrt{(2\pi)^32E(\vec p)}}$.
\item  charged anti-boson: $\frac{1}{\sqrt{(2\pi)^32E(\vec p')}}$.
\end{itemize}
\item  With each outgoing external line we associate
\begin{itemize}
\item  charged boson: $\frac{1}{\sqrt{(2\pi)^32E(\vec p)}}$.
\item  charged anti-boson: $\frac{1}{\sqrt{(2\pi)^32E(\vec p')}}$.
\end{itemize}\end{romanenumerate}

\nowastrona

 \subsubsection{Vacuum energy}

Formally, the vacuum energy can be
computed exactly:
\begin{eqnarray}\nonumber
\cE&:=&\i\log(\Omega|\hat S\Omega)\,=\,\i\log Z(0,0)\\\nonumber
&=&\i\Tr\Big(\log\big({-}\Box{+}m^2{-}\i0\big)-
\log\big({-}(\partial_\mu{+}\i e A_\mu(x))(\partial^\mu{+}\i e A^\mu(x)){+}m^2{-}\i0\big)\Big)
\\\notag
&=&-\i\Tr\Big(
\log\Big(\one+\bigl(
{-}\i e A_\mu(x)\partial^\mu{-}\i e \partial^\mu A_\mu(x)+e^2 A_\mu(x) A^\mu(x)\bigr)D_\fr^{\rm c}\Big)\Big)\\
&=&
\i\sum_\ell\frac{D_\ell}{n_\ell}.\label{paks=}
\end{eqnarray}

Here $D_\ell$ is
the value of  the loop $\ell$ and $n_\ell$ is its {\em symmetry factor}.
Any such a loop is described  by a cyclic 
sequence $(\alpha_1,\dots,\alpha_n)$, where $\alpha_j=1,2$ correspond to $1-$ and $2-$photon vertices. The symmery factor $n_\ell$ is
 the order of the group of the
authomorphisms of this loop. The loop is oriented, hence this group
 is always a subgroup of rotations. In particular, if the loop has $n$ identical vertices, the group is $\zz_n$ and $n_\ell=n$.

\begin{figure}[!h]
\centering
\includegraphics{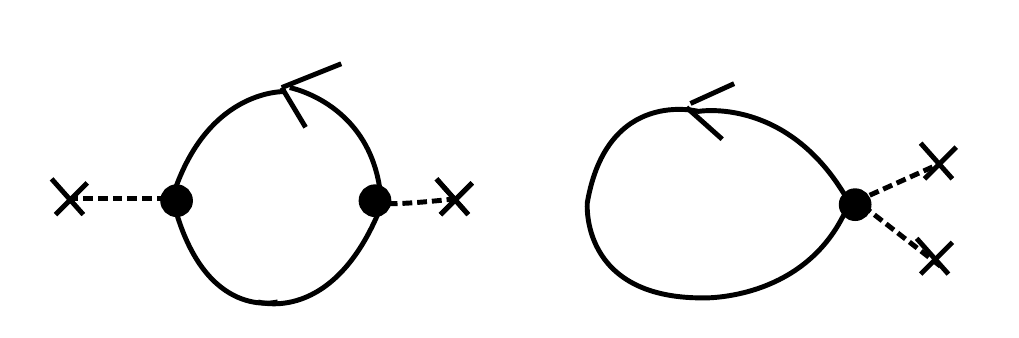}
\caption{Divergent diagrams  for vacuum energy.
\label{diag-9}
}
\end{figure}

Actually, it is better to organize (\ref{paks=}) not in terms of the number of vertices on a loop but in terms of the order wrt $e$.
Using the unitary charge conjugation operator $C$ and $C\Omega=\Omega$ we obtain
\begin{eqnarray*}
(\Omega|\hat S(A)\Omega)&=&(\Omega|C\hat S(A)C^{-1}\Omega)=(\Omega|\hat S(-A)\Omega).\end{eqnarray*}
Therefore,  diagrams   of an odd order in
$e$ vanish. This is the content of  {\em Furry's theorem} for
charged bosons.
Hence  (\ref{paks=}) can be written as
\[\cE=\sum_{n=1}^\infty e^{2n}\cE_n.\]
The expressions for $\cE_n$ obtained from the Feynman rules are convergent for $n\geq3$. $E_2$ is logarithmically divergent, but its physically relevant gauge invariant part is convergent.   $\cE_1$ is quadratically divergent and its gauge-invariant part is logarithmically divergent. It  needs an infinite renormalization, which will be described below.

\subsubsection{Pauli-Villars renormalization}
\label{Pauli-Villars}
The lowest nonzero loop diagrams  are of the
second order in $e$, and hence of the first order in $\alpha=\frac{e^2}{4\pi}$.
There are two kinds of loops of this order: a loop with two
1-photon vertices with symmetry factor $2$ and a loop with a 2-photon vertex with symmetry factor $1$, see the Fig. \ref{diag-9}. The sum of their 
contributions has the form
\beq\cE_1=\int\frac{\d p}{(2\pi)^4} A^\mu(-p)A^\nu(p)\Pi_{\mu\nu}(p).\label{paksq}\eeq
(\ref{paksq}) defines the {\em vacuum energy tensor}  $\Pi_{\mu\nu}(p)$.

We will first
compute $\Pi_{\mu\nu}$  using  the {\em Pauli-Villars regularization}.
The ultraviolet problem is more severe now than it was for the mass-like perturbation, where a single additional fictitious particle sufficed to make the expressions well defined. Now we need two fictitious particles:
\begin{eqnarray*}
m_0^2:=m^2,&&C_0:=1,\\
m_1^2:=m^2+2\Lambda^2,&&C_1:=1,\\
m_2^2:=m^2+\Lambda^2,&&C_2:=-2.
\end{eqnarray*}
Using
\beq \sum_{i=0}^2 C_i=\sum_{i=0}^2 C_im_i^2=0\eeq
we can check that with this choice the  sums used in the following computations are integrable.

In the following formula we have a contribution of the  loop with 2 single-photon vertices and twice the contribution of the loop with a single 2-photon vertex. It is convenient to write the latter  as the sum of two terms, equal to one another. 
\nowastrona

\begin{eqnarray*}
2\Pi_{\mu\nu \Lambda }(p)
&=&\i e^2\int\frac{\d^4
  q}{(2\pi)^4}
\sum_iC_i\Big(\frac{4q_\mu q_\nu}
{\big((q+\frac12p)^2+m_i^2-\i0\big)\big((q-\frac12p)^2+m_i^2-\i0\big)}
\\&&\hspace{7ex}-\frac{g_{\mu\nu}}{\big((q+\frac12p)^2+m_i^2-\i0\big)}
-\frac{g_{\mu\nu}}{\big((q-\frac12p)^2+m_i^2-\i0\big)}\Big)
\\&=&\i e^2\int\frac{\d^4
  q}{(2\pi)^4}
\sum_iC_i\frac{4q_\mu q_\nu-2g_{\mu\nu}(q^2+\frac14p^2+m_i^2)}
{\big((q+\frac12p)^2+m_i^2-\i0\big)\big((q-\frac12p)^2+m_i^2-\i0\big)}
\end{eqnarray*}
\begin{eqnarray*}
&=&-\frac{ e^2}{(4\pi)^2}
\int_0^\infty\d\alpha_1\int_0^\infty\d\alpha_2\sum_iC_i
\Bigg(\frac{(\alpha_1-\alpha_2)^2}{(\alpha_1+\alpha_2)^4}(g_{\mu\nu}p^2-p_\mu
  p_\nu)\\
&&\ \ \ \ \ \ \ \ \ +
2g_{\mu\nu}\left(\frac{\alpha_1\alpha_2}{(\alpha_1+\alpha_2)^4}p^2
-\frac{\i }{(\alpha_1+\alpha_2)^3}+\frac{m_i^2}{(\alpha_1+\alpha_2)^2}\right)\Bigg)
\\&&
\times
\exp\left(-\i(\alpha_1+\alpha_2)m_i^2
-\i\frac{\alpha_1\alpha_2}{\alpha_1+\alpha_2}
  p^2\right)
\\
&=:&
(-g_{\mu\nu} p^2+p_\mu p_\nu)2\Pi_\Lambda^{{\rm gi}}(p^2)+2\Pi^{\rm gd}_{\mu\nu\Lambda}(p^2).
\end{eqnarray*}
\nowastrona
We used the identity (\ref{loop-com}).

\nowastrona

The gauge dependent part of the vacuum energy tensor vanishes:
\begin{eqnarray*}&&
-\Pi_{\mu\nu \Lambda }^{{\rm gd}}(p^2)\\&=&
\sum_iC_i\frac{ e^2}{(4\pi)^2}
\int_0^\infty\d\alpha_1\int_0^\infty\d\alpha_2
\exp\left(-\i(\alpha_1+\alpha_2)m_i^2
-\i\frac{\alpha_1\alpha_2}{\alpha_1+\alpha_2}
  p^2\right)\\&&
\times g_{\mu\nu}
\left(\frac{\alpha_1\alpha_2 p^2}{(\alpha_1+\alpha_2)^4}
-\frac{\i}{(\alpha_1+\alpha_2)^3}+\frac{m_i^2}{(\alpha_1+\alpha_2)^2}\right)\\
&=&
\sum_iC_i\frac{ e^2}{(4\pi)^2}
\rho\partial_\rho\int_0^\infty\d\alpha_1\int_0^\infty\d\alpha_2
\exp\left(-\i\rho\left((\alpha_1+\alpha_2)m_i^2
+\frac{\alpha_1\alpha_2}{\alpha_1+\alpha_2}
  p^2\right)\right)\\&&
\times\frac{\i g_{\mu\nu}}{\rho(\alpha_1+\alpha_2)^3}\Big|_{\rho=1}
\\
&=&
\sum_iC_i\frac{ e^2}{(4\pi)^2}
\rho\partial_\rho\int_0^\infty\d\alpha_1\int_0^\infty\d\alpha_2
\exp\left(-\i\left((\alpha_1+\alpha_2)m_i^2
+\frac{\alpha_1\alpha_2}{\alpha_1+\alpha_2}
  p^2\right)\right)\\&&
\times\frac{\i g_{\mu\nu}}{(\alpha_1+\alpha_2)^3}\ \ \ =\ \  0.
\end{eqnarray*}
To compute the gauge invariant part we
proceed similarly as in Subsubsec. \ref{sec-PV}, see (\ref{inser}), and we obtain
 \begin{eqnarray*}
\Pi_\Lambda ^{{\rm gi}}(p^2)&=&
-\frac{e^2}{2(4\pi)^2}
\int_0^\infty\d\alpha_1\int_0^\infty\d\alpha_2\sum_iC_i
\frac{(\alpha_1-\alpha_2)^2}{(\alpha_1+\alpha_2)^{4}}  
\\ &&
\times \exp\left(-\i(\alpha_1+\alpha_2)m_i^2-
\i\frac{\alpha_1\alpha_2}{\alpha_1+\alpha_2}
  p^2\right)
\\
&=&-\frac{ e^2}{2(4\pi)^2}
\int_0^1\d v\int_0^\infty\frac{\d\rho}{\rho}\sum_iC_i
v^2
 \exp\left(-\i\rho\left(m_i^2
+\frac{(1-v^2)
  p^2}{4}\right)\right)\\
&=&\frac{ e^2}{2(4\pi)^2}
\int_0^1\d v\sum_iC_i
v^2\log\left(m_i^2
+\frac{(1-v^2)p^2}{4}
-\i0\right).
\end{eqnarray*}
\nowastrona
We define
\begin{eqnarray}
\Pi^{\ren}(p^2)&:=&\lim_{\Lambda \to\infty}\big(\Pi_\Lambda ^{{\rm gi}}(p^2)-\Pi_\Lambda ^\gi(0)\big)
\label{piren}\\
&=&\frac{ e^2}{2(4\pi)^2}
\int_0^1\d v
v^2\log\left(1
+\frac{(1-v^2)
  p^2}{4m^2}-\i0\right).
\notag\end{eqnarray}
 Using  (\ref{ide4.1}), and then analytic continuation,
we obtain 
\begin{eqnarray*}&&\Pi^\ren(p^2)\\&=&\frac{e^2}{2\cdot 3(4\pi)^2}\Bigg(
\frac{(p^2+4m^2)^{3/2}}{(p^2)^{3/2}}\log\frac{\sqrt{p^2+4m^2}+\sqrt{p^2}}
{\sqrt{p^2+4m^2}-\sqrt{p^2}}\\&&\hspace{8ex}-\frac{2}{3}-2\Big(\frac{4m^2}{p^2}+1\Big)\Bigg),\ \ \ \ \ \ 0<p^2;\\
&=&\frac{e^2}{2\cdot 3(4\pi)^2}\Bigg(
\frac{(p^2+4m^2)^{3/2}}{(-p^2)^{3/2}}2\arctan\frac{\sqrt{-p^2}}{\sqrt{p^2+4m^2}}
\\&&\hspace{8ex}-\frac{2}{3}-2\Big(\frac{4m^2}{p^2}+1\Big)\Bigg),\ \ \ \ \ \ -4m^2<p^2<0;\\
&=&\frac{e^2}{2\cdot 3(4\pi)^2}\Bigg(
\frac{(-p^2-4m^2)^{3/2}}{(-p^2)^{3/2}}\Big(
\log\frac{\sqrt{-p^2-4m^2}+\sqrt{-p^2}}
{\sqrt{-p^2-4m^2}-\sqrt{-p^2}}-\i\pi\Big)\\&&\hspace{8ex}-\frac{2}{3}-2\Big(\frac{4m^2}{p^2}+1\Big)\Bigg),\ \ \ \ \ \  p^2<-4m^2.
\end{eqnarray*}

\subsubsection{Renormalization of the vacuum energy}

Note that the Fourier tranform of the electromagnetic field is
\beq F_{\mu\nu}(p)=p_\mu A_\nu(p)-p_\nu A_\mu(p).\eeq
Hence
\beq
-\bar{F_{\mu\nu}(p)}F^{\mu\nu}(p)=-p^2|A(p)|^2+|pA(p)|^2.\eeq
Thus  the renormalized 1st order contribution to the vacuum energy is
\beq\cE_1^\ren=-\int\frac{\d p}{(2\pi)^4} 
\Pi^\ren(p^2)\bar{F_{\mu\nu}(p)}F^{\mu\nu}(p).\label{fad}\eeq

We can formally write $\Pi_\infty^\gi(k):=\lim\limits_{\Lambda\to\infty}\Pi_\Lambda^\gi(k)$
(which is typically  infinite). 
Note that the renormalized scattering operator $\hat S_\ren$ is a well defined unitary operator:
\begin{eqnarray}
\hat S_\ren&=&\e^{-\i\Pi_\infty^\gi(0)\int F_{\mu\nu}(x)F^{\mu\nu}(x)\d x} \hat S.\label{formo2}\end{eqnarray}
However, there is no correctly defined renormalized Hamiltonian. Formally, the correct Lagrangian density is obtained by replacing $\cL(x)$ with
\[\cL_\ren(x)=\cL(x)-\Pi_\infty^\gi(0) F_{\mu\nu}(x)F^{\mu\nu}(x).\]


\subsubsection{Method of dispersion relations}

There exists an alternative method to renormalize and compute the
vacuum energy. We start with computing just the imaginary part
of the gauge invariant vaccum energy function, which does not require a regularization, so that we
obtain $\Im\Pi^\ren(p^2)$ from the very beginning:
\begin{eqnarray}\notag
\Im\Pi^{\ren}(p^2)&=&\frac{ e^2}{2(4\pi)^2}
\int_0^1\d v
v^2(-\pi)\theta\left(-1
-\frac{(1-v^2)
  p^2}{4m^2}\right)\\
&=&-\frac{ e^2}{2\cdot 3(4\pi)^2}\frac{\pi}{(-p^2)^{3/2}}
\left|-p^2-4m^2\right|_+^{\frac32}.
\label{imago}\end{eqnarray}
As in (\ref{pdf}), using  $\Pi^\ren(0)=0$ and Thm \ref{dysp1}
 we obtain
\begin{eqnarray}
\Pi^\ren(p^2)&=&
\frac{1}{\pi}\int_{-\infty}^{-4m^2}\d s \Im\Pi^\ren(s)
\left(\frac{1}{s-p^2}-\frac{1}{s}\right).
\label{dispe}\end{eqnarray}

Note that (\ref{imago})  is nonzero only for
$p^2<-4m^2$, and then it is negative.
For such $p$ we can find a coordinate system with $p=(p^0,\vec 0)$. Then
\[-g_{\mu\nu}p^2+p_\mu p_\nu=p_0^2(g_{\mu\nu}+\delta_{\mu0}\delta_{0\nu})\]
and
\beq
-\bar{F_{\mu\nu}(p^0,\vec 0)}F^{\mu\nu}(p^0,\vec0)=p_0^2|\vec A(p^0,\vec0)|^2.\eeq
Thus the imaginary part of (\ref{fad}) is negative (and is
 responsible for the decay).

\subsubsection{Dimensional renormalization}

We present an alternative computation of $\Pi_{\mu\nu}^\ren$ based on the dimensional regularization. We use the Euclidean formalism. 
\begin{eqnarray}\notag
2\Pi_{\mu\nu}^\E(p)
&=&- e^2\int\frac{\d^4
  q}{(2\pi)^4}
\Big(\frac{4q_\mu q_\nu}
{((q+\frac12p)^2+m^2)((q-\frac12p)^2+m^2)}
\\&&\hspace{7ex}-\frac{2g_{\mu\nu}}{q^2+m^2}\Big)\notag
\\&=&- e^2\int\frac{\d^4
  q}{(2\pi)^4}
\frac{4q_\mu q_\nu-2g_{\mu\nu}(q^2+\frac14p^2+m^2)}
{((q+\frac12p)^2+m^2)((q-\frac12p)^2+m^2)}
\notag\\
&=&- \frac{e^2}{2}\int_{-1}^1\d v\int\frac{\d^4
  q}{(2\pi)^4}
\frac{4q_\mu q_\nu-2g_{\mu\nu}(q^2+\frac14p^2+m^2)}
{(q^2+\frac{p^2}{4}+m^2+vqp)^2}
\notag
\\
&\hspace{-37ex}=&\hspace{-20ex}{-} e^2\int_{0}^1{\d v}\int\frac{\d^4
  q}{(2\pi)^4}
\frac{4q_\mu q_\nu-2g_{\mu\nu}(q^2+\frac14p^2+m^2)+v^2(p_\mu p_\nu-g_{\mu\nu}\frac{p^2}{2})}
{(q^2+\frac{p^2}{4}(1-v^2)+m^2)^2},\label{pdf4}
\end{eqnarray}
where we used the Feynman identity (\ref{dim-feyn}), replaced $q+\frac{vp}{2}$ with $q$, used the symmetry $v\to -v$ to remove $\int_{-1}^1 \d v v$ and replace $\frac12\int_{-1}^1\d v$ with  $\int_0^1\d v$.
 After this preparation, we use the dimensional regularization:
\begin{eqnarray}
\int\frac{\d^4 q}{(2\pi)^4}\ \hbox{ is replaced by }\
\frac{\mu^{4-d}\Omega_d}{(2\pi)^d}\int_0^\infty|q|^{d-1}\d|q|,\label{replace1}\\
\int\frac{q_\mu q_\nu\d^4 q}{(2\pi)^4}\ \hbox{ is replaced by }\
\frac{\mu^{4-d}\Omega_d}{d(2\pi)^d}g_{\mu\nu}\int_0^\infty|q|^{d+1}\d|q|,\label{replace2}
\end{eqnarray}
where $\Omega_d$ is given by (\ref{dim1}). Thus (\ref{pdf4}) is replaced by
\begin{eqnarray}&&
\Pi_{\mu\nu}^{\E,d}(p)\,=\,-e^2\frac{\mu^{4-d}\Omega_d}{(2\pi)^d}\int_0^1\d v\int_0^\infty|q|^{d-1}\d|q|\notag\\
&&\times\frac{\big((4/d-2)g_{\mu\nu}q^2-2g_{\mu\nu}(\frac14p^2+m^2)+v^2(p_\mu p_\nu-g_{\mu\nu}\frac{p^2}{2}\big)
}
{\big(q^2+\frac{p^2}{4}(1-v^2)+m^2\big)^2}\notag\\
&=&-\frac{e^2}{(4\pi)^2}\int_0^1\d v\Big(\frac{\mu^24\pi}{
\frac{p^2}{4}(1-v^2)+m^2}\Big)^{2-d/2}\Gamma(2-d/2)\notag\\
&&\times
\Bigg(2g_{\mu\nu}\Big(\frac{p^2}{4}(1-v^2)+m^2\Big)
-2g_{\mu\nu}\Big(\frac14p^2+m^2\Big)+v^2\Big(p_\mu p_\nu-g_{\mu\nu}\frac{p^2}{2}\Big)\Bigg)\notag\\
&=&-\frac{e^2}{(4\pi)^2}\int_0^1\d v\Big(\frac{\mu^24\pi}{
\frac{p^2}{4}(1-v^2)+m^2}\Big)^{2-d/2}\Gamma(2-d/2)v^2(p_\mu p_\nu-g_{\mu\nu}p^2)
\notag\\
&\simeq&-\frac{e^2}{(4\pi)^2}\int_0^1\d v
\Big(-\gamma +\log\mu^24\pi-\log\Big(
\frac{p^2}{4}(1-v^2)+m^2\Big)\Big)v^2(p_\mu p_\nu-g_{\mu\nu}p^2)\notag\\&&
-\frac{e^2}{3(4\pi)^2(2-d/2)}(p_\mu p_\nu-g_{\mu\nu}p^2)
\label{pdf5}\end{eqnarray}
We can now renormalize (\ref{pdf5}):
\begin{eqnarray*}
&&\Pi^{\E,\ren}(p^2)(p_\mu p_\nu-g_{\mu\nu}p^2)\\&=&\lim_{d\to4}\Big(\Pi_{\mu\nu}^{\E,d}(p^2)
-\Pi_{\mu\nu}^{\E,d}(0)\Big)\\
&=&\frac{1}{2(4\pi)^2}\int_0^1\d v v^2
\log\Big(1+
\frac{p^2}{4m^2}(1-v^2)\Big)(p_\mu p_\nu-g_{\mu\nu}p^2).
\end{eqnarray*}
This coincides  with the Wick rotated 
result obtained by the Pauli-Villars method.

\subsubsection{Abstract gauge covariance}

Let us adopt for a moment an abstract setting.
Let $\rr\ni t\mapsto \hat H(t)$ be a time-dependent Hamiltonian
generating the dynamics
\[\hat U(t_+,t_-):=\Texp\Big(-\i\int_{t_-}^{t_+} \hat H(s)\d s\Big).\]
\nowastrona
Let $t\mapsto \hat W(t)$ be a family of unitary operators that have the interpretation of time dependent
{\em gauge transformations}. We will assume that  $\hat W(t)$ converges to identity as $t\to\pm\infty$ and is generated
by a time dependent family of self-adjoint
operators $t\mapsto \hat R(t)$, so that
\[\hat W(t):=\Texp\Big(-\i\int_{-\infty}^t \hat R(s)\d s\Big).\]
Then
\[\hat W(t_+)\hat U(t_+,t_-)\hat W^*(t_-)=\Texp
\Big(-\i\int_{t_-}^{t_+} \hat H_R(s)\d s\Big),\]
where the {\em gauge-transformed Hamiltonian} is
\beq \hat H_R(t):=\hat W(t)\hat H(t)\hat W^*(t)+\hat R(t).\label{gau0}\eeq

\subsubsection{Ward identities}

Let us go back to the setting of quantized charged scalar fields.
The gauge invariance implies strong conditions on the scattering
operator and Green's functions. 

Let $\hat S(A)$ denote the scattering operator for the external 4-potential $A$.
Let $\chi$ be a Schwartz function on $\rr^{1,3}$. It is easy to see that the scattering operator
is gauge-invariant:
\beq \hat S(A)=\hat S(A+\partial\chi).\label{wardo}\eeq
Differentiating this identity w.r.t. $\chi$ and setting $\chi=0$ we
obtain  the
{\em Ward(-Takahashi) identities for the scattering operator} in the position representation:
\[\partial_{y_\mu}\frac{\partial}{\partial_{A_\mu(y)}}\hat S(A)=0.\]
In the momentum representation these identities read
\[p_\mu\frac{\partial}{\partial_{A_\mu(p)}}\hat S(A)=0.\]

\nowastrona

We will write $\langle \hat\psi^*(x'_1)\cdots\hat\psi^*(x_N')\hat\psi(x_N)\cdots\hat\psi(x_{1})\rangle_A$ to express the dependence of
Green's functions on the external 4-potential $A$. 
We have
\begin{eqnarray}\label{wardo1}
&&\langle
\hat\psi^*(x'_1)\cdots\hat\psi^*(x_N')\hat\psi(x_N)\cdots\hat\psi(x_{1})\rangle_{A+\partial\chi}\\
&=&
\langle \hat\psi^*(x'_1)\cdots\hat\psi^*(x_N')\hat\psi(x_N)\cdots\hat\psi(x_{1})\rangle_A\e^{\i
  e\chi(x'_1)+\dots+\i e\chi(x_N')-\i e\chi(x_N)-\dots-\i e\chi(x_1)}.\notag
\end{eqnarray} \nowastrona
By differentiating with respect to $\chi(y)$ and setting $\chi=0$
we obtain 
the {\em Ward(-Takahashi) identities for Green's functions}
in the position representation:
\begin{eqnarray*}
&&\partial_{y_\mu}\frac{\partial}{\partial A_\mu(y)}\langle
\hat\psi^*(x'_1)\cdots\hat\psi^*(x_N')\hat\psi(x_N)\cdots\hat\psi(x_{1})\rangle_A\\
&=&\Bigg(\i\sum_{j=1}^N\delta(y-x_j')-\i\sum_{j=1}^N\delta(y-x_j)\Bigg)
\langle
\hat\psi^*(x'_1)\cdots\hat\psi^*(x_N')\hat\psi(x_N)\cdots\hat\psi(x_{1})\rangle_A
.\end{eqnarray*}

In the  momentum representation these
identities read
\begin{eqnarray*}
&&q_\mu\frac{\partial}{\partial A_\mu(q)}\langle
\hat\psi^*(p_1')\cdots\hat\psi^*(p_N')\hat\psi(p_N)\cdots\hat\psi(p_{1})\rangle_A\\
&=& \sum_{j=1}^N\langle\hat\psi^*(p_1')\cdots\hat\psi^*(p_j'-q)\cdots\hat\psi^*(p_N')\hat\psi(p_N)\cdots\hat\psi(p_{1})\rangle_A+\\
&&-
\sum_{j=1}^N \langle\hat\psi^*(p_1')\cdots\hat\psi^*(p_N')\hat\psi(p_N)\cdots\hat\psi(p_j+q)\cdots\hat\psi(p_{1})\rangle_A
.\end{eqnarray*}

(\ref{wardo}) and (\ref{wardo1}) are essentially obvious if we use the path integral expressions. It is instructive 
to  derive these  statements also in the Hamiltonian formalism. This
derivation is not fully rigorous, since transformations cannot be
implemented, and in general the dynamics does not have a well defined
Hamiltonian.

 Formally, we define the gauge transformation as a unitary operator
\begin{eqnarray}\nonumber
\hat W(\chi,t)&:=&
\exp\left(-\i e\int\d \vec x
\:\chi(t,\vec x)\hat\cQ(\vec x)\right)
\\&=&\exp\left(-\i e\int_{-\infty}^t\d s\int \d \vec x\:
\dot\chi(s,\vec x)\hat \cQ(\vec x)\right)\label{nmun}
\\&=&\Texp\left(-\i e\int_{-\infty}^t\d s\int \d \vec x\:
\dot\chi(s,\vec x)\hat \cQ(\vec x)\right).\nonumber
\end{eqnarray}

\nowastrona
To see the second  identity it is enough to note that $[\hat \cQ(\vec x),
\hat \cQ(\vec y)]=0$, hence we can replace $\Texp$ with $\exp$ in
(\ref{nmun}). 
Clearly,
\begin{eqnarray*}
\hat W(\chi,t)\hat \psi(\vec x)\hat W(\chi,t)^*&=&\e^{\i\chi(t,\vec
  x)}\hat 
\psi(\vec x),\\
\hat W(\chi,t)\hat \eta(\vec x)\hat W(\chi,t)^*&=&\e^{\i\chi(t,\vec x)}\hat \eta(\vec x).
\end{eqnarray*}
Let $\hat H(A,t)$ denote 
(\ref{nmun3}), that is  the Schr\"odinger picture Hamiltonian. Let
$\hat U(A,t_+,t_-)$  be the corresponding dynamics.
\begin{eqnarray*}
&&\hat W(\chi,t)\hat H(t,A)\hat W(\chi,t)^*+e\int\dot\chi(t,\vec x)\hat\cQ(\vec x)\d\vec x
\\
&=&\int\d\vec x
\Bigl(\hat\eta^*(\vec x)\hat\eta(\vec x)-\i e\bigl(A_0(t,\vec x)+\dot\chi(t,\vec x)\bigl){:}\bigl(\hat \psi^*(\vec
x)\hat \eta(\vec x)-
\hat\eta^*(\vec x)\hat\psi(\vec x)\bigr){:}\\
\nonumber
&&+(\partial_i-\i eA_i(t,\vec x))\e^{\i e\chi(t,\vec x)}\hat \psi^*(\vec x)(\partial_i+\i eA_i(t,\vec x))
\e^{-\i e\chi(t,\vec x)}\hat \psi(\vec x)\\
&&+m^2\hat\psi^*(\vec  x)\hat\psi(\vec x)\Bigr)\\
&=&\hat H(t,A+\partial\chi).\end{eqnarray*}
Therefore, by  (\ref{gau0}), we have the following identity, which expresses the gauge covariance:
\begin{eqnarray}\label{first}
\hat W(\chi,t_+)\hat U(A,t_+,t_-)\hat W^*(\chi,t_-)&=&
\hat U(A+\partial \chi,t_+,t_-).
\end{eqnarray}
Using that  $\lim\limits_{t\to\pm\infty}\hat W(\chi,t)=\one$,  we obtain
\begin{eqnarray*}
\hat S(A+\partial\chi)&=&\lim_{t_+,-t_-\to\infty}
\e^{\i t_+ \hat H_0}\hat U(A+\partial\chi,t_+,t_-)
\e^{-\i t_-\hat H_0}\\
&=&\lim_{t_+,-t_-\to\infty}
\e^{\i t_+ \hat H_0}\hat W(\chi,t_+)\hat U(A,t_+,t_-)
\hat W(\chi,t_-)^*\e^{-\i t_-\hat H_0}\\
&=& \hat S(A),\end{eqnarray*}
which implies
(\ref{wardo}).
(\ref{wardo1}) 
is a consequence of (\ref{first}).


\nowastrona

\nowastrona

\subsubsection{Energy shift}

Suppose that the 4-potential does not depend on time and is given by
a Schwartz function $\rr^3\ni \vec x\mapsto A(\vec x)= [A_\mu(\vec x)]$. We assume
that $A_0^2\leq m^2$. The naive
(Weyl ordered) 
 Hamiltonian is
\begin{eqnarray}\nonumber
\hat H&=&\int\d\vec x
\Bigl(\hat\eta^*(\vec x)\hat\eta(\vec x)
+\i eA_0(\vec x){:}\bigl(\hat \psi^*(\vec
x)\hat \eta(\vec x)-
\hat\eta^*(\vec x)\hat\psi(\vec x)\bigr){:}\\
\nonumber
&&+(\partial_i-\i eA_i(\vec x))\hat \psi^*(\vec x)(\partial_i+\i eA_i(\vec x))
\hat \psi(\vec x)
\\
&&+m^2\hat\psi^*(\vec  x)\hat\psi(\vec x)
\Bigr).
\label{nmun3-}\end{eqnarray}
It can be compared with the Weyl ordered free Hamiltonian (\ref{nmun3free}).
We can apply the formula (\ref{bogola}) to compute
the naive energy shift (the difference between the ground state
energies of $\hat H$ and $\hat H_\fr$):
\begin{eqnarray}\notag&&\Tr\Big(\sqrt{-\big(\vec\partial+\i e\vec
    A)^2+m^2-e^2A_0^2}
-\sqrt{-\vec\partial^2+m^2}\Big)\\
&=:&\sum_{n=1}^\infty e^{2n}E_{n}.\notag
\end{eqnarray}
In the above sum all the terms with $n\geq2$ are well defined. The
term with $n=1$ needs renormalization. The renormalized energy
shift is
\begin{eqnarray}E^\ren
&=&-e^2 \int\Pi^\ren(\vec p^2)\bar{F_{\mu\nu}(\vec p)}F^{\mu\nu}(\vec p)
\frac{\d\vec p}{(2\pi)^3}+
\sum_{n=2}^\infty e^{2n}E_{n},\notag
\end{eqnarray}
where $\Pi^\ren  $ was introduced in (\ref{piren}).

\init\section{Dirac fermions}

\nowastrona

In this section we study the {\em Dirac equation}
\[(-\i\gamma^\mu\partial_\mu+m)\psi(x)=0\]
 and its quantization. 
Here, $m\geq0$ and
 $\gamma^\mu$ are {\em Dirac matrices}.


Note that the Dirac equation is complex, and therefore it describes
charged particles. In particular, one can
 consider the Dirac equation in the presence of an external 
electromagnetic 4-potential $A(x)=[A^\mu(x)]$:
\[
\big(\gamma^\mu(-\i\partial_\mu+ eA_\mu(x))+m\big)\psi(x)=0.
\]
\nowastrona

The theory of Dirac fermions is in many ways parallel to the theory of charged scalar bosons described in Sect. \ref{Charged scalar bosons}. 

\subsection{Free Dirac fermions}
\subsubsection{Dirac spinors}

We adopt the following conventions for Dirac matrices
 $\gamma^\mu$, $\mu=0,\dots,3$:
\begin{eqnarray*}
[\gamma^\mu,\gamma^\nu]_+&=&-2g^{\mu\nu},\\
\gamma^{0*}=\gamma^0,&& \ \gamma^{i*}=-\gamma^i,\ \  i=1,2,3
.\end{eqnarray*}

Sometimes we will also
need 
\[\gamma^5:=-\i\gamma^0\gamma^1\gamma^2\gamma^3.\]
It satisfies
\[[\gamma^5,\gamma^\mu]_+=0,\ \ (\gamma^5)^2=\one, \ \  \gamma^{5*}=\gamma^5.\]

All irreducible representations of Dirac matrices
are equivalent and act on the space $\cc^4$.
One of the most common
is the so-called {\em Dirac representation}
\begin{eqnarray*}\gamma^0=\mat{1}{0}{0}{-1},&&
\vec\gamma=\mat{0}{\vec\sigma}{-\vec\sigma}{0},\\
\gamma^5=\mat{0}{1}{1}{0}.&&
\end{eqnarray*}
Here is the {\em Majorana representation}:
\begin{eqnarray*}\gamma^0=\i\mat{0}{-1}{1}{0},&&
\gamma^1=\i\mat{0}{\sigma_1}{\sigma_1}{0},\ \ 
\gamma^2=\i\mat{-1}{0}{0}{1},\ \ 
\gamma^3=\i\mat{0}{\sigma_3}{\sigma_3}{0},\\
\gamma^5=-\mat{0}{\sigma_2}{\sigma_2}{0},&&
\end{eqnarray*}
and the {\em spinor representation}:
\begin{eqnarray*}\gamma^0=\mat{0}{1}{1}{0},&& 
\vec\gamma=\mat{0}{-\vec\sigma}{\vec\sigma}{0},
\\
\gamma^5=\mat{1}{0}{0}{-1}.&&
\end{eqnarray*}
Above we used the {\em Pauli matrices} $\vec\sigma=(\sigma_1,\sigma_2,\sigma_3)$ defined by
\[\sigma_1=\left[\begin{array}{cc}0&1\\1&0\end{array}\right],\ \ \ \ 
\sigma_2=\left[\begin{array}{cc}0&-\i\\\i&0\end{array}\right],\ \ \ \ 
\sigma_3=\left[\begin{array}{cc}1&0\\0&-1\end{array}\right].\]
They satisfy $\sigma_i\sigma_j=2\epsilon_{ijk}\sigma_k$.

Note useful (representation independent) {\em trace identities}:
\begin{eqnarray*}
\Tr \one&=&4,\\
\Tr (a\gamma)(b\gamma)&=&-4ab,\\
\Tr (a\gamma)(b\gamma)(c\gamma)(d\gamma)&=&4(ab)(cd)-4(ac)(bd)+4(ad)(bc).
\end{eqnarray*}

\nowastrona

We also introduce the  {\em spin operators}
\[\sigma^{\mu\nu}:=\frac{\i}{2}[\gamma^\mu,\gamma^\nu].\]
In the Dirac representation
\begin{eqnarray}\nonumber\sigma^{0i}&=&\mat{0}{\i\sigma^i}{\i\sigma^i}{0},\\
\sigma^{ij}&=&\epsilon^{ijk}\mat{\sigma_k}{0}{0}{\sigma_k}.\label{sig}
\end{eqnarray}
The operators $\sigma^{\mu\nu}$ form a representation of the Lie algebra
$so(1,3)=spin(1,3)$. It is the infinitesimal version of the  representation 
\[Spin^\uparrow(1,3)\ni \tilde\Lambda\mapsto \tau(\tilde\Lambda).\]


\nowastrona

\subsubsection{Special solutions and Green's functions}

Note the identity
\begin{eqnarray}\nonumber
-(-\i\gamma\partial+m)(-\i\gamma\partial-m)&=&-\Box+m^2.
\label{D^2}
\end{eqnarray}

Therefore, if 
\[(-\Box+m^2)\zeta(x)=0,\]
then $(\i\gamma^\mu\partial_\mu+m)\zeta(x)$
 is a solution of the homogeneous
Dirac equation:
\[(-\i\gamma^\mu\partial_\mu+m)(\i\gamma^\mu\partial_\mu+m)\zeta(x)=0.\]

In particular, we  have {\em special solutions} of the homogeneous Dirac equation
\begin{eqnarray*}
S^{(\pm)}(x)&=&(\i\gamma\partial+m)D^{(\pm)}(x),\\
S(x)&=&(\i\gamma\partial+m)D(x),\end{eqnarray*}
where $D^{(\pm)}$ and $D$ are the special solutions of the Klein-Gordon equation introduced in Subsubsect. \ref{special}.
We have $\supp S\subset J$.

\nowastrona

If 
\[(-\Box+m^2)\zeta(x)=\delta(x),\]
then $(\i\gamma^\mu\partial_\mu+m)\zeta(x)$ is a {\em Green's function}
of the 
Dirac equation, that is
\[(-\i\gamma\partial+m)(\i\gamma\partial+m)\zeta(x)=\delta(x).\]
In particular, a special role is played by
the  Green functions
\begin{eqnarray*}
S^\pm(x)&=&(\i\gamma\partial+m)D^\pm(x),\\
S^{\rm c}(x)&=&(\i\gamma\partial+m)D^{\rm c}(x),\end{eqnarray*}
where $D^{\pm}$ and $D^{\rm c}$ are the Green's functions of the Klein-Gordon equation introduced  in Subsubsect. \ref{special}.
We have $\supp S^\pm\subset J^\pm$.

\nowastrona

The {\em Dirac propagators} satisfy the identities
\begin{eqnarray*}
S(x)=-S(-x)&=&S^{(+)}(x)+S^{(-)}(x)\\
&=&S^+(x)-S^-(x),\\
S^{(+)}(x)&=&S^{(-)}(-x),\\
S^+(x)=S^-(-x)&=&\theta(x^0)S(x),\\
S^-(x)&=&\theta(-x^0)S(x),\\
S^{\rm c}(x)=S^{\rm c}(-x)&=&\theta(x^0)S^{(-)}(x)-\theta(-x^0)S^{(+)}(x). 
\end{eqnarray*}

Recall that the bosonic causal Green's function in the momentum representation can be written as
\[D^{\rm c}(p)=\frac{1}{p^2+m^2-\i0}.\]
The Dirac causal Green's function  can be written in a similar way:
\begin{eqnarray}\notag
S^{\rm c}(p)&=&\frac{-\gamma p+m}{p^2+m^2-\i0}\\
&=&\frac{1}{\gamma p+m-\i\epsilon},\label{infini}
\end{eqnarray}
where $\i\epsilon$ is the shorthand for $\i0\,\sgn p\gamma$.

\subsubsection{Space of solutions}



\nowastrona

 We  set $\alpha_i=\gamma^0\gamma^i$, $i=1,2,3$, and
$\beta:=\gamma^0$. We obtain matrices satisfying
\begin{eqnarray*}
\beta^2=
\one,\ \ (\alpha_i)^2=\one,&& i=1,\dots,3;\\
\beta\alpha_i+\alpha_i\beta=0,\ \ 
\alpha_i\alpha_j+\alpha_j\alpha_i=0,&& 1\leq i< j\leq 3;\\
\beta^*=\beta,\ \ \alpha_i^*=\alpha_i,&& i=1,\dots,3.
\end{eqnarray*}
In the Dirac representation we have
\[\beta=\mat{1}{0}{0}{-1},\ \ 
\vec\alpha=\mat{0}{\vec\sigma}{\vec\sigma}{0}.\]

Using $\vec\alpha,\beta$ we
can rewrite the Dirac equation in the 
 form of an
evolution equation:
\[\i \partial_t\zeta(t,\vec x)={\mathbb D}\zeta,\ \ {\mathbb D}:=
\vec\alpha \vec p+m\beta.\]
Note that $\dd$ is essentially self-adjoint on $C_{\rm c}^\infty(\rr^{3},\cc^4)$.

\nowastrona

The following
 theorem describes the {\em Cauchy problem} for the Dirac equation:
\bet Let
$\vartheta\in C_{\rm c}^\infty(\rr^{3},\cc^4)$. Then there exists a
unique $\zeta\in C_\sc^\infty(\rr^{1,3})$ that solves the Dirac equation
with initial conditions $\zeta(0,\vec x)=\vartheta(\vec x)$.
It satisfies $\supp\zeta\subset J(\supp \vartheta)$
and 
  is given by
\begin{eqnarray}\label{cauch}
\zeta(t,\vec x)
&=&-\i\int_{\rr^3} S(t,\vec x-\vec y)\beta \vartheta(\vec y)\d\vec{y}.
\end{eqnarray}
\eet

\nowastrona

Let $\cW_\D$ be the space of {\em space-compact solutions of the
Dirac equation}, that is $\zeta\in C_\sc^\infty(\rr^{1,3},\cc^4)$
satisfying $(-\i\gamma^\mu\partial_\mu+m)\zeta=0.$

For $\zeta_1,\zeta_2\in C^\infty(\rr^{1,3},\cc^4)$ set
\beq j^\mu(\zeta_1,\zeta_2,x):=
\bar{\zeta_1(x)}\beta\gamma^\mu\zeta_2(x).\label{prfo}\eeq
We easily check that
\[\partial_\mu
j^\mu(x)=\bar{(-\i\gamma\partial+m)\zeta_1(x)}\beta\zeta_2(x)
-\bar{\zeta_1(x)}\beta(-\i\gamma\partial+m)\zeta_2(x). \]
Therefore, if $\zeta_1,\zeta_2\in\cW_\D$,  then $j^\mu$ is a {\em 
conserved 4-current}:
\[\partial_\mu j^\mu(x)=0.\]
\nowastrona

For $\zeta_1,\zeta_2\in\cW_\D$,  
 the flux of $j^\mu$ 
does not depend on the choice of 
 a Cauchy
 hypersurface $\cS$. It defines a  scalar product on
$\cW_\D$, which will have two  optional symbols:
\[\bar\zeta_1\cdot\zeta_2=(\zeta_1|\zeta_2):=
\int_\cS j^\mu(\zeta_1,\zeta_2,x)\d s_\mu(x).\]
In terms of the Cauchy data this scalar product coincides with the
natural scalar product on $L^2(\rr^3,\cc^4)$:
\[\bar\zeta_1\cdot\zeta_2=
\int \bar{\zeta_1(t,\vec x)}\zeta_2(t,\vec x)\d\vec{x}.\]

The group $\rr^{1,3}\rtimes Spin^\uparrow(1,3)$, acts unitarily on $\cW_{\D}$ by
\[r_{(y,\Lambda)}\zeta(x):=\tau(\tilde\Lambda)\zeta\left((y,\Lambda)^{-1}x\right).\]


\nowastrona

We can also parametrize solutions of the Dirac equation by
{\em space-time functions}. In fact,
for any $f\in C_{\rm c}^\infty(\rr^{1,3},\cc^4)$,
let us write
\[S*f(x):=\int S(x-y)f(y)\d x.\]

\bet\ben\item
 For any $f\in C_{\rm c}^\infty(\rr^{1,3},\cc^4)$,
 $S*f\in\cW_\D$.\item
 Every element of
$\cW_\D$ is of this form.
\item $\bar{S*f_1}\cdot S*f_2
=\int\int
 \bar{f_1(x)}\beta S(x-y)f_2(y)\d x\d y.$
\item If $\supp f_2\times\supp f_2$, then 
\[\bar{S*f_1}\cdot S*f_2=0.\]
\een\eet



\nowastrona

\subsubsection{Classical fields}

We will also consider the space dual to $\cW_\D$, denoted $\cW_\D^\t$.
 In particular, for $x\in\rr^{1,3}$,
$\psi(x),  \psi^*(x)$ will denote the functionals on $\cW_\D$ with values in
 $\cc^4$, called
{\em classical Dirac fields}, given by
\[ 
 \langle \psi(x)|\zeta\rangle:=\zeta(x),\ \ \ 
\langle\psi^*(x)|\zeta\rangle:=\bar{\zeta(x)}
.\]
By (\ref{cauch}),
\[\psi(t,\vec x)=-\i\int S(t,\vec x-\vec y)\beta\psi(0,\vec y)\d\vec{y}.\]
It is convenient to introduce the {\em Dirac conjugate} of the field $\psi$:
\[\tilde\psi(x):=\beta\psi^*(x).\]
(In a large part of the physics literature, $\tilde\psi$ is denoted $\bar\psi$.)

On $\cW_{\D}^\t$ we have the group action  $\rr^{1,3}\rtimes Spin^\uparrow(1,3)\ni
(y,\tilde\Lambda)\mapsto r_{(y,\tilde\Lambda)}^{\t-1}$:
\[r_{(y,\tilde\Lambda)}^{\t-1}\psi(x)=\tau(\tilde\Lambda^{-1})\psi( \Lambda x+y).\]

\nowastrona

\subsubsection{Smeared fields}

We can use the scalar product to pair  solutions.
For $\zeta\in\cW_\D$, the corresponding {\em spatially smeared fields}
are
 the functionals on $\cW_\D$ given by
\begin{eqnarray*}
\langle\psi\lpar\zeta\rpar|\rho\rangle&:=&\bar\zeta\cdot\rho,\\
\langle\psi^*\lpar\zeta\rpar|\rho\rangle&:=&\zeta\cdot \bar\rho,\ \ \ 
\rho\in\cW_\D.
\end{eqnarray*}
Clearly, for any $t$
\begin{eqnarray*}
\psi\lpar\zeta\rpar&=&\int\bar{\zeta(t,\vec x)}\psi(t,\vec x)\d\vec x,\\
\psi^*\lpar\zeta\rpar&=&\int\zeta(t,\vec x)\psi^*(t,\vec x)\d\vec x.
\end{eqnarray*}

\nowastrona

For  $f\in C_{\rm c}^\infty(\rr^{1,3},\cc^4)$, the corresponding {\em
  space-time smeared fields} are given by
\begin{eqnarray*}
\psi[f]&:=&\int \bar{f(x) }\psi(x)\d x=\psi\lpar S*f\rpar,\\
\psi^*[f]&:=&\int f(x) \psi^*(x)\d x=\psi^*\lpar S*f\rpar.
\end{eqnarray*}

\nowastrona

\subsubsection{Diagonalization of the equations of motion}

Let us use the Dirac representation, denoting elements of $\cc^4$ with $\left[\begin{array}{c}\zeta_\uparrow\\ \zeta_\downarrow\end{array}\right]$,
where $\zeta_\uparrow,\zeta_\downarrow\in\cc^2$.  Ater the space-time Fourier transformation
the Dirac equation becomes
\begin{eqnarray*}
-p^0\zeta_\uparrow+\vec\sigma\vec p\zeta_\downarrow+m\zeta_\uparrow&=&0,\\
p^0\zeta_\downarrow-\vec\sigma\vec p\zeta_\uparrow+m\zeta_\downarrow&=&0.
\end{eqnarray*}
This can be rewritten as
\begin{eqnarray*}
\zeta_\uparrow&=&-\frac{\vec\sigma\vec p}{-p^0+m}\zeta_\downarrow,\\
\zeta_\downarrow&=&\frac{\vec\sigma\vec p}{p^0+m}\zeta_\uparrow.\end{eqnarray*}
Using $(\vec\sigma\vec p)^2=\vec p^2$ we obtain
\[-(p^0)^2+\vec p^2+m^2=0.\]

Set $E(\vec p):=\sqrt{\vec p^2+m^2}$, so that $p=(\pm E(\vec p),\vec p)$.
Define
\[\chi_+:=\left[\begin{array}{c}1\\0\end{array}\right],\ \ \ \chi_-:=
\left[\begin{array}{c}0\\ 1\end{array}\right].
\]Traditionally, one often introduces the following  spinors:
\begin{eqnarray}\notag
u(p,\pm1/2)&=&
\frac{\sqrt{E+m}}{\sqrt{2E}}\left[\begin{array}{c}\chi_\pm
\\\frac{\vec{\sigma}\vec{p}}{E+m}\chi_\pm\end{array}\right],\ \ p^0=E(\vec
p)>0;\\
u(p,\pm1/2)&=&
\frac{\sqrt{E+m}}{\sqrt{2E}}\left[\begin{array}{c}
\frac{\mp\vec{\sigma}\vec{p}}{E+m}\chi_\pm\\
\pm\chi_\pm\end{array}\right],\ \ p^0=-E(\vec p)<0.
\label{planewave}\end{eqnarray}
Note that
\begin{eqnarray*}
\left(u(p,s)|u(p,s')\right)&=&\delta_{s,s'},\\
\left(u(p,s)|u(-p,s')\right)&=&0.
\end{eqnarray*}

\nowastrona
The basic {\em plane waves} are defined as
\begin{eqnarray*}
|p,s)=\frac{1}{\sqrt{(2\pi)^3}}u(p,s)\e^{\i px}.\end{eqnarray*}
By writing $(p,s|$, as usual, we will imply the complex conjugation.
We have
 \begin{eqnarray*}
(p,s|p',s')&=&\delta(\vec p-\vec p')\delta_{s,s'},\ \   \ \sgn
(p^0p^{'0})>0 ,
\\
(p,s|p',s')&=&0,\ \ \ \ \ \ \ \ \ \ \ \  \ \sgn(
p^0p^{'0})<0 .
\end{eqnarray*}



\nowastrona

Note that plane waves
 diagonalize simultaneously
the  Dirac Hamiltonian  $\dd$, the      momentum $\vec
p=-\i\vec\partial$
and the scalar product:
\begin{eqnarray*}
\dd|p,s)&=&p^0|p,s),\\
-\i\vec\partial|p,s)&=&\vec p|p,s),
\end{eqnarray*}
\[\bar\zeta_1\cdot\zeta_2=\sum_{s}
\int\big((\zeta_1|p,s)
  (p,s|\zeta_2)
+(\zeta_1|-p,-s)(-p,-s|\zeta_2)\big)
\d\vec{p}.
\]
In addition, positive frequency plane waves diagonalize the ``upper spin in the 3rd direction'' and 
negative frequency plane waves diagonalize the ``lower spin operator in the 3rd direction'': 
\begin{eqnarray*}
\frac12
\left[\begin{array}{cc}\sigma_3&0\\0&0\end{array}\right] |p,s)&=&s|p,s),\ \ \sgn p^0>0,\\
\frac12\left[\begin{array}{cc}0&0\\0&\sigma_3\end{array}\right] |p,s)&=&s|p,s),\ \ \sgn p^0<0.
\end{eqnarray*}

\subsubsection{Plane wave functionals}

{\em Plane wave functionals} are  the functionals defined by plane
waves. One could doubt whether they deserve a special notation. In
the bosonic case the situation was slightly less trivial, because the
pairing was given by the symplectic form. For fermions the pairing is
given  by the scalar product, hence it is straightforward. Anyway,
special notation for plane wave functionals is partly motivated as a
preparation for quantization.

Let $p\in\rr^{1,3}$ with $p^0>0$.
 Anticipating the quantization, 
we will use  different notation for
positive and negative frequencies: 
\begin{eqnarray}\label{poij1}
a(p,s)&:=&\psi\lpar |p,s)\rpar\\ \notag
&=&
\int\frac{\d\vec{x}}{\sqrt{(2\pi)^3}}
\bar{u(p,s)}\e^{-\i \vec p \vec x}\psi(0,\vec x),\\ \label{poij2}
b^*(p,s)&:=&\psi\lpar |-p,-s)\rpar\\ \notag
&=&
\int\frac{\d\vec{x}}{\sqrt{(2\pi)^3}}
\bar{u(-p,-s)}\e^{\i \vec p \vec x}\psi(0,\vec x).
\end{eqnarray}
We have
\begin{eqnarray*}\psi(x)&=&\sum_{s}
\int\frac{\d\vec{p}}{\sqrt{(2\pi)^3}} \left(u(p,s)\e^{\i px} 
a( p,s)+u(-p,-s)\e^{-\i px} 
b^*(p,s)\right)
\\
&=&
\sum_{s}
\int\d\vec{p} \big(|p,s)
a( p,s)+|-p,-s)b^*(p,s)\big).
\end{eqnarray*}

\nowastrona

\nowastrona

\subsubsection{Positive and negative frequency subspaces}
\label{posi-dirac}

We define
\begin{eqnarray*}
\cW_\D^{(+)}&:=&\{\zeta\in\cW_\D\ :\ (p,s|\zeta)=0,\ \ p^0<0\},\\
\cW_\D^{(-)}&:=&\{\zeta\in\cW_\D\ :\ (p,s|\zeta)=0,\ \ p^0>0\}.\end{eqnarray*}
Every $\zeta\in\cW_\D$ can be uniquely decomposed as
$\zeta=\zeta^{(+)}+\zeta^{(-)}$ with $\zeta^{(\pm)}\in
\cW_\D^{(\pm)}$.

On   $\cW_\D^{(+)}$ we keep the old scalar product:
\begin{eqnarray*}
(\zeta_1^{(+)}|\zeta_2^{(+)})
&=&
\sum_s\int(\zeta_1^{(+)}|p,s)(p,s|\zeta_2^{(+)})\d\vec{p}.
\end{eqnarray*}
We set $\cZ_\D^{(+)}$ to be the completion of 
$\cW_\D^{(+)}$ in this scalar product. 

Instead of $\cW_\D^{(-)}$ for quantization we will use the
corresponding
 complex
conjugate space denoted $\bar\cW_\D^{(-)}$ and equipped
with the scalar product
\begin{eqnarray*}
(\bar\zeta_1^{(-)}|\bar\zeta_2^{(-)}):=
\bar{(\zeta_1^{(-)}|\zeta_2^{(-)})}&=&
\sum_s\int(\bar\zeta_1^{(-)}|\bar{-p,-s})
  (\bar{-p,-s}|\bar\zeta_2^{(-)})\d\vec{p}.
\end{eqnarray*}
We set $\cZ_\D^{(-)}$ to be the completion of 
$\bar\cW_\D^{(-)}$ in this scalar product.

The action of $\rr^{1,3}\rtimes  Pin^\uparrow(1,3)$  leaves $\cZ_\D^{(+)}$
and $\cZ_\D^{(-)}$  invariant.

\nowastrona
\subsubsection{Spin averaging}

 $\frac{1}{2m}(\mp p\gamma+m)$ are the projections onto
the positive and negative energy states resp. 
With $E=E(\vec p)=p^0>0$, 
we have the identities
\begin{eqnarray*}
\sum_s u(p,s)\tilde u(p,s)&=&\frac{1}{2E}
\left[\begin{array}{cc}
E+m&-\vec\sigma \vec p\\\vec\sigma\vec
p&-E+m\end{array}\right]\\&=&\frac{ -p\gamma+m}{2E}
 =\ \frac{m}{E}\Lambda_+ ,\\
\sum_s u(-p,-s)\tilde u(-p,-s)
&=&\frac{1}{2E}
\left[\begin{array}{cc}
E-m&-\vec\sigma\vec p\\\vec\sigma\vec p&-E-m\end{array}\right]\\
&=&\frac{-p\gamma-m}{2E}\ = \  -\frac{m}{E}\Lambda_-.
\end{eqnarray*}

\nowastrona
In the following 
{\em spin averaging identities} due to H.B.C.Casimir,
which are useful in computations of scattering
cross-sections, 
the trace involves only the spin degrees of freedom:
\begin{eqnarray*}\sum_{s^+,s^-}
\left|\tilde u(p^+,s^+)B u(p^-,s^-)\right|^2
&=&\frac{\Tr
\tilde B(- p^+\gamma+m) B(- p^-\gamma+m)}{4E^+E^-}
,\\
\sum_{s^+,s^-}
\left|\tilde u(-p^+,-s^+)B u(-p^-,-s^-)\right|^2
&=&\frac{\Tr
\tilde B(-p^+\gamma-m) B(- p^-\gamma-m)}{4E^+E^-},\\
\sum_{s^+,s^-}\left|\tilde u(-p^+,-s^+)B u(p^-,s^-)\right|^2
&=&\frac{\Tr
\tilde B(- p^+\gamma-m) B(- p^-\gamma+m)}{4E^+E^-}
,\\\sum_{s^+,s^-}\left|\tilde u(p^+,s^+)B u(-p^-,-s^-)\right|^2
&=&\frac{\Tr
\tilde B(- p^+\gamma+m) B(- p^-\gamma-m)}{4E^+E^-}
,\end{eqnarray*}
where $B$ is an arbitrary operator on the spinor space and
$\tilde B= \beta B^*\beta$
is its pseudo-Hermitian conjugate.

If we specify $B=\beta$, then
\begin{eqnarray*}\sum_{s^+,s^-}
\left|\bar{u(p^+,s^+)} u(p^-,s^-)\right|^2
&=&
\sum_{s^+,s^-}
\left|\bar{u(-p^+,-s^+)} u(-p^-,-s^-)\right|^2\\=\
\frac{E^+E^-+\vec{p^+}\vec{p^-}+m^2}{E^+E^-}&=&
\frac{(E^++E^-)^2-|\vec{ p_+}-\vec
{ p_-}|^2}{2E^+E^-}
,\\
\sum_{s^+,s^-}\left|\bar{u(-p^+,-s^+)} u(p^-,s^-)\right|^2
&=&
\sum_{s^+,s^-}\left|\bar{u(p^+,s^+)} u(-p^-,-s^-)\right|^2\\=\
\frac{E^+E^-+\vec{p^+}\vec{p^-}-m^2}{E^+E^-}&=&
\frac{-(E^+-E^-)^2+|\vec{ p_+}+\vec
{ p_-}|^2}{2E^+E^-}.
\end{eqnarray*}

\nowastrona

\subsubsection{Quantization}
\label{sec-q-dir}

We would like to describe the quantization of the Dirac
equation. As usual, we will use the ``hat'' to denote  quantized objects. 

We will use the formalism of quantization of charged fermionic systems \cite{DeGe}.

\nowastrona
We want to construct $(\cH, 
 \hat H,\Omega)$ satisfying the standard requirements of QM (1)-(3)  and a
distribution \beq
\rr^{1,3}\ni x\mapsto \hat\psi(x)\eeq
with values  in $\cc^4\otimes B(\cH)$
such that the following conditions are true:
 \ben\item $
(-\i\gamma\partial+m)\hat\psi(x)=0$;
\item
$[\hat\psi_a(0,\vec x),\hat\psi_b^*(0,\vec y)]_+=\delta_{ab}\delta(\vec x-\vec
  y),\ \ \ [\hat\psi_a(0,\vec x),\hat\psi_b(0,\vec y)]_+=0$;
\item 
$\e^{\i t\hat H}\hat\psi(x^0,\vec x)\e^{-\i t\hat H}=\hat\psi(x^0+t,\vec x);$
\item 
$\Omega$ is cyclic for $\hat\psi(x)$,  $\hat\psi^*(x)$.\een

\nowastrona

The above problem has a  solution unique up to a unitary equivalence,
which we describe below.

We set \[\cH:=\Gamma_\a(\cZ_\D^{(+)}\oplus\cZ_\D^{(-)}).\]
Creation/annihilation operators  for the particle space $\cZ_\D^{(+)}\simeq L^2(\rr^3,\cc^2)$ are denoted with the letter $a$ and for the antiparticle space $\cZ_\D^{(-)}\simeq L^2(\rr^3,\cc^2)$ with the letter $b$. Thus, for 
 $p$ on the
mass shell and $s=\pm\frac12$,  using physicist's notation on the left and mathematician's on the right, creation operators for particles/antiparticles are written as
\begin{eqnarray}
 \hat a^*(p,s)&=&\hat a^*\big(|p,s)\big),\label{qrt2=1}\\
\hat b^*(p,s)&=&\hat b^*\big(\bar{|-p,-s)}\big).\label{qrt2+1}\end{eqnarray}
 $\Omega$ is the Fock
vacuum. 
\nowastrona
The quantum field is
\begin{eqnarray*}
\hat\psi(x)&:=&
\sum_s\int\frac{\d\vec{p} }{\sqrt{(2\pi)^3}}
\left(u( p,s)\e^{\i px}\hat a( p,s)+
u( -p,-s)\e^{-\i px}\hat b^*( p,s)\right)
.\end{eqnarray*}
The  quantum Hamiltonian and  momentum are
\begin{eqnarray}
\hat H&=&\int\sum_s \left(\hat a^*( p,s)\hat a(p,s)+\hat b^*( p,s)\hat b(p,s)\right)E(\vec
p)
\d\vec{p},\label{dirac1}\\
\vec {\hat P}&=&\int\sum_s \left(\hat a^*( p,s)\hat a(p,s)+\hat b^*( p,s)\hat
  b(p,s)\right)
\vec p\d\vec{p}.\label{dirac2}
\end{eqnarray}
We also have the {\em charge operator}
\begin{eqnarray}\label{qno}
\hat Q&:=&
\sum_s\int \left(\hat a^*(\vec p,s)\hat a(\vec p,s)
-\hat b^*(\vec p,s)\hat b(\vec p,s)\right)\d\vec{p}.\label{dirac3}\end{eqnarray}

The whole group
$\rr^{1,3}\rtimes  Spin^\uparrow(1,3)$ acts unitarily on $\cH$. 
Moreover, if we set
 $\tilde{\hat\psi}(x):=\beta\hat\psi^*(x)$, then
\beq [\hat\psi_a(x),\tilde{\hat
 \psi}_b(y)]_+=S_{ab}(x-y),\ \ \ 
[\hat\psi_a(x),\hat\psi_b(y)]_+=0.\label{poij}\eeq
%
%

\nowastrona

We have
\begin{eqnarray*}
(\Omega|\hat\psi_a(x)\tilde{\hat\psi}_b(y)\Omega)
&=& S_{ab}^{(+)}(x-y),\\
(\Omega|\T(\hat\psi_a(x)\tilde {\hat\psi}_b(y))\Omega)
&=&S_{ab}^{\rm c}(x-y).
\end{eqnarray*}

\nowastrona

For $f \in C_{\rm
  c}^\infty(\cO,\cc^4)$ we set
\begin{eqnarray*}\hat\psi[f]&:=&\int\bar{f(x)}\hat\psi(x)
\d x,\\
\hat\psi^*[f]&:=&\int\bar{f(x)}\hat\psi^*(x)
\d x.\end{eqnarray*}
We obtain an operator valued distribution satisfying  the Wightman
  axioms  with
$\cD:=\Gamma_\a^\fin(\cZ_\D^{(+)}\oplus \cZ_\D^{(-)})$.

For an open  set $\cO\subset \rr^{1,3}$ the field algebra  is defined as
\[\fF(\cO):=\{\hat\psi^*[f],\hat\psi[f]\ :\ f\in C_{\rm
  c}^\infty(\cO,\cc^4)\}''.\] The observable algebra $\fA(\cO)$ is the subalgebra of
$\fF(\cO)$ fixed by the automorphism 
\[B\mapsto\e^{\i\theta\hat
  Q}B\e^{-\i\theta\hat
 Q},\]
where $\hat Q$ will be defined in (\ref{qno}).
The nets of algebras $\fF(\cO)$ and $\fA(\cO)$, $\cO\subset\rr^{1,3}$,
 satisfy the Haag-Kastler axioms.

\nowastrona
\subsubsection{Quantization in terms of smeared fields}

There exists an alternative equivalent formulation of the quantization
program, which uses the smeared fields instead of point fields.
Instead of
(\ref{distro}) we look for an antilinear function
\[\cW_\D\ni\zeta\mapsto\hat\psi\lpar\zeta\rpar\]
with values in bounded operators such that
\ben\item
$
[\hat\psi\lpar\zeta_1\rpar,\hat\psi^*\lpar\zeta_2\rpar]_+
=\bar\zeta_1\cdot\zeta_2,$\hspace{5ex}
${}[\hat\psi\lpar\zeta_1\rpar,\hat\psi\lpar\zeta_2\rpar]_+
=0.$
\item
$\hat\psi\lpar r_{(t,\vec0)}\zeta\rpar
=\e^{\i t\hat H}\hat\psi\lpar\zeta\rpar\e^{-\i t\hat H}.$
\item 
$\Omega$ is cyclic for
$ \hat\psi\lpar\zeta\rpar$, $ \hat\psi^*\lpar\zeta\rpar$.\een

\nowastrona

One can pass between these two kinds of  quantization  by
\beq
\hat\psi\lpar\zeta\rpar =\int\bar{\zeta(t,\vec x)}\hat\psi(t,\vec x)\d\vec x.
\eeq

\nowastrona
\subsubsection{Dirac sea quantization}

When we quantized a fermionic field we demanded
that
the quantum Hamiltonian $\hat H$ be positive. In the bosonic case this
condition can be  dropped if we start from a positive classical
Hamiltonian $H$. Usually  this suffices to guarantee  the positivity of
$\hat H$. (If we start from a classical Hamiltonian that is not
positive definite, the bosonic quantum counterpart has no chances of
being positive).

Suppose now that we drop the positivity requirement of $\hat H$ in the
fermionic case. Then we have  many possible quantizations. Among them
one is distinguished -- it is just the usual {\em  second
  quantization}. It means that we
 consider the  antisymmetric Fock space
$\Gamma_\a(\cW_\D^\cpl)$, where
 $\cW_\D^\cpl$ denotes the completion of
  $\cW_\D$ in its natural
scalar product.

The Hilbert space $\cW_\D^\cpl$ is equipped with a distinguished family of commuting self-adjoint
operators: the Dirac operator $\dd$ and the momentum operator
$-\i\vec\partial$. We can second quantize them using the operation
  $\d\Gamma$ 
obtaining the operators on $\Gamma_\a(\cW_\D^\cpl)$, the Hamiltonian and the momentum
\begin{eqnarray}H&=&\d\Gamma(\dd),\label{naive1}\\
\vec P&=&\d\Gamma(-\i\vec\partial).\label{naive2}
\end{eqnarray}
The number operator will be rebaptized as the {\em charge} and  denoted
\[Q=\d\Gamma(\one).\]
(Let us stress that we do not use ``hats'' in the above notation).

Let us reinterpret $\psi^*(x)$/$\psi(x)$ (without ``hats'')
 as the creation/annihilation
operators on the space  $\Gamma_\a(\cW_\D^\cpl)$. 
As in  (\ref{poij}), they satisfy
\beq [\psi_a(x),\tilde{
 \psi}_b(y)]_+=S_{ab}(x-y),\ \ \ 
[\psi_a(x),\psi_b(y)]_+=0.\label{look}\eeq 
The plane wave functionals $a(p,s)$, $a^*(p,s)$, $b^*(p,s)$, $b(p,s)$ defined as in
 (\ref{poij1}) and   (\ref{poij2}) in terms of $\psi(x)$, $\psi^*(x)$, can be used to diagonalize the Hamiltonian,  momentum and charge
\begin{eqnarray}
 H&=&\int\sum_s \left( a^*( p,s) a(p,s)-
 b( p,s) b^*(p,s)\right)E(\vec
p)
\d\vec{p},\label{diracsea1}\\
\vec {P}&=&\int\sum_s \left( a^*( p,s) a(p,s)- b( p,s)
  b^*(p,s)\right)
\vec p\d\vec{p},\label{diracsea2}\\
Q&=&\int\sum_s \left( a^*( p,s) a(p,s)+ b( p,s)
  b^*(p,s)\right)
\vec p\d\vec{p}.\label{diracsea3}
\end{eqnarray}

The vacuum of  $\Gamma_\a(\cW_\D^\cpl)$ is annihilated by $\psi(x)$, hence also by $a(p,s)$ and $b^*(p,s)$.
 It is
 the state of the lowest charge
possible.
 Therefore, it 
 will be called
the {\em bottom of the Dirac sea}. We
will call the above described procedure the {\em Dirac  sea quantization}.

The reader should compare the formulas for $H$
(\ref{diracsea1}), $\vec P$ (\ref{diracsea2}) and $Q$ (\ref{diracsea3}) with 
$\hat H$ (\ref{dirac1}), $\vec{\hat P}$ (\ref{dirac2}) and $\hat Q$ (\ref{dirac3}). They  differ only by  the order of a part of field operators. So formally they coincide modulo  an (infinite) additive constant.

The usual quantization, called the {\em positive energy
  quantization} 
and the Dirac sea quantization are just
two inequivalent representations of canonical anticommutation relations.
If $\cW_\D$ had
a finite dimension (which can be accomplished by applying both an
infrared and ultraviolet cutoff), then the Dirac sea quantization
would be unitarily equivalent with the positive energy quantization by the
procedure invented by Dirac and called often {\em filling the Dirac
  sea}. The Hamiltonians $H$ and $\hat H$, and as we see later, the
charges $Q$ and $\hat Q$ would differ by a finite constant. The
momenta $\vec P$ and $\vec{\hat P} $ would  coincide.

\subsubsection{Fermionic Hamiltonian formalism}

Bosonic quantum fields can be interpreted as a quantization of a
classical system. In the Hamiltonian (on-shell) formalism this system is 
described by an appropriate symplectic space. 
In the charged case, the symplectic space can be viewed as a complex
space and instead of the symplectic structure it is natural to consider an
appropriate Hermitian form. The spaces $\cY_\KG$ and $\cW_\KG$ were
examples of such spaces.
Symmetries
are described by symplectic transformations. The dynamics is generated
by a
(classical) Hamiltonian -- a function on the symplectic space.

An important element of the Hamiltonian formalism is the
``algebra of classical observables'' -- the  commutative algebra of
 functions on the symplectic space equipped with the Poisson
bracket. 
One can ask whether there exists an
analogous structure behind fermionic quantum fields.

Clearly, the space $\cW_\D$, which is equipped with a scalar product, is
the obvious fermionic analog of a (complex) symplectic space from the
bosonic case. 
 The fermionic
analog of the ``algebra of classical observables'' considered in the
literature, eg.  \cite{Sr}, is the $\zz_2$-graded algebra of operators on
$\Gamma_\a(\cW_\D^\cpl)$ equipped with the  {\em graded commutator}.

 The space $\Gamma_\a(\cW_\D^\cpl)$
is equipped with the {\em fermionic parity} operator, which we denote by $I:=(-\one)^Q$. 
An  operator $A$  satisfying $IAI=\pm A$ will be called
even/odd. Operators that are 
either even or odd will be called {\em homogeneous}. If
$A$ is homogeneous
we will write $|A|=0$ if $A$ is even and
$|A|=1$ if $A$ is odd.
The analog of the Poisson bracket is the  graded commutator:
\beq \{A,B\}:=AB-(-1)^{|A|\,|B|}BA.\label{poio}\eeq

Note that $\psi(x)$,  $\psi^*(x)$ are odd operators and for such operators
$\{\cdot,\cdot\}$ coincides with the anticommutator. Thus, to make (\ref{look}) look ``classical'', we can replace $[\cdot,\cdot]_+$ with $\{\cdot,\cdot\}$ in this identity.

Note that the  ``classical'' version of the Dirac theory has a quantum character.
In particular, the ``classical fermionic algebra''  is an algebra of operators on a Hilbert space and
symmetries are unitary. Nevertheless, one has a far reaching analogy with
the usual commutative classical mechanics.

\subsubsection{Fermionic Lagrangian  formalism}

The  Lagrangian formalism in the bosonic case    involves the commutative
algebra of functions on the space-time (the ``off-shell
formalism'').
In the literature one can also find its fermionic analog.
 The 
 fermionic
Lagrangian formalism involves the {\em Grassmann algebra} generated by
{\em anticommuting} functions on space-time. This algebra is generated
by anticommuting fields
$\rr^{1,3}\ni x\mapsto \psi(x),\psi^*(x)$.
(Thus, the anticommutators of the off-shell $\psi(x)$, $\psi^*(y)$ are always zero, unlike in the on-shell formalism).

Note that every Grassmann algebra, besides multiplication, is equipped
with the  {\em integral} (called sometimes the {\em Berezin
  integral}),
 the {\em left} and the {\em right
 derivative}. We will use the left derivative as the standard one (see eg. \cite{DeGe}).

The
  {\em Lagrangian density} is an even
element of this Grassmann algebra:
\begin{eqnarray*}
\cL(x)&=&-\12\left(\tilde\psi(x)\gamma^\mu(-\i\partial_\mu)\psi(x)+\i\partial_\mu\tilde\psi(x)\,\gamma^\mu\psi(x)\right)-m\tilde\psi(x)\psi(x),
\end{eqnarray*}
where as usual  $\tilde\psi(x)=\beta\psi^*(x)$. 
The Euler-Lagrange equations 
\beq
\partial_{\tilde\psi}\cL-
\partial_\mu\frac{\partial \cL}{\partial\tilde\psi_{,\mu}}=0,\ \ 
\partial_{\psi}\cL-
\partial_\mu\frac{\partial \cL}{\partial\psi_{,\mu}}=0
\label{euler-lagrange1c}\eeq
 yield the Dirac equation.

One can define the {\em stress-energy tensor}
\begin{eqnarray}\nonumber
\cT^{\mu\nu}(x)&:=&-\frac{\partial \cL(x)}{\partial\psi_{,\mu}(x)}\partial^\nu\psi(x)-
\frac{\partial \cL(x)}{\partial\tilde\psi_{,\mu}(x)}\partial^\nu\tilde\psi(x)
+g^{\mu\nu}\cL(x)\\
&=&
\frac{1}{2}\Big(\tilde\psi(x)\gamma^\mu(-\i\partial^\nu)\psi(x)
+\i\partial^\nu\tilde\psi(x)\gamma^\mu\psi(x)\Big)\notag\\&&
-
g^{\mu\nu}\Big(\frac{1}{2}\big(\tilde\psi(x)\gamma(-\i\partial)\psi(x)+
\i\partial\tilde\psi(x)
\gamma\psi(x)\big)+m\tilde\psi(x)
\psi(x)\Big).\notag
\end{eqnarray}
It is conserved on shell
\[\partial^\mu \cT_{\mu\nu}(x)=0.\]
The components of the stress-energy tensor with the first temporal coordinate
are called the {\em Hamiltonian density} and {\em momentum
  density}.
\begin{eqnarray*}
\cH(x)& :=& \cT^{00}(x)\\
&=&\frac{1}{2}\big(\psi^*(x)\vec\alpha(-\i\vec\partial)\psi(x)
+\i\vec\partial\psi^*(x)\vec\alpha\psi(x)\Big)+m\psi^*(x)\beta\psi(x),
\label{diraco1}\\
\cP^i(x)&:=&\cT^{0i}(x)\\
&=&
-\frac{1}{2}\big(\psi^*(x)(-\i\partial^i)\psi(x)+\i\partial^i\psi^*(x)\psi(x)\big).\label{diraco2}
\end{eqnarray*}

Note that in 
(\ref{diraco1}) and (\ref{diraco2}) we put $\psi^*$ on the left and $\psi$ on the right. This is the Wick ordering for the Dirac sea quantization, which can be called the {\em charge Wick ordering}. 
The {\em Hamiltonian and  momentum} defined from these
densities coincide with the operators defined by the Dirac sea second
quantization (\ref{naive1}), (\ref{naive2}):
\begin{eqnarray*}
H&=&\int \cH(t,\vec x)\d\vec x,\\
 \vec P&=&\int \vec\cP(t,\vec x)\d\vec x.\end{eqnarray*}

\nowastrona

\subsubsection{Classical 4-current}

The Lagrangian is invariant w.r.t. the $U(1)$ symmetry $\psi\mapsto\e^{-\i\theta}\psi$. The Noether 4-current associated to this symmetry is
the {\em 4-current}, defined as
\begin{eqnarray*}\cJ^\mu(x)&:=&
\i\Big(\tilde\psi(x)\frac{\partial\cL(x)}{\partial\tilde\psi_{,\mu}}
-\frac{\partial\cL(x)}{\partial\psi_{,\mu}}\psi(x)\Big)\\
&=&
\tilde\psi(x)\gamma^\mu\psi(x)
.\end{eqnarray*} 
It is conserved on shell and self-adjoint:
\begin{eqnarray}
\partial_\mu \cJ^\mu(x)&=&0,\label{po1}\\
\cJ^\mu(x)^*&=&\cJ^\mu(x).\label{po2}\end{eqnarray}

The  sesquilinear form  given by $\cJ$ coincides with (\ref{prfo}):
\begin{eqnarray*}
\bar\zeta_1\cJ^\mu(x)\zeta_2&=&  j^\mu(\bar\zeta_1,\zeta_2,x)\\
\\
&=& \bar{\zeta_1(x)}\beta\gamma^\mu\zeta_2(x),\  \ \zeta_1,\zeta_2\in\cW_\D.\end{eqnarray*}

The {\em current} or the spatial part  of 4-current can be expressed in terms of the
$\alpha$ matrices:
\[\vec\cJ(x)= \psi^*(x)\vec\alpha\psi(x).\]
The 0th component of the 4-current is called the {\em charge density}
\[\cQ(x):=\cJ^0(x)=
\psi^*(x)\psi(x).\] 
The {\em  charge} is
\begin{eqnarray*}
Q&:=&\int \cQ(t,\vec x)\d\vec x\\
&=&
\sum_s\int \big( a^*(\vec p,s) a(\vec p,s)
+ b(\vec p,s)b^*(\vec p,s)\big)\d\vec{p}.\end{eqnarray*}

$x\mapsto \cQ(t,\vec x)$ is a well defined distribution with values in
operators on
 space $\Gamma_\a(\cW_\D^\cpl)$. We have the relations
\begin{eqnarray}\nonumber
\{  \cQ(t,\vec x), \psi(t,\vec y)\}&=&- \psi(t,\vec y)\delta(\vec
x-\vec y),\\ {}\nonumber
\{  \cQ(t,\vec x), \psi^*(t,\vec y)\}&=& \psi^*(t,\vec y)\delta(\vec
x- \vec y),\\{} \label{westill-}
\{  \cQ(t,\vec x),  \cQ(t,\vec y)\}&=&0,
\end{eqnarray}
where the  bracket coincides now with the commutator, since
$\cQ$ is even.

For $\chi\in C_{\rm c}^\infty(\rr^3,\rr)$, let $\alpha_\chi$ denote the
 $*$-automorphism of the algebra of operators on $\cW_\D$
 defined by
\begin{eqnarray}\notag\alpha_\chi(\psi(0,\vec x))&:=&\e^{-\i\chi(\vec
    x)}\psi(0,\vec x).
\label{gaugo5}\end{eqnarray}
Obviously, 
\begin{eqnarray}\notag\alpha_\chi(\psi^*(0,\vec x))&=&\e^{\i\chi(\vec
    x)}\psi^*(0,\vec x).
\label{gaugo15}\end{eqnarray}
$\alpha_\chi$ is called
the {\em gauge transformation} at time $t=0$ corresponding to $\chi$.
Set
\beq
 Q(\chi)=\int\chi(\vec x)\cQ(0,\vec x) \d\vec
x.\label{illo-5}\eeq It can be used to implement the corresponding  gauge
transformation:
 \[\alpha_\chi(B)=\e^{\i  Q(\chi)}B\e^{-\i  Q(\chi)}.\]


\subsubsection{Quantum 4-current}

Let us try to introduce the {\em quantum 4-current density}  as an operator valued distribution on $\Gamma_\a(\cZ_\D^{(+)}\oplus\cZ_\D^{(-)})$
by the antisymmetric quantization of the classical expression
\begin{eqnarray}
\cJ^\mu(x)&:=&\frac12\big(\hat\psi^*(x)\beta\gamma^\mu\hat\psi(x)
-\psi(x)\bar\beta\bar\gamma^\mu\psi^*(x)\big).\end{eqnarray}
(See Subsubsect. 
\ref{Weyl and Wick quantization} for the definition of antisymmetric quantization. Note that $(\beta\gamma^\mu)^*=\beta\gamma^\mu$, and hence $\bar\beta\bar\gamma^\mu$ is the transpose of $\beta\gamma^\mu$). The charge conjugation $C$, which we introduce later on in Subsubsect. 
\ref{discro}, 
satisfies $C\Omega=\Omega$ and $C\hat\cJ^\mu(x)C^*=-\cJ^\mu(x)$. Therefore,
$(\Omega|\cJ^\mu(x)\Omega)=0$. Hence
\begin{eqnarray*}
\hat \cJ^\mu(x)
&=&{:}
\tilde{\hat \psi}(x)\gamma^\mu\hat\psi(x){:}
.\end{eqnarray*}
Formally, we can check the quantum versions of the relations (\ref{po1}) the (\ref{po2}).
We have
\[\vec{\hat\cJ}(x)= {:}\hat\psi^*(x)\vec\alpha\hat\psi(x){:},\]
and the 0th component of the 4-current is called the {\em charge density}
\[\hat\cQ(x):=\hat\cJ_0(x)=
{:}\hat\psi^*(x)\hat\psi(x){:}.\] 

Formally, the charge density satisfies
\begin{eqnarray}\nonumber
[\hat \cQ(t,\vec x),\hat\psi(t,\vec y)]&=&-\hat\psi(t,\vec y)\delta(\vec
x-\vec y),\\ {}\nonumber
[\hat \cQ(t,\vec x),\hat\psi^*(t,\vec y)]&=&\hat\psi^*(t,\vec y)\delta(\vec
x- \vec y),\\{} \label{westill}
[\hat \cQ(t,\vec x),\hat \cQ(t,\vec y)]&=&0
.\end{eqnarray}

For $\chi\in C_{\rm c}^\infty(\rr^3)$ let $\alpha_\chi$ denote the gauge transformation at time $t=0$ defined as a $*$-automorphism
of the algebra generated by fields satisfying (\ref{gaugo-q}), and hence also (\ref{gaugo1-q}). Assume that $\chi\neq0$. Let us check whether $\alpha_\chi$ is unitarily implementable.

On the level of annihilation operators
we have
\begin{eqnarray*}
\alpha_\chi(\hat a(p))&=&
\sum_{s_1}\int\int\frac{\d\vec
  x\d\vec p_1}{(2\pi)^{3}}\bar{u(p,s)}u(p_1,s_1)\e^{\i(\vec p_1-\vec p)\vec x-\i e\chi(\vec x)}\hat a(p_1)\\
&&\!\!\!\!\!+\sum_{s_1}
\int\int\frac{\d\vec
  x\d\vec p_1}{(2\pi)^{3}}\bar{u(p,s)}u(-p_1,-s_1)
\e^{-\i(\vec p_1+\vec p)\vec x-\i e\chi(\vec x)}\hat b^*(p_1).
\end{eqnarray*}
Let $q_\chi(\vec p,s;\vec p_1,s_1)$ denote the integral kernel on the second line above.
We need to check whether it is square integrable.
Now
\begin{eqnarray}
\sum_{s,s_1}|\bar{u(p,s)}u(-p_1,-s_1)|^2&=&\frac{|\vec p+\vec
  p_1|^2+\big(E(\vec p)-E(\vec p_1)\big)^2}{2E(\vec p)E(\vec p_1)} .\label{eqna2}
\end{eqnarray}
After integrating in $\vec x$ we obtain fast decay in
 $\vec p+\vec p_1$, which allows us to 
control
the numerator of (\ref{eqna2}).
We obtain
\[\int|q_\chi(\vec p,\vec p_1)|^2\d\vec p\sim \frac{C}{E(\vec p_1)^{2}},\]
which is not  integrable.
 Therefore,  $\alpha_\chi$ is not implementable by the {\em Shale-Stinespring
criterion}, see Thm \ref{shale}.

Formally, with
\beq
\hat Q(\chi):=\int\chi(\vec x) \hat \cQ(0,\vec x) \d\vec
x,\label{illo1}\eeq $\e^{\i e\hat Q(\chi)}$ implements the gauge
transformation:
 \[\alpha_\chi(B)=\e^{\i e\hat Q(\chi)}B\e^{-\i e\hat Q(\chi)}.\]
But we know that nontrivial gauge transformations are not implementable.
Thus for nonzero $\chi$ (\ref{illo1}) cannot be defined as a closable operator.

However, the (quantum) charge 
\begin{eqnarray}
\hat Q&:=&\int \hat \cQ(t,\vec x)\d\vec x\end{eqnarray}
is a well defined self-adjoint operator, which we already discussed before.


For further reference let us express the charge density 
in terms of creation and annihilation operators:
\begin{eqnarray*}\hat \cQ(x)&=&
\int\int \frac{\d \vec p_1\d \vec
  p_2}{(2\pi)^{3}}\bar{u(p_1,s_1)}u(p_2,s_2)
\e^{-\i x p_1+\i x p_2}\hat a^*(p_1,s_1)\hat a(p_2,s_2)\\&&
-\int\int \frac{\d \vec p_1\d \vec
  p_2}{(2\pi)^{3}}\bar{u(-p_1,-s_1)}u(-p_2,-s_2)
\e^{\i x p_1-\i x p_2}\hat b^*(p_2,s_2)\hat b(p_1,s_1)\\&&
+\int\int \frac{\d \vec p_1\d \vec
  p_2}{(2\pi)^{3}}\bar{u(p_1,s_1)}u(-p_2,-s_2)
\e^{-\i x p_1-\i x p_2}\hat a^*(p_1,s_1)\hat b^*(p_2,s_2)\\&&
+\int\int \frac{\d \vec p_1\d \vec
  p_2}{(2\pi)^{3}}\bar{u(-p_1,-s_1)}u(p_2,s_2)
\e^{\i x p_1+\i x p_2}\hat b(p_1,s_1)\hat a(p_2,s_2).
\end{eqnarray*}
To obtain $\vec{\hat\cJ}(x)$ one inserts $\vec \alpha$ between
$\bar{u(\cdot,\cdot)}$ and $u(\cdot,\cdot)$.

\subsection{Dirac fermions in an external 4-potential}

\subsubsection{Dirac equation in an external 4-potential}

Let \beq \rr^{1,3}\ni x\mapsto A(x)=[A_\mu(x)]\in\rr^{1,3}\label{source3}\eeq be a given
 function. In most of this subsection we assume that (\ref{source3}) is Schwartz.
The 
Dirac equation in an {\em external 4-potential} $A$ is
\beq
\big(\gamma^\mu(-\i\partial_\mu+ eA_\mu(x))+m\big)\psi(x)=0.
\label{sfa1d}\eeq 

If $\psi$ satisfies (\ref{sfa1d}) and $\rr^{1,3}\ni x\mapsto \chi(x)\in\rr$ is
an arbitrary smooth function, then $\e^{\i e\chi}\psi$ satisfies (\ref{sfa1d}) with $A$
replaced with $A+\partial\chi$. 

Note the identity
\begin{eqnarray}\nonumber
&&-\big(\gamma^\mu(-\i\partial_\mu+ eA_\mu(x))+m\big)
\big(\gamma^\mu(-\i\partial_\mu+eA_\mu(x))-m\big)\\
&=&-(\partial_\mu+\i eA_\mu(x))(\partial^\mu+\i eA^\mu(x))+m^2+\frac{e}{2}\sigma^{\mu\nu}F_{\mu\nu}(x).
\label{kleigo}\end{eqnarray}

Let $D^{\pm}(x,y)$ denote the retarded/advanced Green's function of (\ref{kleigo}). Then
\begin{eqnarray*}
S^{\pm}(x,y)
&:=&\big(\gamma^\mu(-\i\partial_{x^\mu}+eA_\mu(x))-m\big)D(x,y)
\end{eqnarray*}
is the 
 retarded/advanced Green's function of (\ref{sfa1d}), that is,  the unique
solution of
\beq\big(\gamma^\mu(-\i\partial_\mu+
eA_\mu(x))+m\big)S^\pm(x,y)=\delta(x-y)\label{sfa-cd}\eeq  
satisfying
\[\supp S^\pm\subset\{x,y\ :\ x\in J^\pm(y)\}.\]

We set
\[S(x,y):=S^+(x,y)-S^-(x,y).\]
Clearly,
\[\supp S\subset\{x,y\ :\ x\in J(y)\}.\]


\nowastrona


%

\nowastrona

We would like to introduce a field $\rr^{1,3}\ni
x\mapsto \psi(x)$ satisfying (\ref{sfa1d}).
If we  assume that it acts on $\cW_\D$ and coincides with
the free field $\psi_\fr(x)$
 at $x^0=0$, such a field is given by
\begin{eqnarray}
\psi(t,\vec x)
&=&-
\i\int_{\rr^3}S(t,\vec x;0,\vec y)\beta\psi_\fr(0,\vec y)
\d\vec{y}.\label{deco1}
\end{eqnarray}

\nowastrona

\subsubsection{Lagrangian and Hamiltonian formalism}

(\ref{sfa1d})  can be obtained as the Euler-Lagrange of a
variational problem.
The {\em Lagrangian density} can be taken as
\begin{eqnarray*}
\cL(x)&=&-
\frac{1}{2}\left(\tilde\psi(x)\gamma^\mu(-\i\partial_\mu)\psi(x)
+\i\partial_\mu\tilde\psi(x)\gamma^\mu\psi(x)
\right)\\
&&- \tilde\psi(x)eA_\mu(x)\gamma^\mu\psi(x)-m \tilde\psi(x)\psi(x).
\end{eqnarray*}
The Euler-Lagrange equations (\ref{euler-lagrange1c})
yield (\ref{sfa1d}).

We can introduce the {\em Hamiltonian density}
\begin{eqnarray*}
\cH(x)&=&\dot\psi(x)\frac{\partial\cL(x)}{\partial{\dot\psi(x)}}
+\dot\psi^*(x)\frac{\partial\cL(x)}{\partial{\dot\psi^*(x)}}-\cL(x)\\
&=&
\frac12\big(\psi^*( x)\vec\alpha(-\i\vec\partial)\psi(x)+\i\vec\partial\psi^*(x)\vec\alpha\psi(x)\big)\\
&&+\psi^*(x)\bigl(
e\vec\gamma\vec A(x)+m\beta+eA_0( x)\bigr)\psi( x).
\end{eqnarray*}
 The  {\em Hamiltonian} 
\[H(t)=\int \cH(t,\vec x)\d\vec x
\]
can be interpreted as a self-adjoint operator on $\Gamma_\a(\cW_\D^\cpl)$
that
 generates the ``classical'' dynamics
\begin{eqnarray*}
\dot\psi(t,\vec x)=\i\{H(t),\psi(t,\vec x)\},
\end{eqnarray*}
where now $\{\cdot,\cdot\}$ has the meaning of the commutator.


\subsubsection{Classical discrete symmetries}
\label{Classical discrete symmetries1}

Let $\kappa$ be a unitary $4\times4$ matrix satisfying
\[\kappa\bar\kappa=\one,\ \ \ \kappa\gamma^\mu\kappa^{-1}=-\bar\gamma^\mu,\]
where the bar denotes the complex conjugation. In particular, $\kappa\beta\kappa^{-1}=-\bar\beta$.
Note also that 
\[\kappa\bar{\kappa\bar u}=u,\ \ \ u\in\cc^4.\]

Choose $\xi_C\in\cc$, $|\xi_C|=1$. If $\zeta$ solves the Dirac equation with the 4-potential $A$, then so does
$\xi_C\kappa\bar\zeta$ with the 4-potential $-A$. Thus replacing
\begin{eqnarray*}
&&\psi(x),\psi^*(x),A(x)\\
&\hbox{with }&\bar\xi_C\kappa\psi^*(x),\xi_C\bar\kappa\psi(x),-A(x)\end{eqnarray*}
is a symmetry of the Dirac  equation with external
4-potentials
 (\ref{sfa1d}). It is called {\em charge conjugation} and denoted $\cC$.

The matrix $\kappa$ depends on a representation. In the Majorana
representation it is the identity. In the Dirac and spinor representation
it can be chosen to be $\gamma^2$ multiplied by an arbitrary phase
factor. In fact, in these representations $\bar\gamma^\mu=\gamma^\mu$,
except for $\mu=2$ satisfying $\bar\gamma^2=-\gamma^2$. 
When we consider the Dirac representation, we will adopt the convention
\[\kappa:=\i\gamma^2.\]
Then $\bar\kappa=\kappa=\kappa^*$.
The spinor
basis that we chose in (\ref{planewave}) is compatible with $\kappa$:
\beq \kappa\bar{u(p,s)}=u(-p,-s).\eeq

Choose $\xi_P\in\{1,-1\}$. Recall that ${\rm P}$ denotes the space
inversion. Replacing  \begin{eqnarray*}
&&\psi(x),\psi^*(x),\big(A_0(x),\vec A(x)\big)\\
&\hbox{with }&
\xi_P\gamma^0\psi({\rm P}x),\ \ \xi_P\gamma^0\psi^*({\rm P}x),\ \ \big(A_0({\rm P}x),-\vec A({\rm
  P}x)\big)\end{eqnarray*} is a symmetry of  (\ref{sfa1d})  called {\em parity}  and denoted $\cP$.

Choose $\xi_T\in\cc$, $|\xi_T|=1$. Recall that ${\rm T}$ denotes the time
reflection. Replacing (in the Dirac representation) \begin{eqnarray*}
&&\psi(x),\psi^*(x),\big(A_0(x),\vec A(x)\big)\\
&\hbox{with }&\bar\xi_T\gamma^1\gamma^3\psi^*({\rm T}x),\ \ \xi_T\gamma^1\gamma^3
\psi({\rm T}x),\ \ \big(A_0({\rm T}x),-\vec A({\rm
  T}x)\big)\end{eqnarray*} is a symmetry of  (\ref{sfa1d})  called {\em time reversal}  and denoted $\cT$.

The symmetry that is guaranteed by the CPT Theorem consists in replacing 
 \begin{eqnarray*}
&&\psi(x),\ \ \psi^*(x),\ \  A(x)\\
&\hbox{with }&
\i\gamma^5\psi(-x),\ \ -\i\gamma^5\psi^*(-x),\ \ -A(-x).\end{eqnarray*}
It is  denoted $\cX$. (Note that $\i\gamma^5=\gamma^0\gamma^1\gamma^2\gamma^3$). 

Assume that
$\bar\xi_C\xi_P\xi_T=\i$. Then \[\cX=\cC\cP\cT\]
 and we have the relations
\begin{eqnarray*}&&\cC^2=\cP^2=-\cT^2=-\cX^2=\id,\\
&&\cC\cP+\cP\cC=\cC\cT+\cT\cC=0,\\
&&\cX\cP+\cP\cX=\cX\cT+\cT\cX=0,\\
&&\cC\cX-\cX\cC=\cP\cT-\cT\cP=0.
\end{eqnarray*}

To understand better these relations, let us notice that the automorphisms $\cP$, $\cC\cT$ and $\cX$ anticommute and
\[\cP^2=(\cC\cT)^2=-\cX^2=\id,\]
where $\id$ denotes the identity.  Thus together with  $Spin^\uparrow(1,3)$ they represent the group $Pin_+(1,3)$, see Subsubsect.
\ref{Double coverings}.

 Besides,
\[(\cP\cT)^2=-\id\]
and $\cP\cT$ commutes with
$\cP$, $\cC\cT$, $\cX$. Thus it behaves as $\i\cdot\id$.
Thus the group generated by  $Spin^\uparrow(1,3)$, $\cC$, $\cP$ and $\cT$ is $Pin_\ext(1,3)$, see Subsubsect. \ref{Complex Lorentz groups}.

\subsubsection{Quantization}

We are looking for a quantum field satisfying
\beq
\big(\gamma^\mu(-\i\partial_\mu+ eA_\mu(x))+m\big)\hat\psi(x)=0
\label{sfa1aa}\eeq
such that
\[\hat\psi(\vec x):=\hat\psi(0,\vec x)=\hat\psi_\fr(0,\vec x).\]
Clearly the solution is obtained by decorating (\ref{deco1}) with hats.

As in the bosonic case, we ask whether the fields are implemented by a 
 a unitary dynamics. Equivalently, we want to check if the classical dynamics generated by
$H_\Int(t)$ satisfies the Shale-Stinespring criterion.

Arguments parallel to those of Subsubsect. \ref{shale2} show that the classical scattering operator is unitarily implementable. 

An analysis similar to that of Subsect. \ref{Quantization-bosons} shows
that  the dynamics from $t_-$ to $t_+$ is implementable on the Fock space
iff the spatial part of the 4-potential is the same at the initial and final time:
\beq\vec A(t_+,\vec x)=\vec A(t_-,\vec x),\ \ \vec x\in\rr^3.\label{pqiq-fermion}\eeq

\subsubsection{Quantum Hamiltonian}

Formally, we can also obtain the quantum field from a unitary dynamics:
\[\hat\psi(t,\vec x):=\Texp\left(-\i\int_t^0 \hat H(s)\d s\right)\hat \psi(0,\vec
x)\Texp\left(-\i\int_0^t \hat H(s)\d s\right), \]
\nowastrona
\noindent
where the   Schr\"odinger picture Hamiltonian $\hat H(t)$ and the corresponding 
interaction picture Hamiltonian are
\begin{eqnarray}\nonumber
\hat H(t)&=&\int\d\vec x
{:}\bigl(\hat\psi^*(\vec x)(\vec\alpha(-\i\vec\partial+e\vec A(t,\vec
x))+m\beta+eA_0(t,\vec x))\hat\psi(\vec x)
\bigr){:},
\label{nmun3a}
\\\nonumber
\hat H_\Int(t)
&=&\int\d\vec x e A_\mu(t,\vec x)\hat \cJ^\mu_\fr(t,\vec x)
.\label{nmun4a}
\end{eqnarray}

\nowastrona

Note that unlike in the case of charged bosons we use the Wick ordering. 
This is because $\hat H(t)$ differs from $\hat H_\fr$ by a term involving the 4-current $\hat J_\fr^\mu(t,\vec x)$, which is automatically  Wick ordered. Therefore, we can assume that both $\hat H(t)$ and $\hat H_\fr$ are Wick ordered, which was impossible for charged bosons.

\subsubsection{Quantized discrete symmetries}
\label{discro}
The discrete symmetries considered in
Subsubsect. \ref{Classical discrete symmetries1} remain true when we
decorate
 the fields
with ``hats''. Thus on the level of quantum
observables
the discrete symmetries are the same as in the classical
case.

Let us now  discuss the implementation of these
symmetries by unitary or antiunitary operators on the Hilbert
space $\Gamma_\a(\cZ_\D^{(+)}\oplus\cZ_\D^{(-)})$.
 We will discuss this for free fields, that is, for $A=0$.
As in the bosonic case, this will imply some properties of
 the scattering operator  $\hat S(A)$.

First consider the charge conjugation. We  define the following  unitary operator on
$\cZ_\D^{(+)}\oplus\cZ_\D^{(-)}$
\[\chi(g_1,\bar g_2):=(\xi_C\bar\kappa  g_2 ,\bar\xi_C\kappa\bar g_1).\]
We check that
\[\chi|p,s)=\xi_C\bar{|-p,-s)},\ \ \ \chi\bar{|-p,-s)}=\bar\xi_C|p,s).\]
We set
$C:=\Gamma(\chi)$. We have $C^2=\one$,
\[ C\hat\psi_\fr(x)C^{-1}=\bar\xi_C\kappa\hat\psi_\fr^*(x),\ \ C\hat\psi_\fr^*(x)C^{-1}=\xi_C\bar\kappa\hat\psi_\fr(x),\]
\[C\hat\cQ_\fr(x)C^{-1}=-\hat\cQ_\fr(x),\ \ \ 
C\vec{\hat\cJ}_\fr(x)C^{-1}=-\vec{\hat\cJ}_\fr(x),\]
\[C\hat S(A)C^{-1}=\hat S(-A).\]

Define the following unitary operator on
$\cZ_\D^{(+)}\oplus\cZ_\D^{(-)}$:
\[\pi \big( g_1,\bar g_2\big):=\big(\xi_P \gamma^0g_1\circ{\rm P},
\xi_P\gamma^0\bar g_2\circ{\rm P}\big).\]
We check that
\[\pi|E,\vec p,s)=\xi_P|E,-\vec p,s),\ \ \pi|{-}E,{-}\vec p,s)=\xi_P|{-}E,\vec p,s).\]
Set $P:=\Gamma(\pi)$.
 We have $P^2=\one$,
\[ P\hat\psi_\fr(x)P^{-1}=\xi_P\gamma^0\hat\psi_\fr({\rm P}x),\ \ P\hat\psi_\fr^*(x)P^{-1}=\xi_P\gamma^0\hat\psi_\fr^*({\rm P}x),\]
\[P\hat\cQ_\fr(x)P^{-1}=\hat\cQ_\fr({\rm P}x),\ \ \ 
P\vec{\hat\cJ}_\fr(x)P^{-1}=-\vec{\hat\cJ}_\fr({\rm P}x),\]
\[P\hat S(A_0,\vec A)P^{-1}=\hat S(A_0\circ{\rm P},-\vec A\circ{\rm P}).\]

Define
(in the Dirac representation)
 the following  antiunitary operator on
$\cZ_\D^{(+)}\oplus\cZ_\D^{(-)}$:
\[\tau(g_1,\bar g_2):=\big(\xi_T\gamma^1\gamma^3\bar
      {g}_1\circ{\rm T},\bar\xi_T\gamma^1\gamma^3 {g}_2\circ{\rm T}\big).\]
We check that
\[\tau|E,\vec p,s)=\xi_T|E,-\vec p,-s),\ \ \tau|{-}E,{-}\vec p,s)=\bar\xi_T|{-}E,\vec p,-s).\]
Set $T:=\Gamma(\tau)$.
 We have $T^2=-\one$,
\[ T\hat\psi_\fr(x)T^{-1}=\bar\xi_T\gamma^1\gamma^3\hat\psi_\fr^*({\rm T}x),\ \ T\hat\psi_\fr^*(x)T^{-1}=\xi_T\gamma^1\gamma^3\hat\psi_\fr({\rm T}x),\]
\[T\hat\cQ_\fr(x)T^{-1}=\hat\cQ_\fr({\rm T}x),\ \ \ 
T\vec{\hat\cJ}_\fr(x)T^{-1}=-\vec{\hat\cJ}_\fr({\rm T}x),\]
\[T\hat S(A_0,\vec A)T^{-1}=\bar{\hat S(A_0\circ{\rm T},-\vec A\circ{\rm T})}.\]

\subsubsection{$2N$-point Green's functions}

We consider again a Dirac field in an external  4-potential $[A^\mu(x)]$.
For $y_{N},\dots y_1$, $x_{N},\dots,x_1$, the {\em $2N$ point 
 Green's function} are defined
as follows:
\begin{eqnarray*}
&&\big\langle\tilde{\hat\psi}(y_1)\cdots\tilde{\hat\psi}(y_N)\hat\psi(x_{N})\cdots\hat\psi(x_{1})\big\rangle\\&:=&
\left(\Omega^+|\T\left(
\tilde{\hat \psi}(y_1)\cdots\tilde{\hat \psi}(y_N){\hat
  \psi}(x_{N})\cdots{\hat
\psi}(x_{1})\right)
\Omega^-\right).
\end{eqnarray*}

One can organize Green's functions in terms of the {\em generating function}:
\begin{eqnarray*}&&Z(g,\tilde g)\\
&:=&\sum_{n=0}^\infty\int\cdots\int 
\frac{(-1)^N}{(N!)^2}\big\langle\tilde{\hat\psi}(y_1)\cdots\tilde{\hat\psi}(y_N)\hat\psi(x_{N})\cdots\hat\psi(x_{1})\big\rangle\\
&&\times g(y_1)\cdots g(y_N)\tilde g(x_N)\cdots\tilde g(x_1)
\d y_1\cdots\d y_N\d x_N\cdots\d x_1\\
&=&
\left(\Omega\Big|\Texp\left(-\i\int_{-\infty}^{\infty} \hat H_\Int(t)\d t-\i\int g(x)\tilde{\hat \psi}_\fr(x)\d x-\i\int \tilde g(x)\hat \psi_\fr(x)\d x\right)\Omega\right)
,\end{eqnarray*}
where $\rr^{1,3}\ni x\mapsto g(x),\tilde g(x)\in\cc^4$ are Grassmann variables
anticommuting with $\hat\psi(x)$, $\tilde{\hat\psi}(x)$.

One can retrieve Green's functions from the generating function:
\begin{eqnarray*}&&
\big\langle\tilde{\hat\psi}(y_1)\cdots\tilde{\hat\psi}(y_N)\hat\psi(x_N)\cdots\hat\psi(x_1)\big\rangle\\&=&
(-1)^{N}\frac{\partial^{2N}}{\partial g(y_1)\cdots\partial g(y_N)\partial \tilde g(x_N)\cdots \partial \tilde g(x_1)}
Z(g,\tilde g)\Big|_{g=\tilde g=0}.\end{eqnarray*}

We introduce also the {\em amputated Green's function}
\begin{eqnarray*}&&
\big\langle\tilde{\hat\psi}(p_1')\dots\tilde{\hat\psi}(p_N')\hat\psi(p_N)\cdots\hat\psi(p_1)\big\rangle_\amp\\
&:=&
\bigl(\gamma p_1'+m\bigr)\cdots\bigl(\gamma p_{N}'+m\bigr)
\bigl(\gamma p_{N}+m\bigr)\cdots
\bigl(\gamma p_1+m\bigr)\\&&
\times \big\langle\tilde{\hat\psi}(p_1')\cdots\tilde{\hat\psi}(p_N')\hat\psi(p_N)\cdots\hat\psi(p_1)\big\rangle.\end{eqnarray*}

Introduce {\em many particle plane waves}
\begin{eqnarray*}&&
|\bar{-p_{N'}',-s_{N'}'};\dots;\bar{-p_1',-s_1'};p_{N},s_{N};\dots;p_1,s_1)\\&:=&\hat b^*(p_{N'}',s_{N'}')\cdots
\hat b^*(p_1',s_1')\hat a^*(p_{N},s_{N})\cdots\hat a^*(p_1,s_1)\Omega,\end{eqnarray*}
where all $p_i$, $p_i^{\prime}$ are  on shell.
{\em Scattering amplitudes} are the
 matrix elements of the scattering operator $\hat S$ between
plane waves.
One can compute scattering amplitudes from the amputated Green's functions:
\begin{eqnarray*}&&
\left(\bar{-p_{n^{+\prime}}^{+\prime},-s_{n^{+\prime}}^{+\prime}}
;\dots;p_{n^+}^+,s_{n^+}^+;\dots
|\,\hat S\, 
|
\bar{-p_{n^{-\prime}}^{-\prime},-s_{n^{-\prime}}^{-\prime}};
\dots;
p_{n^-}^-,s_{n^-}^-;\dots\right)\\[3ex]
&=& 
\frac{\cdots\tilde u(p_{n^+}^+,s_{n^+}^+)\cdots
u(-p_{n^{+\prime}}^{+\prime},-s_{n^{+\prime}}^{+\prime})
\tilde  u(-p_{n^{-\prime}}^{-\prime},-s_{n^{-\prime}}^{-\prime})\cdots
 u(p_{n^-}^{-},s_{n^-}^{-})\cdots
}
{\sqrt{(2\pi)^{3(n^++n^{+\prime}+n^{-\prime}+n^-)}}}
\\[2ex]
&&\times
\big\langle\cdots\hat\psi(p_{n^+}^+)\cdots\tilde{\hat\psi}(p_{n^{+\prime}}^{+\prime})
\psi(-p_{n^{-\prime}}^{-\prime})\cdots
\tilde{\hat\psi}(-p_{n^-}^{-})\cdots\big\rangle_\amp
.\end{eqnarray*}


The scattering operator and Green's functions satisfy  the Ward identities analogous to those satisfied by charged bosons.

\subsubsection{Path integral formulation}

We have the following formula for the generating function:

\begin{eqnarray}&&Z(g,\tilde g)\label{patho3}\\
&=&\det\big(\gamma^\mu(-\i\partial_\mu+ eA_\mu(x))+m\big)
\big(-\i\gamma^\mu\partial_\mu+m-\i\epsilon\big)^{-1}
\notag\\
&&\times\exp\left(\i \tilde g
\big(\gamma^\mu(-\i\partial_\mu+eA_\mu(x))+m-\i\epsilon\big)^{-1}
g\right)\notag\\
&=&\det\left(\one+
\gamma_\mu eA^\mu S_\fr^{\rm c}
\right)\notag\\
&&\times\exp\left(\i \tilde g S_\fr^{\rm c}
\Bigl(\one+\gamma_\mu eA^\mu S_\fr^{\rm c}
\Bigr)^{-1}g\right)
,\notag\end{eqnarray}
where $\epsilon$ has the same meaning as in (\ref{infini}).

In terms of path integrals this can be formally written as
\begin{eqnarray*}
&&\frac{\int\lpi_y\d\tilde\psi(x)\lpi_{y'}\d\psi(y)
\exp\left(\i\int\big(\cL(x)-g(x)\tilde\psi(x)-\tilde g(x)\psi(x)\big)\d x\right)}
{\int\lpi_y\d\tilde\psi(y)\lpi_{y'}\d\psi(y')
\exp\left(\i\int\cL_\fr(x)\d x\right)}.
\end{eqnarray*}

\nowastrona
\subsubsection{Feynman rules}

The Feynman  rules are very similar to those for charged bosons, except that there are no two-photon vertices.
Here are the Feynman rules  for
 Green's functions.

\ben \item In the $n$th order we draw all possible topologically distinct
Feynman diagrams with $n$ vertices and external lines. All the
charged lines have a natural arrow.
\item  To each  vertex we associate the factor
    $-\i e\gamma^\mu A_\mu(p^+-p^-)$.
\item To each line we associate the propagator
 $-\i S_\fr^{\rm c}(p)=-\i\frac{-p\gamma+m}{p^2+m^2-\i0}$
\item For internal lines we integrate over the variables with the measure
$\frac{\d^4 p}{(2\pi)^4}$.
\item If two diagrams differ only by an exchange of two fermionic lines, there
  is an additional factor $(-1)$ for one of them. This implies, in particular,
  that loops have an additional factor $(-1)$.
\een
\nowastrona


\nowastrona

To compute  scattering amplitudes with $N^-$ incoming and $N^+$
outgoing particles we draw the same diagrams as
for $N^-+N^+$-point Green's functions.
The rules are changed only
concerning the external lines.

\nowastrona

\begin{romanenumerate}\item
 With each incoming external line we associate
\begin{itemize}
\item  fermion: $\frac{1}{\sqrt{(2\pi)^3}}u(p,s)$.
\item  anti-fermion: $\frac{1}{\sqrt{(2\pi)^3}}\tilde u(-p,-s)$.
\end{itemize}
\item  With each outgoing external line we associate
\begin{itemize}
\item  fermion:  $\frac{1}{\sqrt{(2\pi)^3}}\tilde u(p,s)$.
\item  anti-fermion: $\frac{1}{\sqrt{(2\pi)^3}} u(-p,-s)$.
\end{itemize}
Each incoming and outgoing  antifermion has an additional factor
$(-1)$. (This follows from the rule (5) above).
\end{romanenumerate}
\nowastrona
 \subsubsection{Vacuum energy}

Formally, the  vacuum energy can be
computed exactly:
\begin{eqnarray}\nonumber
\cE&:=&\i\log(\Omega|\hat S\Omega)\,=\,\i\log Z(0,0)\\\nonumber
&=&\i\Tr\Big(\log
\big(\gamma^\mu({-}\i\partial_\mu{+} eA_\mu(x))+m-\i\epsilon\big)
-\log\big({-}\i\gamma^\mu\partial_\mu+m-\i\epsilon\big)\Big)
\\\nonumber
&=&\i\Tr
\log\left(\one+
\gamma^\mu eA_\mu S_\fr^{\rm c}
\right)
\\
&=&
\i\sum_{n=1}^\infty\frac{D_n}{n}.
\label{paks==}\end{eqnarray}

Here $D_n$ is
the value of  the loop with $n$ vertices.
Note that $n$ in the denominator is the order of the group of the authomorphisms of a loop with $n$ vertices, which is $\zz_n$.

{\em Furry's theorem}, proven as in the bosonic case, says that diagrams  for charged fermions of the odd order in
$e$
 vanish.
Hence  (\ref{paks==}) can be written as
\[\cE=\sum_{n=1}^\infty e^{2n}\cE_n,\]
where $e^{2n}\cE_n=\i\frac{D_{2n}}{2n}$.

Just as in he bosonic case, the expressions for $\cE_n$ obtained from the Feynman rules are convergent for $n\geq3$. The gauge invariant part of $\cE_2$ is convergent. The computation of $\cE_1$ will be described below -- it  needs an infinite renormalization.

There exists a close relationship between the fermionic and bosonic vacuum energy. To see it, note that using  $\gamma^5\gamma_\mu(\gamma^5)^{-1}=-\gamma_\mu$, we obtain
\begin{eqnarray}
\cE
&=&\i\Tr\Big(\log
\big(-\gamma^\mu(-\i\partial_\mu+eA_\mu(x))+m-\i\epsilon\big)\notag\\
&&\hspace{5ex}-\log\big(\i\gamma^\mu\partial_\mu+m-\i\epsilon\big)\Big).
\label{paks=+}\end{eqnarray}
We add up $\frac12$(\ref{paks==}) and  $\frac12$(\ref{paks=+}) and use identity
(\ref{kleigo}). We obtain
\begin{eqnarray}\nonumber
\cE&=&\frac\i2\Tr\Big(-\log\big(-(\partial_\mu+\i eA_\mu(x))(\partial^\mu+\i eA^\mu(x))+m^2+\frac{e}{2}\sigma^{\mu\nu}F_{\mu\nu}(x)-\i0\big)\\
&&\hspace{3ex}+\log
\big(-\Box+m^2-\i0\big)\Big)\notag \\
&=&\frac\i2\Tr\log\Big(\one+\Big(\i e\partial_\mu A^\mu(x)+
\i eA^\mu(x)\partial_\mu +e^2 A_\mu(x)A^\mu(x)\notag\\
&&\hspace{10ex}+\frac{ e}{2}\sigma^{\mu\nu} F_{\mu\nu}(x)\Big)D_\fr^{\rm c}\Big).
\label{kleigo+}\end{eqnarray}
We can compare (\ref{kleigo+}) with a similar expression in the bosonic case (\ref{paks=}).

\subsubsection{Pauli-Villars renormalization}

A single electron loop with two vertices coming from a 4-potential $A^\mu$ leads
to a contribution of the form
\[\cE_1= \int\frac{\d p}{(2\pi)^4} A^\mu(-p)A^\nu(p)\Pi_{\mu\nu}(p).\]
Unfortunately, computed naively, $\Pi_{\mu\nu}(p)$ is divergent.

We will 
compute it  using  the  Pauli-Villars regularization, similarly as for charged bosons, see Subsubsect. \ref{Pauli-Villars}:

\nowastrona

\begin{eqnarray*}
2\Pi_{\mu\nu \Lambda }(p)&=&-\sum_i C_i\i e^2\int\frac{\d^4
  q}{(2\pi)^4}
\frac{\Tr\gamma_\mu\left((q+\frac12 p)\gamma+m_i\right)\gamma_\nu\left((q-\frac12p)\gamma+m_i\right)}
{\left((q+\frac12p)^2+m_i^2-\i0\right)\left((q-\frac12p)^2+m_i^2-\i0\right)}
\\
&=&-\sum_iC_i\i e^2\int\frac{4\d^4
  q}{(2\pi)^4}
\frac{2q_\mu q_\nu-\frac12p_\mu p_\nu-
g_{\mu\nu}(q^2-\frac14p^2+m_i^2)}
{((q+\frac12p)^2+m_i^2-\i0)((q-\frac12p)^2+m_i^2-\i0)}
\end{eqnarray*}
\begin{eqnarray*}
&=&\sum_iC_i\frac{ e^2}{(4\pi)^2}
\int_0^\infty\d\alpha_1\int_0^\infty\d\alpha_2
\Bigg(\frac{8\alpha_1\alpha_2}{(\alpha_1+\alpha_2)^4}(p_\mu
  p_\nu-g_{\mu\nu}p^2)\\
&&\ \ \ \ \ \ \ \ \ +
4g_{\mu\nu}\left(\frac{\alpha_1\alpha_2}{(\alpha_1+\alpha_2)^4}p^2
+\frac{\i }{(\alpha_1+\alpha_2)^3}+\frac{m_i^2}{(\alpha_1+\alpha_2)^2}\right)\Bigg)
\\&&
\times
\exp\left(-\i(\alpha_1+\alpha_2)m_i^2
-\i\frac{\alpha_1\alpha_2}{\alpha_1+\alpha_2}
  p^2\right)
\\
&=:&
(-g_{\mu\nu} p^2+p_\mu p_\nu)2\Pi_\Lambda^{{\rm gi}}(p^2)+2\Pi^{\rm gd}_{\mu\nu\Lambda}(p).
\end{eqnarray*}
\nowastrona
We used the identity (\ref{loop-com}).

\nowastrona

The gauge dependent part of the vacuum energy tensor 
up to a coefficient is the same as for charged bosons and vanishes.
We apply the same substitutions and use the same identities as in the charged boson case:
\begin{eqnarray*}
\Pi_\Lambda ^{{\rm gi}}(p^2)&=&
-\frac{e^2}{(4\pi)^2}\sum_iC_i
\int_0^\infty\d\alpha_1\int_0^\infty\d\alpha_2
\frac{4\alpha_1\alpha_2}{(\alpha_1+\alpha_2)^{4}}  
\\ &&
\times \exp\left(-\i(\alpha_1+\alpha_2)m_i^2-
\i\frac{\alpha_1\alpha_2}{\alpha_1+\alpha_2}
  p^2\right)
\\
&=&-\frac{ e^2}{(4\pi)^2}
\sum_iC_i
\int_0^1\d v\int_0^\infty\frac{\d\rho}{\rho}
(1-v^2)
\\&&
\times \exp\left(-\i\rho\left(m_i^2
+\frac{(1-v^2)
  p^2}{4}\right)\right)\\
&=&\frac{ e^2}{(4\pi)^2}
\sum_iC_i
\int_0^1\d v
(1-v^2)\log\left(m_i^2
+\frac{(1-v^2)
  p^2}{4}-\i0\right)\\
&=&\frac{ e^2}{(4\pi)^2}
\sum_iC_i
\left(\int_0^1\d v
(1-v^2)\log\left(1
+\frac{(1-v^2)
  p^2}{4m_i^2}-\i0\right)+\frac{1}{3}\log m_i^2\right).
\end{eqnarray*}
\nowastrona
Define
\begin{eqnarray}
\Pi^{\ren}(p^2)&:=&
\lim_{\Lambda \to\infty}\big(\Pi_\Lambda ^\gi(p^2)-\Pi_\Lambda ^\gi(0)\big)
\label{piren1}
\\\notag
&=&\frac{ e^2}{(4\pi)^2}
\int_0^1\d v
(1-v^2)\log\left(1
+\frac{(1-v^2)
  p^2}{4m^2}-\i0\right).
\end{eqnarray}

Denote the vacuum  energy function for neutral bosons, introduced in (\ref{piren1+}), by  $\pi_{\rm b}^\ren$.
 Let $\Pi_{\rm b}^\ren$ denote the vacuum energy function for 
charged bosons
(\ref{piren}) and $\Pi_{\rm f}^\ren$
for charged fermions
(\ref{piren1}). Let us note the following identity:
\beq 2\Pi_{\rm b}^\ren(p^2)+\Pi_{\rm f}^\ren(p^2)=4e^2\pi^\ren(p^2).\eeq
This identity can be also derived from  (\ref{kleigo+}), (\ref{paks=}) and (\ref{paks}).

\nowastrona
\subsubsection{Method of dispersion
  relations}

The imaginary
part of the gauge invariant vacuum energy function can be computed without regularization:
\begin{eqnarray*}
\Im\Pi^\ren(p^2)&=&
\Im\frac{e^2}{(4\pi)^2}
\int_0^1\d v
(1-v^2)\log\left(m^2
+\frac{(1-v^2)  p^2}{4}-\i0\right)\\&=&
\frac{ e^2}{(4\pi)^2}
\int_0^1\d v
(1-v^2)(-\pi)\theta\left(
-\frac{(1-v^2)  p^2}{4}-m^2\right)\\&=&
-\frac{4e^2\pi}{3(4\pi)^2}                         \frac{(-p^2+2m^2)}{(-p^2)^{3/2}}
\left|-p^2-4m^2\right|_+^{\frac12}, \ \ \ \ p^2\in\rr.
\end{eqnarray*}
The full vacuum energy tensor can be obtained by using the once
subtracted dispersion relations, as in (\ref{dispe}).
\nowastrona

\subsubsection{Dimensional renormalization}

We can also use dimensional regularization to compute $\Pi_{\mu\nu}^\ren$.  We use the Euclidean formalism. 
\begin{eqnarray}\notag
2\Pi_{\mu\nu}^\E(p)
&=& e^2\Tr\one\int\frac{\d^4
  q}{(2\pi)^4}
\frac{2q_\mu q_\nu-\frac12p_\mu p_\nu-g_{\mu\nu}(q^2-\frac14 p^2+m^2)}
{((q+\frac12p)^2+m^2)((q-\frac12p)^2+m^2)}
\\&=& e^2\Tr\one\int_{-1}^1\d v\int\frac{\d^4
  q}{(2\pi)^4}
\frac{2q_\mu q_\nu-\frac12p_\mu p_\nu-g_{\mu\nu}(q^2-\frac14 p^2+m^2)}
{(q^2+\frac{p^2}{4}+m^2+vqp)^2}
\notag
\\\notag
&=& e^2\Tr\one\int_{0}^1{\d v}\int\frac{\d^4
  q}{(2\pi)^4}\\&&\hspace{-10ex}\times
\frac
{2q_\mu q_\nu-\frac12p_\mu p_\nu-g_{\mu\nu}(q^2-\frac14 p^2+m^2)+v^2(\frac12p_\mu p_\nu-g_{\mu\nu}\frac{p^2}{4})}
{(q^2+\frac{p^2}{4}(1-v^2)+m^2)^2}.\label{pdf4+}
\end{eqnarray}
Then we use the dimensional regularization. Besides the rules (\ref{replace1}) and (\ref{replace2}) we have a new rule:
\begin{eqnarray}
\Tr\one\ \hbox{ is replaced by }\ 2^{d/2}.
\label{newrule}\end{eqnarray}
 Thus (\ref{pdf4+}) is replaced by
\begin{eqnarray}&&
\Pi_{\mu\nu}^{\E,d}(p)\,=\,e^2\frac{2^{d/2}\mu^{4-d}\Omega_d}{(2\pi)^d}\int_0^1\d v\int_0^\infty|q|^{d-1}\d|q|\notag\\
&&\times\frac{\big((2/d-1)g_{\mu\nu}q^2
-\frac12p_\mu p_\nu-g_{\mu\nu}(-\frac14 p^2+m^2)+v^2(\frac12p_\mu p_\nu-g_{\mu\nu}\frac{p^2}{4})}
{\big(q^2+\frac{p^2}{4}(1-v^2)+m^2\big)^2}\notag\\
&=&\frac{4e^2}{(4\pi)^2}\int_0^1\d v\Big(\frac{\mu^22\pi}{
\frac{p^2}{4}(1-v^2)+m^2}\Big)^{2-d/2}\Gamma(2-d/2)\notag\\
&&\times
\Bigg(g_{\mu\nu}\Big(\frac{p^2}{4}(1-v^2)+m^2\Big)
-\frac12p_\mu p_\nu-g_{\mu\nu}\Big(-\frac14 p^2+m^2\Big)+v^2\Big(\frac12p_\mu p_\nu-g_{\mu\nu}\frac{p^2}{4}\Big)\Bigg)
\notag\\
&=&\frac{2e^2}{(4\pi)^2}\int_0^1\d v\Big(\frac{\mu^22\pi}{
\frac{p^2}{4}(1-v^2)+m^2}\Big)^{2-d/2}\Gamma(2-d/2)(v^2-1)(p_\mu p_\nu-g_{\mu\nu}p^2)
\notag\\
&\simeq&\frac{2e^2}{(4\pi)^2}\int_0^1\d v
\Big(-\gamma +\log(\mu^22\pi)-\log\Big(
\frac{p^2}{4}(1-v^2)+m^2\Big)\Big)(v^2-1)(p_\mu p_\nu-g_{\mu\nu}p^2)\notag\\&&
+\frac{4e^2}{3(4\pi)^2(2-d/2)}(p_\mu p_\nu-g_{\mu\nu}p^2).
\label{pdf5a}\end{eqnarray}
We can now renormalize (\ref{pdf5a}):
\begin{eqnarray*}
&&\Pi^{\E,\ren}(p^2)(p_\mu p_\nu-g_{\mu\nu}p^2)\\
&=&\lim_{d\to4}\Big(\Pi_{\mu\nu}^{\E,d}(p^2)
-\Pi_{\mu\nu}^{\E,d}(0)\Big)\\
&=&\frac{1}{(4\pi)^2}\int_0^1\d v(1- v^2)
\log\Big(1+
\frac{p^2}{4m^2}(1-v^2)\Big)(p_\mu p_\nu-g_{\mu\nu}p^2).
\end{eqnarray*}
Again, this coincides  with the Wick rotated 
result obtained by the Pauli-Villars method.

\ber
In the above computations we first try to  eliminate gamma matrices. The only remnant of gamma matrices is $\tr\one$, where $\one$ is the identity on the space of Dirac spinors, to which we apply the rule (\ref{newrule}). However, we would have obtained the same final result if we used eg. the rule  $\Tr\one=4$, since at the end we apply the normalization condition $\Pi_{\mu\nu}^\ren(0)=0$. We use the condition 
 (\ref{newrule}), since it is the usual choice in the literature. 

Note, however, that in more complicated situations the dimensional renormalization can be problematic, especially for fermions in the presence of $\gamma^5$. \eer

\subsubsection{Energy shift}

Suppose that the 4-potential does not depend on time and is given by
a Schwartz function $\rr^3\ni \vec x\mapsto A(\vec x)=[A_\mu(\vec x)]$.

The free Hamiltonian is
\begin{eqnarray}\nonumber
\hat H_\fr&=&
\int\d\vec x
{:}\hat\psi^*(\vec x)\big(\vec\alpha(-\i\vec\partial)+m\beta\big)\hat\psi(\vec x){:}
.\end{eqnarray}
 The naive interacting
 Hamiltonian is
\begin{eqnarray}\nonumber
\hat H&=&\int\d\vec x
{:}\hat\psi^*(\vec x)\big(\vec\alpha(-\i\vec\partial+e\vec A(\vec
x))+m\beta+eA_0(\vec x)\big)\hat\psi(\vec x){:}
.
\label{nmun3a+}
\end{eqnarray}
We apply (\ref{bogol1a}) to compute the difference between the ground state
energies of $\hat H$ and $\hat H_\fr$, obtaining 
\begin{eqnarray}\notag&&\Tr\Big(-\big|\vec\alpha(-\i\vec\partial+e\vec A(\vec
x))+m\beta+eA_0(\vec x)\big|+\big|\vec\alpha(-\i\vec\partial)+m\beta)\big|\Big)\\
&=&\sum_{n=1}^\infty e^{2n}E_{n}(A).\label{termi}
\end{eqnarray}

Note that we could have assumed that $\hat H_\fr$ and $\hat H$ are given by the antisymmetric quantization, used the formula (\ref{bogola}), and we would have obtained the same result for the energy shift.
Indeed, formally, the Wick and Weyl quantized versions of $\hat H_\fr$ and $\hat H$ differ by the same (infinite) constant (which was not true in the bosonic case).

All the terms in (\ref{termi}) with $n\geq2$ are well defined. The
term with $n=1$ needs renormalization. The renormalized energy
shift is
\begin{eqnarray}E^\ren
&=&e^2\int \Pi^\ren(\vec p^2)
\bar{F_{\mu\nu}(\vec p)}F^{\mu\nu}(\vec p)
\frac{\d\vec p}{(2\pi)^3}
+
\sum_{n=2}^\infty e^{2n}E_{n}(A),\notag
\end{eqnarray}
where $\Pi^\ren  $ was introduced in (\ref{piren1}).

\init\section{Majorana fermions}

In this section we consider again  the {\em Dirac equation}
\[(-\i\gamma^\mu\partial_\mu+m)\phi(x)=0.\]
We will quantize the space of its solutions satisfying 
the {\em Majorana condition}. We obtain a formalism that describes
{\em neutral} fermions. 

In the bosonic case we first treated the neutral case and only then the charged case. In the fermionic case it is convenient to reverse the order.

\nowastrona
\subsection{Free Majorana fermions}

\subsubsection{Charge conjugation}
Consider a representation of Dirac matrices $\gamma^\mu$.
Let $\kappa$ be a unitary $4\times4$ matrix described in Subsubsect.
\ref{Classical discrete symmetries1}.
We say that  $u\in\cc^4$ is {\em neutral} or satisfies the {\em
  Majorana condition}
 if $u=\kappa\bar u$.

Recall that in the Majorana
representation $\kappa$ can be taken to be the identity.  
In the Dirac and spinor representation
$\kappa:=\i\gamma_2.$

\subsubsection{Space of solutions}
If a function $\zeta$  satisfies the Dirac equation
\[(-\i\gamma^\mu\partial_\mu+m)\zeta(x)=0,\] then
$\kappa\bar\zeta$ also satisfies the Dirac equation. Therefore, we can
restrict the Dirac equation to 
functions $\zeta$ satisfying the
{\em Majorana condition}
\beq 
\kappa\bar \zeta=\zeta.\label{majorana}
\eeq
The space of smooth space compact solutions of the Dirac equation satisfying
(\ref{majorana}) will be denoted $\cY_\D$.
Note that it is a real vector space equipped with a nondegenerate
scalar product
\[\bar\zeta_1\cdot\zeta_2=\int\bar{\zeta_1(t,\vec x)}\zeta_2(t,\vec
x)\d\vec x.\]

In the Majorana representation the space $\cY_\D$ consists simply of real
functions. However, we will most often use the Dirac representation,
where the Majorana condition is less trivial.

Let $\phi(x)$ be the  linear functional on $\cY_\D$ defined by
\[\langle \phi(x)|\zeta\rangle=\zeta(x).\]
The complexification of $\cY_\D$, that is $\cc\cY_\D$, can be
identified with $\cW_\D$. We can extend $\phi(x)$ to $\cc\cY_\D$ by complex linearity. The subspace $\cY_\D$ is then determined by the condition
\beq \kappa\phi^*(x)=\phi(x),\label{major}\eeq
where $*$ is the complex conjugation as defined in (\ref{sense}).

\subsubsection{Smeared fields}

Smeared fields are defined very similarly as for Dirac fields. Note
that in spite of the similarity of the formulas, the objects are
different: they act on the real space $\cY_\D$, and not on the complex
space $\cW_\D$.

For $\zeta\in\cW_\D$, the corresponding {\em spatially smeared field} is
 the functional on $\cY_\D$ given by
\begin{eqnarray*}
\langle\phi\lpar\zeta\rpar|\rho\rangle&:=&\bar\zeta\cdot\rho,\ \ \
\rho\in\cY_\D.
\end{eqnarray*}
Clearly, for any $t$
\begin{eqnarray*}
\phi\lpar\zeta\rpar&=&\int\bar{\zeta(t,\vec x)}\phi(t,\vec x)\d\vec x.
\end{eqnarray*}

\nowastrona

For  $f\in C_{\rm c}^\infty(\rr^{1,3},\cc^4)$ such that $\kappa\bar
f=f$,
 the corresponding {\em
  space-time smeared field} is given by
\begin{eqnarray*}
\phi[f]&:=&\int \bar{f(x)} \phi(x)\d x=\phi\lpar S*f\rpar.
\end{eqnarray*}

\nowastrona

\subsubsection{Plane waves}

Since we
consider neutral fields, the generic name for the momentum variable is again $k$, instead of $p$.

Recall that
in the Dirac representation  we defined the plane waves $u(k,s)$ given by (\ref{planewave}). These plane waves are
compatible with the Majorana condition in the
following sense:
\beq \kappa\bar{u(k,s)}=u(-k,-s).\eeq
We can introduce the plane wave functionals, where $k^0>0$,
\begin{eqnarray*}
a(k,s)&:=&\phi\lpar |k,s)\rpar\\ \notag
&=&
\int\frac{\d\vec{x}}{\sqrt{(2\pi)^3}}
\bar{u(k,s)}\e^{-\i \vec k \vec x}\phi(0,\vec x).
\end{eqnarray*}
Note that
\begin{eqnarray*}
a^*(k,s)&:=&\phi\lpar |-k,-s)\rpar\\ \notag
&=&
\int\frac{\d\vec{x}}{\sqrt{(2\pi)^3}}
\bar{u(-k,-s)}\e^{\i \vec k \vec x}\phi(0,\vec x).
\end{eqnarray*}
We have
\begin{eqnarray*}\phi(x)&=&\sum_{s}
\int\frac{\d\vec{k}}{\sqrt{(2\pi)^3}} \left(u(k,s)\e^{\i kx} 
a( k,s)+u(-k,-s)\e^{-\i kx} 
a^*(k,s)\right)
\\
&=&
\sum_{s}
\int\d\vec{k} \big(|k,s)
a( k,s)+|-k,-s)a^*(k,s)\big).
\end{eqnarray*}

\subsubsection{Quantization}
\label{sec-q-dir-}

To quantize the Dirac equation with the Majorana condition we use the
formalism of quantization of  neutral  fermionic systems \cite{DeGe}.

\nowastrona
We want to construct $(\cH,\hat H,\Omega)$ satisfying the standard requirements of QM (1)-(3) and
a distribution \beq
\rr^{1,3}\ni x\mapsto \hat\phi(x),\eeq
with values  in $\cc^4\otimes B(\cH)$, satisfying the Majorana condition
\beq \kappa\hat\phi^*(x)=\phi(x),\label{majorq}\eeq
and such that the following conditions are true:
 \ben\item $
(-\i\gamma\partial+m)\hat\phi(x)=0;$
\item
$[\phi_a(0,\vec x),\phi_b^*(0,\vec y)]_+=2\delta_{ab}\delta(\vec x-\vec
  y)$;
\item 
$\e^{\i t\hat H}\hat\phi(x^0,\vec x)\e^{-\i t\hat H}=\hat\phi(x^0+t,\vec x)$;
\item 
$\Omega$ is cyclic for $\hat\phi(x)$.\een

\nowastrona

The above problem has an essentially unique  solution,
which we describe below.

Let $\cZ_\D\simeq L^2(\rr^3,\cc^2)$ denote the fermionic positive frequency Hilbert space defined in Subsubsect. \ref{posi-dirac}.
We set $\cH:=\Gamma_\a(\cZ_\D)$.
Creation/annihilation operators on $\cZ_\D$ will be denoted $\hat
a^*$/$\hat a$.
In particular, for $k$ on shell and $s=\pm\frac12$, we have creation operators, written below in both physicist's and mathematician's notation:
\beq \hat a^*(k,s)=\hat a^*\big(|k,s)\big).\label{qrt2-}\eeq
\nowastrona
The quantum field is
\begin{eqnarray*}
\hat\phi(x)&:=&
\sum_s\int\frac{\d\vec{k}}{\sqrt{(2\pi)^3}}
\left(u( k,s)\e^{\i kx}\hat a( k,s)+
u( -k,-s)\e^{-\i kx}\hat a^*( k,s)\right)
.\end{eqnarray*}
The quantum Hamiltonian and  momentum are
\begin{eqnarray*}
\hat H&:=&\int\sum_s \hat a^*( k,s)\hat a(k,s)\varepsilon(\vec
k)
\d\vec{k},\\
\vec {\hat P}&:=&\int\sum_s \hat a^*( k,s)\hat a(k,s)
\vec k\d\vec{k}.
\end{eqnarray*}
The whole
$\rr^{1,3}\rtimes  Spin^\uparrow(1,3)$ acts unitarily on $\cH$. 
Moreover, if we set
 $\tilde{\hat\phi}(x):=\beta\hat\phi^*(x)$, then
\beq [\hat\phi_a(x),\tilde{\hat
 \phi}_b(y)]_+=2S_{ab}(x-y).\label{poij=}\eeq
%
%

\nowastrona

We have
\begin{eqnarray*}
(\Omega|\hat\phi_a(x)\tilde{\hat\phi}_b(y)\Omega)
&=& 2S_{ab}^{(+)}(x-y),\\
(\Omega|\T(\hat\phi_a(x)\tilde {\hat\phi}_b(y))\Omega)
&=&2S_{ab}^{\rm c}(x-y).
\end{eqnarray*}

\nowastrona

For  $f\in C_{\rm c}^\infty(\rr^{1,3},\cc^4)$ such that $\kappa\bar
f=f$, we set
\begin{eqnarray*}\hat\phi[f]&:=&\int\bar{f(x)}\hat\phi(x)
\d x.\end{eqnarray*}
If we use the Majorana representation, so that $\kappa=\one$, we obtain an operator valued distribution satisfying
 the  Wightman
  axioms  with
$\cD:=\Gamma_\a^\fin(\cZ_\D)$.

For an open  set $\cO\subset \rr^{1,3}$ the field algebra  is defined as
\[\fF(\cO):=\{\hat\phi[f]\ :\ f\in C_{\rm
  c}^\infty(\cO,\cc^4), \ \kappa \bar f=f\}''.\] The observable
algebra $\fA(\cO)$ is the even subalgebra of
$\fF(\cO)$.
The nets of algebras $\fF(\cO)$ and $\fA(\cO)$, $\cO\subset\rr^{1,3}$,
satisfy the Haag-Kastler axioms.

\nowastrona
\subsubsection{Quantization in terms of smeared fields}

There exists an alternative equivalent formulation of the quantization
program, which uses the smeared fields instead of point fields.
We look for a linear function
\[\cY_\D\ni\zeta\mapsto\hat\phi\lpar\zeta\rpar\]
with values in bounded self-adjoint operators such that
\ben\item
$
[\hat\phi\lpar\zeta_1\rpar,\hat\phi\lpar\zeta_2\rpar]_+
=2\bar\zeta_1\cdot\zeta_2$;
\item
$\hat\phi\lpar r_{(t,\vec0)}\zeta\rpar
=\e^{\i t\hat H}\hat\phi\lpar\zeta\rpar\e^{-\i t\hat H}$;
\item 
$\Omega$ is cyclic for
$ \hat\phi\lpar\zeta\rpar$.\een

\nowastrona

One can pass between these two versions of the quantization  by
\beq
\hat\phi\lpar\zeta\rpar =\int\bar{\zeta(t,\vec x)}\hat\phi(t,\vec x)\d\vec x.
\eeq

\subsection{Majorana fermions with a mass-like perturbation}

\subsubsection{Classical fields}

The meaning of the expression a {\em mass-like perturbation} is slightly different for fermions, where we perturb $m$,  and for bosons,
where  we perturb $m^2$.

``Classical'' Majorana fields with a mass-like perturbation satisfy the Majorana condition (\ref{major}) and the  equation
\beq\label{poisson8=}
(-\i\gamma_\mu\partial^\mu +m)\phi(x)=-\sigma(x)\phi(x),\eeq
where we assume that $\rr^{1,3}\ni x\mapsto \sigma(x)$ is a given real Schwartz
function.

Let us define the corresponding retarded and advanced propagators as
the unique distributional solutions of
\beq
(-\i\gamma_\mu\partial^\mu+\sigma(x))S^\pm(x,y)=\delta(x-y)\eeq
satisfying
\[\supp S^\pm\subset\{x,y\ :\ x\in J^\pm(y)\}.\]
We also set
\[S(x,y):=S^+(x,y)-S^-(x,y).\]
Clearly
\[\supp S\subset\{x,y\ :\ x\in J(y)\}.\]
 
The ``classical'' Majorana field coinciding with the free field at
time $t=0$ is defined as
\begin{eqnarray}
\phi(t,\vec x)&=&\int S(t,\vec x,0,\vec y)\beta\phi_\fr(0,\vec
y)\d \vec y.\notag 
\label{poio22}
\end{eqnarray}

\subsubsection{Lagrangian and Hamiltonian formalism}
The  Lagrangian density that yields  (\ref{poisson8=}) is
\begin{eqnarray*}
\cL(x)&=&
-\frac{1}{2}\left(\tilde\phi(x)\gamma^\mu(-\i\partial_\mu)\phi(x)
+
\i\partial_\mu\tilde\phi(x)\gamma^\mu\phi(x)
\right)+ \tilde\phi(x)\big(m+\sigma(x)\big)\phi(x),
\end{eqnarray*}
where $\phi(x)$ are off-shell fields satisfying the Majorana condition (\ref{major}).

We can introduce the {\em Hamiltonian density}
\begin{eqnarray*}
\cH(x)&=&
\dot\phi(x)\frac{\cL(x)}{\partial_{\dot\phi(x)}}-\cL(x)\\
&=&
\frac12\Big(\phi^*( x)\vec\alpha(-\i\vec\partial)\phi(x)+\i\vec\partial\phi^*(x)\phi(x)\Big)
+\phi^*(x)\big(m+\sigma(x)\big)\beta\Big)\phi( x),
\end{eqnarray*}
and the  {\em Hamiltonian} 
\[H(t)=\int \cH(t,\vec x)\d\vec x.
\]

\subsubsection{Quantum fields}
The quantum fields should satisfy the Majorana condition (\ref{majorq}), the equation
\beq\label{poisson8=+}
(-\i\gamma_\mu\partial^\mu +m)\hat\phi(x)=-\sigma(x)\hat\phi(x),\eeq
and they should coincide with the free fields at time $t=0$:
\[\hat\phi(\vec x):=\hat\phi(0,\vec x)=\hat\phi_\fr(0,\vec x).\]
The quantization amounts to putting ``hats'' onto
(\ref{poio22}).

We  write the  Schr\"odinger picture Hamiltonian as
\[\hat H(t):=
\int{:}
\hat\phi^*(\vec x)\Big(\alpha_i\i\partial_i
+\big(m+\sigma(t,\vec x)\big)\beta\Big)\hat \phi(\vec  x){:}\d\vec x.\]

The  interaction picture Hamiltonian is
\begin{eqnarray}
\hat H_\Int(t)&=&\frac12\int\sigma(t,\vec x){:}\hat\phi_\fr^*(t,\vec
x)\beta\hat\phi_\fr(t,\vec
x)
{:}\d \vec x.\nonumber\end{eqnarray}

As usual, we define the scattering operator,
scattering amplitudes,
Green's functions, amputated Green's functions and  the generating
function.

\subsubsection{Path integral formulation}

The generating function (and hence all the other quantities introduced
above) can be computed exactly. It equals
\begin{eqnarray}\nonumber
Z(f)&=&\det\left(\frac{\big(-\i\gamma\partial+m+\sigma\big)}{\big(-\i\gamma\partial+m+\sigma-\i0\big)}\exp\Bigl(-\sigma\frac{1}{-\i\gamma\partial+m-\i0}\Bigr)\right)^{\frac12}\\\nonumber
&&\times\exp \left(\frac{\i}{2}\bar f(-\i\partial\gamma+m+\sigma-\i0)^{-1}f\right)\\\nonumber
&=&\det\left(\bigl(\one+\sigma S_\fr^{\rm c}\bigr)\exp\bigl(-\sigma S_\fr^{\rm c}\bigr)\right)\\
&&\times\exp\left(\frac{\i}{2}\bar f S_\fr^{\rm c}\left(\one+\sigma S_\fr^{\rm c}\right)^{-1} f\right)
.\label{causapro2=}
\end{eqnarray}

(\ref{causapro2=})
can be expressed in terms of path integrals:
\begin{eqnarray}\label{causapro1=}
C\int\lpi_x\d\phi(x)\exp\left(\i\int\big(\cL(x)-f(x)\phi(x)\big)\d x\right).
 &&\nonumber\end{eqnarray}
Here, $C$ is a normalization constant, which  does not depend on $f$.
As usual, the formula (\ref{causapro1=}) is only symbolic, the full information is contained  in
(\ref{causapro2=}).

One can derive Feynman rules fully analogous to the Feynman rules of
bosonic mass-like perturbations.

\subsubsection{Vacuum energy}

The logarithm of the vacuum-to-vacuum scattering amplitude can be
computed exactly:
\begin{eqnarray}\nonumber\cE\ =\ 
\i\log(\Omega|S\Omega)&=&\i\log Z(0)\\\nonumber
&=&\frac{\i}{2}\Tr\Big(\log(1+\sigma S^{\rm c})+\sigma S^{\rm c}\Big)\\
&=&\i\sum_{n=2}^\infty\frac{(-1)^{n+1}}{2n}\Tr(\sigma S^{\rm c})^n
=:\sum_{n=2}^\infty\cE_n\notag
.\end{eqnarray}
Note that $\cE_n=-\i\frac{D_n}{2n}$, where
 $D_n=(-1)^n\Tr(\sigma S^{\rm c})^n$ is
the value of  the loop with $n$ vertices, similarly to the bosonic case (\ref{paks}) except for a different sign.

\subsubsection{Renormalization of the vacuum energy}

The $n$th order contribution to the vacuum energy has the form
\begin{eqnarray}\cE_n&=&\int\pi(k_1,\dots,k_n)\notag\\&&
\times\sigma(k_1)\cdots\sigma(k_{n-1})\sigma(-k_1\cdots-k_{n-1})\frac{\d k_1}{(2\pi)^4}\cdots\frac{\d k_{n-1}}{(2\pi)^4}.\label{specia}
\end{eqnarray}
For $n=2,3,4$, $\cE_n$ are divergent and need renormalization.

Using the Pauli-Villars method we define for $n=1,2,3$ the renormalized vacuum energy functions
\begin{eqnarray*} \pi^\ren(k_1,\dots,k_{n-1})&:=&\lim_{\Lambda\to\infty}
\big(\pi_\Lambda(k_1,\dots,k_{n-1})-\pi_\Lambda(0,\dots,0)\big).
\end{eqnarray*}
Thus
\begin{eqnarray*}&&\cE_n^\ren\\&=&\int\pi^\ren(k_1,\dots,k_{n-1})\sigma(k_1)\cdots\sigma(k_{n-1})\sigma(-k_1\cdots-k_{n-1})\frac{\d k_1}{(2\pi)^4}\cdots\frac{\d k_{n-1}}{(2\pi)^4}\\
&=&\lim_{\Lambda\to\infty}\Bigg(
\int\pi_\Lambda(k_1,\dots,k_{n-1})\sigma(k_1)\cdots\sigma(k_{n-1})\sigma(-k_1\cdots-k_{n-1})\frac{\d
k_1}{(2\pi)^4}\cdots\frac{\d k_{n-1}}{(2\pi)^4}\\
&&-\pi_\Lambda(0,\dots,0)\int\sigma(x)^n\d x\Bigg).
\end{eqnarray*}
The renormalized scattering operator $\hat S_\ren$ is a well defined unitary operator. Formally, we have
\begin{eqnarray*}
\hat S_\ren&=&\e^{\big(\i\pi_\infty(0)\int \kappa(x)^2\d x
+\i\pi_\infty(0,0)\int \kappa(x)^3\d x
+\i\pi_\infty(0,0,0)\int \kappa(x)^4\d x
\big)} \hat S.\label{formo0}\\
\cL_\ren(x)&=&\cL(x)+\pi_\infty(0) \kappa(x)^2
+\pi_\infty(0,0)\kappa(x)^3+\pi_\infty(0,0,0) \kappa(x)^4.
\label{formo-}\end{eqnarray*}


\subsubsection{Pauli-Villars renormalization of the 2nd order term}

 $\cE_3$ and $\cE_4$ are logaritmically divergent. Below we  present computations only for $\cE_2$, which is quadratically divergent.  As a special case of (\ref{specia}) for $n=2$ we write
\[\cE_2=\int|\sigma(k)|^2\pi(k^2)
\frac{\d k}{(2\pi)^4}.\]
Using  the {\em Pauli-Villars regularization}, as in Subsubsect. \ref{Pauli-Villars}, we compute:
\nowastrona

\begin{eqnarray*}
&&4\pi_\Lambda(k^2)\\
&=&-\i\int\frac{\d^4
  q}{(2\pi)^4}\sum_iC_i\tr
\frac{\big(-(q+\frac12k)\gamma+m_i\big)\big(-(q-\frac12k)\gamma+m_i\big)}
{((q+\frac12k)^2+m_i^2-\i0)((q-\frac12k)^2+m_i^2-\i0)}
\\
&=&-\i\int\frac{\d^4
  q}{(2\pi)^4}\sum_iC_i
\frac{\big(-4q^2+k^2+4m_i^2\big)}
{((q+\frac12k)^2+m_i^2-\i0)((q-\frac12k)^2+m_i^2-\i0)}
\\
&=&
-\frac{1}{(4\pi)^2}\int_0^\infty\d\alpha_1\int_0^\infty\d\alpha_2\sum_iC_i\Bigg(
\frac{4\alpha_1\alpha_2k^2}{(\alpha_1+\alpha_2)^2}+\frac{4m^2}{(\alpha_1+\alpha_2)^2}+\frac{2\i}{(\alpha_1+\alpha_2)^3}\Bigg)\\&&\times
\exp\left(-\i(\alpha_1+\alpha_2)m_i^2
-\i\frac{\alpha_1\alpha_2}{\alpha_1+\alpha_2}k^2 \right)\\
&=&
-\frac{1}{(4\pi)^2}
\int_0^1\d v\int_0^\infty\frac{\d\rho}{\rho}
\sum_iC_i\Bigg((1-v^2)k^2+4m^2+\frac{2\i}{\rho}\Bigg)\\
&&\times\exp\left(-\i\rho\left( m_i^2
+\frac{1-v^2}{4}
  k^2\right)\right)\\
&=&\frac{1}{(4\pi)^22}
\int_0^1\d v
\sum_iC_i\Big((1-v^2)k^2+4m_i^2\Big)\log\left(m_i^2
+\frac{(1-v^2)}{4}
  k^2-\i0\right).
\end{eqnarray*}
\nowastrona
Note that at the end we use  (\ref{ide3}) besides (\ref{ide3a}), because of the quadratic divergence.

Finally, the {\em renormalized vacuum energy function} is defined as
\begin{eqnarray}
\pi^\ren(k^2)
&=&\lim_{\Lambda\to\infty}\Big(\pi_\Lambda(k^2)-\pi_\Lambda(0)\Big)\label{piren10}
\\
&=&
-\frac{m^2}{(4\pi)^2}\int_0^1\Big(\frac{1}{2}+\frac{(1-v^2)k^2}{8m^2}\Big)
\log\Big(1+\frac{(1-v^2)k^2}{4m^2}-\i0\Big)\d v.\notag
\end{eqnarray}

\appendix
\init\section{Appendix}

\subsection{Second quantization}
\subsubsection{Fock spaces}

Let $\cZ$ be a Hilbert space.
Let $S_n$ denote the {\em permutation group of $n$ elements} and
 $\sigma\in S_n$.
$\Theta(\sigma)$ is defined as the unique operator in 
$B(\otimes^n\cZ)$
such that
\[\Theta(\sigma)g_1\otimes\cdots\otimes g_n
=g_{\sigma^{-1} (1)}\otimes\cdots\otimes
g_{\sigma^{-1} (n)},\ \ g_1,\dots,g_n\in\cZ.\]
$\Theta(\sigma)$ is unitary. We define the
{\em symmetrization/antisymmetrization projections}
\begin{eqnarray*}
\Theta_\s^n:=\frac{1}{n!}\sum_{\sigma\in S_n}\Theta(\sigma),&&
\Theta_\a^n:=\frac{1}{n!}\sum_{\sigma\in S_n}\sgn\sigma\Theta(\sigma)
.\end{eqnarray*}
In what follows we will consider in parallel the symmetric/antisymmetric, or bosonic/fermionic case.  To facilitate notation we will write $\sa$ for {\em either $\s$ or $\a$}.

$\Theta_\sa^n$ are orthogonal projections.
The {\em   $n$-particle bosonic/fermionic space}
is defined as
\beq{\otimes}_\sa^n\cZ:=\Theta_\sa^n{\otimes}^n\cZ.\label{fock1}\eeq
The {\em  bosonic/fermionic Fock space}  is 
 \beq\Gamma_\sa(\cZ):=\loplus_{n=0}^\infty{\otimes}_\sa^n\cZ.\label{fock2}\eeq
The {\em vacuum vector} is $\Omega:=1\in\otimes_\sa^0\cZ=\cc$.

We use the convention saying that  the tensor products and direct sums used in (\ref{fock1}) and (\ref{fock2}) 
are completed in their natural topology, so that  $n$-particle spaces and  Fock space are Hilbert spaces. Sometimes we may want a similar construction without the completion (in particular, if $\cZ$ is not a Hilbert space). Then we will speak about {\em algebraic $n$-particle spaces} or {\em algebraic Fock spaces}.

\subsubsection{Creation/annihilation operators}

\nowastrona

For  $g\in\cZ$ we define the {\em creation operator}
\[\hat a^*(g)\Psi:=\Theta_\sa^{n+1}
\sqrt{n+1} g\otimes\Psi,\ \ \Psi\in\otimes_\sa^n\cZ,\]
and the  {\em annihilation operator}
$\hat a(g):=\left(\hat a^*(g)\right)^*$.

\medskip
Above we used a compact notation
 for
creation/annihilation operators popular among
mathematicians. Physicists commonly prefer  
 another notation, which is longer and less canonical, but often more flexible. In order to
introduce it, we need to
 fix an
 identification of $\cZ$ with $L^2(\Xi)$ of some measure space $\Xi$
 with its elements called generically $\xi$ and the measure called $\d\xi$.
For instance,  $\Xi$ can be $\rr^d$ with the Lebesgue measure.
 Every $g\in \cZ$ can be represented as a function $\Xi\ni \xi\mapsto g(\xi)$. Then
\begin{eqnarray}\hat a^*(g)&=&\int g(\xi)\hat a^*(\xi)\d \xi,\label{nota1}\\
 \ \ \hat a(g)&=&\int \bar
{g(\xi)}\hat a(\xi)\d \xi.\label{nota}\end{eqnarray}
We will call the notation on the left of (\ref{nota1}) and (\ref{nota})
{\em ``mathematician's notation''}
and on the right {\em ``physicist's notation''}.

Sometimes one  introduces formal symbols $|\xi)$ treated as vectors, possibly nonnormalizable, such that for $g\in L^2(\Xi)$ we can write 
\[g=\int|\xi)g(\xi)\d\xi,\ \ \ g(\xi)=(\xi|g).\]
We have the following dictionary  between  creation operators written in the ``physicist's notation'' (on the left) and the ``mathematician's notation''
(on the right):
\begin{eqnarray} \hat a^*(\xi )&=&\hat a^*\big(|\xi )\big),\label{qrt2}\\
\int(\xi |g)\hat a^*(\xi )\d \xi &=&\hat a^*(g).\label{qrt4}
\end{eqnarray}

Let $[\cdot,\cdot]_-$, resp.
 $[\cdot,\cdot]_+$ denote the commutator, resp. anticommutator.
Bosonic/fermionic creation and annihilation operators satisfy the canonical commutation/anticommutation relations, which in the ``mathematician's notation'' read
\begin{eqnarray*}
[\hat a^*(f),\hat a^*(g)]_\mp=[\hat a(f),\hat a(g)]_\mp&=&0,\\{}
[\hat a(f),\hat a^*(g)]_\mp&=&(f|g)=\int\bar{f(\xi)}g(\xi)\d\xi,
\end{eqnarray*}
and in the ``physicist's notation'', at least for $\Xi=\rr^d$, have the form
\begin{eqnarray*}
[\hat a^*(\xi),\hat a^*(\xi')]_\mp=[\hat a(\xi),\hat a(\xi')]_\mp&=&0,\\{}
[\hat a(\xi),\hat a^*(\xi')]_\mp&=&\delta(\xi-\xi').
\end{eqnarray*}

\nowastrona
\subsubsection{Weyl/antisymmetric and Wick quantization}
\label{Weyl and Wick quantization}

Let
\beq
(\xi_1,\cdots\xi_m,\xi_n',\cdots,\xi_1')\mapsto 
b(\xi_1,\cdots\xi_m,\xi_n',\cdots,\xi_1')\label{kerno}\eeq
be a complex function, symmetric/antisymmetric separately wrt the first $m$
and the last $n$ arguments. Let us introduce the following expression:
\begin{eqnarray}\label{kerno1}&&
\int\cdots\int b(\xi_1,\cdots\xi_m,\xi_n',\cdots,\xi_1')\\
&&
\times a^*(\xi_1)\cdots a^*(\xi_m)a(\xi_n')\cdots a(\xi_1')
\d\xi_1\cdots\d\xi_m\d\xi_n'\cdots\d\xi_1',\notag\end{eqnarray}
where $a(\xi)$ and $a^*(\xi)$ are commuting/anticommuting symbols.

In the symmetric case (\ref{kerno1}) can be interpreted as 
a {\em polynomial on $\cZ\oplus\bar\cZ$}. Indeed, if we interpret the symbols $a(\xi)$ as the evaluations of $g\in\cZ= L^2(\rr^d)$:
\[\langle a(\xi)|g\rangle:=g(\xi),\ \ \langle a^*(\xi)|g\rangle:=\bar{g(\xi)},\]
then (\ref{kerno1}) has the meaning of a polynomial function.
It is common to use the name a 
{\em polynomial}  for (\ref{kerno1}) also in the antisymmetric case.


\nowastrona

The {\em Wick quantization of  (\ref{kerno1})} is the operator on the Fock space given by  the same expression, except 
that we put the ``hats'' on $a$ and $a^*$. Note that 
 the creation operators are on the left and annihilation operators are on the right:
\begin{eqnarray*}
&&\int b(\xi_1,\cdots\xi_m,\xi_n',\cdots,\xi_1')\\
&&\times\hat a^*(\xi_1)\cdots \hat a^*(\xi_m)\hat a(\xi_n')\cdots \hat a(\xi_1')\d
\xi_1,\cdots\xi_n\d \xi_1'\cdots\d\xi_m'.
\end{eqnarray*}

In practice we often have some fields, say $\varphi_1(\xi)$, $\varphi_2(\xi)$, that can be written as linear combinations of $a(\xi)$ and $a^*(\xi)$, eg.
\begin{eqnarray*}
\varphi_i(\xi)&=&\int A_i(\xi)a(\xi)+
\int B_i(\xi)a^*(\xi).\label{pada61}
\end{eqnarray*}
Their quantizations are denoted by ``hats'':
\begin{eqnarray*}
\hat\varphi_i(\xi)&=&\int A_i(\xi)\hat a(\xi)+
\int B_i(\xi)\hat a^*(\xi).
\end{eqnarray*}

 Suppose we have a  polynomial
\beq
\sum_{i_1,\dots,i_m}\int\cdots\int c_{i_1,\dots,i_m}(\xi_1,\cdots\xi_m)
\varphi_{i_1}(\xi_1)\cdots \varphi_{i_m}(\xi_m)
\d\xi_1\cdots\d\xi_m.\label{kerno9}\eeq
We assume that the coefficients $c_{i_1,\dots,i_m}(\xi_1,\cdots\xi_m)$ are symmetric/antisymmetric. The most natural quantization of  (\ref{kerno9})
 is the operator on the Fock space given by  the same expression, where we just put ``hats'' on the fields. It is called the {\em Weyl quantization  } in the bosonic case. In the fermionic case this quantization seems to have no established name, although it would be tempting to call it the {\em fermionic Weyl quantization}. Following \cite{DeGe}, we will call it the {\em antisymmetric quantization}.

By inserting (\ref{pada61}), we obtain a polynomial expressed in terms of $a(\xi)$ and $a^*(\xi)$. Its Wick quantization has the traditional notation where the expression decorated with hats is put between double dots:
\begin{eqnarray*}&&
{:}\sum_{i_1,\dots,i_m}\int\cdots\int c_{i_1,\dots,i_m}(\xi_1,\cdots\xi_m)
\hat\varphi(\xi_1)\cdots \hat\varphi(\xi_m)\d\xi_1\cdots\d\xi_m{:}.\end{eqnarray*}

For 1st order polynomials their Wick quantization obviously coincides with their Weyl/antisymmetric quantization:
\begin{eqnarray*}
{:}\int f(\xi)\hat\varphi(\xi)\d\xi{:}&=&\int f(\xi)\hat\varphi(\xi)\d\xi.
\end{eqnarray*}

We will often use Wick quantizations of second degree polynomials.
For instance, let $c(\xi,\xi')$ be a symmetric/antisymmetric function. Then
the Wick and Weyl/antisymmetric quantizations differ by the vacuum expectation value:
\begin{eqnarray*}&&
{:}\int\int c(\xi,\xi') \hat\varphi(\xi)\hat\varphi(\xi')\d\xi\d\xi'{:}\\
&=&\int\int c(\xi,\xi') \Big(A(\xi)A(\xi) \hat a^*(\xi)\hat a^*(\xi')
+A(\xi) B(\xi') \hat a^*(\xi)\hat a(\xi')\\&&
\pm B(\xi)A(\xi') \hat a^*(\xi')\hat a(\xi)+
B(\xi) B(\xi') \hat a(\xi)\hat a(\xi')\Big)\d\xi\d\xi'
\\
&=&\int\int c(\xi,\xi') \hat\varphi(\xi)\hat\varphi(\xi')\d\xi\d\xi'
-
\int\int c(\xi,\xi')\big(\Omega|\hat \varphi(\xi)\hat\varphi(\xi')\Omega\big)\d\xi\d\xi'.
\end{eqnarray*}

\nowastrona
\subsubsection{Second quantization of operators}

For  a contraction $q$ on $\cZ$ we define the operator $\Gamma(q)$ on
$\Gamma_\sa(\cZ)$ by
\[\Gamma(q)\Big|_{\otimes_\sa^n\cZ}=q\otimes\cdots\otimes
q\Big|_{\otimes_\sa^n\cZ} .\]
$\Gamma(q)$ is called the {\em second quantization of $q$}.

Similarly, for an operator $h$  we define the  operator $\d\Gamma(h)$ by
\[\d\Gamma(h)\Big|_{\otimes_\sa^n\cZ}=h\otimes1^{(n-1)\otimes}+\cdots
+1^{(n-1)\otimes}\otimes h\Big|_{\otimes_\sa^n\cZ}.\]
$\d\Gamma(h)$ is called the {\em (infinitesimal) second quantization of $h$}.

If  $h$ is
 the multiplication operator by $h(\xi)$, then
using physicist's notation we have
\[\d\Gamma(h)=\int h(\xi)\hat a^*(\xi) \hat a(\xi)\d\xi.\]

Note the identity $\Gamma(\e^{\i t h})=\e^{\i t\d\Gamma(h)}$.

\subsubsection{Implementability of Bogoliubov translations}

Consider bosonic creation/annihilation operators. Let $\xi\mapsto f(\xi)$ be a complex function. Set
\begin{eqnarray*}
\hat a_1^*(\xi)&=\hat a^*(\xi)+\bar{f(\xi)},\\
\hat a_1(\xi)&=\hat a(\xi)+f(\xi).
\end{eqnarray*}
A proof of the following well-known fact can be found eg. in \cite{DeGe}.
\bet There exists a unitary operator $U$ on the Fock space such that
\[U\hat a^*(\xi)U^*=\hat a_1^*(\xi),\ \ U\hat a(\xi)U^*=\hat a_1(\xi),\]
iff 
\[\int |f(\xi)|^2\d\xi<\infty.\]
Up to a phase factor 
\[U=\exp\Big(\int \big(\hat a(\xi) f(\xi)-\hat a^*(\xi)\bar{ f(\xi)}\big)\d\xi\Big).\]
\label{transla}\eet

The following formula  is a time-dependent generalization of the well-known
identity $\e^{\i \hat a^*(f)+\i \hat a(f)}=\e^{-\frac{1}{2}(f|f)}\e^{\i \hat a^*(f)}\e^{\i \hat a(f)}$:
\begin{eqnarray}\label{bch}
&&\Texp\Bigg(\i \hat a^*\Big(\int f(t)\d t\Big)+\i \hat a\Big(\int f(t)\d t\Big)\Bigg)\\
&=&\e^{-\int\int(f(t_1)|f(t_2))\theta(t_1-t_2)\d t_1\d t_2}\notag
\e^{\i \hat a^*\big(\int f(t)\d t\big)}
\e^{\i \hat a\big(\int f(t)\d t\big)}
.\notag
\end{eqnarray}

\subsubsection{Implementability of Bogoliubov rotations}

We will treat simultaneously the bosonic and fermionic case.
The upper signs will always correspond to the bosonic case and lower to the fermionic case.

Let  $p$, $q$ be  operators with the integral kernels
$p(\xi,\xi')$, $q(\xi,\xi')$. We assume that $q(\xi,\xi')=\pm q(\xi',\xi)$.
Set
\begin{eqnarray}
\hat a_1^*(\xi)&=&\int\big(p(\xi,\xi')\hat a^*(\xi)+q(\xi,\xi')\hat a(\xi')\big)\d\xi',\\
 \hat a_1(\xi)&=&\int\big(\bar{q(\xi,\xi')}\hat a^*(\xi')+\bar{p(\xi,\xi')}\hat a(\xi)\big)\d\xi'.
\end{eqnarray}
Assume that
\begin{eqnarray*}
p^*p\mp q^\t \bar q=\one,&& p^*q\mp q^\t \bar p=0,\\
 pp^*\mp qq^*=\one,&& pq^\t \mp qp^\t =0,
\end{eqnarray*}
which guarantees that $\hat a_1^*$, $\hat a_1$ satisfy the same commutation/anticommutation relations as $\hat a^*$, $\hat a$.

Here, we use the following notation:
For an operator $p$
we will write
 $p^*$  for its Hermitian conjugate, $p^\#$ for its
transpose of $p$ and $\bar p$ for its complex conjugate.
If the integral kernel of $p$ is $p(\xi,\xi')$, then clearly
\[p^*(\xi,\xi')=\bar{p(\xi',\xi)},
\ \ \ p^\t(\xi,\xi')=p(\xi',\xi),\ \ \ \bar p(\xi,\xi')=\bar{p(\xi,\xi')}.\]

\bet
There exists a unitary $U$ on the Fock space such that
\[U\hat a^*(\xi)U^*=\hat a_1^*(\xi),\ \ U\hat a(\xi)U^*=\hat a_1(\xi),\]
iff $q$ is Hilbert-Schmidt, that means,
\[\int\int|q(\xi,\xi')|^2\d\xi\d\xi'<\infty.\]
\label{shale}\eet

The above theorem is called the {\em Shale criterion} \cite{Sh} in  the bosonic and the {\em Shale-Stinespring criterion} \cite{ShSt} in the fermionic case. See also eg. \cite{DeGe}.

\subsubsection{Infimum of a van Hove Hamiltonian}

Consider a bosonic Hamiltonian of the form
\beq H:=\int\varepsilon(\xi)\hat a^*(\xi)\hat a(\xi)\d\xi+\int v(\xi)\hat a^*(\xi)\d\xi+
\int \bar{v(\xi)}\hat a(\xi)\d\xi.\label{hov}\eeq
Such Hamiltonians are sometimes called {\em van Hove Hamiltonians} \cite{De,DeGe}.
Assume that $\varepsilon$ is positive. We would like to compute the infimum of the spectrum of $H$, denoted $\inf H$.

By completing the square we can rewrite (\ref{hov}) as
\beq\int\varepsilon(\xi)\Big(\hat a^*(\xi)+\frac{\bar{v(\xi)}}{\varepsilon(\xi)}\Big)
\Big(\hat a(\xi)+\frac{v(\xi)}{\varepsilon(\xi)}\Big)\d\xi-\int \frac{|v(\xi)|^2}{\varepsilon(\xi)}\d\xi
.\label{hov1}\eeq
It is easy to see that the infimum  of the first term in (\ref{hov1}) is zero. Hence 
\beq \inf H=-\int \frac{|v(\xi)|^2}{\varepsilon(\xi)}\d\xi\label{hov2}.\eeq

\subsubsection{Infimum of a Bogoliubov Hamiltonian}
Consider a bosonic or fermionic Hamiltonian
\begin{eqnarray}\notag
H&:=&\int h(\xi,\xi')\big(\hat a^*(\xi)\hat a(\xi')\pm \hat a(\xi) \hat a^*(\xi')\big)\d\xi\d\xi'\\
&&+\int\big(g(\xi,\xi')\hat a^*(\xi)\hat a^*(\xi')\pm \bar{g(\xi,\xi')}\hat a(\xi)\hat a(\xi')\big)\d\xi\d\xi'.\label{bogol}
\end{eqnarray}
We assume that $\bar{h(\xi,\xi')}= h(\xi',\xi)$, $g(\xi,\xi')=\pm g(\xi',\xi)$.
We will call (\ref{bogol}) {\em Bogoliubov Hamiltonians}. Note that (\ref{bogol}) is the Weyl/antisymmetric quantization of the corresponding classical quadratic Hamiltonian. In the case of an infinite number of degrees of freedom 
it is often ill defined, but even then it is useful to consider such formal expressions.

 We have the following formula for the infimum of $H$ \cite{DeGe}:
\beq \inf
H=\pm\frac{1}{2}\Tr\left[\begin{array}{cc}h^2\mp gg^*&\mp hg\pm gh^\#\\
g^*h-h^\#g^*& h^{\#2}\mp g^*g\end{array}\right]^{\frac12}.\label{bogola}\eeq
Here, we  write $h$ for the operator with the integral kernel $h(\xi,\xi')$ and $g$ for the operator with the integral kernel $g(\xi,\xi')$.

Consider the Wick ordered version of (\ref{bogol}):
\begin{eqnarray}\notag
{:}H{:}&:=&2\int h(\xi,\xi')\hat a^*(\xi)\hat a(\xi')\d\xi\d\xi'\\
&&+\int\big(g(\xi,\xi')\hat a^*(\xi)\hat a^*(\xi')\pm \bar{g(\xi,\xi')}\hat a(\xi)\hat a(\xi)\big)\d\xi\d\xi'.\label{bogol1}
\end{eqnarray}
(In the case an infinite number of degrees of freedom ${:}H{:}$ has a better chance to be well defined compared with $H$).
The formula for the infimum of ${:}H{:}$ is more complicated, but is more likely to lead to a finite expression \cite{DeGe}:
\beq\inf
{:}H{:}
=\frac{1}{2}\Tr\Bigg(\pm\left[\begin{array}{cc}h^2\mp gg^*&\mp hg\pm gh^\#\\
g^*h-h^\#g^*& h^{\#2}\mp g^*g\end{array}\right]^{\frac12}\mp
\left[\begin{array}{cc}h&0\\0
& h^\#\end{array}\right]\Bigg)
.\label{bogol1a}\eeq

\subsection{Miscellanea}

\subsubsection{Identities for Feynman integrals}

\begin{eqnarray}
\frac{1}{A-\i0}&=&\i\int_0^\infty\d\alpha\exp(-\i\alpha A),\label{ide1}\\
 p_\mu&=&\i\partial_{z_\mu}\exp(-\i pz)\Big|_{z=0},\label{ide0}\\
\int\frac{\d p}{(2\pi)^4}\exp\left( -\i(ap^2+bp)\right)&=&\i\frac{\sgn( a)}{(4\pi)^2a^2}
\exp
\left( \i b^2/4a\right).\label{ide2}
\end{eqnarray}
Using these identities, a typical evaluation of a loop integral goes as follows:
\begin{eqnarray}
&&\frac{\i}{(2\pi)^4}\int\frac{P(q)\d q}
{(a_1q^2+2b_1q+c_1-\i0)\cdots(a_nq^2+2b_nq+c_n-\i0)}\notag\\
&=&\frac{\i^{n+1}}{(2\pi)^4}\int_0^\infty\d \alpha_1\cdots\int_0^\infty\d\alpha_n\int\d q
P(q)\notag\\
&&\times\exp\Big(-\i\alpha_1(a_1q^2+2b_1q+c_1)\cdots-
\i\alpha_n(a_nq^2+2b_qq+c_q)\Big)\notag\\
&=&\frac{\i^{n+1}}{(2\pi)^4}\int_0^\infty\d \alpha_1\cdots\int_0^\infty\d\alpha_n\int\d q
P(\i\partial_z)\exp\Big(-\i(\alpha_1a_1\cdots+\alpha_na_n)q^2\notag\\
&&\ \ -\i(\alpha_1 b_1\cdots+\alpha_n b_n+z)q-\i(\alpha_1 c_1\cdots+\alpha_n c_n)\Big)\Big|_{z=0}\notag\\
&=&-\frac{\i^{n}}{(4\pi)^2}\int_0^\infty\d \alpha_1\cdots\int_0^\infty\d\alpha_n
(\alpha_1 a_1\cdots+\alpha_n a_n)^{-2}P(\i\partial_z)\notag\\
&&\times\exp\Big(\i\frac{(\alpha_1b_1\cdots+\alpha_nb_n+z)^2}
{4(\alpha_1a_1\cdots+\alpha_na_n)}-\i(\alpha_1 c_1\cdots+\alpha_n c_n)\Big)\Big|_{z=0}.\label{loop-com}
\end{eqnarray}

If $\sum C_i=0$, then
\beq
\int_0^\infty\sum_iC_i
\frac{\d\rho}{\rho}\e^{-\i\rho A_i}=-\sum_i C_i\log( A_i-\i0).\label{ide3}\eeq
If in addition $\sum C_i A_i=0$, then
\beq
\int_0^\infty\sum_iC_i
\frac{\d\rho}{\rho^2}\e^{-\i\rho A_i}=-\sum_i C_iA_i\log( A_i-\i0).\label{ide3a}\eeq

\begin{eqnarray}\int\log(A^2-w^2)\d w&=& w\log(A^2-w^2)-2w\notag\\
&&+A\log\frac{(A+w)}{(A-w)},\ \ \ \ \ \ \ 0<w<A;\label{ide4}\\
\int w^2\log(A^2-w^2)\d w&=& \frac{w^3}{3}\log(A^2-w^2)
-\frac{2w^3}{9}-\frac{2A^2w}{3}\notag\\
&&+\frac{A^3}{3}\log\frac{(A+w)}{(A-w)}, \ \ \ \ \ \ \  0<w<A.\label{ide4.1}
\end{eqnarray}

\subsubsection{Identities for the dimensional regularization}

The {\em Feynman identity}:
\beq\frac{1}{AB}=\frac12\int_{-1}^1\frac{\d v}{\big(\frac12(A+B)+\frac12(A-B)v\big)^2}.\label{dim-feyn}
\eeq

The behavior of $\Gamma$ around $0$:
\beq\Gamma(2-d/2)\simeq\frac{1}{2-d/2}-\gamma.\label{dim0}\eeq

The area of the unit $d-1$-dimensional sphere:
\beq\Omega_d=\frac{2\pi^{d/2}}{\Gamma(d/2)}.\label{dim1}\eeq

Integrals, which can be reduced to special cases of the Euler integral:
\begin{eqnarray}
\int_0^\infty\frac{t^{d-1}}{(t^2+A^2)^2}\d t&=&\frac12(A^2)^{-2+d/2}\Gamma(d/2)\Gamma(2-d/2),\label{dim2}\\
\int_0^\infty\frac{t^{d+1}}{(t^2+A^2)^2}\d t&=&\frac12(A^2)^{-1+d/2}\Gamma(1+d/2)\Gamma(1-d/2)\notag\\
&\hspace{-11ex}=&\hspace{-7ex}\frac12(A^2)^{-1+d/2}\Gamma(d/2)\Gamma(2-d/2)(-1+2/d)^{-1}.
\label{dim3}\end{eqnarray}

Typical integrals:
\begin{eqnarray}
&&\frac{\mu^{4-d}\Omega_d}{(2\pi)^d}\int_0^\infty\frac{|q|^{d-1}}
{\big(q^2+A^2\big)^2}\d|q|\notag\\
&=&\frac{1}{(4\pi)^2}\Big(\frac{\mu^24\pi}{A^2}\Big)^{2-d/2}
\Gamma(2-d/2)\notag\\
&\approx&\frac{1}{(4\pi)^2}\Bigg(1+(2-d/2)\log\frac{\mu^24\pi}{A^2}
\Bigg)\Big(\frac{1}{2-d/2}-\gamma\Big)\notag\\
&\approx&
\frac{1}{(4\pi)^2}\Big(-\gamma+\log\frac{\mu^24\pi}{A^2}
+\frac{1}{(2-d/2)}\Big),
\label{dim4}
\\[4ex]
&&\frac{\mu^{4-d}\Omega_d}{(2\pi)^d}\int_0^\infty(-1+2/d)\frac{|q|^{d+1}}
{\big(q^2+A^2\big)^2}\d|q|\notag\\
&=&\frac{A^2}{(4\pi)^2}\Big(\frac{\mu^24\pi}{A^2}\Big)^{2-d/2}
\Gamma(2-d/2)\notag\\
&\approx&\frac{A^2}{(4\pi)^2}\Bigg(1+(2-d/2)\log\frac{\mu^24\pi}{A^2}
\Bigg)\Big(\frac{1}{2-d/2}-\gamma\Big)\notag\\
&\approx&
\frac{A^2}{(4\pi)^2}\Big(-\gamma+\log\frac{\mu^24\pi}{A^2}
+\frac{1}{(2-d/2)}\Big).
\label{dim5}\end{eqnarray}

\subsubsection{Operator identities}
If $A$ is a positive self-adjoint operator, then
\begin{eqnarray}
A^{1/2}&=&\int\frac{A}{(A+\tau^2)}\frac{\d\tau}{2\pi},
\label{sqrt}\\
A^{-1/2}&=&\int\frac{1}{(A+\tau^2)}\frac{\d\tau}{2\pi}\notag\\
&=&-2\int\frac{1}{(A+\tau^2)^2}\tau^2\frac{\d\tau}{2\pi}.
\label{sqrt1}
\end{eqnarray}
In the following identity $\kappa$ is a certain operator. It
is useful when studying $n$th  order loop
diagrams:
\begin{eqnarray}\notag
&&\int\Tr\frac{1}{(A+\tau^2)^2}\kappa
\Big(\frac{1}{(A+\tau^2)}\kappa\Big)^{n-1}
\tau^2\frac{\d\tau}{2\pi}\\
&=&-\frac{1}{2n}\int\Tr
\Big(\frac{1}{(A+\tau^2)}\kappa\Big)^{n}\frac{\d\tau}{2\pi}.
\label{sqrt2}
\end{eqnarray}

\subsubsection{Coulomb and Yukawa potential}

If $\rho\in C_{\rm c}(\rr^3)$, then 
\[\rho=-\Delta f\]
has a unique solution in functions that decay at infinity
 given by
\beq f(\vec x)\,=\, (-\Delta)^{-1}\rho(\vec x)=\int\frac{1}{4\pi|\vec x-\vec y|}\rho(\vec y)\d\vec y.\label{electro}\eeq
For large $|\vec x|$, (\ref{electro}) has the asymptotics
\beq\frac{1}{4\pi|\vec x|}\int\rho(\vec y)\d\vec y+ O\left(\frac{1}{|\vec x|^2}\right).\label{beha}\eeq

More generally
\beq (m^2-\Delta)^{-1}\rho(\vec x)=\int\frac{\e^{-m|\vec x-\vec y|}}{4\pi|\vec x-\vec y|}\rho(\vec y)\d\vec y.\label{electro1}\eeq

\nowastrona

\subsubsection{Vector fields}

Consider a vector field $\rr^3\ni\vec x\mapsto \vec A(\vec x)\in\rr^3$. 
We say that it is
 {\em transversal} if
\[\div\vec A(\vec x)=0.\]
If it is  not necessarily transversal but sufficiently nice,
its {\em transversal part} is defined as
\beq \vec A_\tr(\vec x):=
\vec A(\vec x)+(-\Delta)^{-1}\vec\partial\div\vec A(\vec x).\label{nice}\eeq


We have the identities
\begin{eqnarray}\notag
\int \vec A(\vec x)^2\d\vec x&=&\int \vec A_\tr(\vec x)^2\d\vec x+
\int\big((-\Delta)^{-1/2}\div\vec A(\vec x)\big)^2\d\vec x,\\
\int \big(\vec\partial\vec A(\vec x)\big)^2\d\vec x&=&
\int \big(\vec\partial\vec A_\tr(\vec x)\big)^2\d\vec x
+\int \big(\div\vec A(\vec x)\big)^2\d\vec x,\\
\int \big(\vec\partial\vec A_\tr(\vec x)\big)^2\d\vec x&=&
\frac12\int \big(\rot\vec A(\vec x)\big)^2\d\vec x.
\end{eqnarray}

\subsubsection{Dispersion relations}

The {\em principal value of 
$\frac{1}{\xi}$}, denoted $\cP\frac1\xi$, is the distribution acting on a test function $f$  as
\[\cP\int\frac{f(\xi)}{\xi}\d\xi
:=\lim_{\epsilon\searrow0}\left(
\int_{-\infty}^{-\epsilon}+  \int_{\epsilon}^\infty\right)
\frac{f(\xi)}{\xi}\d\xi.\]
It appears in the {\em Sochocki formula}
\[\frac{1}{\xi\pm\i0}=\lim_{\epsilon\searrow0}\frac{1}{\xi\pm\i\epsilon}=
\mp\i\pi\delta(\xi) +\cP\frac{1}{\xi}.\]

 Let $f$ be holomorphic on $\{\Im z>0\}$ with continuous boundary values at the real line.
 Let $f=f_\R+\i f_\I$ be
its decomposition into the real and imaginary part. The following theorem follows easily from the Cauchy formula and describes what physicists call
{\em  dispersion relations}:
\bet
Assume that $f\in C^1(\rr)$, $\frac{f}{1+| E |}\in L^1(\rr)$ and on the upper half-plane
$\lim\limits_{| E |\to\infty} f( E )=0$.  Then for
 $ E \in\rr$
\begin{eqnarray*}
f_\R( E +\i0)&=&\frac{1}{\pi}\cP\int\frac{f_\I(\xi+\i0)}{\xi- E }\d\xi,\\
f_\I( E +\i0)&=&-\frac{1}{\pi}\cP\int\frac{f_\R(\xi+\i0)}{\xi- E }\d\xi.
\end{eqnarray*}\label{dysp}
\eet

Sometimes a function $f$ does not have  enough decay, and instead we can apply Thm \ref{dysp} to its derivative.
Then by integrating we obtain the so-called {\em once substracted dispersion relations}.

\bet
Assume that $f\in C^2(\rr)$, $\frac{f'}{1+| E |}\in L^1(\rr)$ and on the upper half-plane
$\lim\limits_{| E |\to\infty} f'( E )=0$.  Then for
 $ E \in\rr$
\begin{eqnarray*}
f_\R( E +\i0)&=&f_\R(0+\i0)+\frac{1}{\pi}\cP\int f_\I(\xi+\i0)\left(\frac{1}{\xi- E }-\frac{1}{\xi}\right)\d\xi,\\
f_\I( E +\i0)&=&f_\I(0+\i0)-\frac{1}{\pi}\cP\int f_\R(\xi+\i0)\left(\frac{1}{\xi- E }-\frac{1 }{\xi}\right)\d\xi.
\end{eqnarray*}\label{dysp1}
\eet

\end{document}